\documentclass[fleqn,usenatbib]{mnras}

\usepackage{amsmath,amssymb}    

\newcommand{\kms}{\mathrm{km\ s^{-1}}\,}
\usepackage{natbib}

\usepackage{natbib, aas_macros}
\usepackage{graphicx, color, url}
\usepackage{grffile} 
\usepackage{epstopdf}
\usepackage{float}

\usepackage{hyperref}
\usepackage{etoolbox}
\makeatletter
\patchcmd\@combinedblfloats{\box\@outputbox}{\unvbox\@outputbox}{}{%
   \errmessage{\noexpand\@combinedblfloats could not be patched}%
}%
\makeatother

\title[Modeling the Milky Way as a Dry Galaxy]{Modeling the Milky Way as a Dry Galaxy}
\author[M. S. Fujii et al.]{M. S. Fujii$^{1}$\thanks{E-mail:
fujii@astron.s.u-tokyo.ac.jp (MSF)}, J. B\'edorf$^{2}$, J. Baba$^{3}$, and S. Portegies Zwart$^{2}$\\
$^{1}$Department of Astronomy, Graduate School of Science, The University of Tokyo, 7-3-1 Hongo, Bunkyo-ku, Tokyo, 113-0033, Japan\\
$^{2}$Leiden Observatory, Leiden University, NL-2300RA Leiden, The Netherlands\\
$^{3}$National Astronomical Observatory of Japan, Mitaka-shi, Tokyo 181-8588, Japan}

\date{Accepted . Received ; in original form }

\usepackage{pgfplots}

\begin{document}

\def\eps@scaling{1.0}%
\newcommand\epsscale[1]{\def\eps@scaling{#1}}%
\newcommand\plotone[1]{%
 \centering 
 \leavevmode  
 \includegraphics[scale={\eps@scaling}]{#1}%
}%

\label{firstpage}
\pagerange{\pageref{firstpage}--\pageref{lastpage}} \pubyear{2002}
\maketitle



\begin{abstract}
We construct a model for the Milky Way Galaxy composed of a stellar disc and bulge embedded in 
a dark-matter halo. All components are modelled as $N$-body systems 
with up to 8 billion equal-mass particles and integrated up to 
an age of 10\,Gyr.
We find that net angular-momentum of the dark-matter halo with a
spin parameter of $\lambda=0.06$ is required to form a
relatively short bar 
($\sim 4$\,kpc) with a high pattern speed (40--50\,km\,s$^{-1}$).
By comparing our model with observations of the Milky Way Galaxy, we conclude that 
a disc mass of $\sim 3.7\times10^{10}M_{\odot}$ and an initial bulge 
scale length and velocity of $\sim 1$\,kpc and $\sim 300$\,km\,s$^{-1}$, respectively,
fit best to the observations.
The disc-to-total mass fraction 
($f_{\rm d}$) 
appears to be an important 
parameter for the evolution of the Galaxy and models with 
$f_{\rm d}\sim 0.45$ are most similar to the Milky Way Galaxy.
In addition, we compare the velocity distribution in the solar 
neighbourhood in our simulations with observations in the Milky Way
Galaxy. 
In our simulations the observed gap in the velocity distribution,
which is expected to be caused by the outer Lindblad resonance 
(the so-called Hercules stream), appears to be a time-dependent structure.
The velocity distribution changes on a time scale of 20--30\,Myr and therefore it is 
difficult to estimate the pattern speed of the bar from the 
shape of the local velocity distribution alone. 
\end{abstract}




\section{Introduction}
Simulating the Milky Way (MW) Galaxy as an $N$-body system without gas is 
an important step in understanding its structure, kinematics,
and dynamics. 
The complex structures of the Galactic disc such as the bar
and spiral structures are composed of a bunch of individual 
stars orbiting around the Galactic center. 
We can examine the self-consistent evolution of the disc, bulge, and 
halo using $N$-body simulations. Especially when using models in which the 
dark-matter halo is composed of particles as opposed to an analytic
potential, we are able to follow the evolution of the bar as it emits the discs 
angular momentum to the live halo \citep{2002ApJ...569L..83A,2009ApJ...697..293D}.

Self-consistent MW models with a `live' dark-matter halo 
have been proposed by several previous studies
\citep{2005ApJ...631..838W,1997A&A...327..983F}.
\citet{2005ApJ...631..838W}
proposed a self-consistent equilibrium model of the MW. 
They constructed the distribution functions of the disc 
with a bulge and halo from integrals of motions 
by iteratively solving the Poisson equation~\citep{1995MNRAS.277.1341K}.
The resulting model is compared with observational properties
such as the rotation curve and the line-of-sight velocity-dispersion
profile of the bulge region~\citep{2002ApJ...574..740T}. 
They proposed two models for the MW; one disc-dominated model
and one halo-dominated model, both of which with a relatively massive
bulge ($M_{\rm b}>10^{10}M_{\odot}$). 
In \citet{2010ApJ...720L..72S} the authors found, using $N$-body 
simulations, that a model with a small bulge (less than $\sim 8$\,\% of 
the disc mass) fits the bulge kinematics 
data, observed by the Bulge Radial Velocity Assay 
(BRAVA)~\citep{2008ApJ...688.1060H,2012AJ....143...57K}, better than
the previously predicted massive classical bulge.
Apart from the bulge kinematics, there is additional available
observational data of the MW such as the velocity dispersion 
of the disc stars and surface density of the disc measured 
in the solar neighbourhood \citep{2016ARA&A..54..529B}. 

Here we construct an improved self-gravitating model 
for the MW that takes these observations into account. We perform 
a range of $N$-body simulations using models that are setup using 
the methods described in~\citet{1995MNRAS.277.1341K} and \citet{2005ApJ...631..838W} (see \S\,\ref{Sec:NbodySimulations}).

One of the difficulties in finding the initial conditions for a best
fitting Galaxy model is that we cannot predict the
outcome of the simulation before actually having performed the run.
The chaotic gravitational dynamics of the particles (see
\cite{PORTEGIESZWART2018160} for a brief overview) 
prevents making a preliminary assessment of the simulation results as
function of the many initial parameters.
Although many automated optimization strategies exist, we decided to do this by manually mapping the parameter space and guiding our next run based on the analyzed data. This is a rather labour intensive procedure, but allows us to converge more efficiently than any automated procedure.

We perform the simulations with at least one million particles in the Galactic disc.
Such a large number of particles 
is necessary in order to 
obtain a reliable result at the end of the simulation~\citep{2011ApJ...730..109F}. 
To simulate our models we use our recently developed tree-code, 
{\tt Bonsai}, which utilizes massive parallel GPU computing systems~\citep{2014hpcn.conf...54B}. 
Using {\tt Bonsai}, we are able to perform a large number of $N$-body 
simulations, with a high enough resolution. Our largest model contains 
eight billion particles for the galaxy and the dark-matter halo combined.
By comparing the properties 
of the simulated model with recent MW observations, we find 
the best matching configuration parameters.

We further ``observe'' the largest simulations in order to 
compare the results with the MW Galaxy
to investigate the velocity distribution in the local neighbourhood.
It is expected that spiral structures can be seen in the local velocity 
distribution
as well as the imprint of resonances such as the Hercules stream, 
although the origin of the velocity structures is still
debatable~\citep{2000AJ....119..800D,2011MNRAS.417..762Q,2014A&A...563A..60A,
  2017MNRAS.466L.113M,2017ApJ...840L...2P,2018MNRAS.474...95H,2018MNRAS.477.3945H,
  2018arXiv180401920H}.
In this paper, we propose a set of best matching parameters for the MW
Galaxy and make a comparison with observations.
In Section 2, we describe the details of our models and $N$-body 
simulations. Our best-fitting models are presented in Section 3.
In Section 4, we show detailed analyses of the best-fitting
models. The results are summarized in Section 5.

\section{$N$-body simulations}\label{Sec:NbodySimulations}

To find a best-fitting MW model we performed a series of
$N$-body simulations. We simulate a live dark-matter halo with an embedded stellar disc.
The simulations have up to $\sim 8$ billion particles.

The initial conditions are generated using {\tt GalactICS}
and the simulations are performed using the parallel GPU 
tree-code, {\tt Bonsai}~\citep{2012JCoPh.231.2825B, 2014hpcn.conf...54B}, which is part of the Astrophysical
Multipurpose Software Environment \citep[AMUSE, ][]{2013CoPhC.183..456P, 2013A&A...557A..84P,AMUSE}. 
In this section details of the models, parameters, and simulations are described.
Since we tested more than 50 models we only cover the most important models in the main text,
the detailed parameters for all models are summarized in Appendix~\ref{Sect:AllModel}.

\subsection{Initial conditions}

\subsubsection{Dark-Matter Halo}

In the initial condition generator, {\tt GalactICS}~\citep{1995MNRAS.277.1341K,2005ApJ...631..838W}, 
the dark-matter halo is modeled using the NFW density profile~\citep{1997ApJ...490..493N}:

\begin{eqnarray}
\rho_{\rm NFW}(r) = \frac{\rho_{\rm h}}{(r/a_{\rm h})(1+r/a_{\rm h})^3},
\end{eqnarray}

with the following potential:

\begin{eqnarray}
\Phi_{\rm NFW} = -\sigma_{\rm h}^2\frac{\log (1+r/a_{\rm h})}{r/a_{\rm h}},
\end{eqnarray}
where $a_{\rm h}$ is the scale radius, $\rho_{\rm h}\equiv\sigma^2/(4{\rm \pi} G a_{\rm h}^2)$
is the characteristic density, and $\sigma_{\rm h}$ is the characteristic 
velocity dispersion. The gravitational constant, $G$, is set to be unity.
Since the NFW profile has an infinite extent the mass  
distribution is truncated by a halo tidal radius using
an energy cutoff $E_{\rm h}\equiv\epsilon_{\rm h}\sigma_{\rm h}^2$,
where $\epsilon _{\rm h}$ is the truncation parameter with $0<\epsilon _{\rm h}<1$.
Setting $\epsilon _{\rm h}=0$ yields a full NFW profile
\citep[see][for details]{2005ApJ...631..838W}. 

We therefore have $a_{\rm h}$, $\sigma_{\rm h}$, $\epsilon_{\rm h}$, and $\alpha_{\rm h}$
as the parameters that configure the dark-matter halo model.
We summarize the values of these parameters for our models in 
Tables \ref{tb:params} and \ref{tb:A1}.
Here $a_{\rm h}$ and $\sigma_{\rm h}$ are chosen such that the models match 
the observed circular rotation velocity at the Sun's location, 
$V_{\rm circ, \odot} = 238\pm15$\,km\,s$^{-1}$ \citep{2016ARA&A..54..529B}. 
The Galactic virial radius is estimated to be $r_{\rm vir}=282\pm30$\,kpc
\citep{2016ARA&A..54..529B}, and 
from cosmological simulations, we know that dark-matter halos with 
a mass of $10^{12}M_{\odot}$ typically have a concentration parameter of
$c\equiv r_{\rm vir}/a_{\rm h}=10$--17 \citep{2002ApJ...573..597K}. 
These give $a_{\rm h}\sim 15$--30\,kpc.
Recently it was suggested that $c\sim 10$ ($a_{\rm h}\sim25$\,kpc) 
\citep{2015MNRAS.452.1217C}, but theoretical modeling, based on observations,
suggest smaller values such as $a_{\rm h}=19.0 \pm 4.9$\,kpc \citep{2017MNRAS.465...76M} and 
$a_{\rm h}=14.39^{+1.30}_{-1.15}$\,kpc \citep{2016MNRAS.463.2623H}.
Our choice for $a_{\rm h}$ is on the lower end of the above 
ranges, namely 10--22\,kpc.

To reproduce the observed circular rotational velocity at the location of 
the Sun, 
we 
set $\sigma_{\rm h}=380$--500\,km\,s$^{-1}$. For $\epsilon_{\rm h}$, we use 0.7--0.85 
to get an halo outer radius, $r_{\rm h}$, that is close to the 
observed Galactic virial radius, $r_{\rm vir}=282\pm30$\,kpc
\citep{2016ARA&A..54..529B}. The above settings result in a halo 
mass, $M_{\rm h}$, of $\sim6$--$19\times 10^{11}M_{\odot}$ 
(see Tables \ref{tb:masses} and \ref{tb:A2}).
In these tables we further see that the halo mass is most sensitive
to the value of $a_{\rm h}$ if we fix the circular rotation velocity at the 
location of the Sun;
if $a_{\rm h}$ is small then $M_{\rm h}$ is also small. 

On the other hand, 
the MW's virial mass is estimated to 
be $M_{\rm vir}=1.3\pm 0.3 \times 10^{12}M_{\odot}$~\citep{2016ARA&A..54..529B}. 
In this work, we are interested in the halo's inner region 
where the Galactic disc is located, because the outer region only has 
a limited effect on the disc evolution. 
We, therefore, consider the total halo mass and outer radius,
which are configured using the truncation parameter ($\epsilon_{\rm h}$), 
of less significance in this work.

The final parameter that controls the halo configuration is the spin parameter $\alpha_{\rm h}$. 
This parameter controls the sign of the angular momentum along the symmetry 
axis, $J_{z}$.
When $\alpha _{\rm h}=0.5$, the number of halo particles with a
positive and negative $J_{z}$ is equal, and therefore there is
no spin. If $\alpha _{\rm h}>0.5$, the halo rotates in 
the same direction as the disc.
A developing galactic bar transfers a part of its angular momentum into the 
halo~\citep{2002ApJ...569L..83A,2009ApJ...697..293D}. As a consequence,
if the halo has an initial spin, the final bar becomes shorter~\citep{2014ApJ...783L..18L,2018MNRAS.477.1451F}.
\citet{2008ApJ...679.1239W} showed that galaxy models, similar to the MW, tend to 
develop bars that are too long when compared to observations.
We, therefore, give the halo a spin in the same direction as the disc,
where $\alpha_{\rm h}=0.8$ is our standard value.
For comparison, we also used models without halo 
spin ($\alpha_{\rm h}=0.5$) and a weaker than default spin ($\alpha_{\rm h}=0.65$).

The halo spin is commonly characterized using the spin parameter defined by 
\citet{1969ApJ...155..393P,1971A&A....11..377P}:
\begin{eqnarray}
\lambda = \frac{J|E|^{1/2}}{GM_{\rm h}^{5/2}},
\end{eqnarray}
where $J$ is the magnitude of the angular momentum vector and $E$ is the total 
energy.
For our models, $\alpha_{\rm h}=0.8$ (0.65) corresponds to $\lambda\sim0.06$ (0.03).
For MW size galaxies, the halo spin is suggested to 
be $\lambda=0.03$--0.05 \citep{2002ApJ...573..597K} and
 $\lambda=0.03$--0.04 \citep{2007MNRAS.376..215B}
from cosmological $N$-body simulations.
Observationally, 
using the Sloan Digital Sky Survey Data Release 7 
\citep[SDSS DR7; ][]{2013ApJ...775...19C}, 
the halo spin is estimated to be $\lambda = 0.039$ for barred galaxies, but 
$\lambda = 0.061$ for galaxies where the bar is short.
These values are consistent with our spinning halo models.

\subsubsection{Galactic disc}

For the disc we use the surface density distribution given by
\begin{eqnarray}
\Sigma (R) = \Sigma_{0} {\rm e}^{-R/R_{\rm d}},
\end{eqnarray}
where $\Sigma_{0}$ is the central surface density and 
$R_{\rm d}$ is the disc scale length. 
In {\tt GalactICS}, $\Sigma_{0}$ is a function of the 
disc mass ($M_{\rm d,0}$).
The vertical structure is given by ${\rm sech}^2(z/z_{\rm d})$,
where $z_{\rm d}$ is the scale height of the disc.
The radial velocity dispersion is assumed to follow 
$\sigma_{R}^2(R)=\sigma_{R0}^2\exp(-R/R_{\rm d})$, where 
$\sigma_{\rm R0}$ is the radial velocity dispersion at the center
of the disc. 
For the disc model this gives the following free parameters, $M_{\rm d,0}$, $R_{\rm d}$, 
$z_{\rm d}$, and $\sigma_{R0}$. 
We use $R_{\rm d}=2.6$\,kpc as our standard setting, to match the observed 
value of $2.6\pm0.5$\,kpc \citep{2016ARA&A..54..529B}.
For $z_{\rm d}$, we adopt $z_{\rm d}=0.2$ or 0.3\,kpc.
This is slightly smaller than the observed scale height of the Galactic 
thin disc, $0.30\pm0.05$\,kpc \citep{2016ARA&A..54..529B}. 
However, when the bar and spiral arms develop, 
dynamical heating causes thickening of the disc.

For the total disc mass, we use $M_{\rm d}=3.1$--$4.1\times 10^{10}M_{\odot}$, 
consistent with the
observed disc mass; $3.5\pm 1\times10^{10}M_{\odot}$ for the thin disc,
plus $6\pm3 \times10^9M_{\odot}$ for the thick disc \citep{2016ARA&A..54..529B}.
As with the halo, the disc is infinite and must be truncated. This is controlled by the 
truncation radius, $R_{\rm out}$, and the truncation sharpness, $\delta R_{\rm d}$.
We set $R_{\rm out}=30$\,kpc and $\delta R_{\rm d}=0.8$\,kpc, 
which is similar to the values in \citet{2005ApJ...631..838W}.
To configure the final parameter, $\sigma_{R0}$, we aim on getting an initial
Toomre $Q$ value~\citep{1964ApJ...139.1217T} at the reference 
radius ($2.5R_{\rm d}$), $Q_0$, of $\sim 1.2$.
For our models this leads to $\sigma_{R0}\sim$ 70--105\,km\,s$^{-1}$,
where $\sigma_{R0}$ becomes larger when the disc mass increases and $Q_0$ 
is kept constant.
The values we adopt for these parameters are summarized in 
Tables \ref{tb:params} and \ref{tb:A1}, and the resulting disc mass 
($M_{\rm d}$), which is slightly larger than the parameter $M_{\rm d,0}$, 
and outer radius ($R_{\rm d, t}$) is summarized in 
Tables~\ref{tb:masses} and \ref{tb:A2}.

In these Tables, we also summarize the disk-to-total mass fraction ($f_{\rm d}$),
which is measured for the mass within $2.2R_{\rm d}$. 
In \citet{2018MNRAS.477.1451F}, we showed that $f_{\rm d}$ is a critical parameter
to control the bar formation epoch. We therefore add $f_{\rm d}$ to the model
parameters.

\subsubsection{Bulge}

The bulge component is based on the Hernquist model \citep{1990ApJ...356..359H},
but, as with the disc and halo, the distribution function is modified to allow truncation.
The model is described as 
\begin{eqnarray}
\rho_{\rm H} = \frac{\rho_{\rm b}}{(r/a_{\rm b})(1+r/a_{\rm b})^3}
\end{eqnarray}
and
\begin{eqnarray}
\Phi_{\rm H} = \frac{\sigma_{\rm b}^2}{1+r/a_{\rm b}},
\end{eqnarray}
where $a_{\rm b}$ is the scale radius, 
$\rho_{\rm b}=\sigma_{\rm b}^2/(2{\rm \pi} Ga_{\rm b}^2)$ is the characteristic density,
and $\sigma_{\rm b}$ is the characteristic velocity of the bulge.
This gives $a_{\rm b}$, $\sigma_{\rm b}$, and the truncation 
parameter, $\epsilon_{\rm b}$,  as the free parameters.
We adopt $a_{\rm b}=0.2$--1.2\,kpc and 
$\sigma_{\rm b}=$270--400\,km\,s$^{-1}$. We chose $\epsilon_{\rm b}$ 
such that the outer radius of the bulge  ($r_{\rm b, t}$) is $\sim 1$--3\,kpc.
These settings result in a bulge mass 
of $M_{\rm b}\sim3$--$9\times 10^{10}M_{\odot}$ (see Tables~\ref{tb:masses} and \ref{tb:A2}).
In \citet{2010ApJ...720L..72S}, they showed that even the model without a classical bulge
fits to the observational data. In our previous studies, however, we found that 
models without a classical bulge form too strong bars \citep{2018MNRAS.477.1451F}. 
We therefore add a classical bulge in the initial conditions.
We did not give the bulge a preferential spin.

All parameters and their chosen values, are summarized 
in Table~\ref{tb:params} for our best fitting models and \ref{tb:A1} for the others, 
and the resulting masses and radii are 
given in Tables~\ref{tb:masses} and \ref{tb:A2}.
For our standard resolution, we set the number of particles 
in the disc component to $\sim 8$ million (8M), with this resolution
the results do not strongly depend on the resolution 
(see Appendix~\ref{Sect:Resolution} for details).
In order to avoid numerical heating, we assign the same mass to each of the 
particles, irrespective of the component (disc, bulge or halo) they belong to.
The number of particles for each model is also given in
Tables~\ref{tb:masses} and \ref{tb:A2}.

\begin{table*}
\scriptsize
\raggedright
\caption{Models and parameters\label{tb:params}}
\begin{tabular}{lccccccccccc}
\hline
           &  \multicolumn{4}{l}{Halo} &  \multicolumn{4}{l}{disc} &  \multicolumn{3}{l}{Bulge} \\
Model &  $a_{\rm h}$ & $\sigma_{\rm h}$ & $\epsilon_{\rm h}$ & $\alpha_{\rm h}$& $M_{\rm d,0}$ & $R_{\rm d}$ & $z_{\rm d}$ & $\sigma_{\rm R0}$  & $a_{\rm b}$ & $\sigma_{\rm b}$ & $\epsilon_{\rm b}$ \\ 
   &  (kpc) & ($\kms$) &  &  & $(10^{10}M_{\odot})$ & (kpc) & (kpc) & ($\kms$)  & (kpc) & $(\kms)$\\
\hline \hline
MWa/a5B & 10 & 420 & 0.85 & 0.8 & 3.61 & 2.3 & 0.2 & 94 & 0.75 & 330 & 0.99  \\ 
MWb/b6B & 10 & 380 & 0.83 & 0.8 & 4.1 & 2.6 & 0.2 & 90 & 0.78 & 273 & 0.99  \\ 
MWc0.8/c7B & 12 & 400 & 0.80 & 0.8 & 4.1 & 2.6 & 0.2 & 90 & 1.0 & 280 & 0.97  \\ 
MWc0.65/c0.5 & 12 & 400 & 0.80 & 0.65/0.5 & 4.1 & 2.6 & 0.2 & 90 & 1.0 & 280 & 0.97  \\ 
\hline
\end{tabular}
\newline
{ \scriptsize
The settings of the free parameters used to configure the halo 
(column 2-5), disc (column 6-9) and bulge (10-12). The first column indicates 
the model name as referred to in the text. `xB' in the name indicates the number of halo particles if it is over 1 billion.
The values 0.8, 0.65 and 0.5 for MWc indicate the halo spin parameter. 
}
\end{table*}

\begin{table*}
\scriptsize
\raggedright
\caption{Mass, radius, and the number of particles\label{tb:masses}}
\begin{tabular}{lcccccccccccc}
\hline
Model    & $M_{\rm d}$ & $M_{\rm b}$ & $M_{\rm h}$ & $M_{\rm b}/M_{\rm d}$ & $R_{\rm d, t}$ & $r_{\rm b, t}$ & $r_{\rm h, t}$ & $Q_0$  & $N_{\rm d}$ & $N_{\rm b}$ & $N_{\rm h}$ & $f_{\rm d}$\\ 
   & ($10^{10}M_{\odot}$) & ($10^{10}M_{\odot}$) & ($10^{10}M_{\odot}$) & &  (kpc) & (kpc) & (kpc) &  &   &  & & \\ 
\hline \hline
MWa & 3.73 & 0.542 & 86.8 & 0.15 & 31.6 & 3.16 & 239 & 1.3  & 8.3M  & 1.2M  & 194M  & 0.459\\ 
MWa5B & 3.73 & 0.542 & 86.8 & 0.15 & 31.6 & 3.16 & 239 & 1.3  & 208M  & 30M  & 4.9B & 0.459\\
MWb & 4.23 & 0.312 & 62.3 & 0.07 & 31.6 & 2.69 & 241 & 1.2  & 8.3M  & 0.6M  & 123M  & 0.471\\ 
MWb6B & 4.23 & 0.312 & 62.3 & 0.07 & 31.6 & 2.69 & 241 & 1.2  & 415M  & 33M  & 6.1B  & 0.471\\ 
MWc0.8/c0.65/c0.5 & 4.19 & 0.37 & 76.7 & 0.09 & 31.6 & 3.06 & 233 & 1.2  & 8.3M  & 0.7M  & 151M  & 0.472\\ 
MWc7B          & 4.19 & 0.379  & 76.7 &  0.09  & 31.6 & 3.06 & 233 & 1.2 & 415M & 37M &  7.6B & 0.472\\
\hline
\end{tabular}
\newline
{ \scriptsize
The generated properties for each of our models. The first column indicates the name of the model. Columns 2-4 give the mass of the component (disc, bulge, halo). The 
fifth column gives the bulge to disc mass-ratio. Column 6-8 shows the radius of the component and column 9 Toomre's $Q$ value at $2.5R_{\rm d}$. Columns 10-12 gives the used number of 
particles, per component. Here the exact number for $N_{\rm b}$ and $N_{\rm h}$ is chosen such to produce the observed mass ratio's given that the particles are equal mass. 
Column 13 gives the disc to total mass fraction at $2.2R_{\rm d}$.
}
\end{table*}

\subsection{Simulation code and parameters}

To simulate the models described in the previous section
we use the latest version of {\tt Bonsai}, a parallel, GPU accelerated,
$N$-body tree-code~\citep{2012JCoPh.231.2825B, 2014hpcn.conf...54B}. 
{\tt Bonsai} has been designed to run efficiently on GPU accelerators. 
To achieve the performance required for this project, all particle and tree-structure data is stored in the GPUs on-board memory.
Because the limited amount of GPU memory competes with the desire to run billion-particle models,
the code is able to use multiple GPUs in parallel and has been shown to scale efficiently to 
thousands of GPUs. The version of {\tt Bonsai} used for this research is able to simulate billion-particle 
MW models in reasonable time. Our largest model, with 8 billion particles,
took about one day using 512 GPUs\footnote{{\tt Bonsai} is available at: {\tt https://github.com/treecode/Bonsai}}.

For this work, we have made a number of improvements to the code as described 
in~\citet{2012JCoPh.231.2825B, 2014hpcn.conf...54B}.
The first improvement is related to the post-processing. The most compute intensive post-processing
operations are executed during the simulation itself. At that point, all data is already loaded
in memory and can be processed by using all reserved compute nodes in parallel. 
Although not all the post-processing analysis is handled during the simulation, the most 
compute and memory intensive ones are. For the other post-processing operations, we produce, 
during the simulation, (reduced) data files that can be further processed using a small number
of processors. Without this on-the-fly post-processing the whole analysis phase would take an 
order of magnitude more processing time than the actual simulation.

We also updated the writing of snapshot data to disk. This is now a fully asynchronous 
operation and happens in parallel with the simulation. For our largest models the amount of data 
stored on disk, per snapshot, is on the order of hundreds of gigabytes.
Even on a distributed file-system, this operation takes a large amount of time during which the 
GPUs would have been idle if this was not handled asynchronously.

The simulations described here have been performed on the Piz Daint computer at CSCS in 
Switzerland. This machine has recently been upgraded and outfitted with NVIDIA P100 GPUs. 
So the final improvement is that, apart from the usual bug fixes, we updated {\tt Bonsai} 
to properly support and efficiently use this new GPU generation. 
The architectural upgrades in these GPUs improve the performance of {\tt Bonsai}  by roughly
a factor 2.5 compared to the previously installed GPU generation (NVIDIA K20).

For all simulations, we use a shared time-step of $\sim 0.6$ Myr, a gravitational softening 
length of 10\,pc and as opening angle $\theta=0.4$.
Each simulation runs for 10\,Gyr and has an energy error on the order of $10^{-4}$,
which is sufficient for $N$-body system simulations~\citep{2015ComAC...2....2B}.

\begin{figure*}
\epsscale{.45}
\plotone{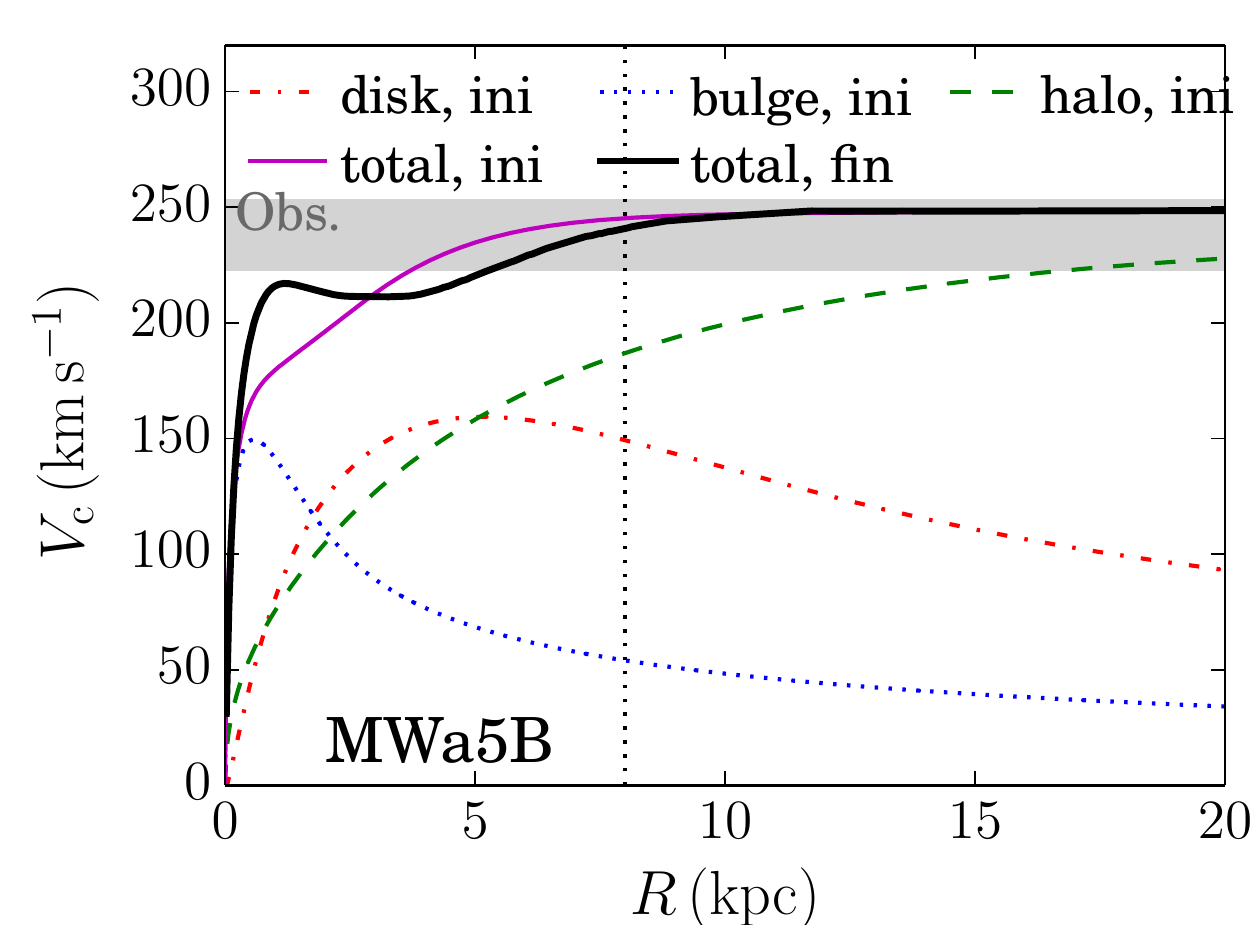}\plotone{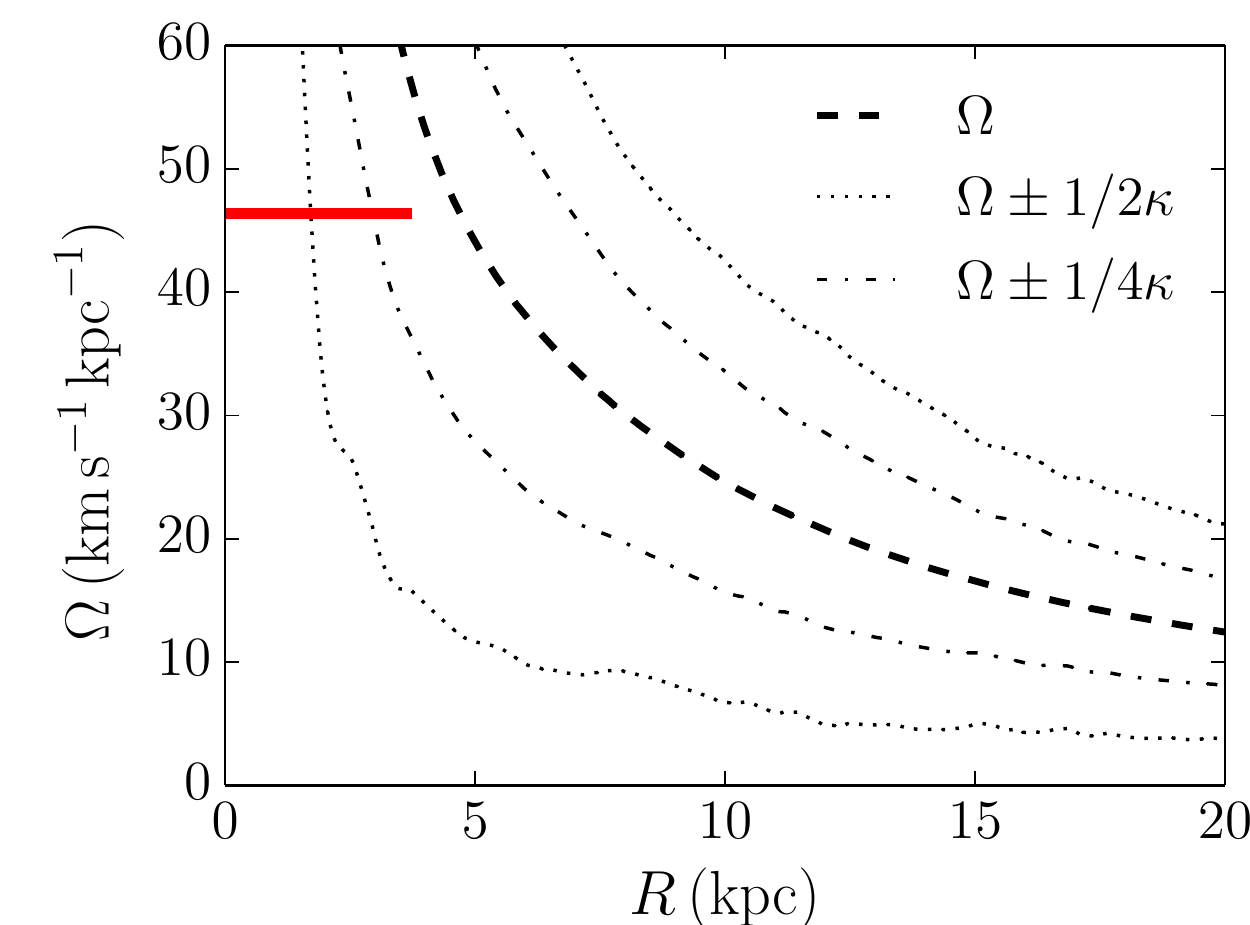}\\
\plotone{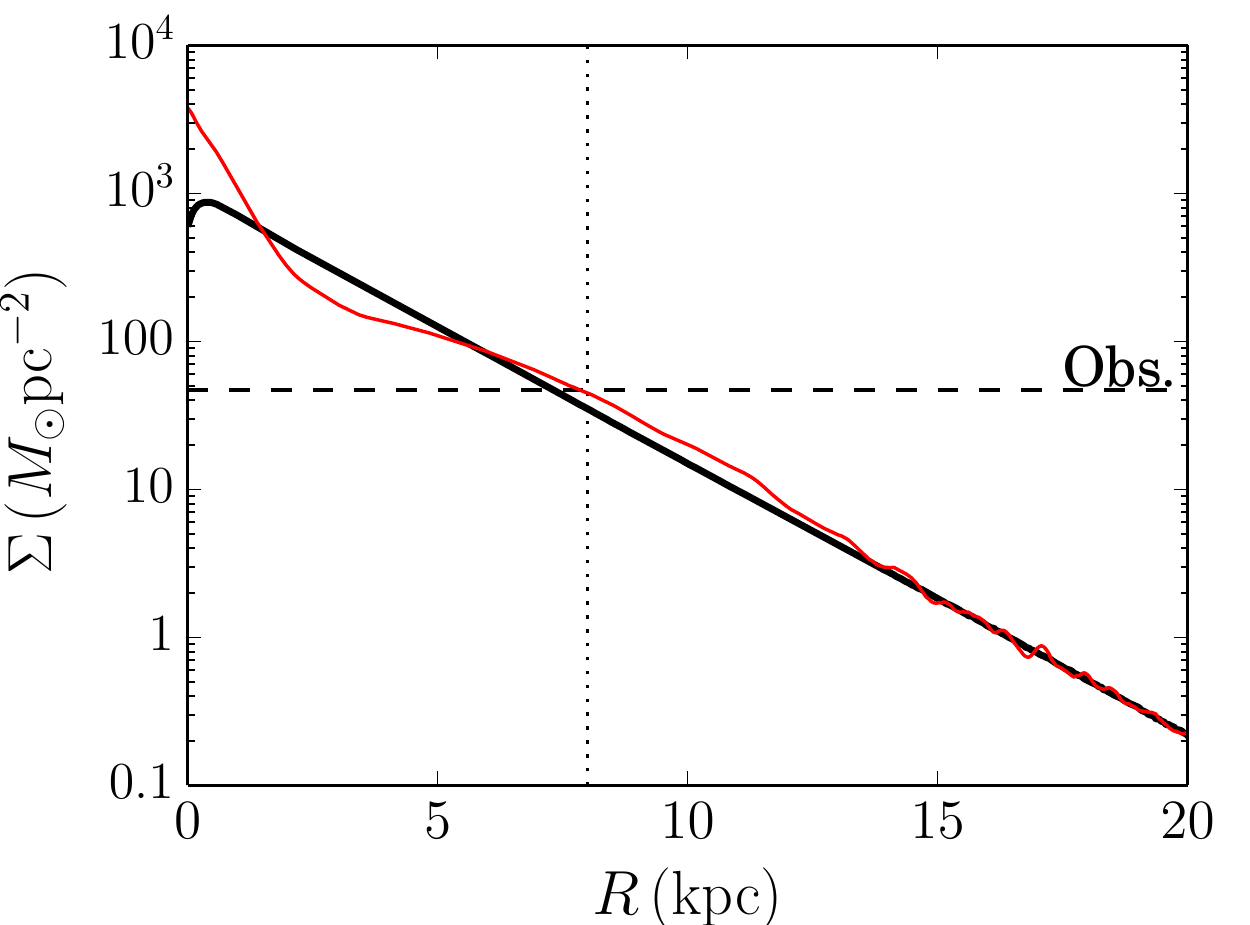}\plotone{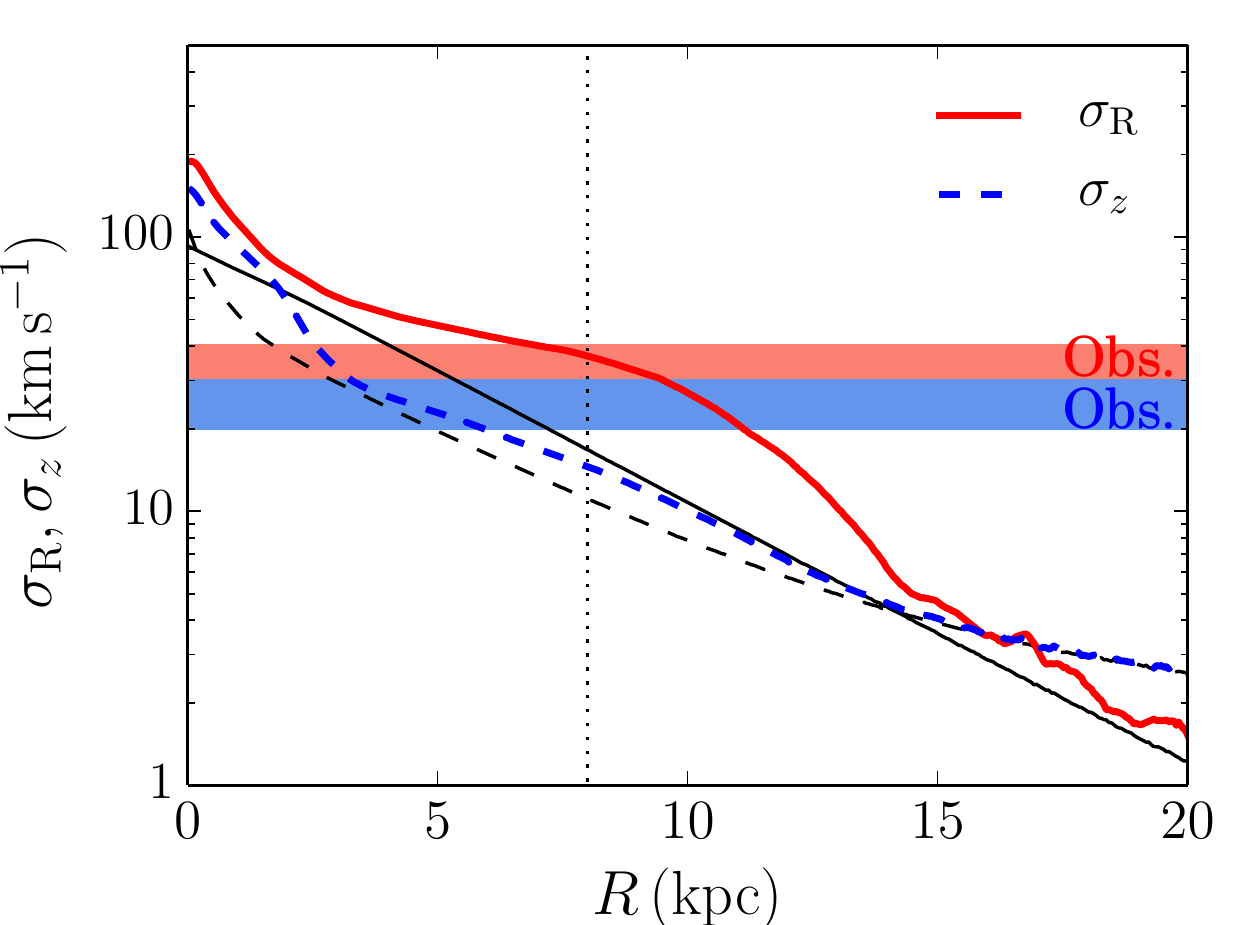}\\
\plotone{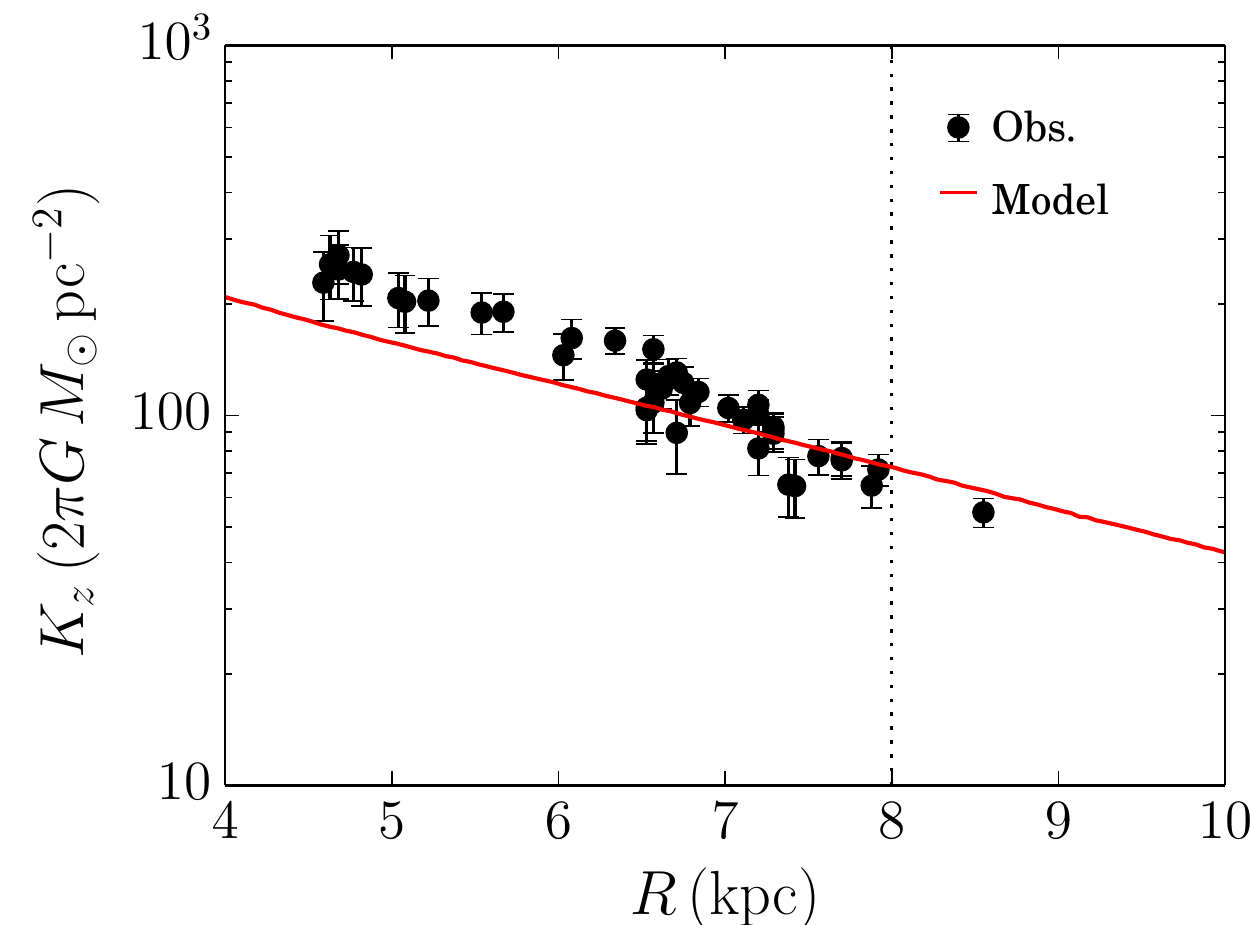}\plotone{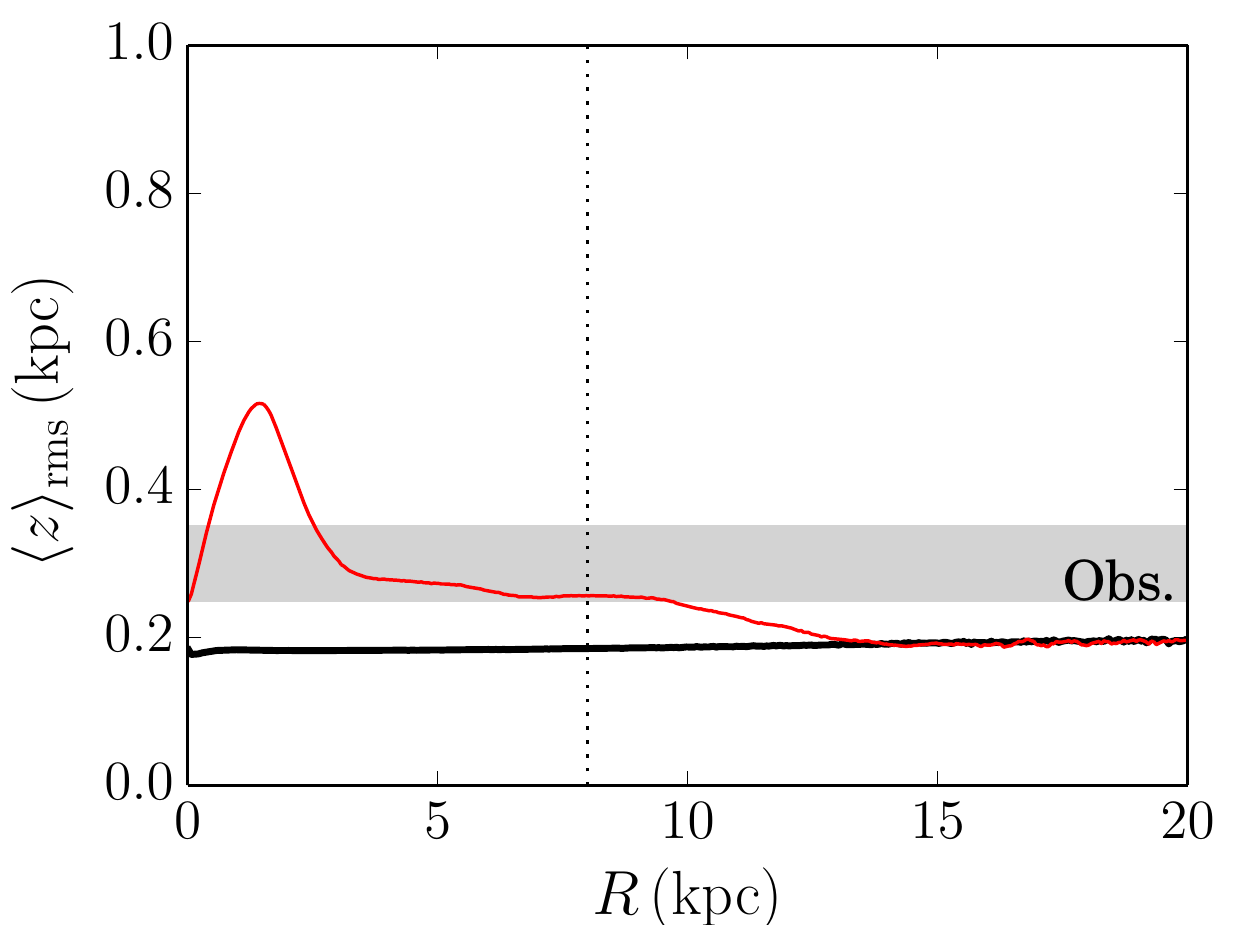}\\
\plotone{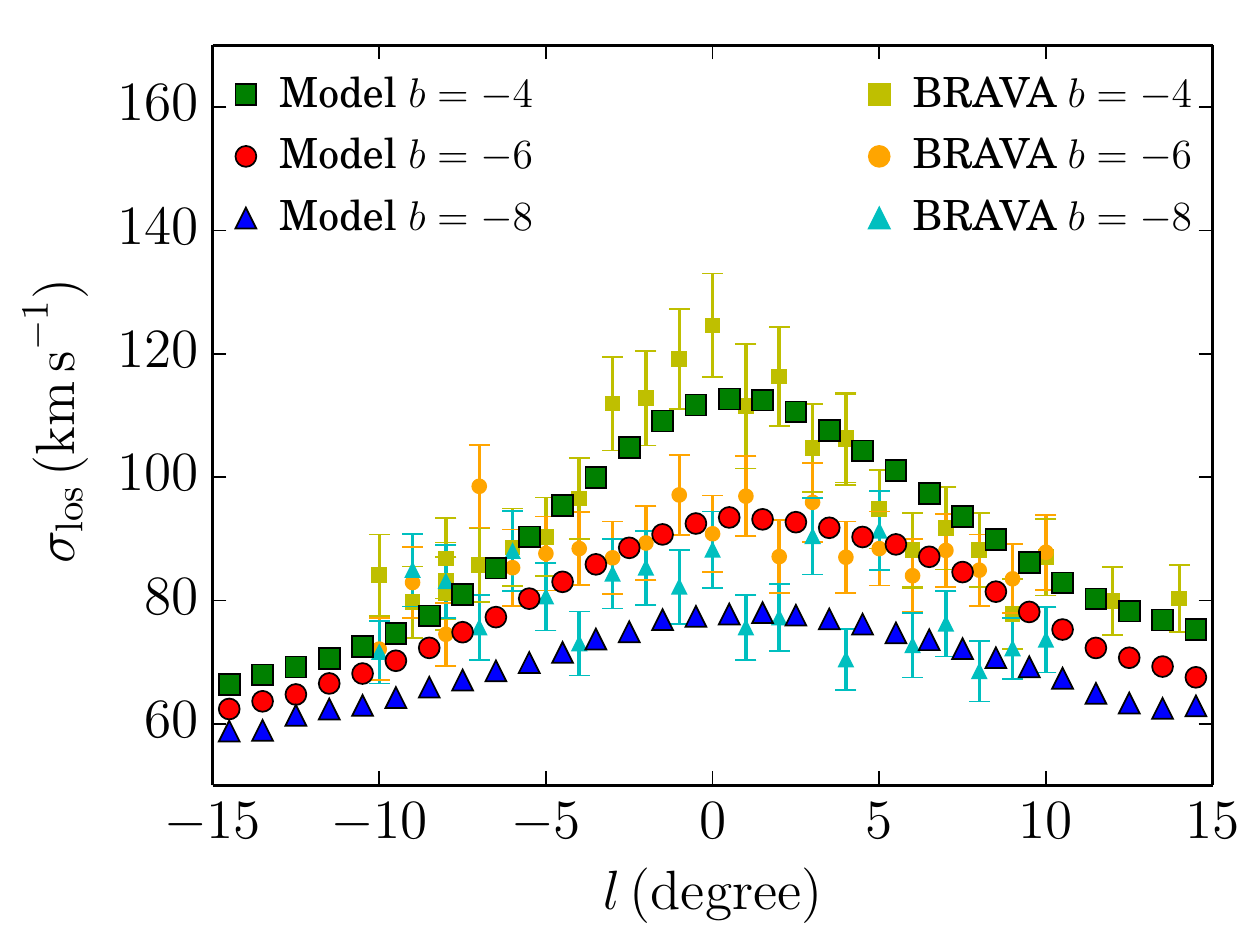}\plotone{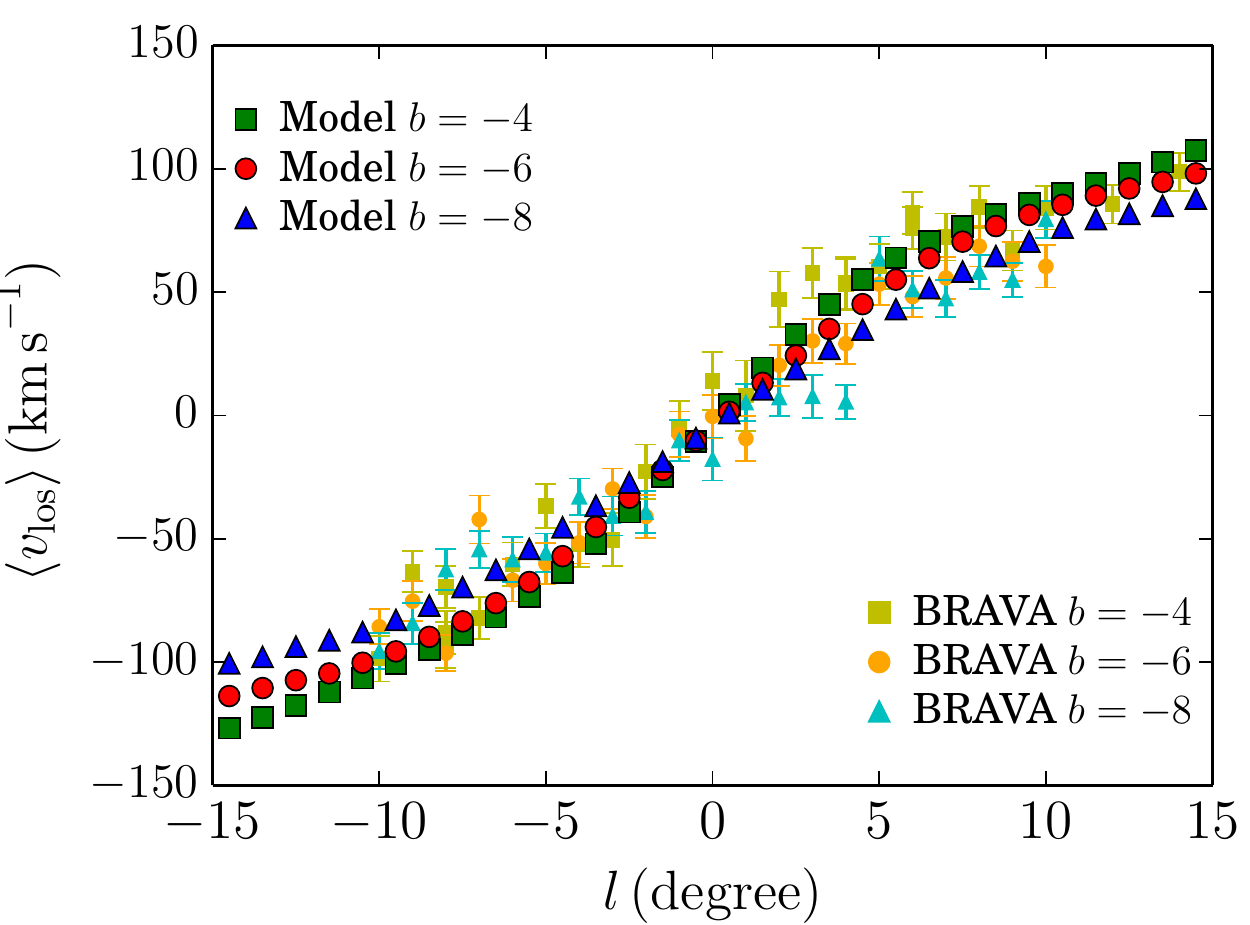}\\
\plotone{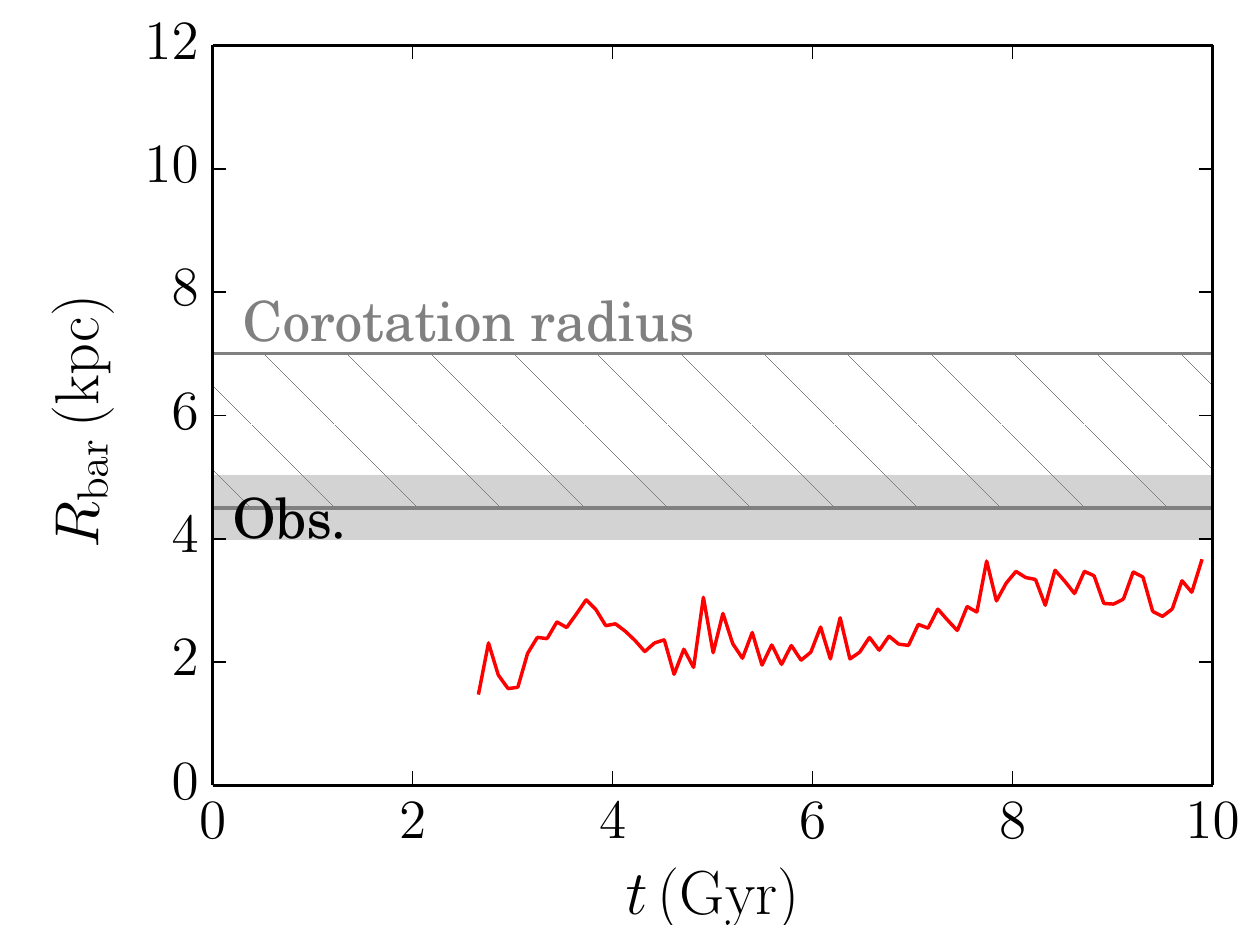}\plotone{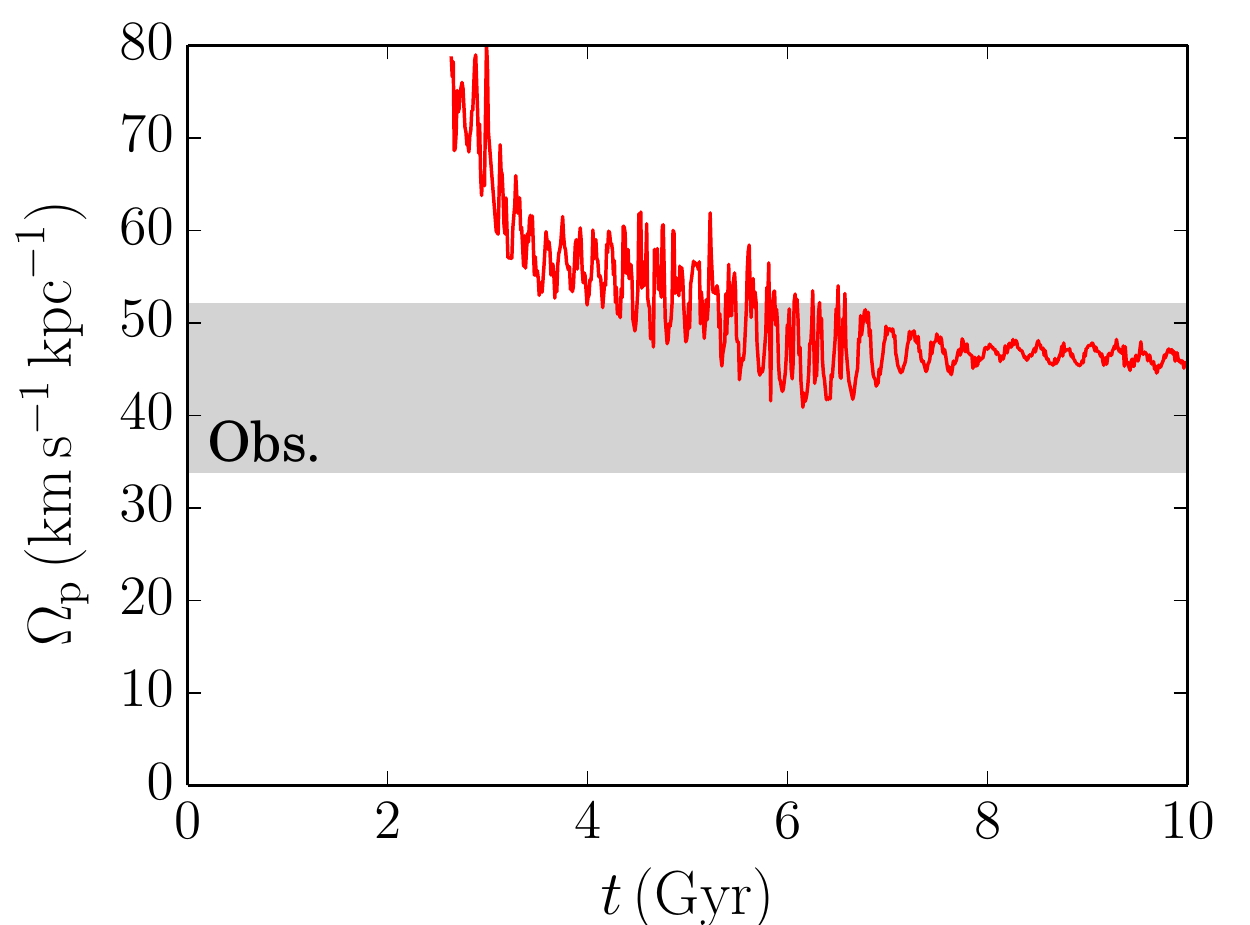}\\
\caption{Results of Model a5B.
(a) Initial and final rotation curves. 
(b) Angular frequencies of the disc (black curves) and bar (red line) at 10\,Gyr. The length of the red line indicate the length of the bar.
(c) Surface densities, and
(d) radial and vertical velocity dispersion of the disc.
(e) Total (disc, bulge, and dark-matter) density within $|z|<1.1$\,kpc ($K_z$) at 10\,Gyr. Symbols with error bars indicate observations~\citep{2013ApJ...779..115B}.
  (f) Scale height of the disc.
Black and color curves in panels (c), (d), and (f) indicate the initial and final (10\,Gyr) distribution, respectively.
 (g) Line-of-sight velocity dispersion and 
(h) mean velocity of the bulge region ($R<3$\,kpc) at 10\,Gyr. Symbols with error bars are BRAVA data.
(i) Bar length and 
(j) pattern speed of the bar as a function of time. 
Vertical dotted lines in panels (a) and (c)--(f) indicate 8\,kpc from the galactic center. \label{fig:a5B}
}
\end{figure*}

\begin{figure*}
\epsscale{.45}
\plotone{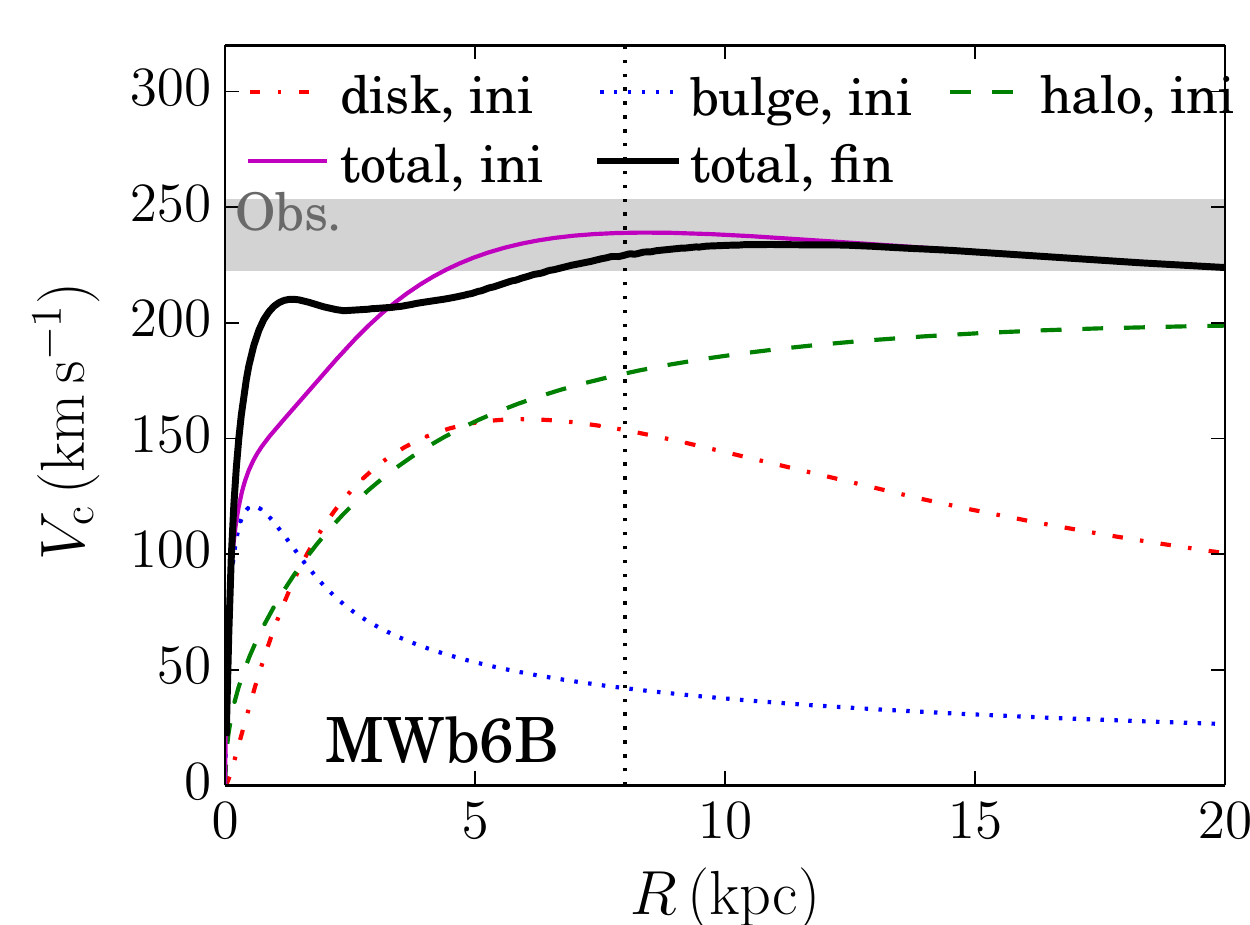}\plotone{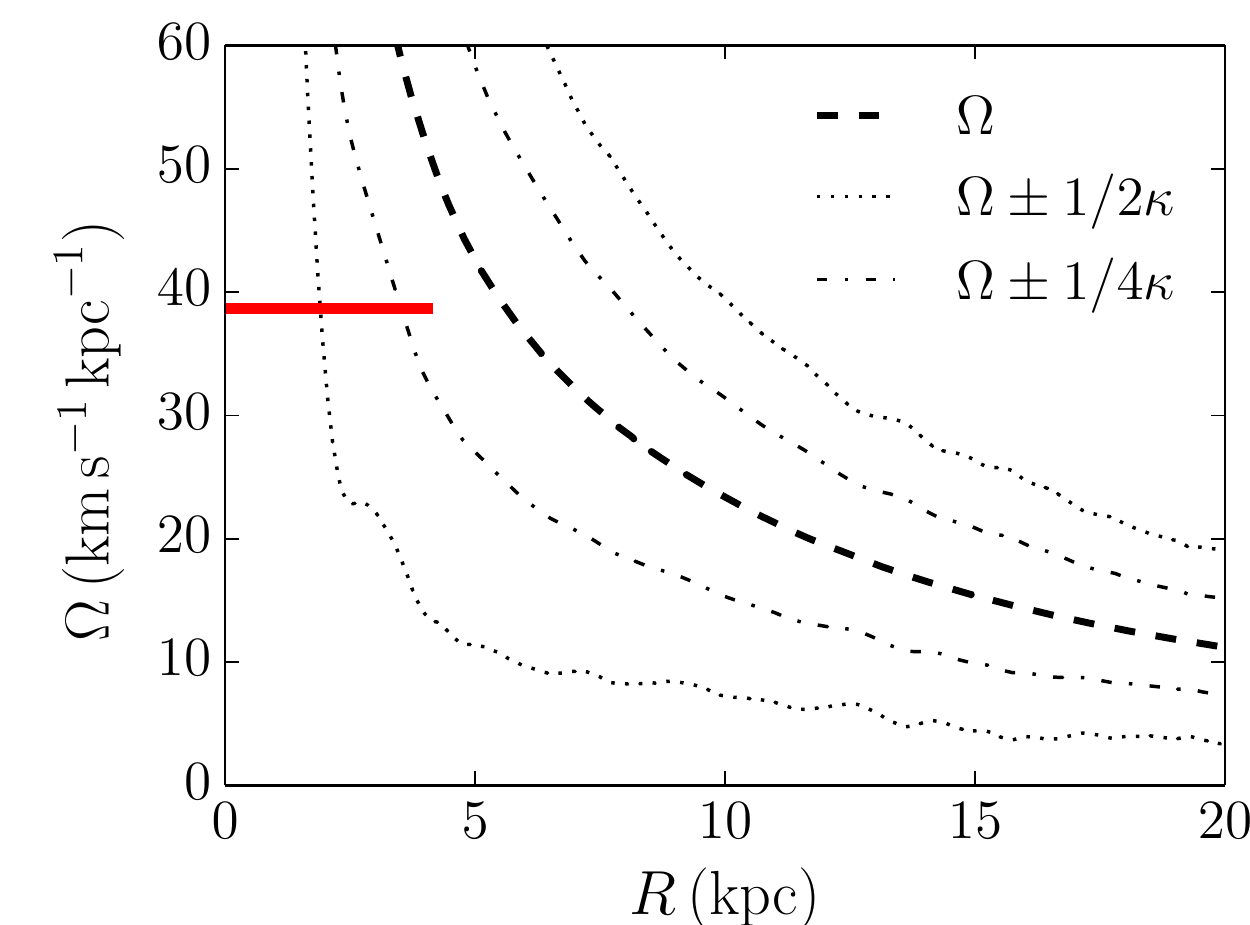}\\
\plotone{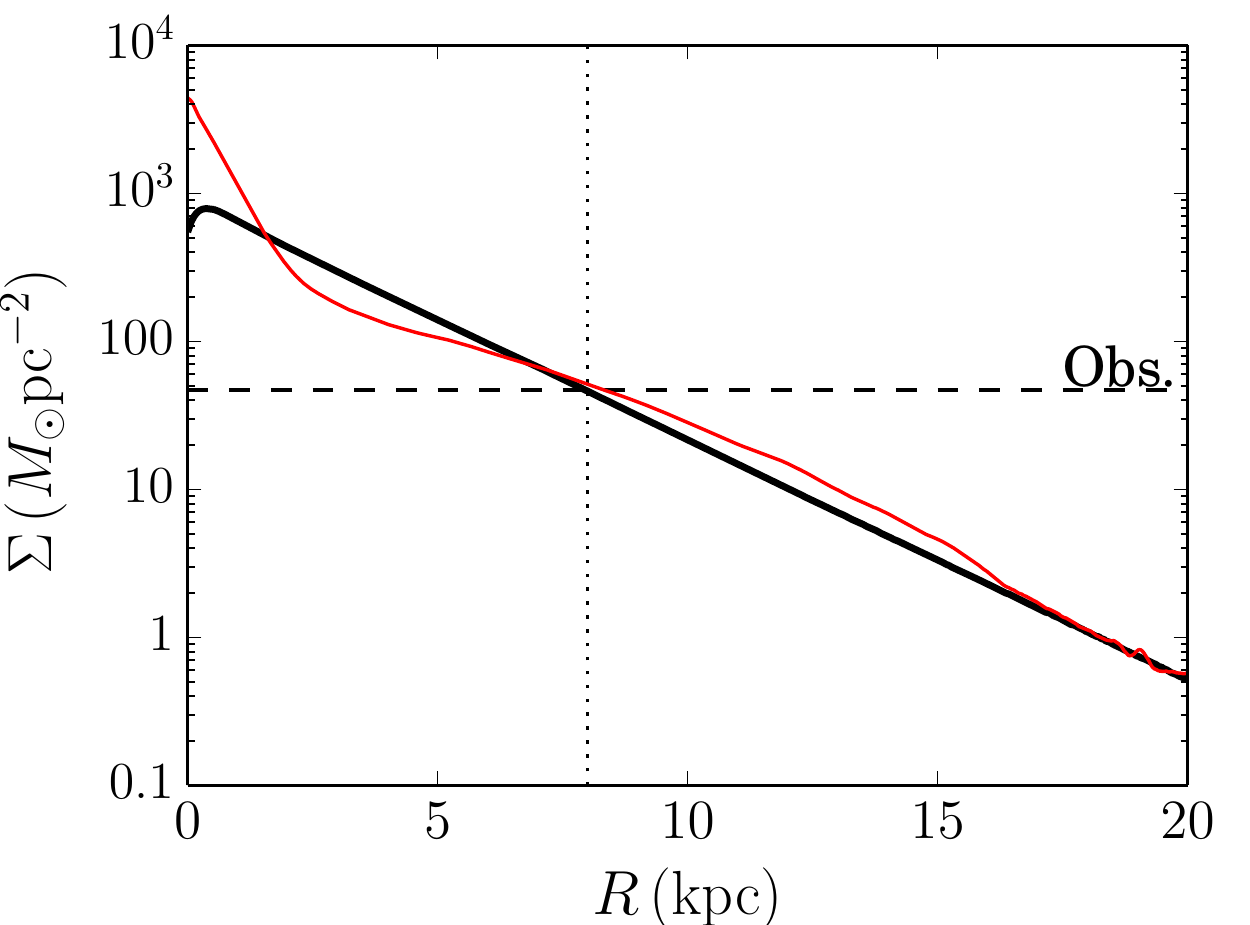}\plotone{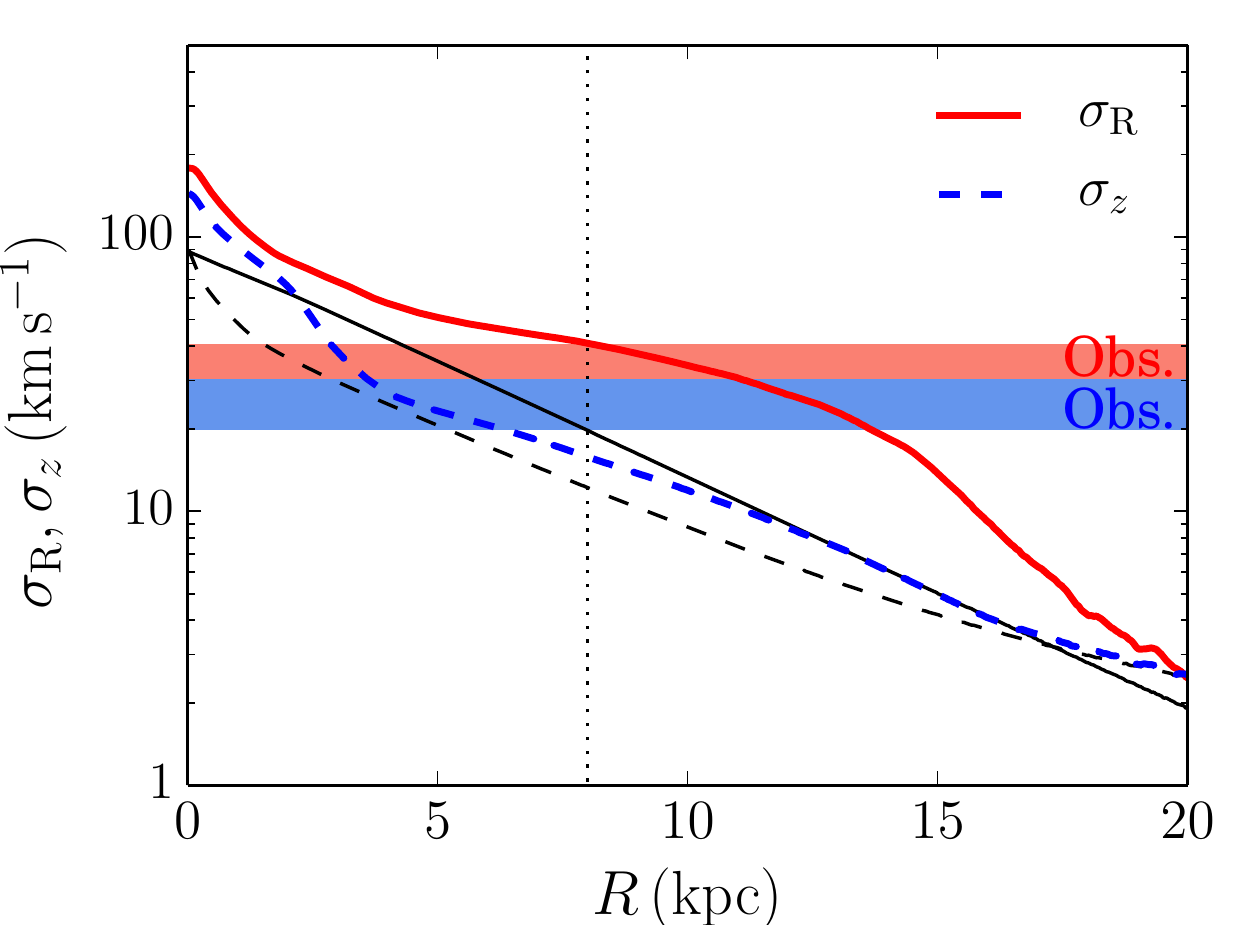}\\
\plotone{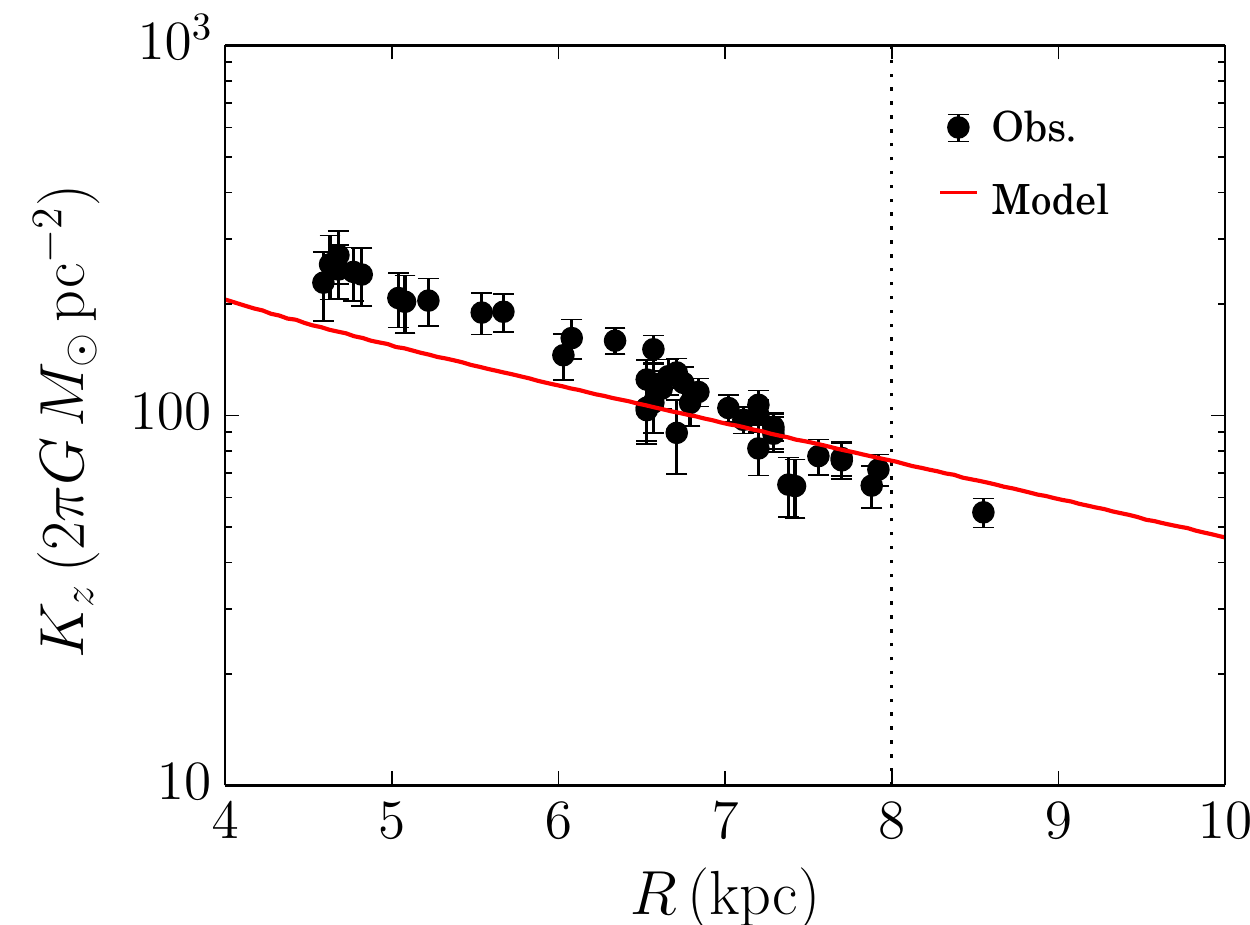}\plotone{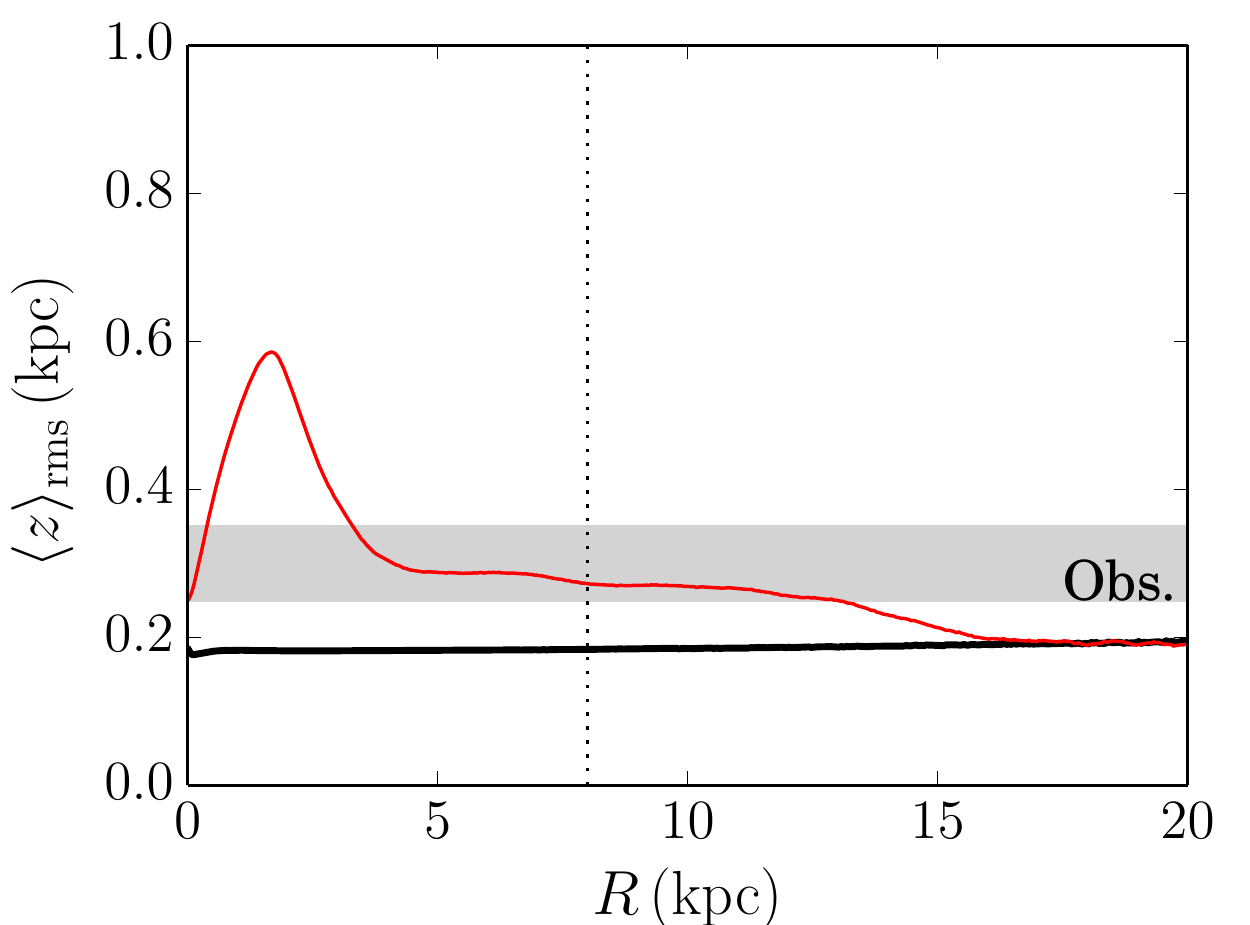}\\
\plotone{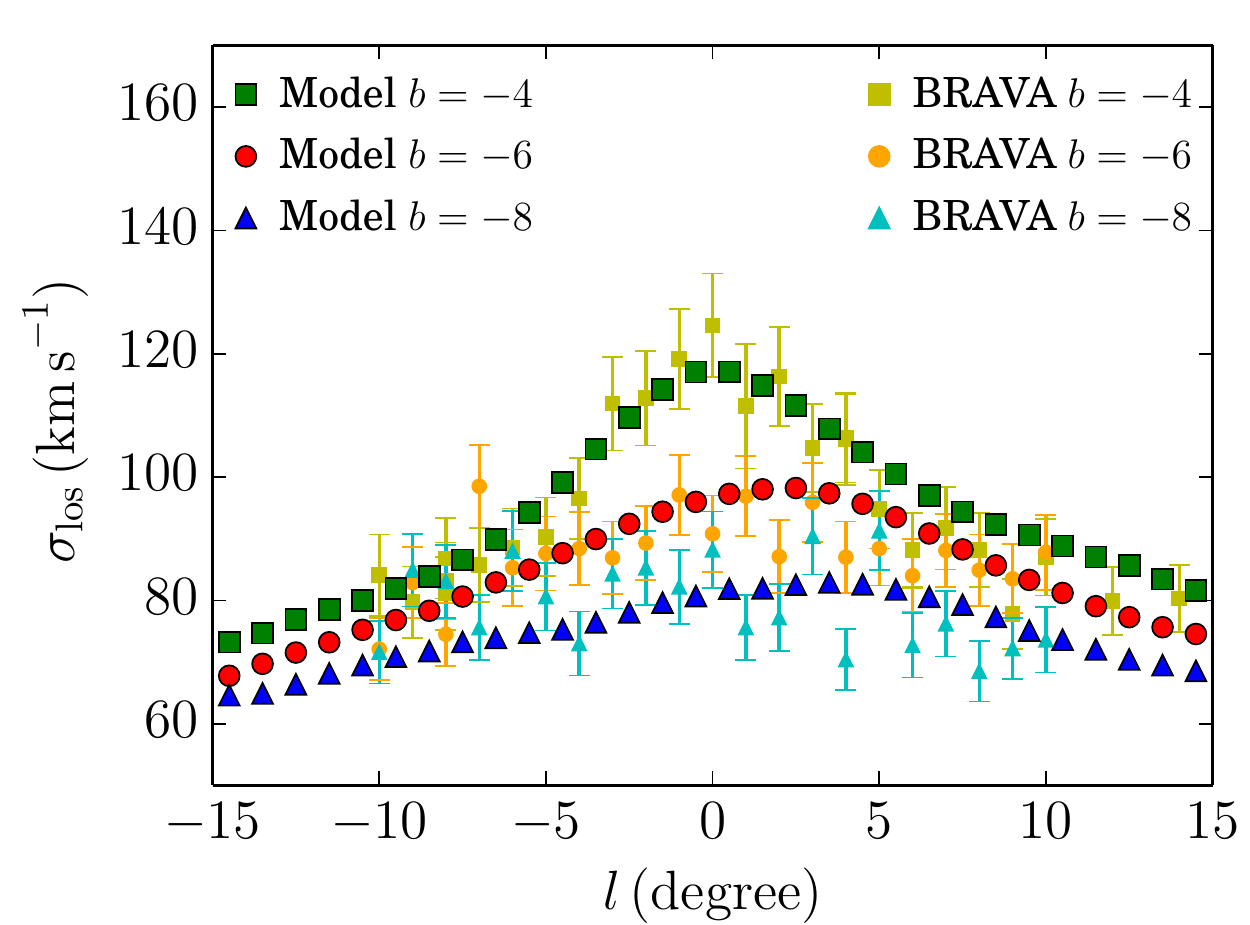}\plotone{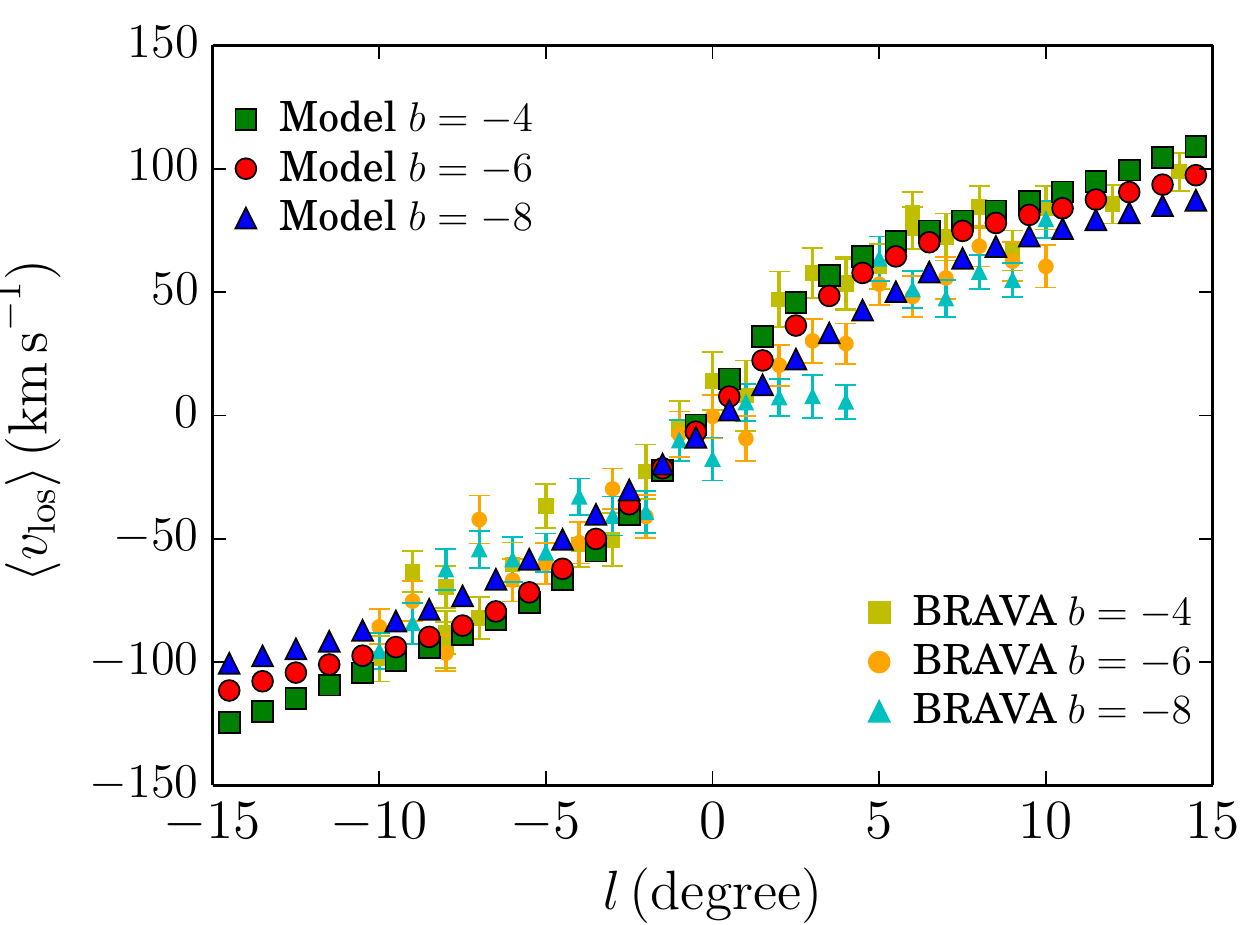}\\
\plotone{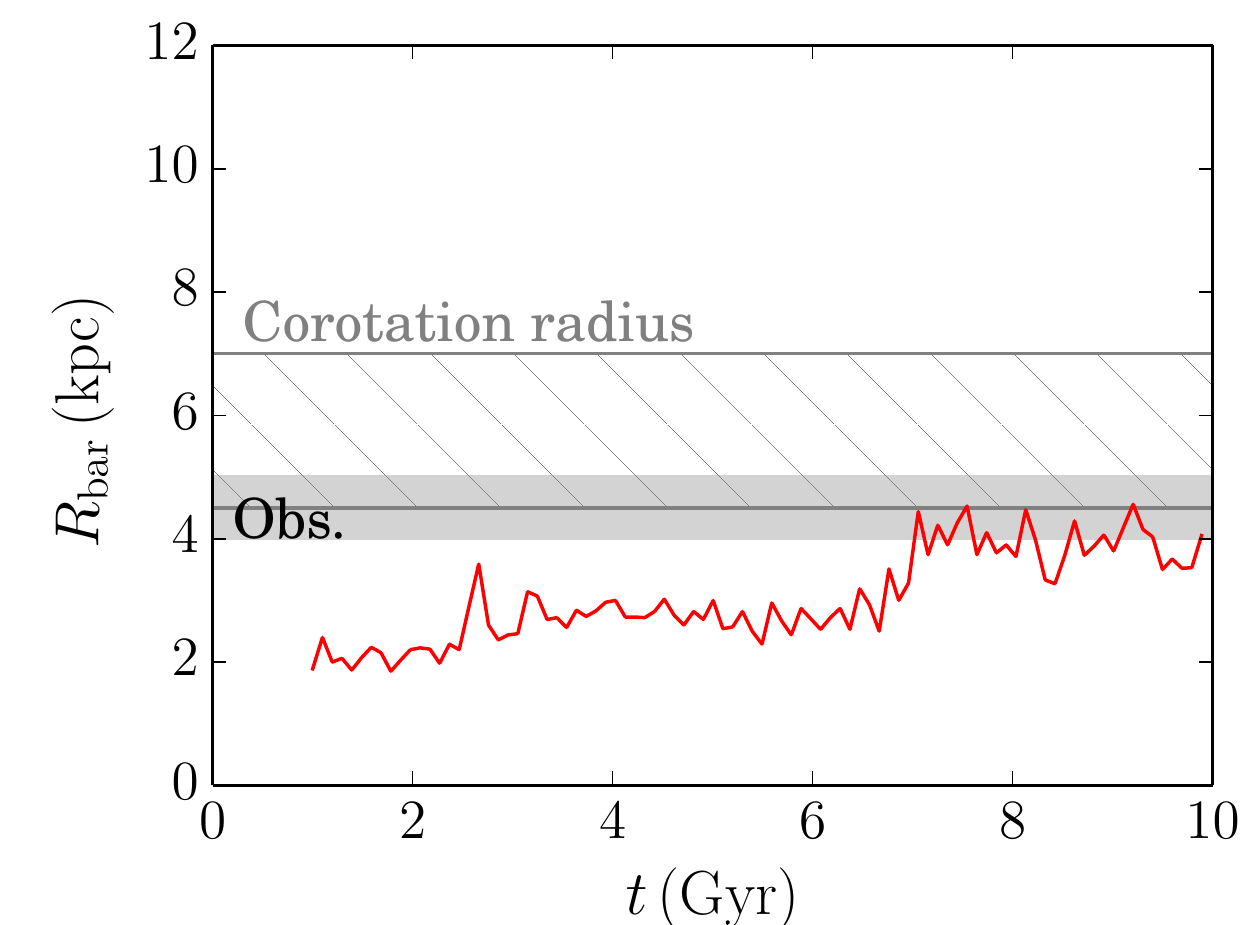}\plotone{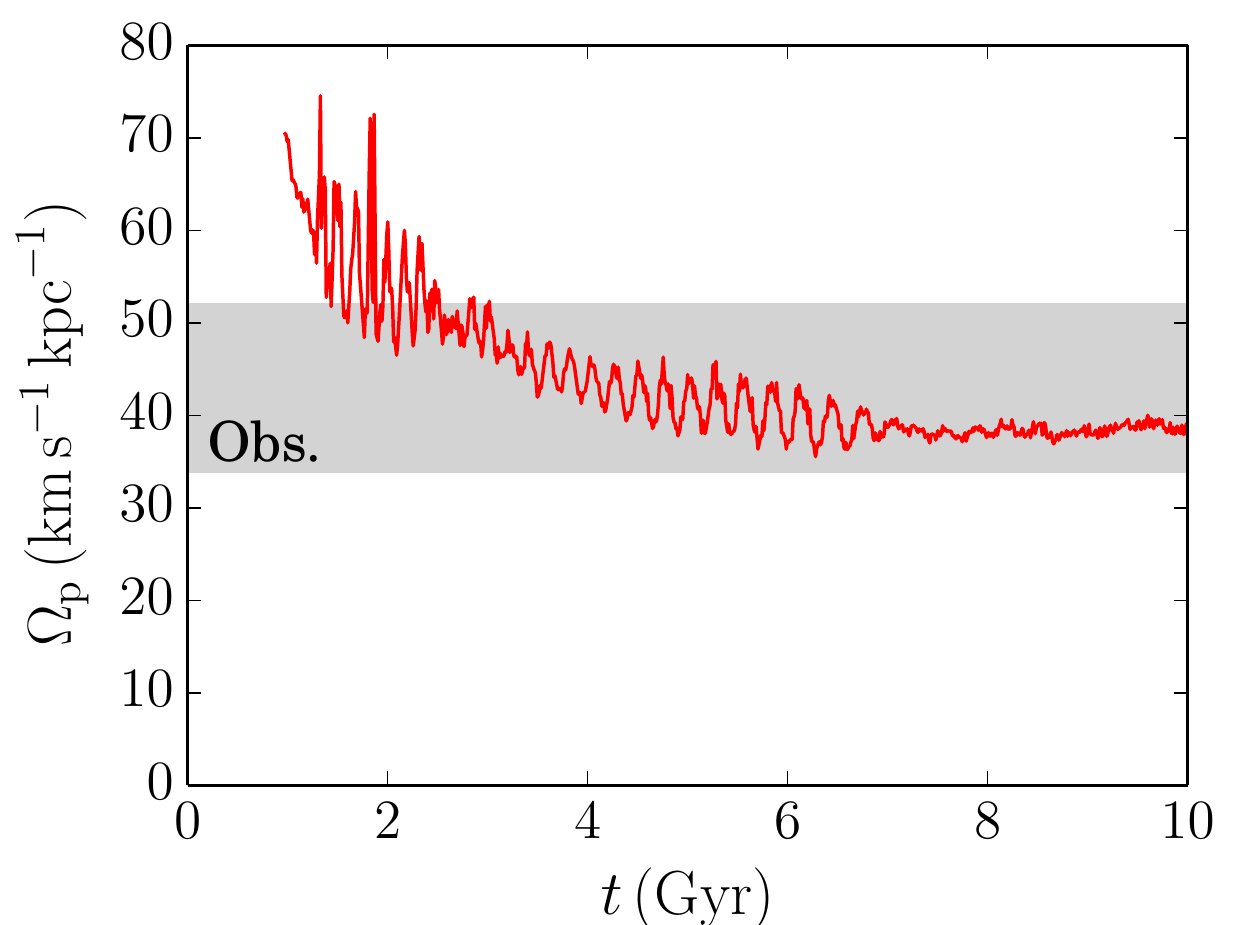}\\
\caption{Same as Fig.\ref{fig:a5B}, but for model MWb6B. \label{fig:b6B}}
\end{figure*}

\begin{figure*}
\epsscale{.45}
\plotone{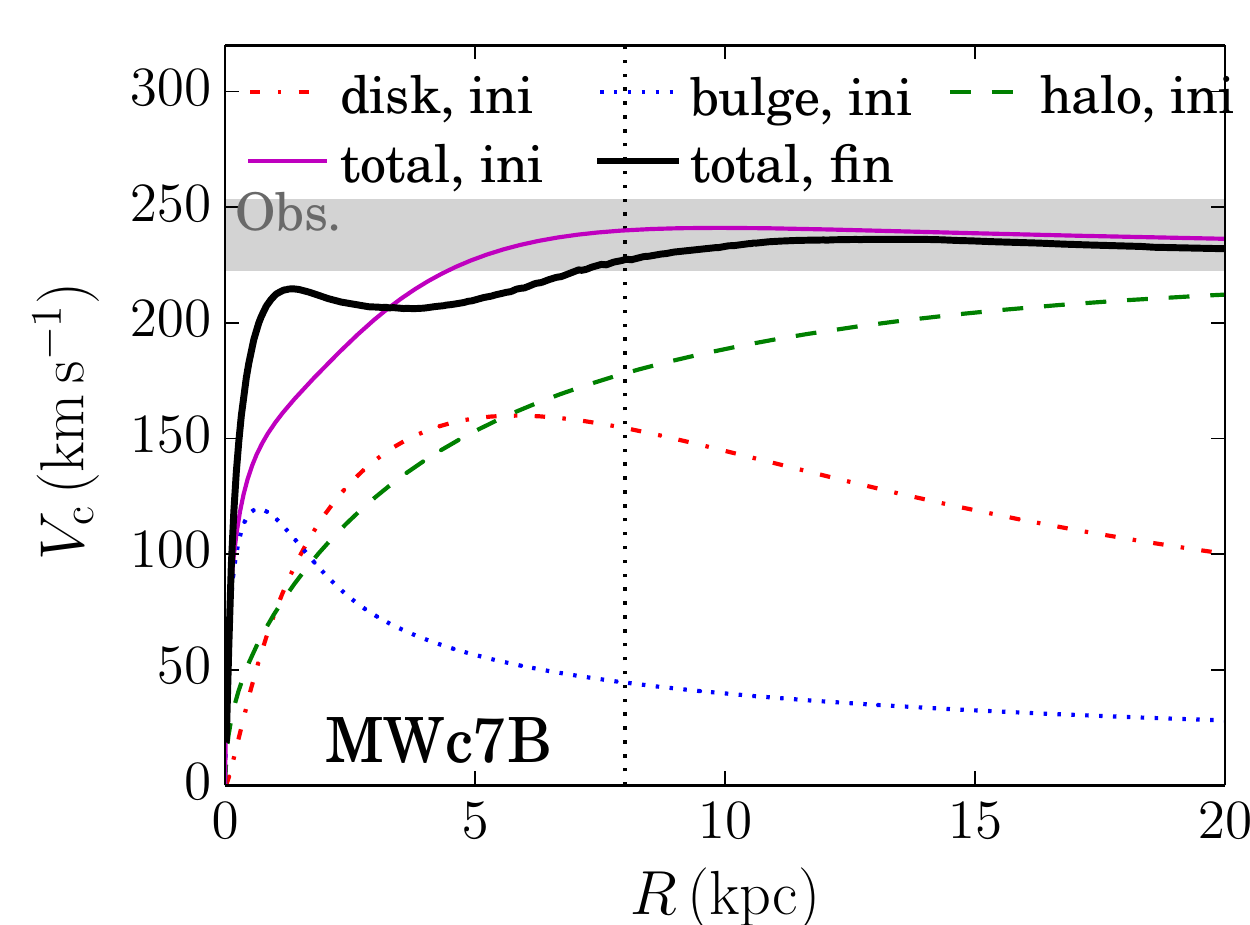}\plotone{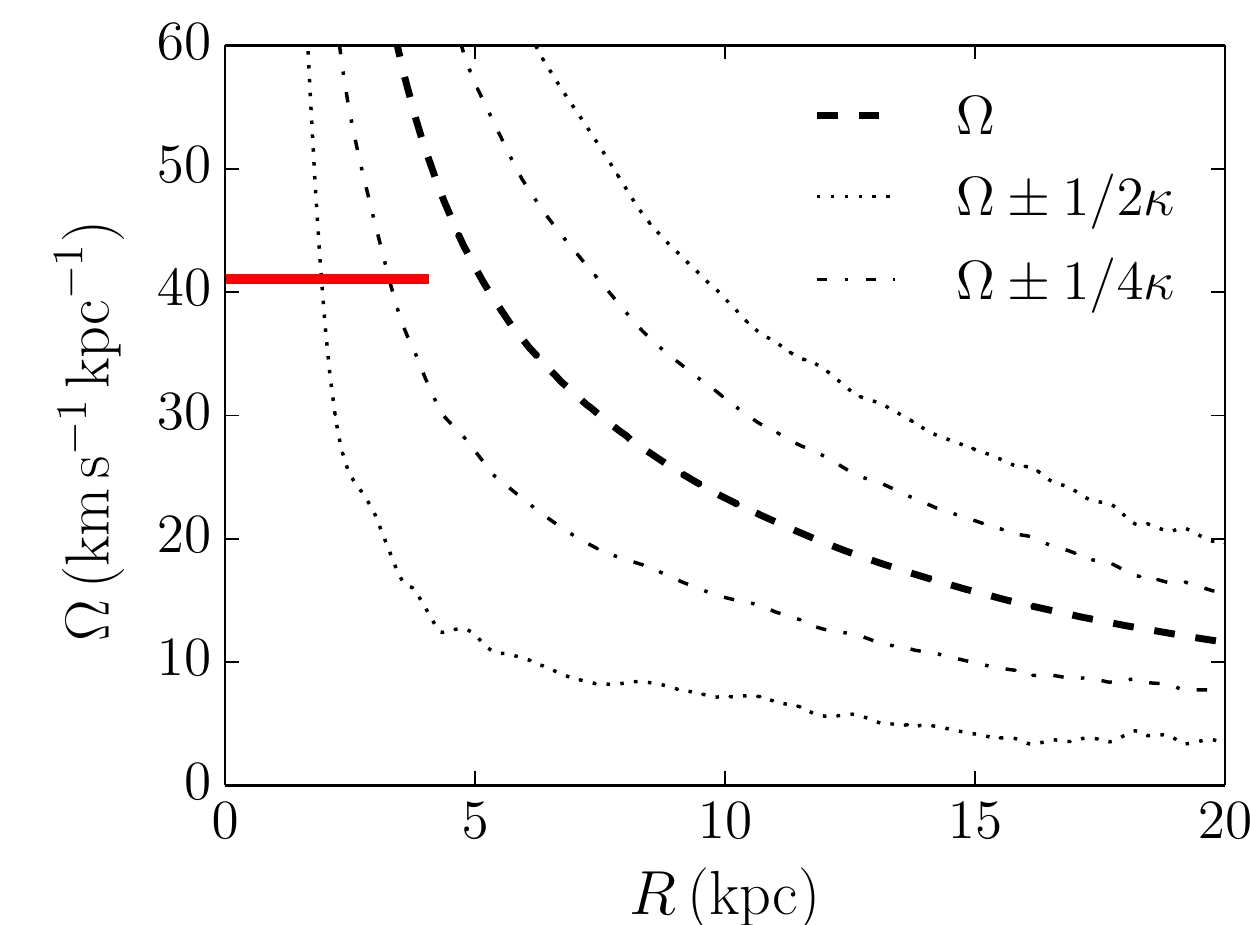}\\
\plotone{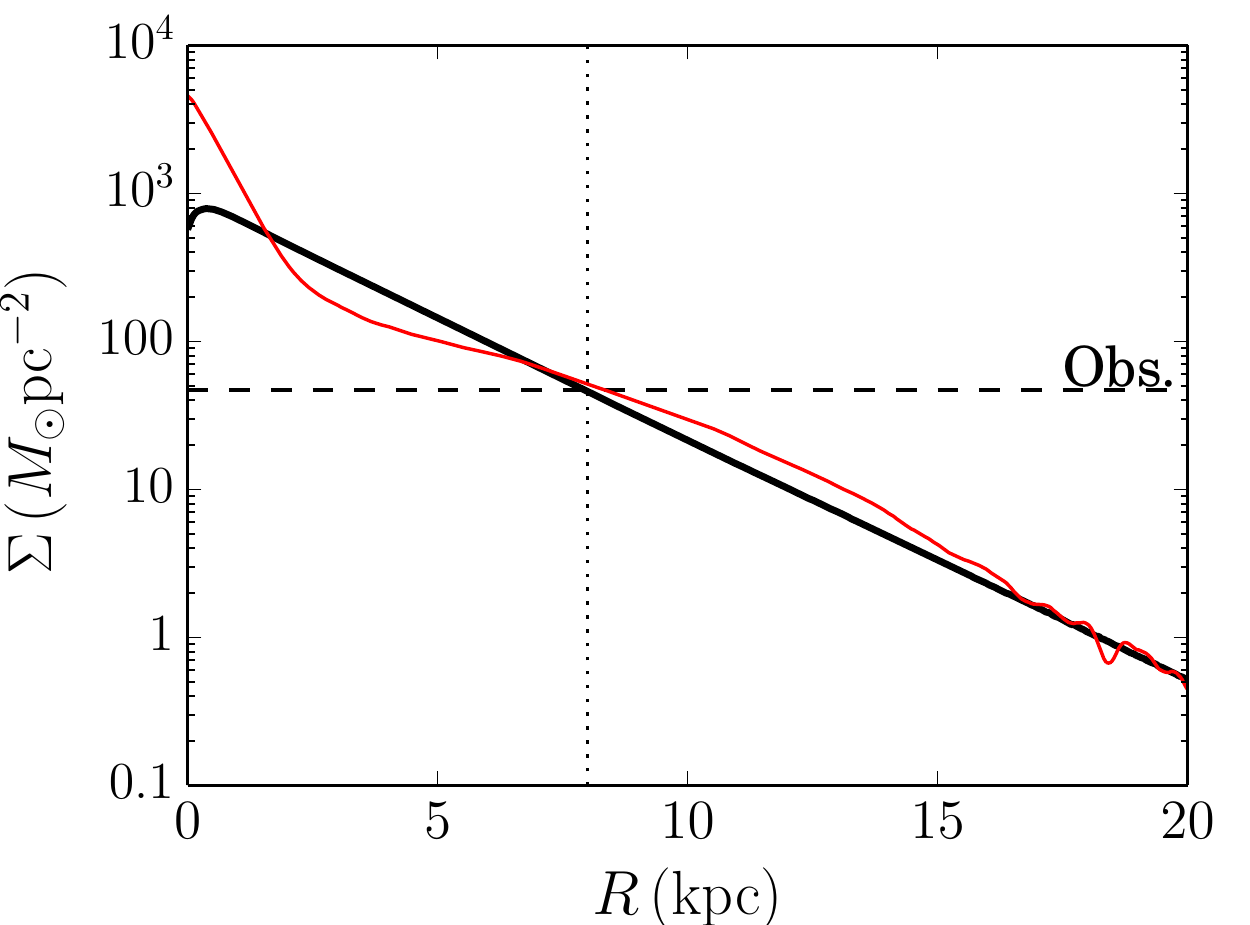}\plotone{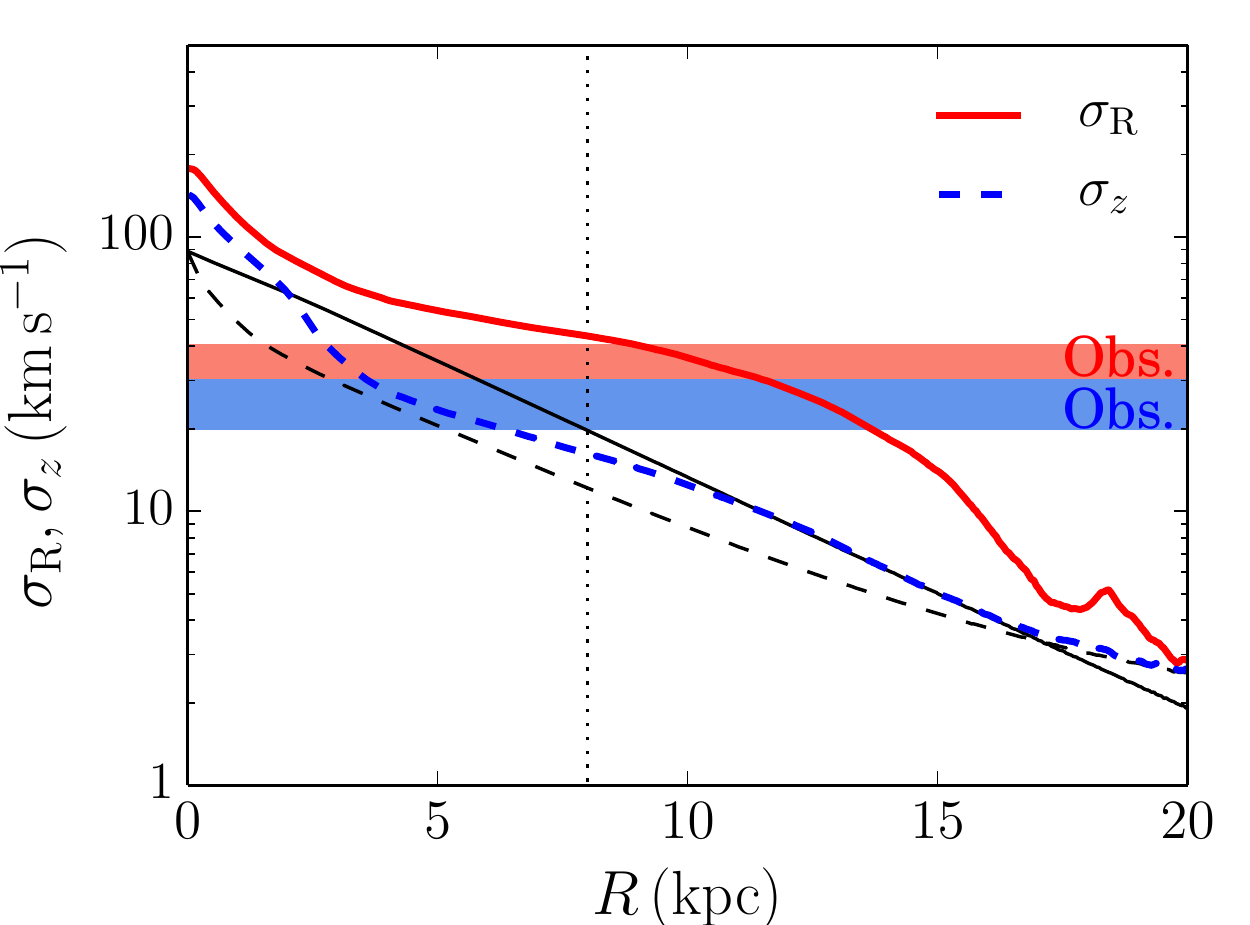}\\
\plotone{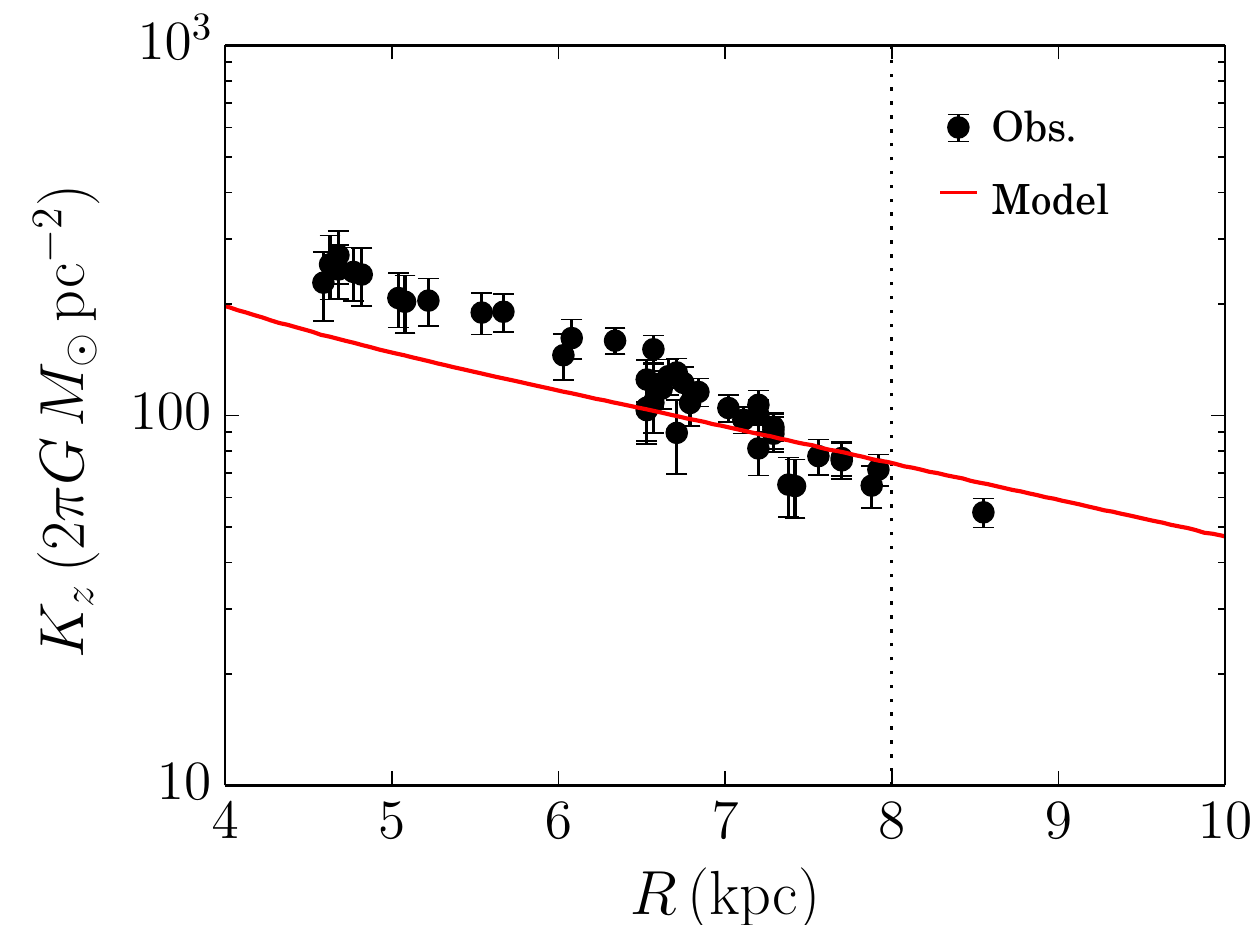}\plotone{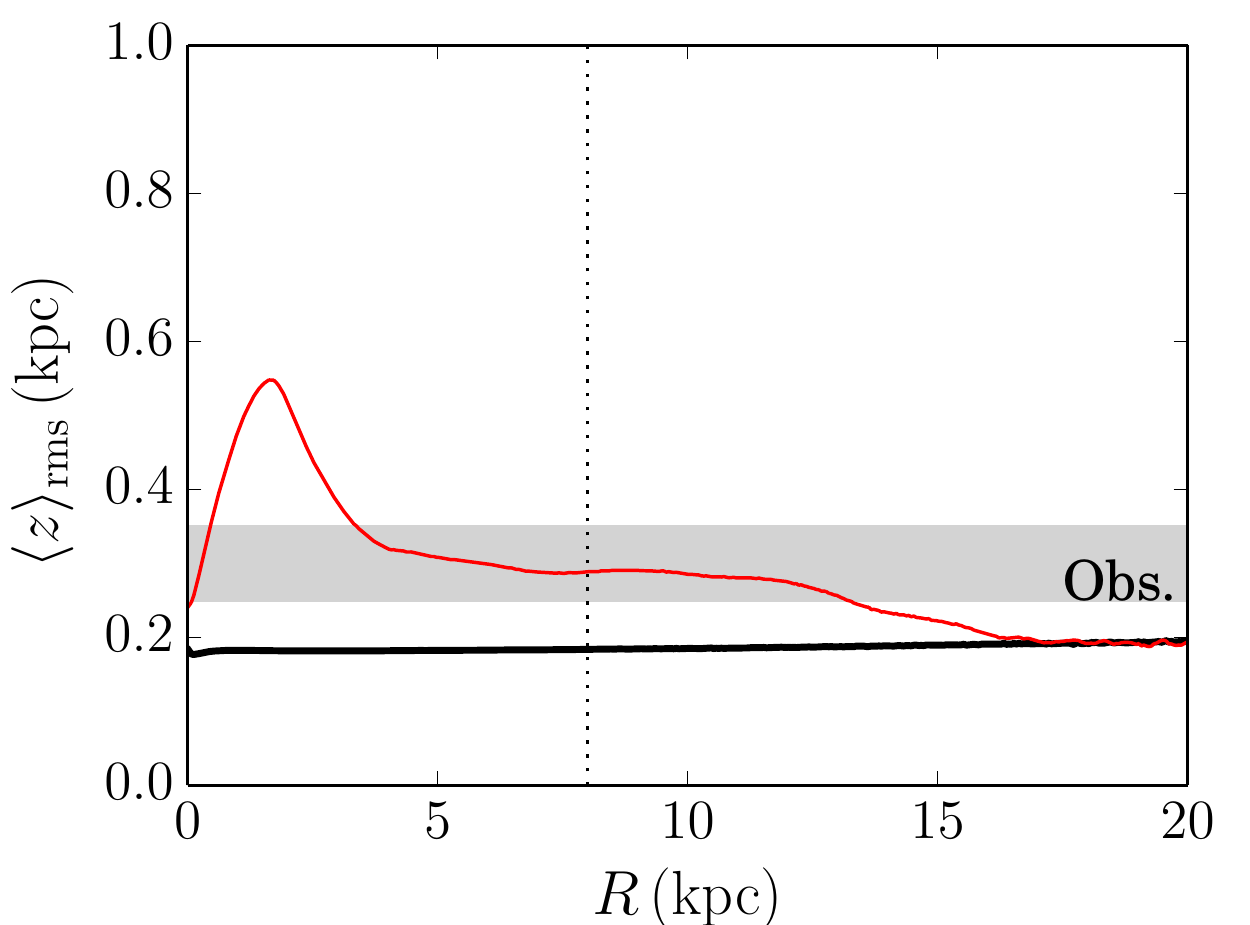}\\
\plotone{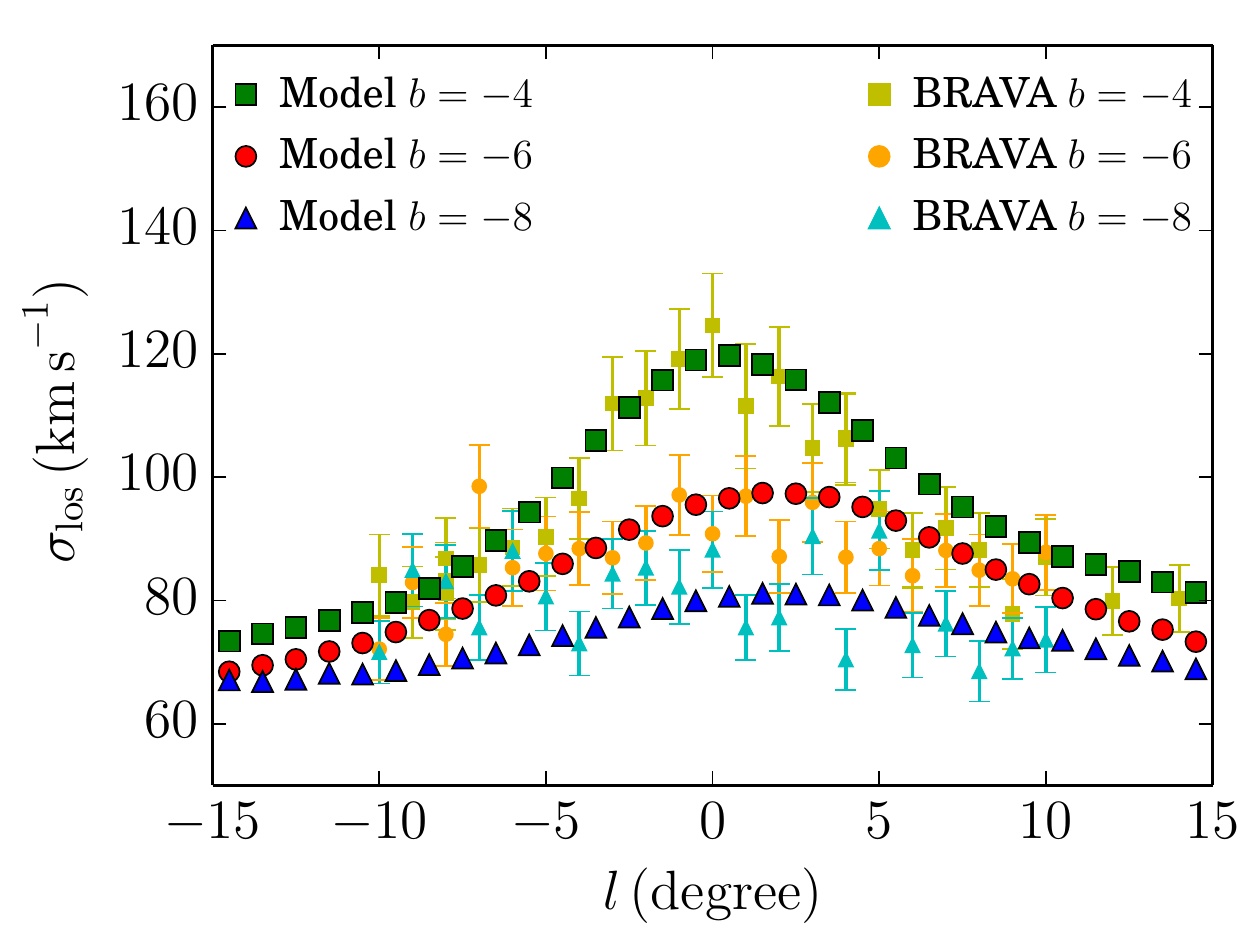}\plotone{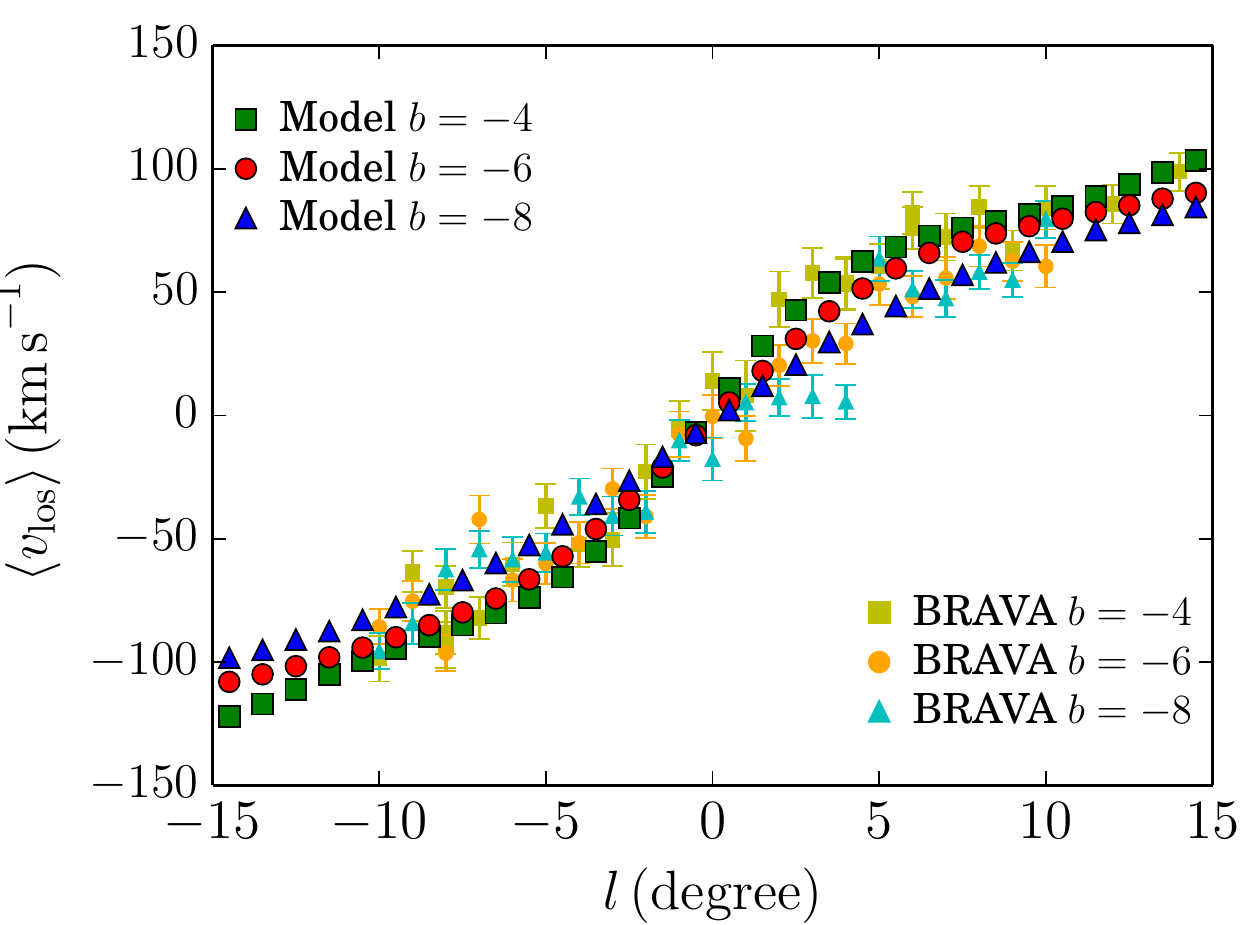}\\
\plotone{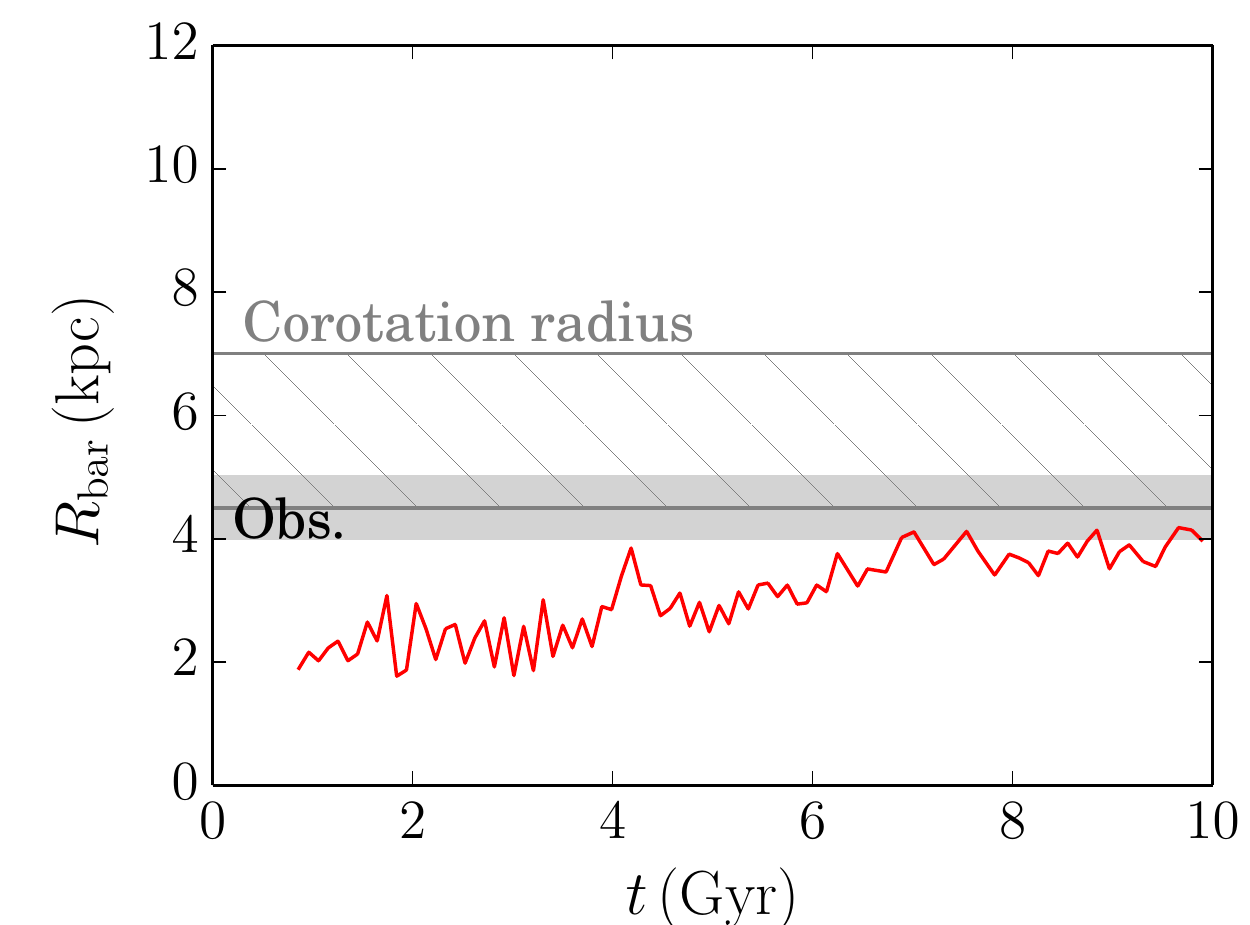}\plotone{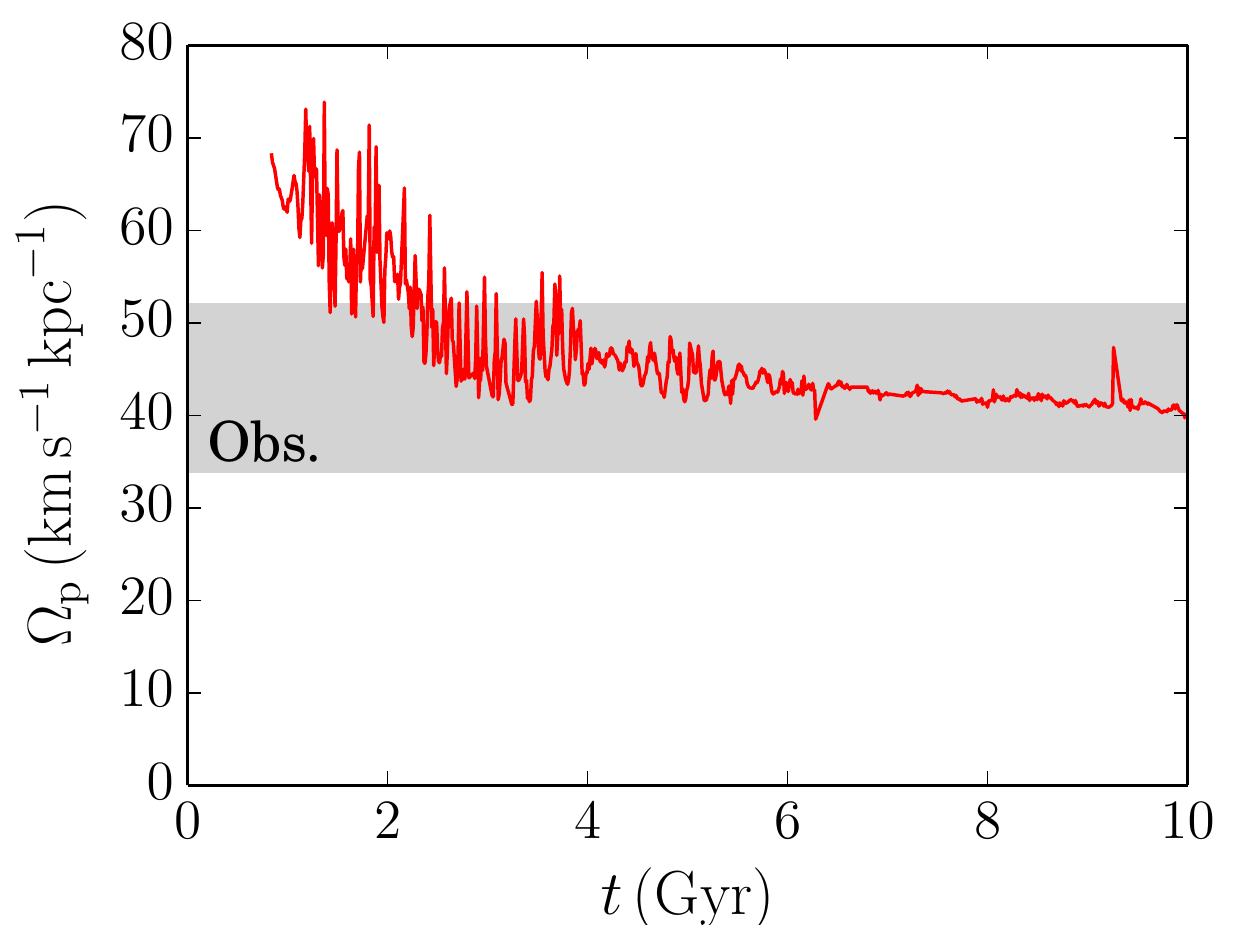}\\
\caption{Same as Fig.\ref{fig:a5B}, but for model MWc7B.\label{fig:c7B} }
\end{figure*}

\begin{figure*}
\epsscale{.45}
\plotone{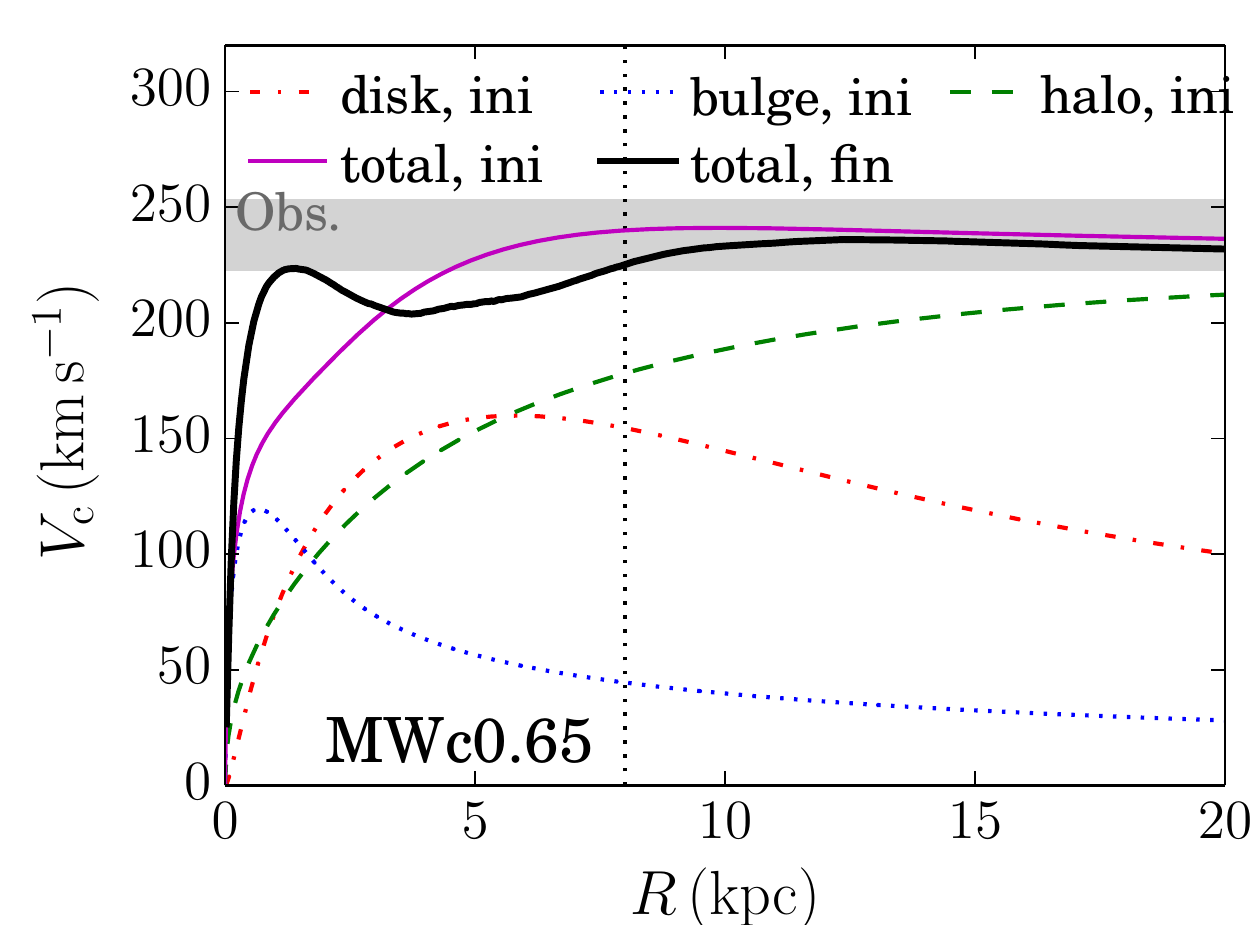}\plotone{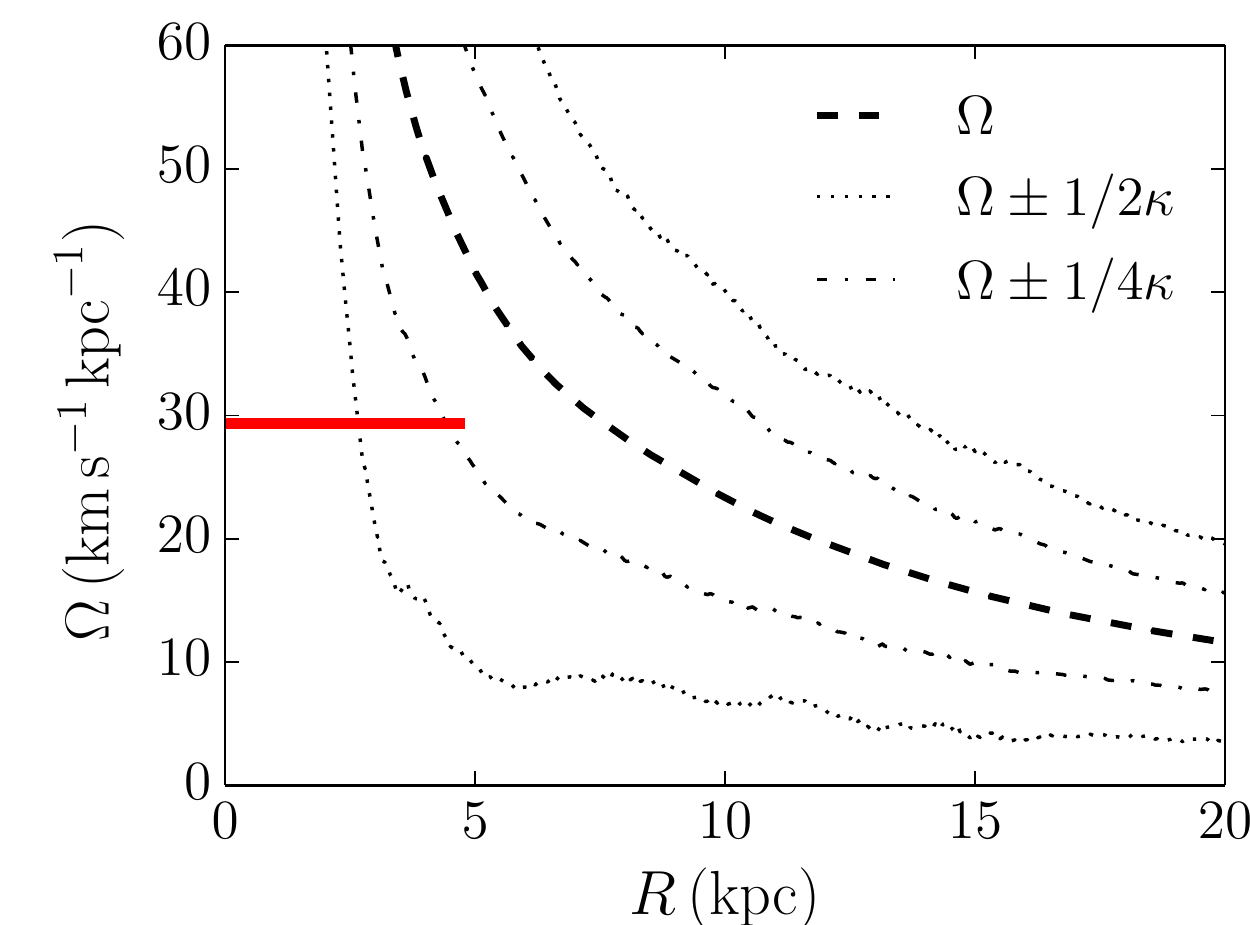}\\
\plotone{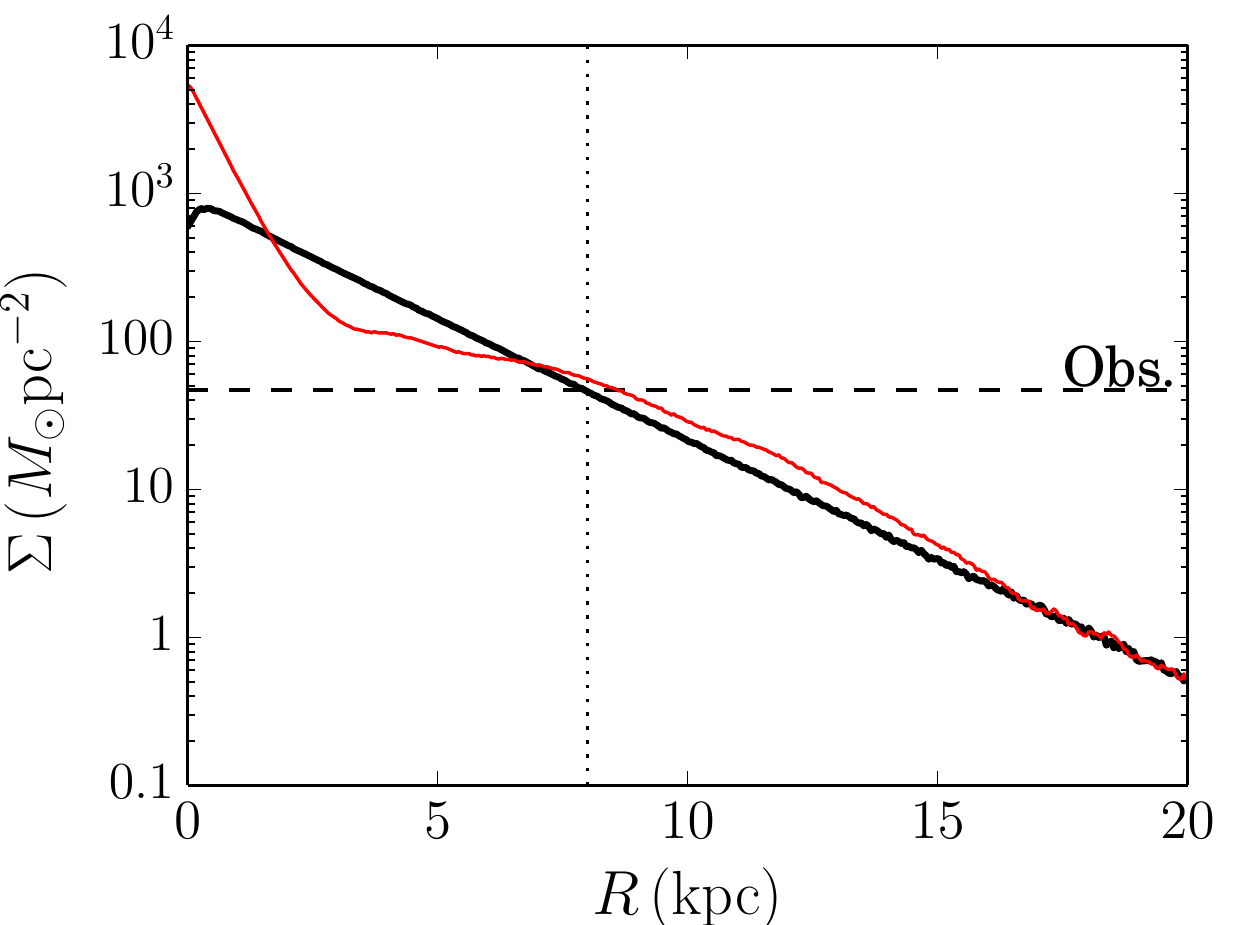}\plotone{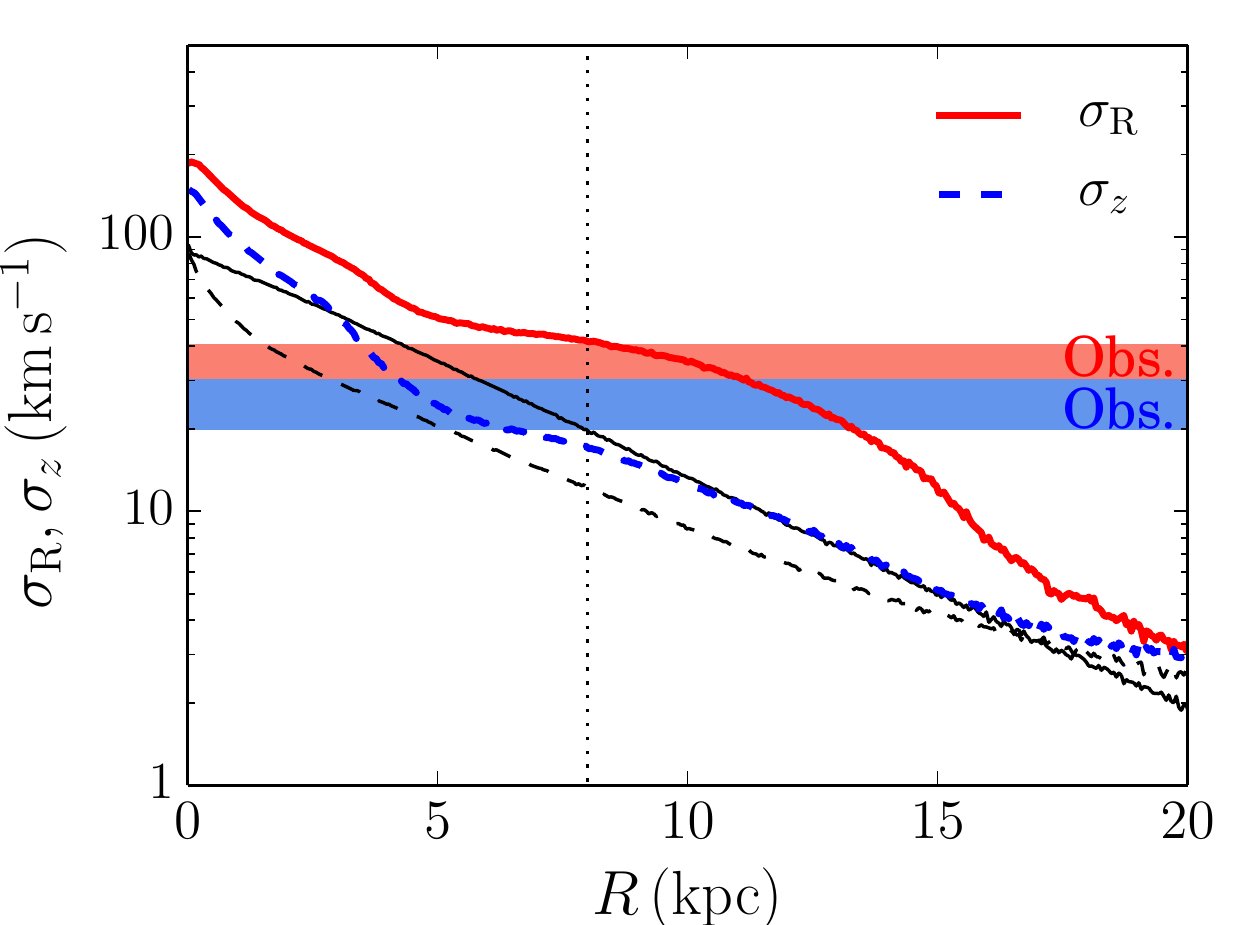}\\
\plotone{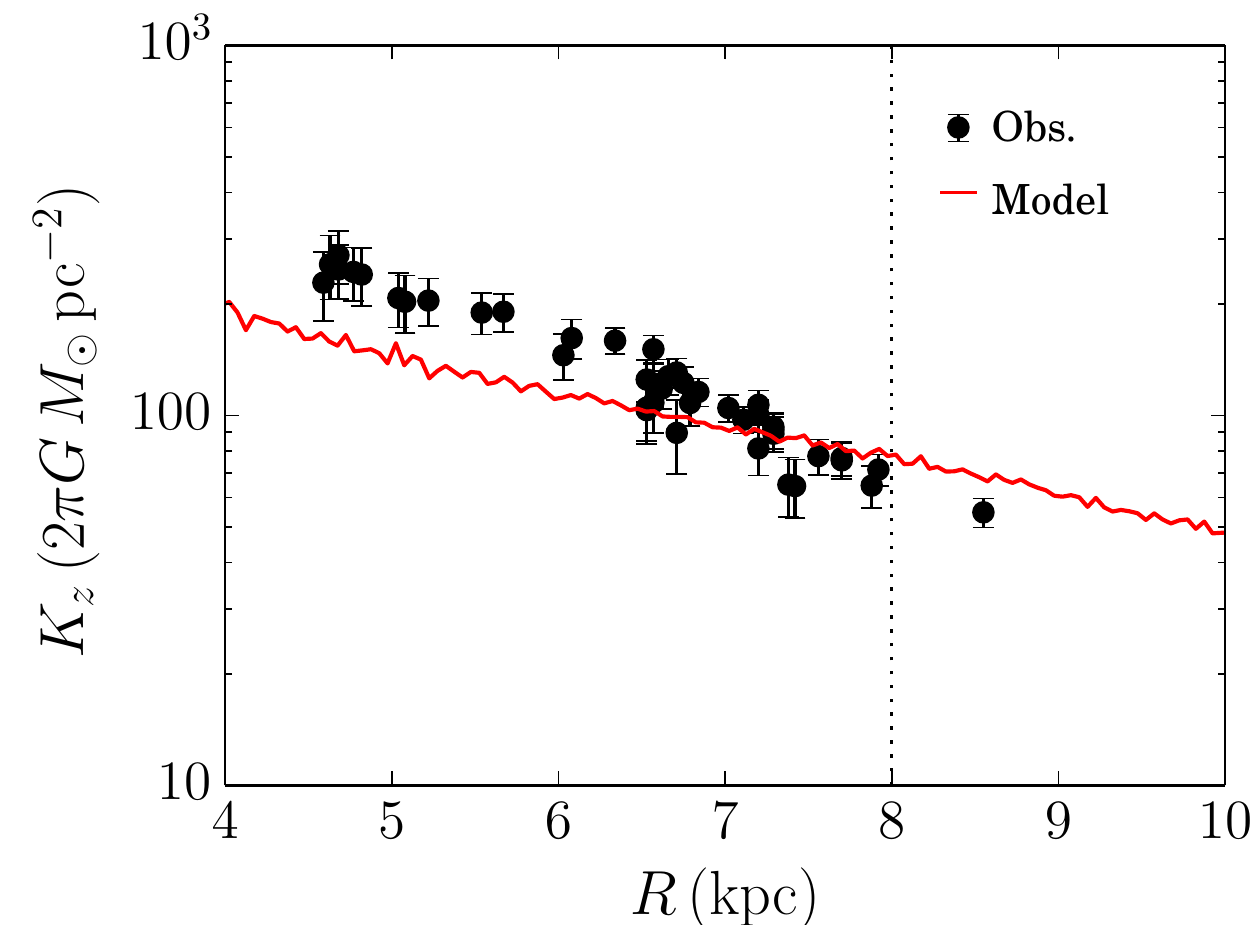}\plotone{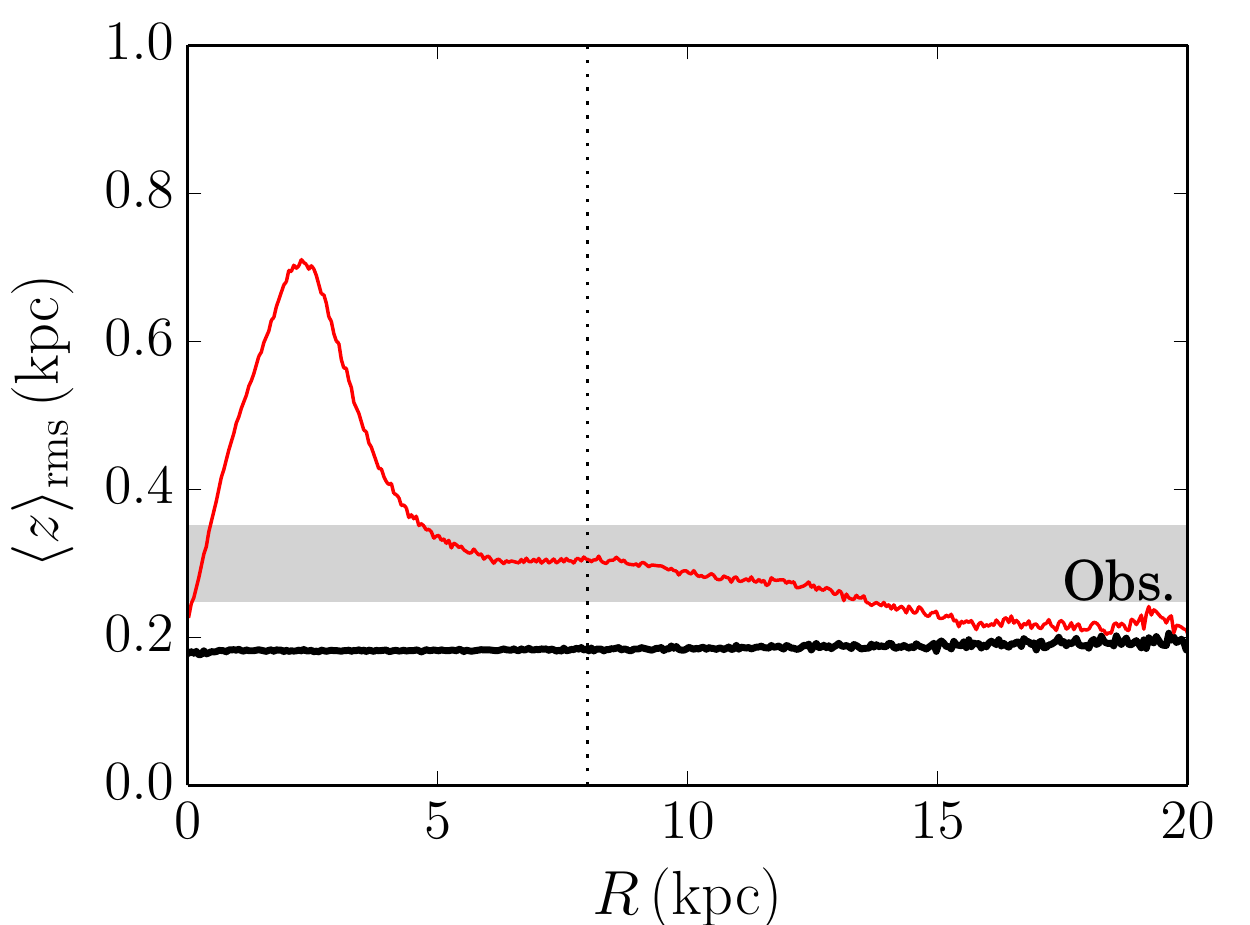}\\
\plotone{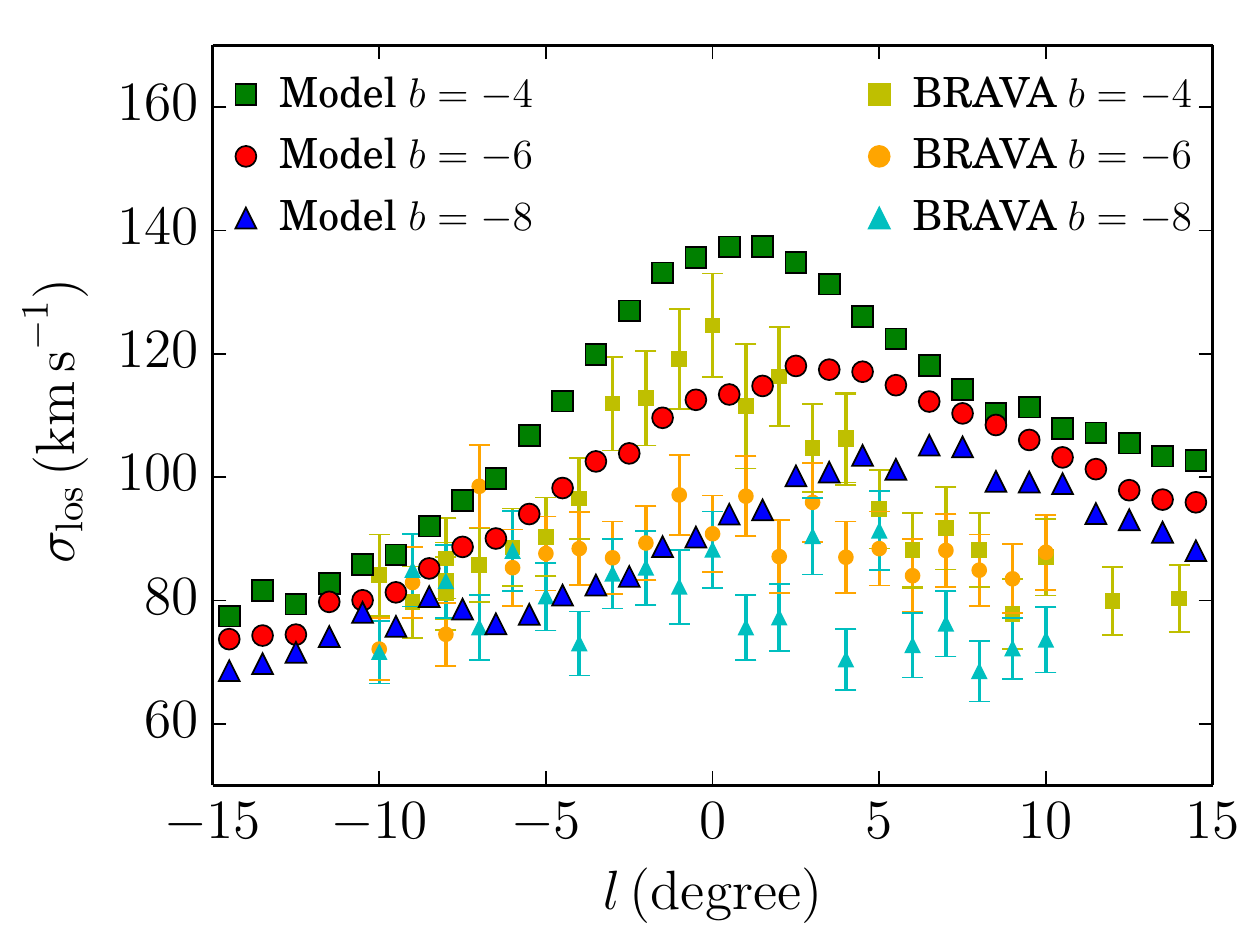}\plotone{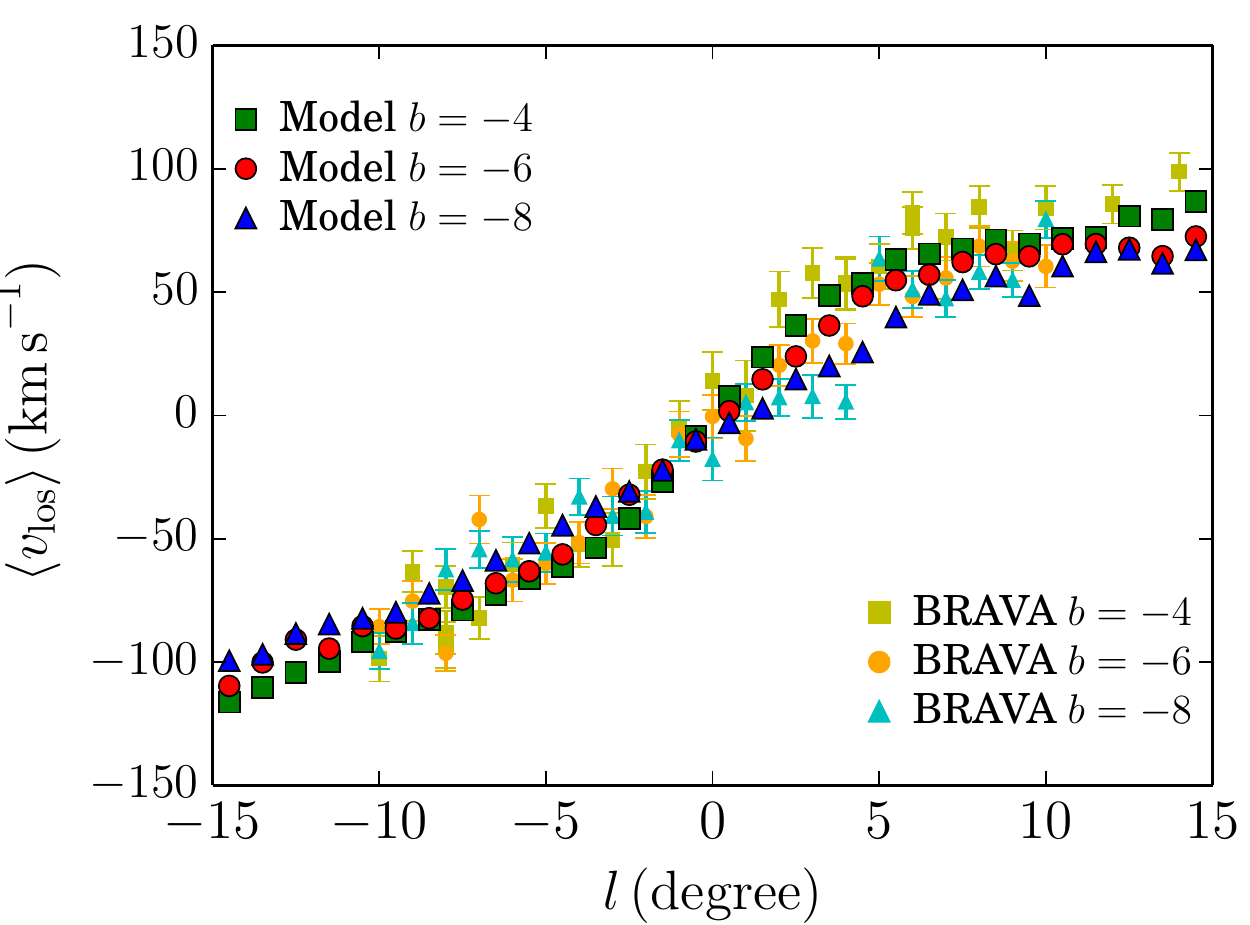}\\
\plotone{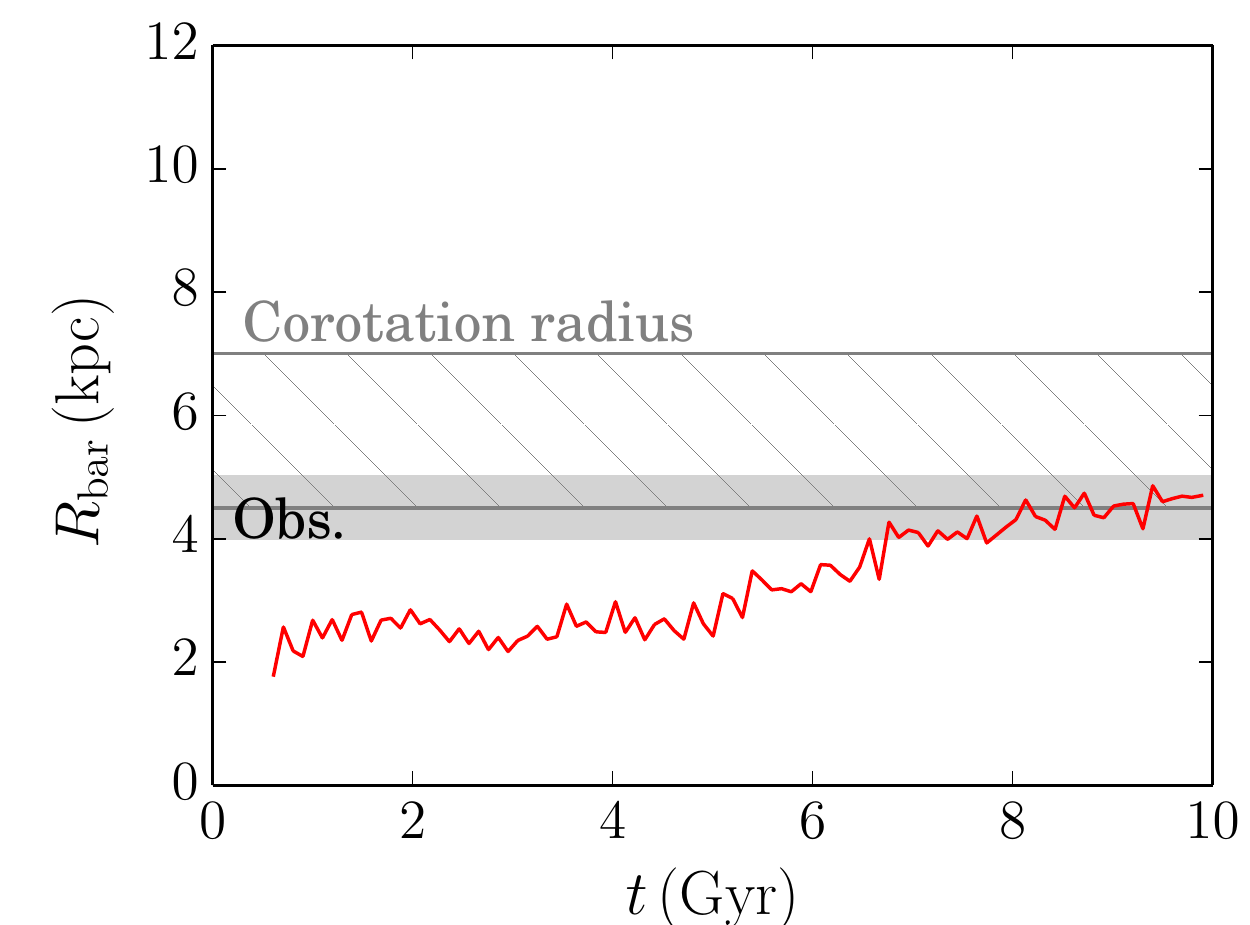}\plotone{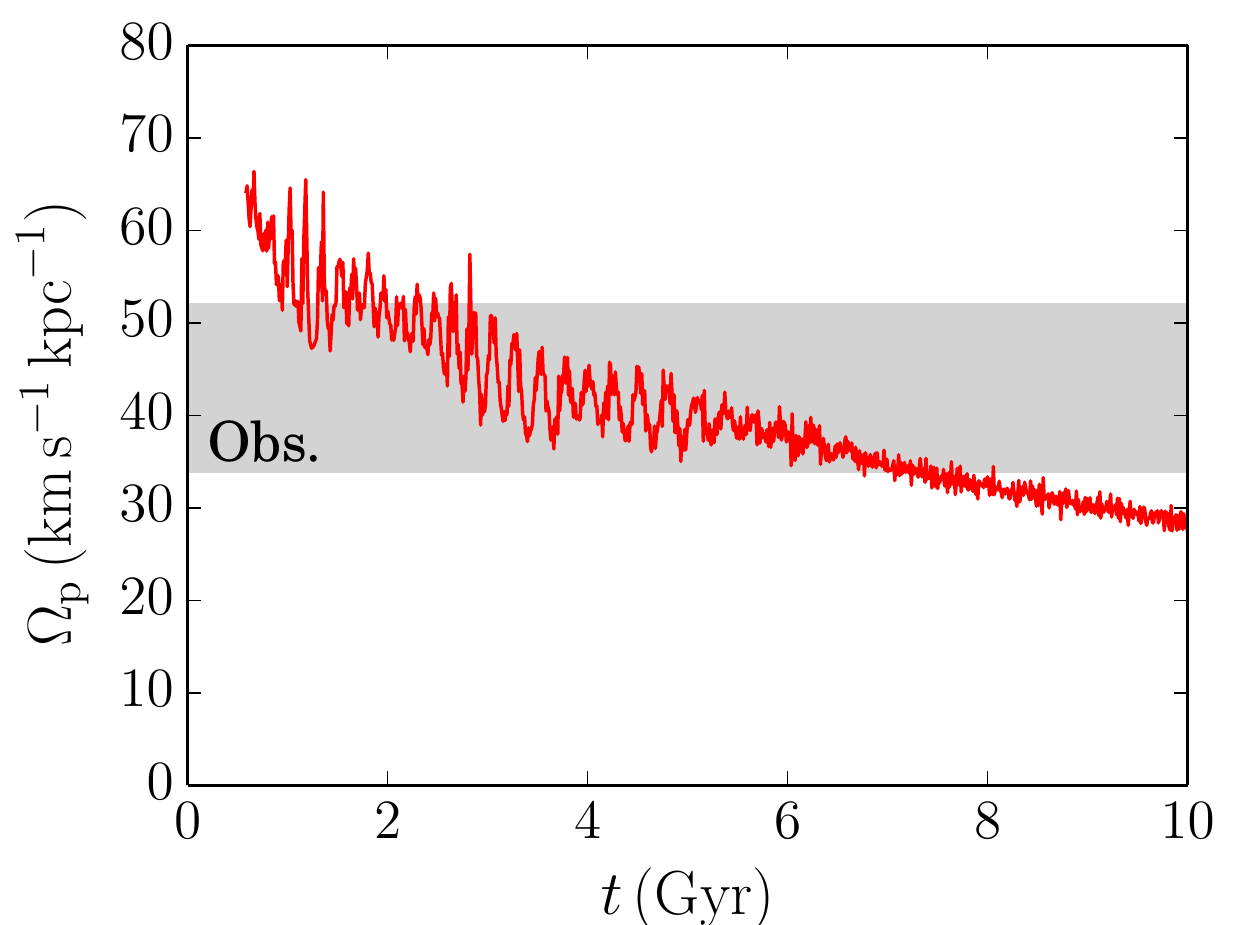}\\
\caption{Same as Fig.\ref{fig:a5B}, but for model MWc0.65.\label{fig:c0.65}}
\end{figure*}

\begin{figure*}
\epsscale{.45}
\plotone{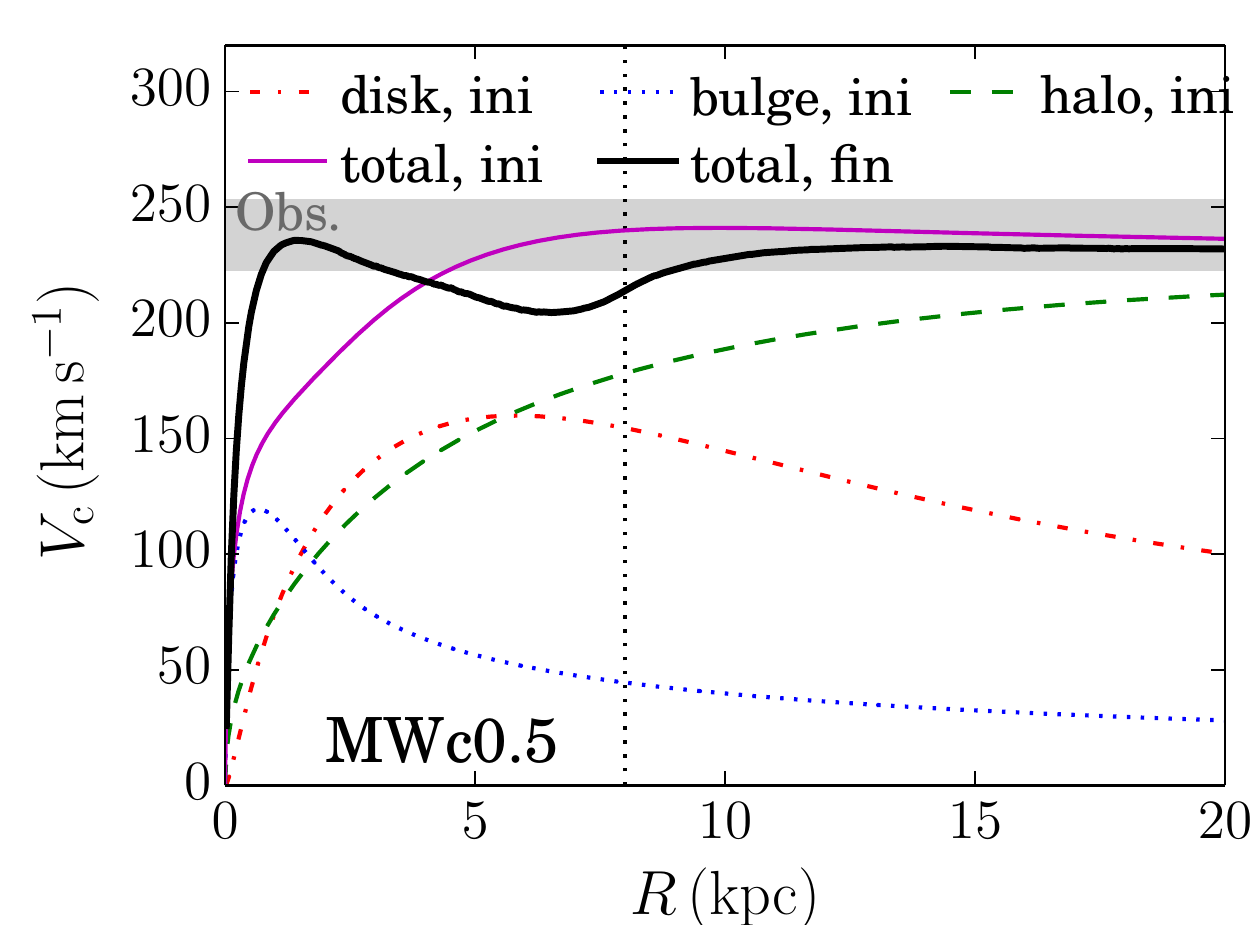}\plotone{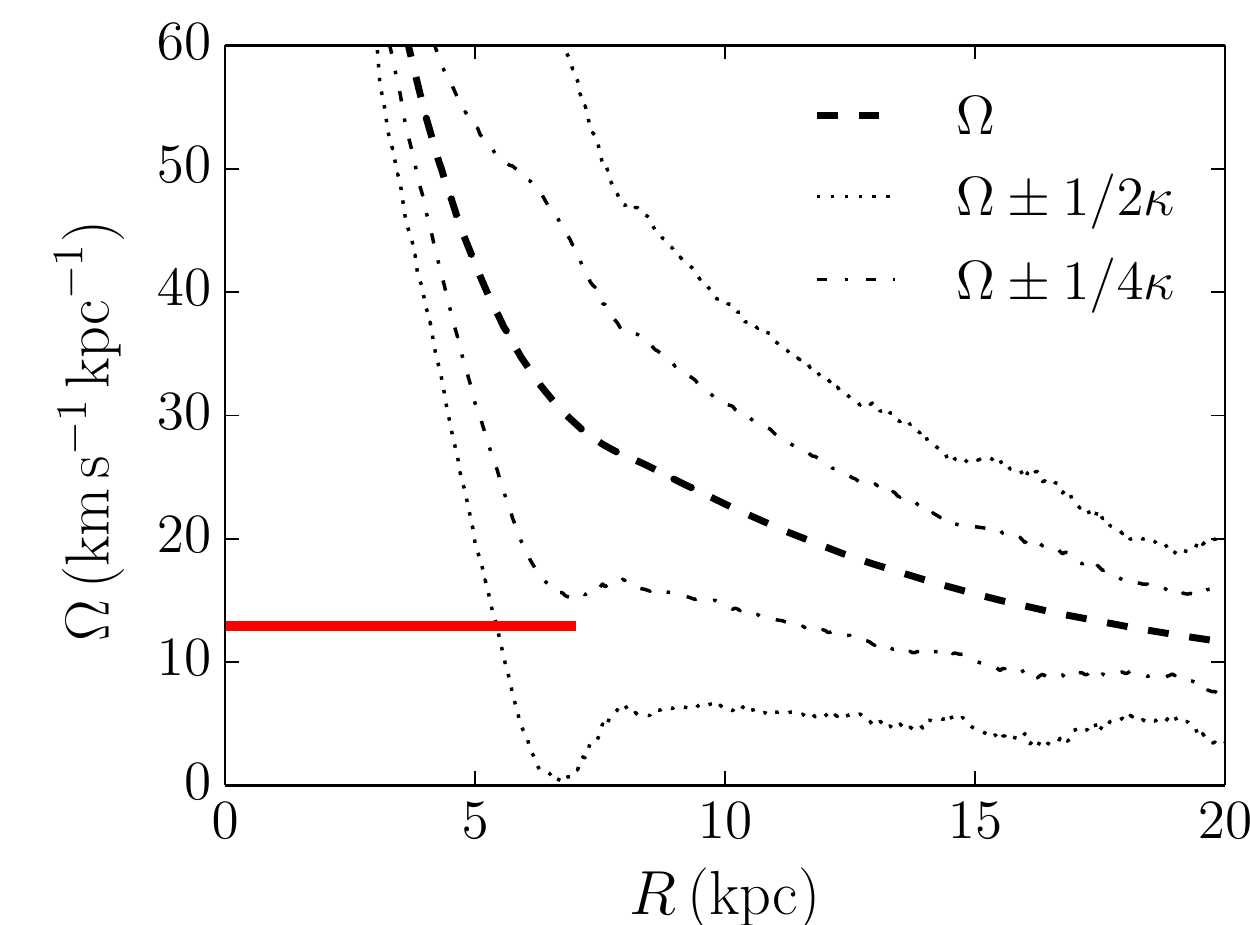}\\
\plotone{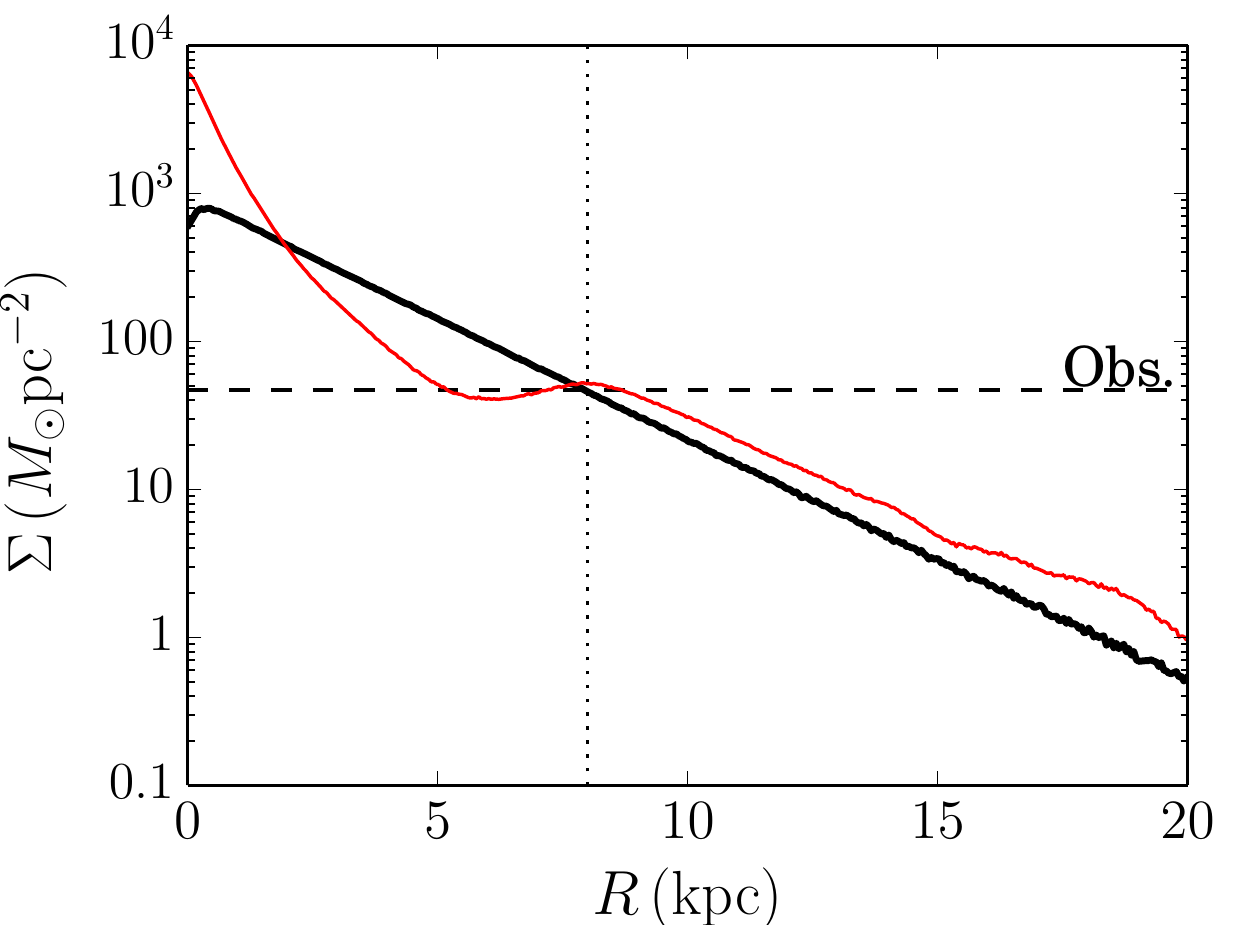}\plotone{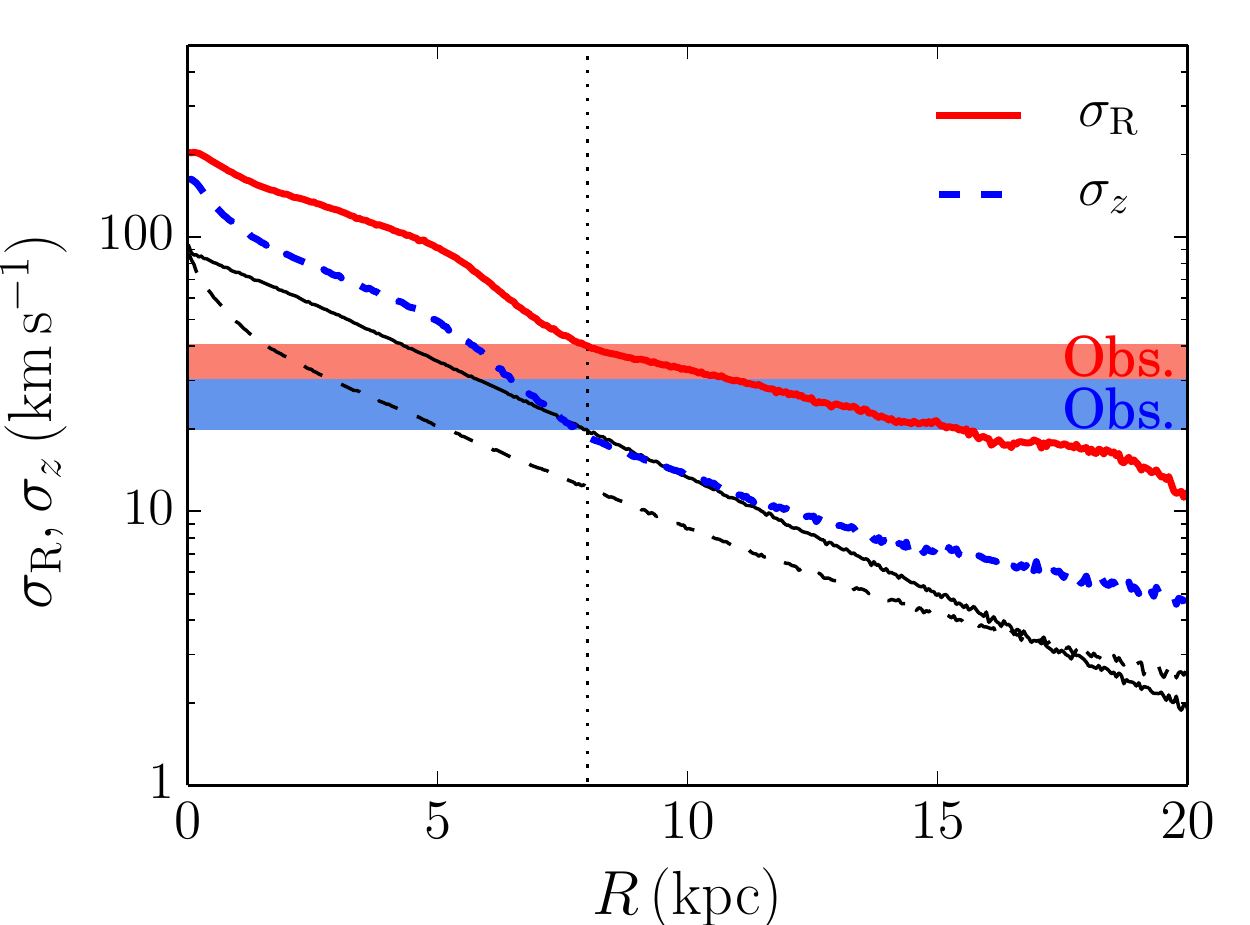}\\
\plotone{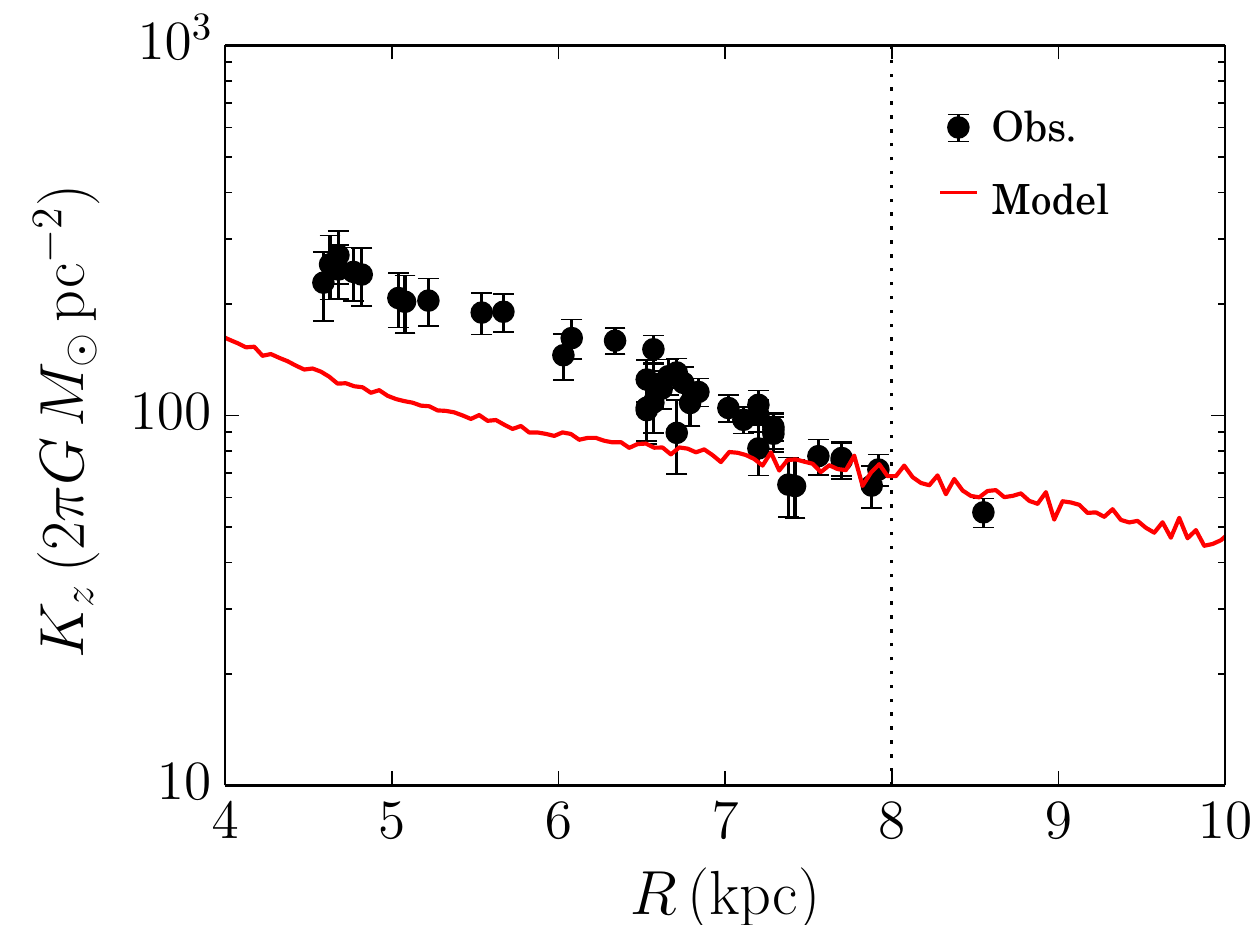}\plotone{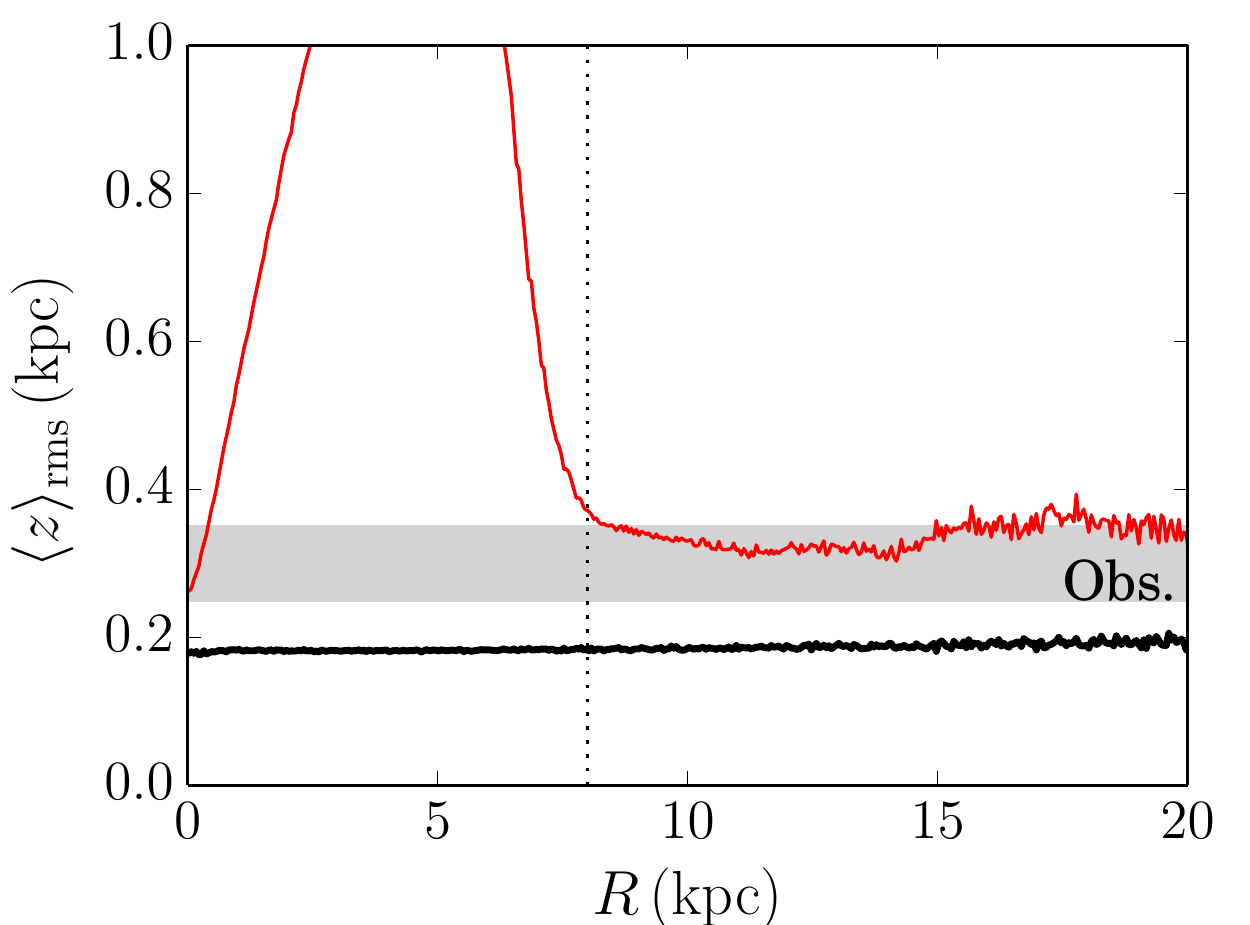}\\
\plotone{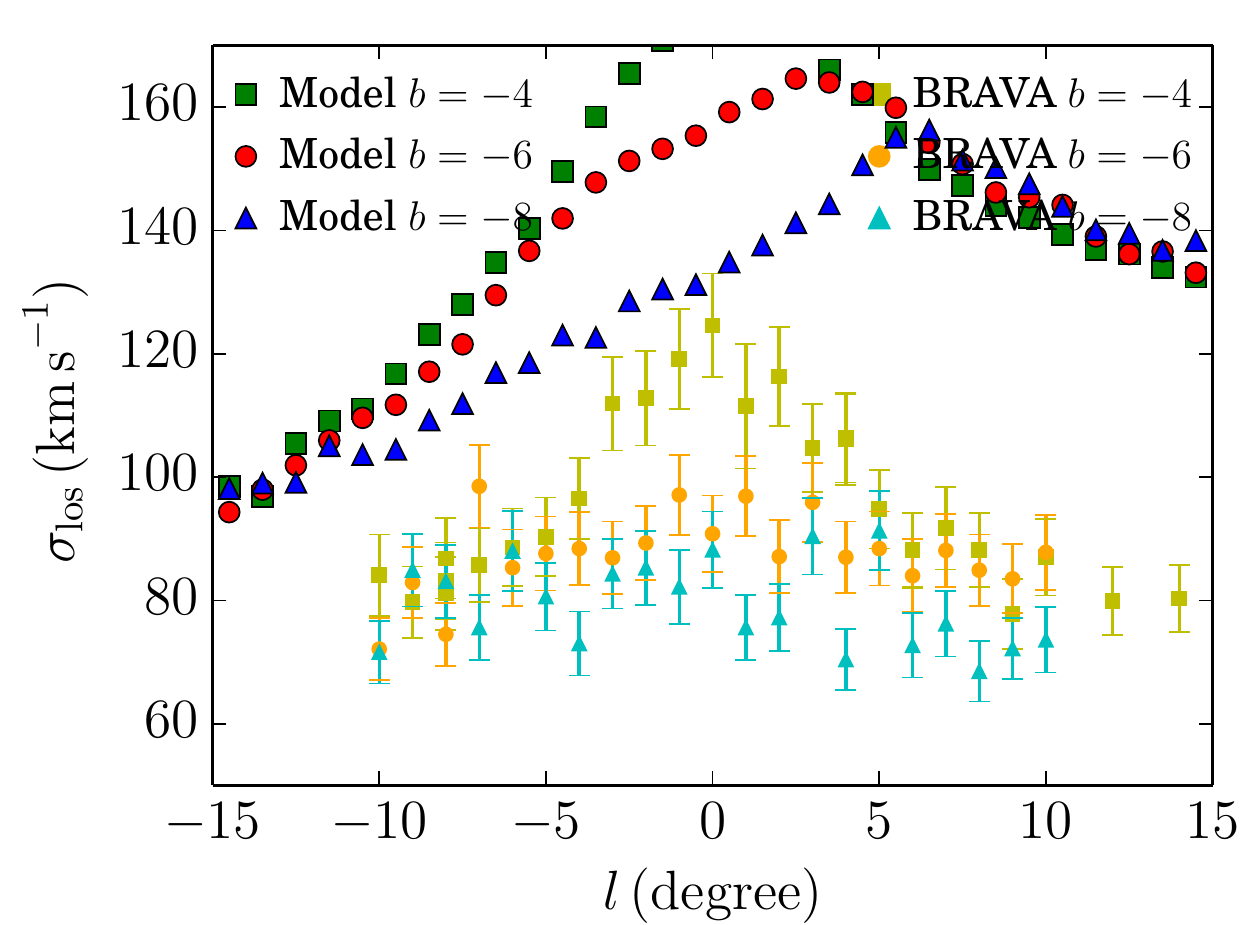}\plotone{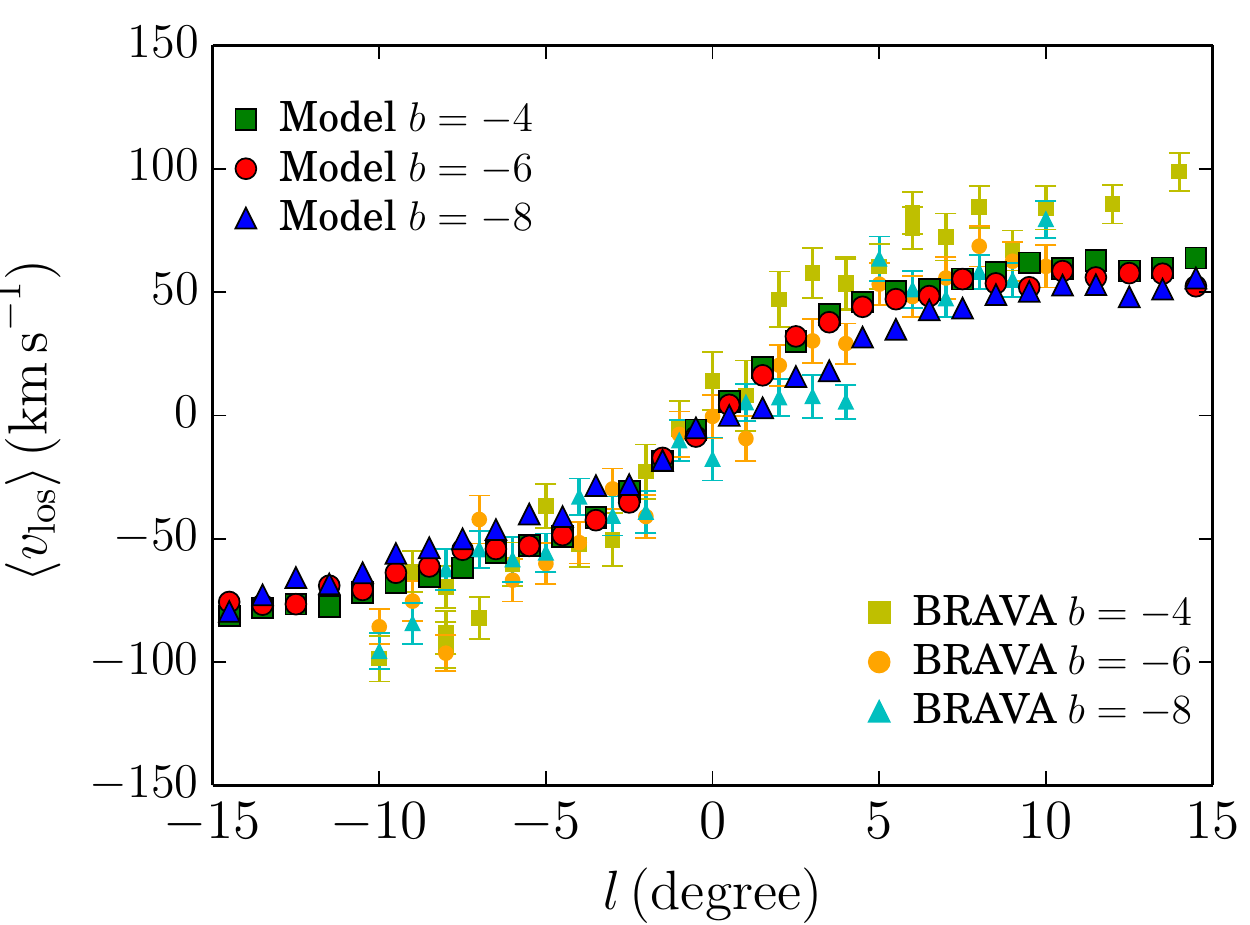}\\
\plotone{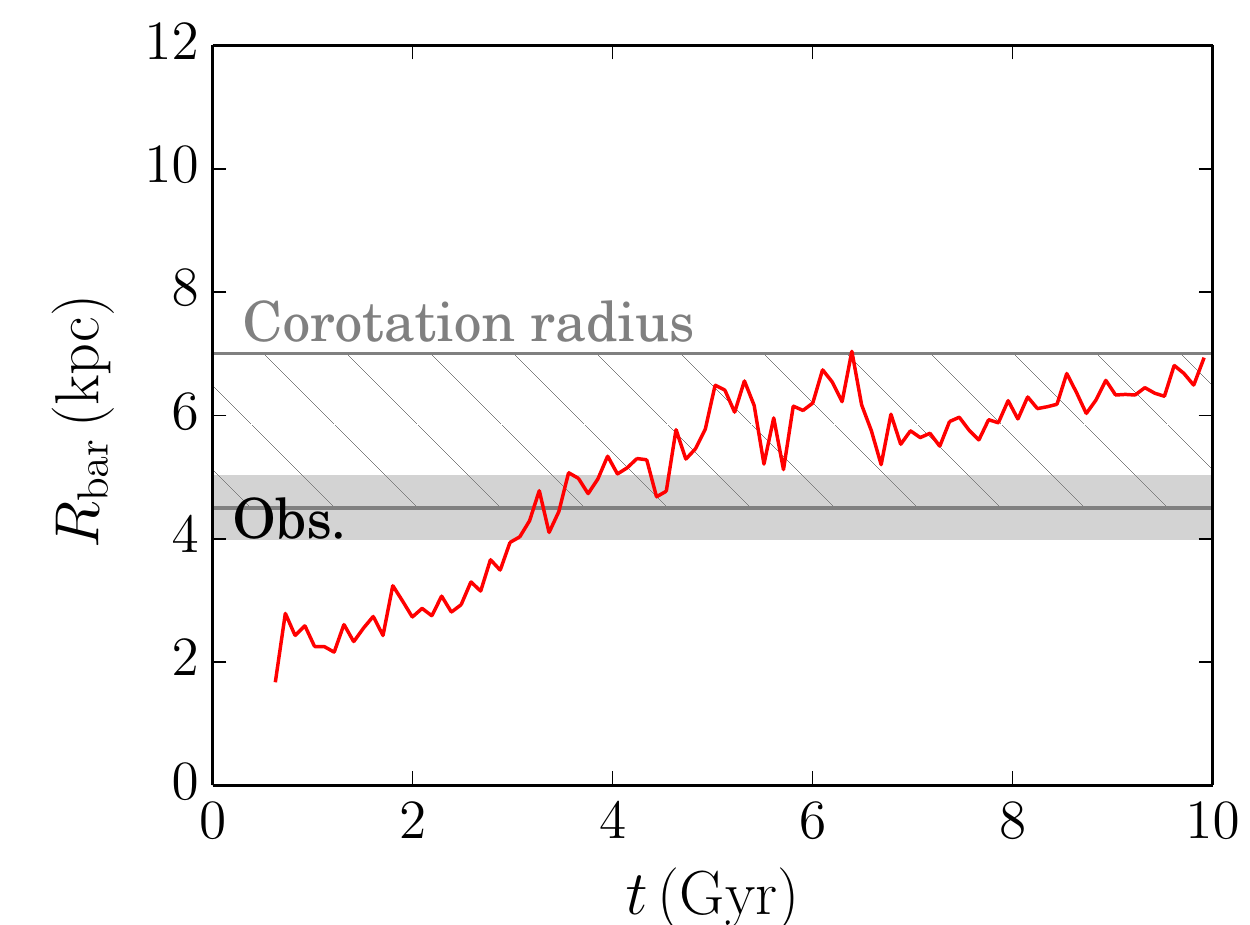}\plotone{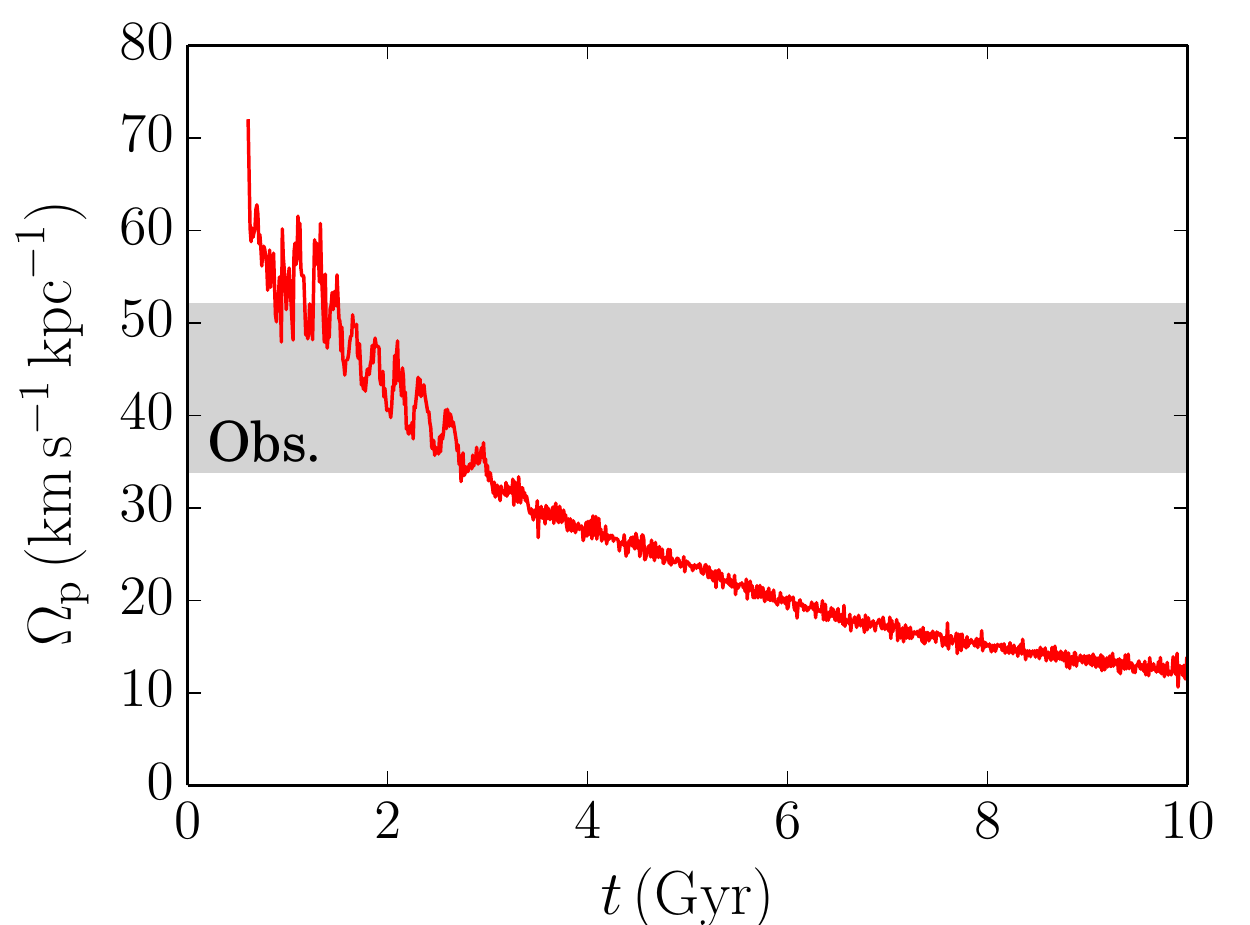}\\
\caption{Same as Fig.\ref{fig:a5B}, but for model MWc0.5.\label{fig:c0.5}}
\end{figure*}

\section{Results}

Each model is simulated for 10\,Gyr after which we compare the resulting disc and bulge 
structure with observations.
The results of our simulations are shown in Figs.~\ref{fig:a5B}--\ref{fig:c0.5}.
The panels in the figures present, from the top-left to bottom-right, 
\renewcommand{\labelenumi}{(\alph{enumi})}

\begin{enumerate}
    \item the initial and final rotation curves, 
    \item angular frequency of the bar and disc at $t=10$\,Gyr,
    \item surface density profile, 
    \item disc radial and vertical velocity dispersion, 
    \item disc and dark-matter density within $|z|<1.1$\,kpc
      ($K_{\rm z}/2\pi G$) \citep{1991ApJ...367L...9K},
    \item disc scale height,
    \item line-of-sight velocity dispersion of the bulge region ($R<3$\,kpc), 
    \item mean line-of-sight velocity of the bulge region,
    \item the time evolution of the bar length,
    \item the time evolution of the bar's pattern speed,  
\end{enumerate}

\begin{figure*}
\raggedright
\epsscale{.35}
\plotone{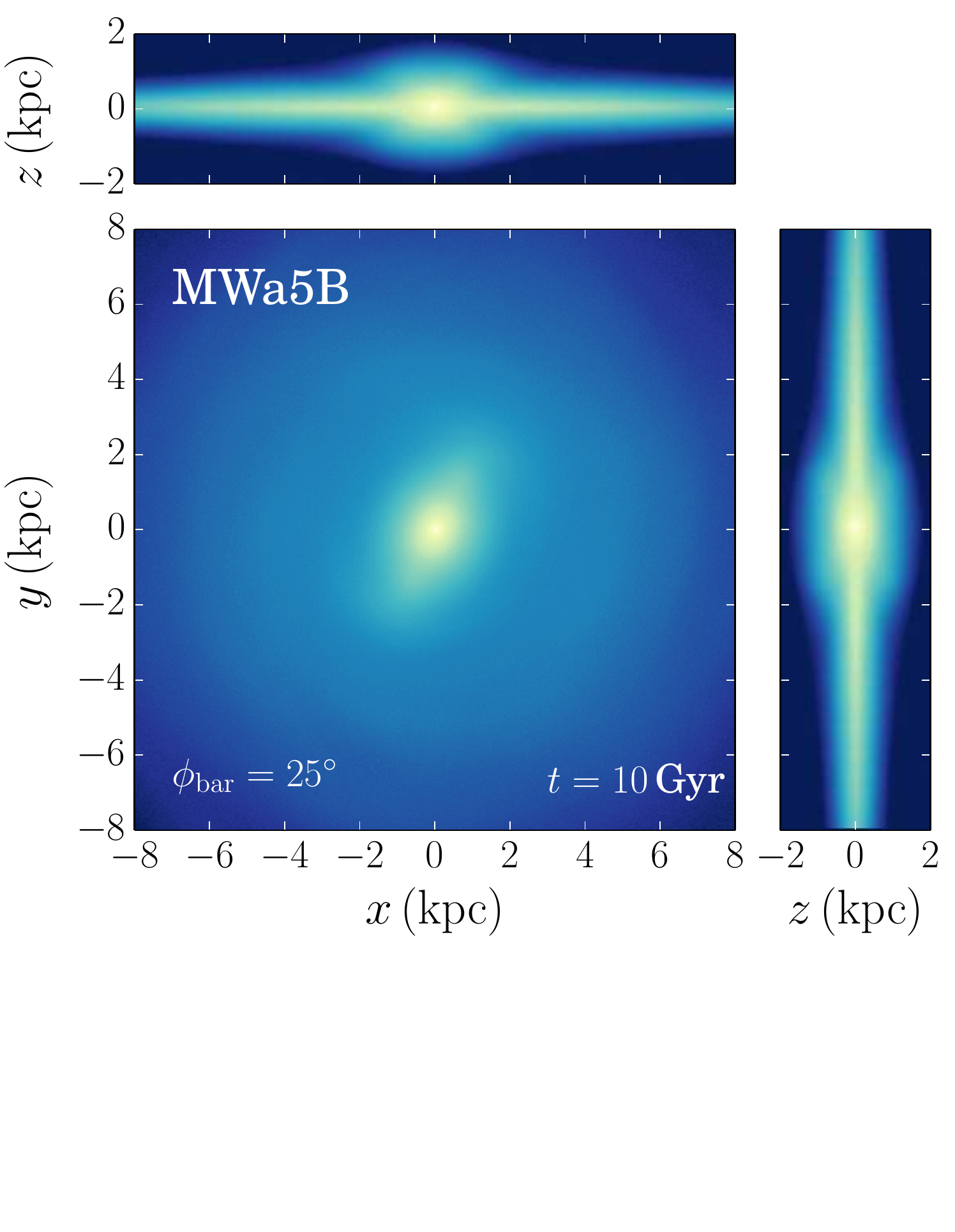}\plotone{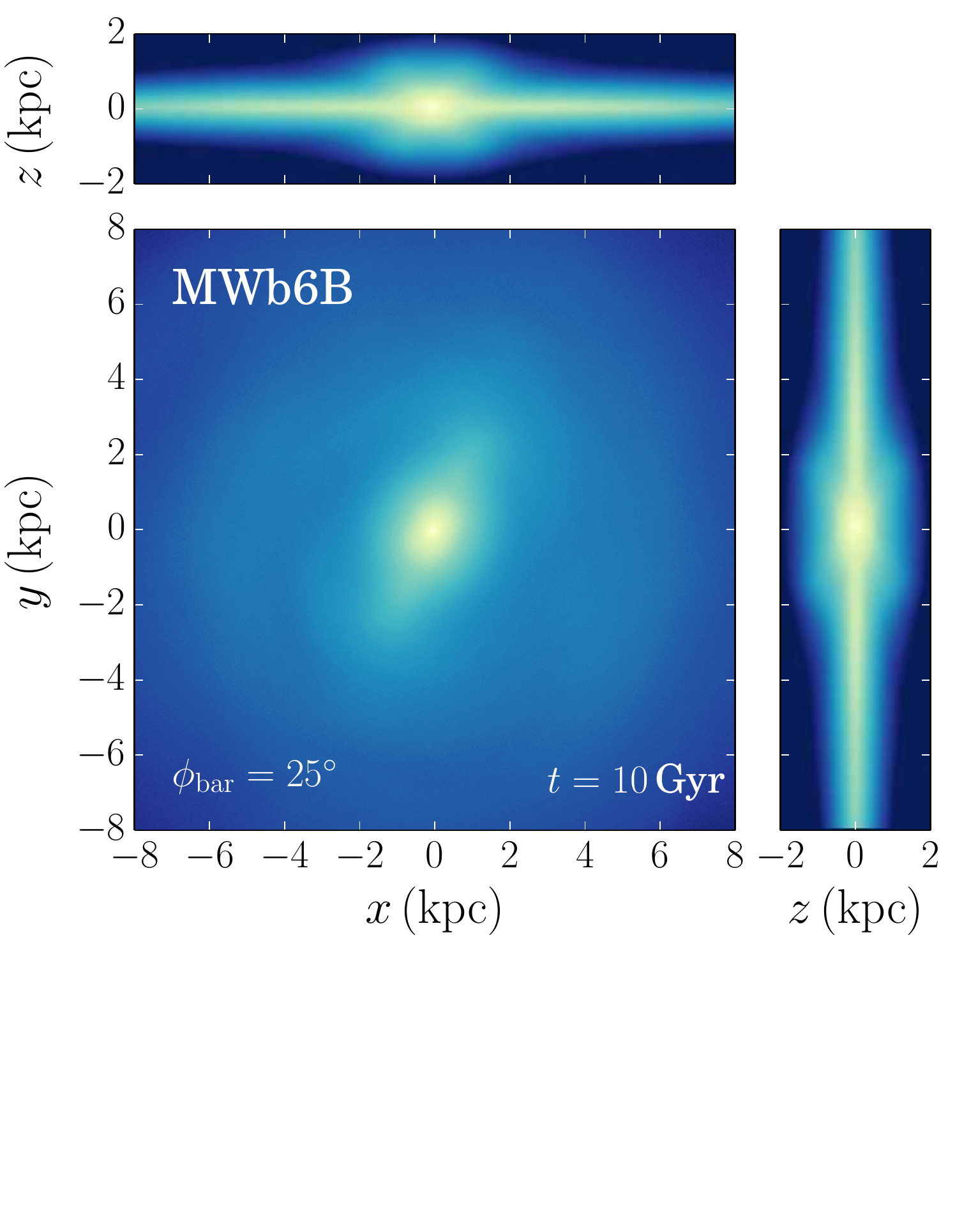}\plotone{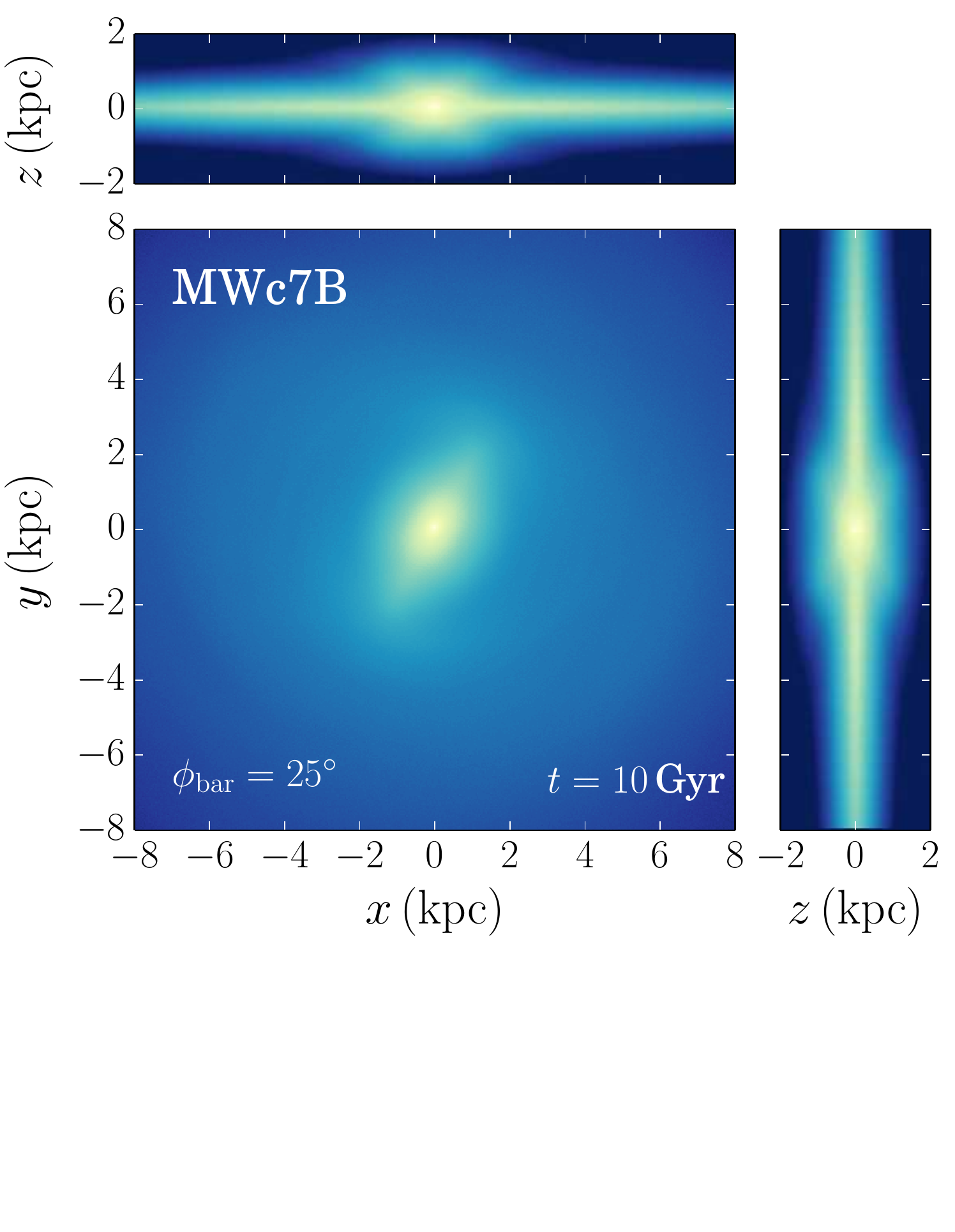}
\vspace{-5.0em} 
\caption{Face- and edge-on view at $t=10$\,Gyr for models MWa5B, MWb6B, and MWc7B (from left to right). 
The Sun is located on the $y$-axis and the bar angle with respect to the Sun-Galactic Center line $\phi_{\rm bar}=25^{\circ}$. \label{fig:snapshots_7B}}
\end{figure*}

We also present 
the face- and edge-on views of models MWa5B, MWb6B, and MWc7B in Figure \ref{fig:snapshots_7B}.
Here, we assume that the bar angle with respect to the Sun-Galactic Center line ($\phi_{\rm bar}$)
is $25^{\circ}$. 
In the edge-on view, we see a weak x-shaped bulge. 
The movies of the face-on view for these models are available as 
online materials.

\subsection{Rotation curves}

In panel (a) of Figs.~\ref{fig:a5B}--\ref{fig:c0.5}, we present the circular velocity ($V_{\rm c}$) of the 
disc at $t=0$ (magenta) and at $t=10$\,Gyr (black). The contributions of 
the individual components, at $t=0$, are shown in red (disc), blue (bulge),
and green (halo).
The grey shaded region indicates the observed circular rotation 
velocity at the Sun's location, 
$238\pm15$\,km\,s$^{-1}$ \citep{2016ARA&A..54..529B}. 
The vertical dotted line marks the distance of 8\,kpc from the centre. 
We measure $V_{\rm c}$ at $t=10$\,Gyr using the centrifugal force calculated 
in the simulation. For the initial conditions, $V_{\rm c}$ is calculated using 
the assumed potential. 

Compared with the initial conditions, the circular velocity at 8\,kpc drops
at the end of the simulation ($t=10$\,Gyr) for nearly all the models shown here.
The size and depth of the dip in the rotation curve correspond to the strength of the bar
(see models MWc7B, MWc0.65, and MWc0.5).
As the simulation progresses the shape of the rotation curve in the inner region changes, 
and the final curve has a peak at a few kpc from the Galactic centre. 
Compared to the rotation curve observed in the inner region such as in
\citet{2012PASJ...64...75S}, our peak is less high.
The higher peak in the observations is, however, obtained by measurements of the H1 gas velocity.
The clumps of gas in the inner region of the Galactic disc could have a higher
line-of-sight velocity due to their motion along the bar, and the actual circular velocity
might be lower than that obtained from the gas velocity~\citep{2010PASJ...62.1413B}.
We, therefore, discuss the circular velocity at 8\,kpc, but not the shape of the curve in the inner region.
The measurements of the circular velocity at 8\,kpc ($V_{\rm c,8kpc}$) for $t=10$\,Gyr are summarized in 
Tables~\ref{tb:results} and \ref{tb:A3}.

\subsection{Halo spin}

\begin{figure*}
\raggedright
\epsscale{.35}
\plotone{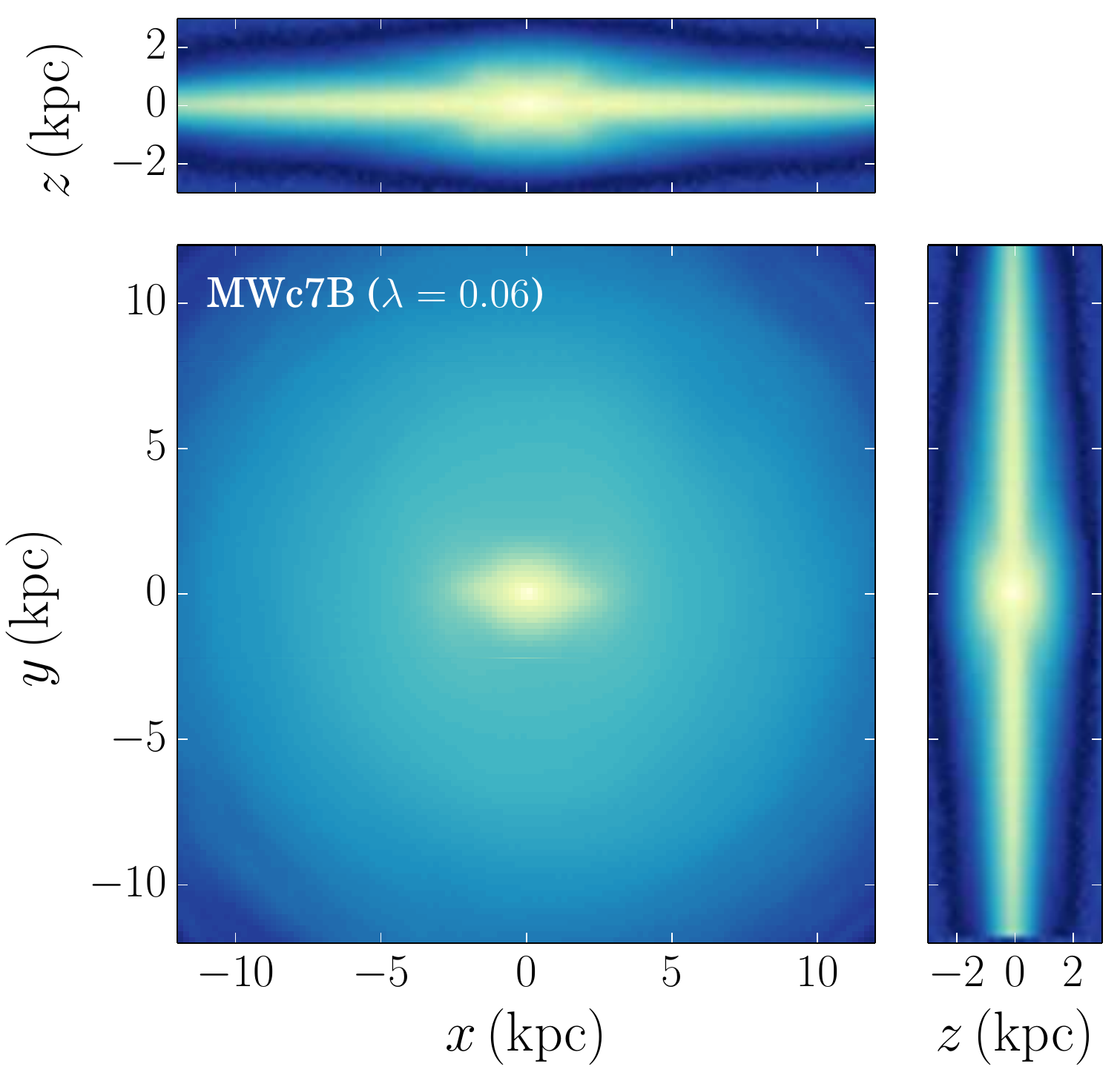}
\plotone{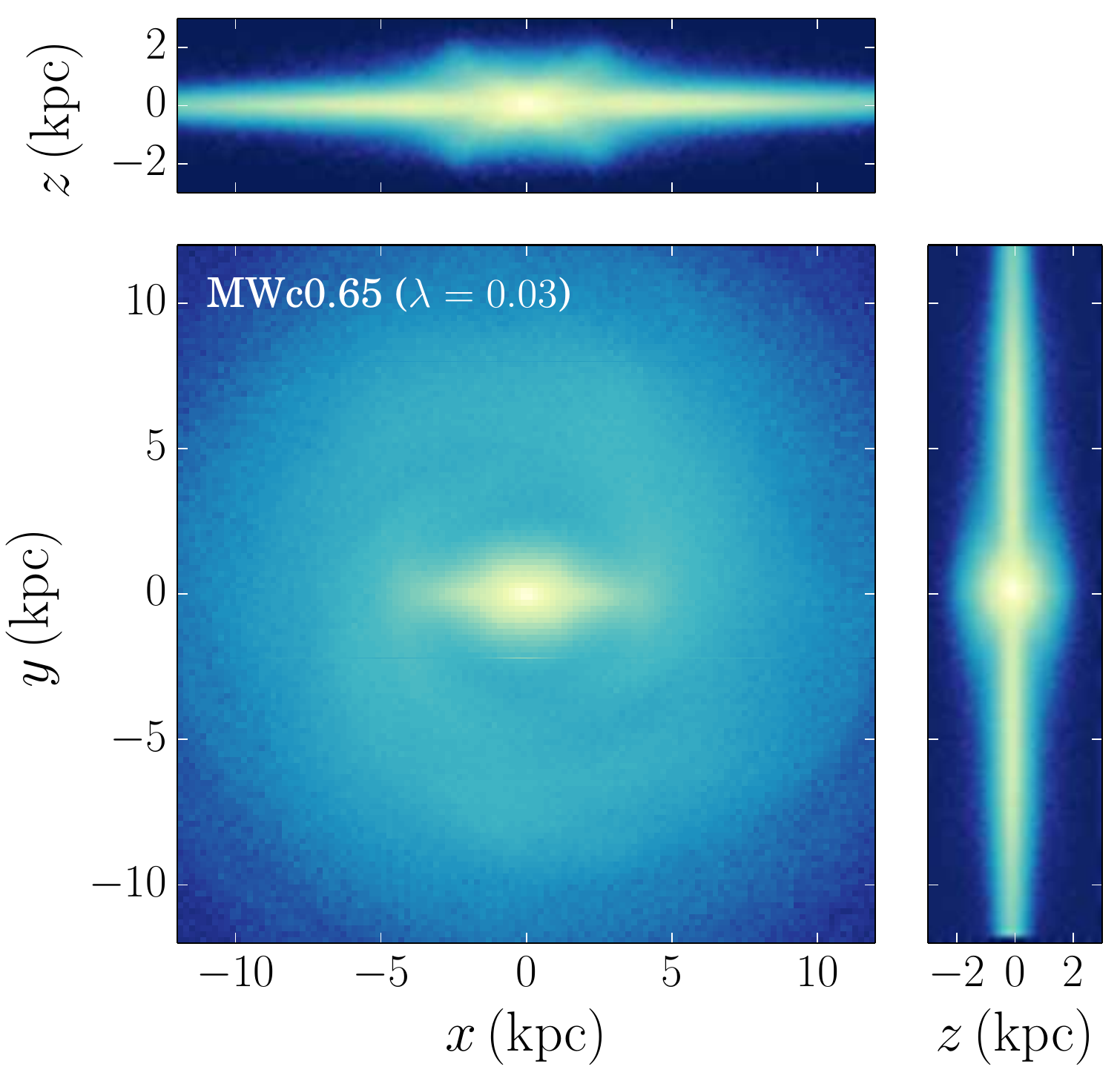}
\plotone{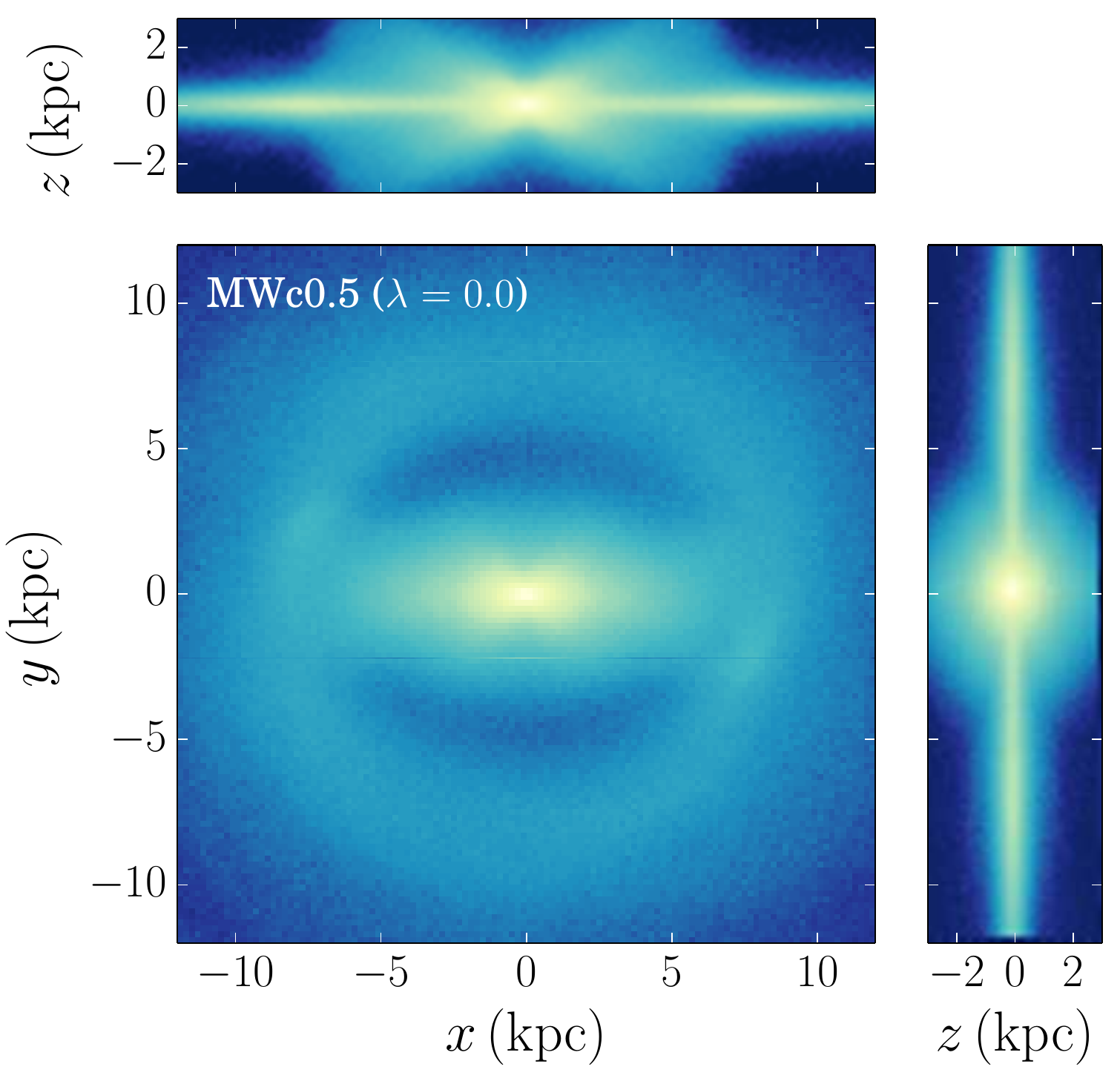}\\
\caption{Surface density maps at $t=10$\,Gyr for models MWc7B, MWc0.65 and MWc0.5 (left to right). \label{fig:snap_c}}
\end{figure*}

We find that halo spin is one of the most crucial parameters in order to reproduce the 
MW Galaxy. 
The halo spin parameter strongly affects the bar evolution.
In general, bars grow longer by transferring their angular momentum to 
the halo, and the pattern speed slows down
\citep{2002ApJ...569L..83A}. If the halo initially has a spin then
the pattern speed of the bar increases and the bar length becomes 
shorter as the halo spin 
increases~\citep{2014ApJ...783L..18L,2018MNRAS.477.1451F}. 
For model MWc, we adopt a range of different halo spin parameters where we keep the other parameters fixed.
As is shown in Figs.~\ref{fig:c7B}--\ref{fig:c0.5}, the bars in the lesser ($\alpha_{\rm h}=0.65$) and 
no spin ($\alpha_{\rm h}=0.5$) cases evolve 
much stronger compared to the case with $\alpha_{\rm h}=0.8$.
In Fig.~\ref{fig:snap_c}, we present the surface density of these models at $t=10$\,Gyr.
Without halo spin, we see an x-shaped bulge that is much stronger than observed in the MW
\citep{2013MNRAS.435.1874W}. 
The effect of the halo spin can also be seen in the bulge kinematics,
where the line-of-sight velocity dispersion of the bulge region 
becomes much higher than in the models without halo spin. 

Our results indicate that an initial spin parameter of $\alpha_{\rm h}\sim 0.8$, 
which corresponds to $\lambda\sim0.06$, matches the observations best.
This value is consistent with the value observationally estimated for 
disc galaxies with a short bar (smaller than a quarter of the 
disc outer radius) \citep{2013ApJ...775...19C}.
We further investigated a model with a larger halo spin, but the
  result was not significantly different from those of a model with
  $\alpha_{\rm h}\sim 0.8$ (see Appendix C). From this result, we expect
  that a fixed potential case would be similar to the maximum halo spin cases
  because in both cases net angular momentum transfer 
  from disk to halo is forbidden.
We therefore adopt $\alpha_{\rm h}\sim0.8$ for 
most of our models without explicitly indicating it in their name.

\subsection{Comparison with disc observations}

In order to evaluate the differences between models and observations, we use 
the stellar surface density of the disc ($\Sigma_{\rm 8kpc}$), radial velocity dispersion 
($\sigma_{R, {\rm 8kpc}}$), and disc and dark-matter density within $|z|<1.1$\,kpc 
($K_{z, {\rm 8kpc}}$) at 8\,kpc from the Galactic center. 
Hereafter, we assume that the Sun is located at 8\,kpc from the Galactic centre.

In panel (c) we present the disc surface density profile 
at the start of the simulation (black curve) and at the end
of the simulation ($t=10$\,Gyr, red curve). The vertical dotted line indicates the 
location of the Sun, and the horizontal dashed line the surface density
of the Galactic disc observed in the solar neighbourhood; $47.1\pm 3.4 M_{\odot}$\,pc$^{-2}$ 
\citep{2015ApJ...814...13M,2016ARA&A..54..529B}.
Compared to the initial density profile, the central density increases and 
the surface density around the bar end drops slightly at $t=10$\,Gyr. 
The density increases again in the region outside the bar length.
Since we configured our models to form a bar with a length of $\lesssim5$\,kpc,
the surface density at 8\,kpc at $t=10$\,Gyr is always larger than the initial
value. 

\begin{figure}
\includegraphics[width=0.9\columnwidth]{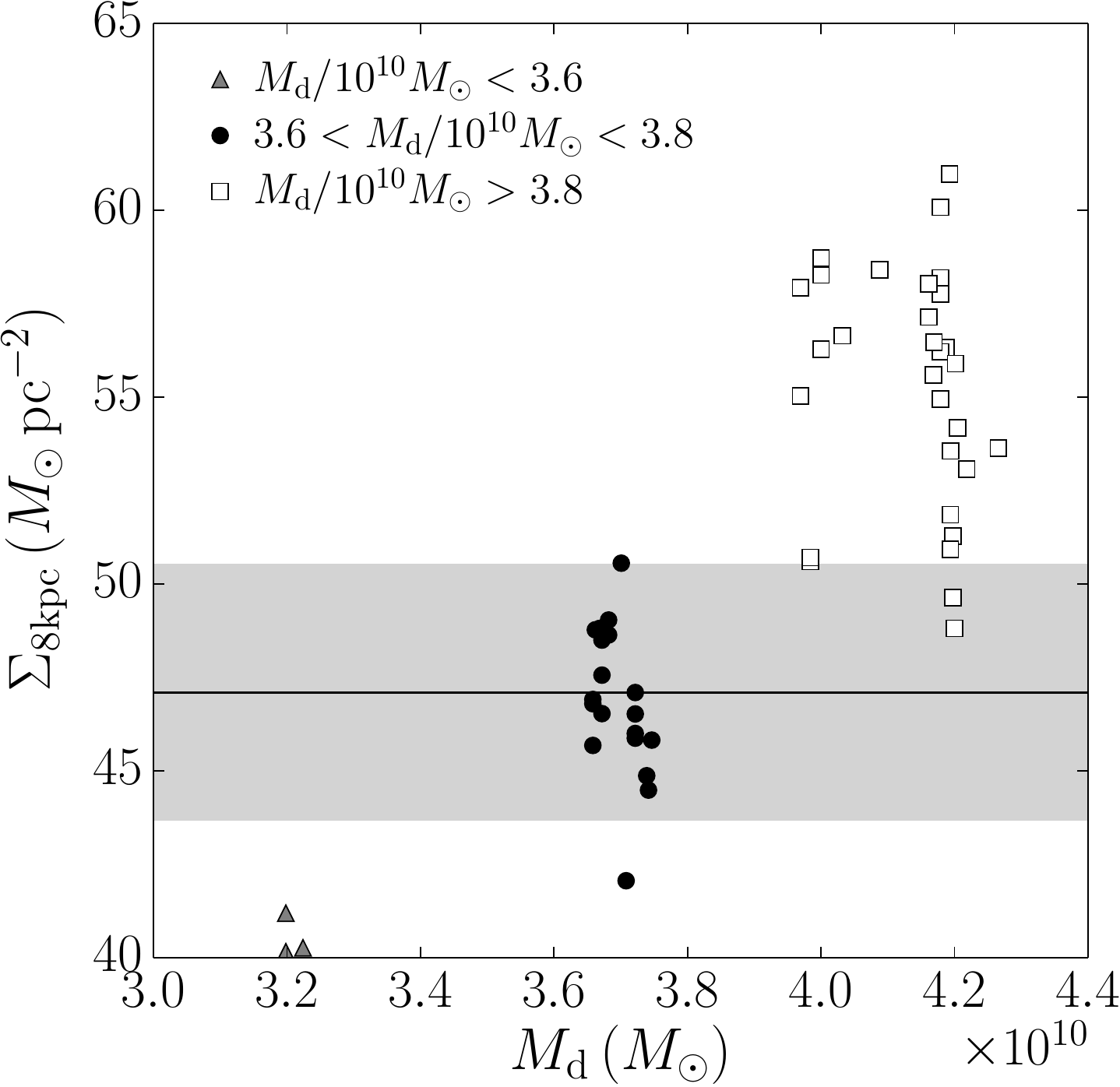}%
\caption{Relation between the disc mass and the disc surface density at 8\,kpc at $t=10$\,Gyr.
Each symbol indicates one model.
The black line and grey shaded region mark the observed value; $47.1\pm 3.4 M_{\odot}$\,pc$^{-2}$
\label{fig:Sigma_Md}.}
\end{figure}

The initial surface density at 8\,kpc depends on both the disc mass ($M_{\rm d}$) and 
the disc scale length ($R_{\rm d}$). The scale length is, however, limited by
observations to $R_{\rm d}\sim2.6$\,kpc \citep{2016ARA&A..54..529B}. 
We tested the correlation between the disc mass
($M_{\rm d}$) and the surface density at 8\,kpc at $t=10$\,Gyr ($\Sigma_{\rm 8kpc}$). 
The results are shown in Fig.~\ref{fig:Sigma_Md}. The correlation coefficient 
between $M_{\rm d}$ and $\Sigma_{\rm 8kpc}$ is 0.82.
Based on this result we set the disc mass of our models to $\sim 3.7\times 10^{10}M_{\odot}$.

\begin{figure}
\includegraphics[width=0.9\columnwidth]{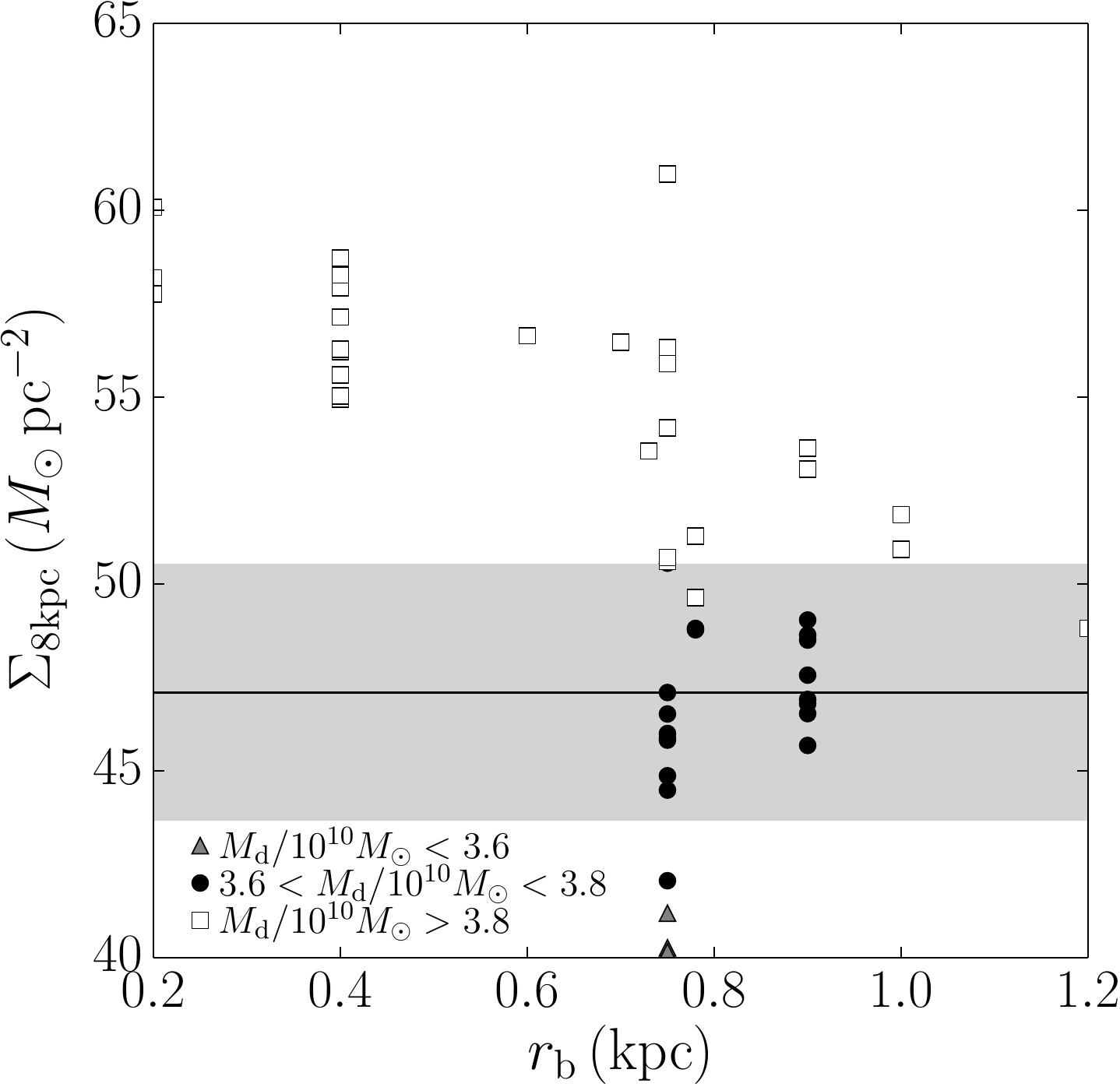}%
\caption{Relation between the bulge scale length and the disc surface density at 8\,kpc at $t=10$\,Gyr.
The black line and grey shaded region mark the observed value; $47.1\pm 3.4 M_{\odot}$\,pc$^{-2}$ 
\label{fig:Sigma_rb}.}
\end{figure}

\begin{figure}
\includegraphics[width=0.9\columnwidth]{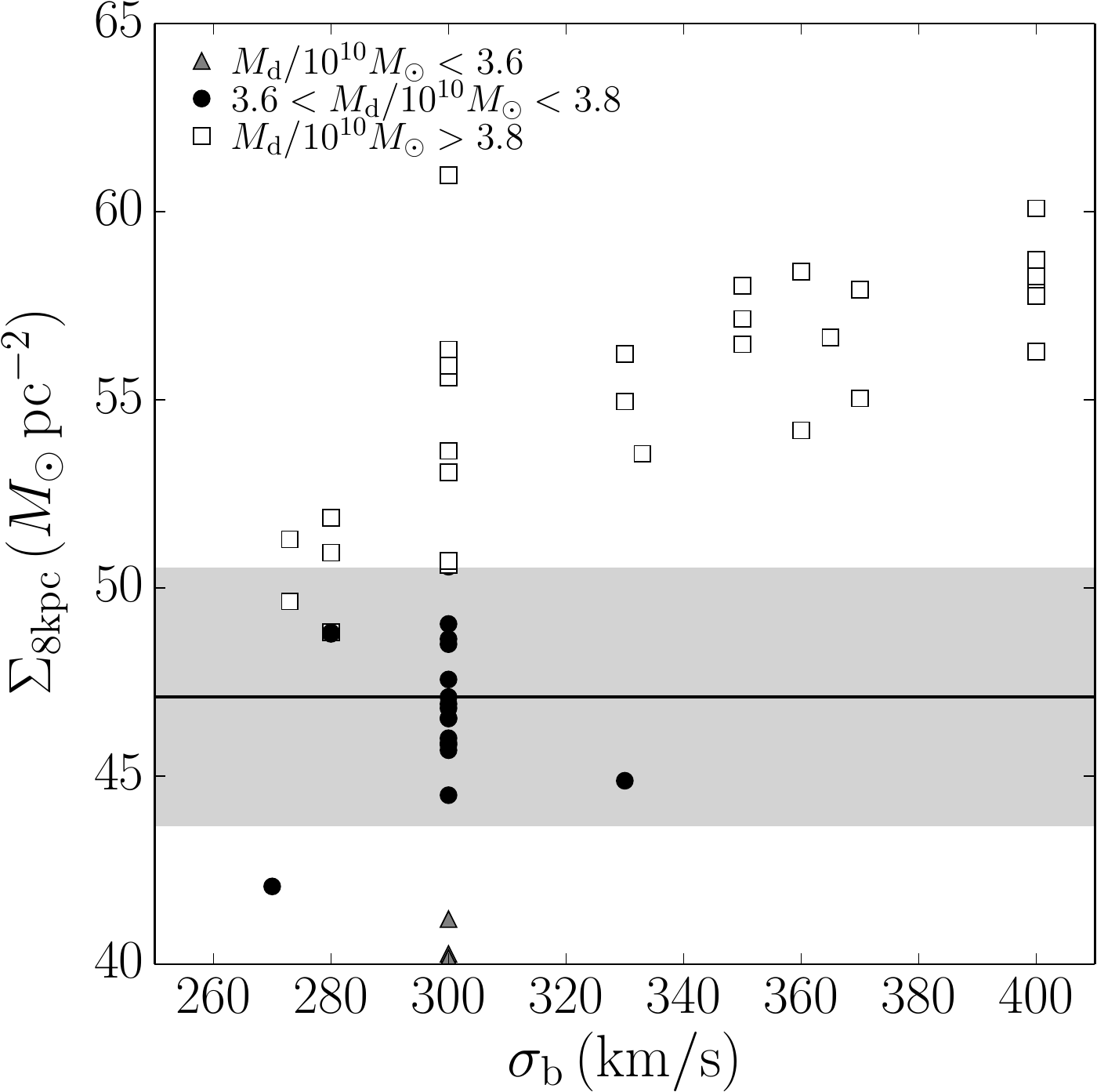}%
\caption{Relation between the bulge characteristic velocity and the disc surface density at 8\,kpc at $t=10$\,Gyr.
The black line and grey shaded region mark the observed value; $47.1\pm 3.4 M_{\odot}$\,pc$^{-2}$ 
\label{fig:Sigma_vb}.}
\end{figure}

We further find a weak correlation between the initial bulge scale length ($r_{\rm b}$) and $\Sigma_{\rm 8kpc}$ 
(with a coefficient of $-0.63$) and between the initial bulge characteristic velocity ($\sigma _{\rm b}$)
and  $\Sigma_{\rm 8kpc}$ (with a coefficient of 0.60), see Figs.~\ref{fig:Sigma_rb} and ~\ref{fig:Sigma_vb}. 
If we focus on models with $M_{\rm d} \sim 3.7M_{\odot}\times 10^{10}$ (black circles in the figures), 
we find that $r_{\rm b}\sim 1$\,kpc and 
$\sigma_{\rm b}\sim 300$\,km\,s$^{-1}$ gives a stellar surface density of the disc (at 8\,kpc)
which is similar to the observations.

Now we investigate the relation between the initial parameters and the final 
velocity dispersion of the disc. 
The radial (red solid line) and vertical (blue dashed line) velocity dispersion of 
the disc at $t=10$\,Gyr are presented in panel (d). 
In the same panel, we present the observed
radial (red shaded) and vertical (blue shaded) velocity dispersion of 
the MW disc in the solar neighbourhood, 
$35\pm 5$ and $25\pm 5$\,km\,s$^{-1}$, respectively \citep{2016ARA&A..54..529B}. 
Black curves in the same panel indicate the initial distributions.
Both the radial ($\sigma_R$) and vertical ($\sigma_z$) velocity dispersion
increase after the bar formation. 
When we look at our simulations we see that the ratio between the radial
and vertical velocity dispersion is always larger than the observed ratio. 
This might be due to the lack of gas in our simulation. The vertical velocity dispersion
can be increased by massive objects in the disc such as giant molecular clouds
rather than spiral arms \citep{1987ApJ...322...59C,1990MNRAS.245..305J,2008gady.book.....B}. 
We, therefore, ignore the vertical 
velocity dispersion when we tune our models to match with observations. 

\begin{figure}
\includegraphics[width=0.9\columnwidth]{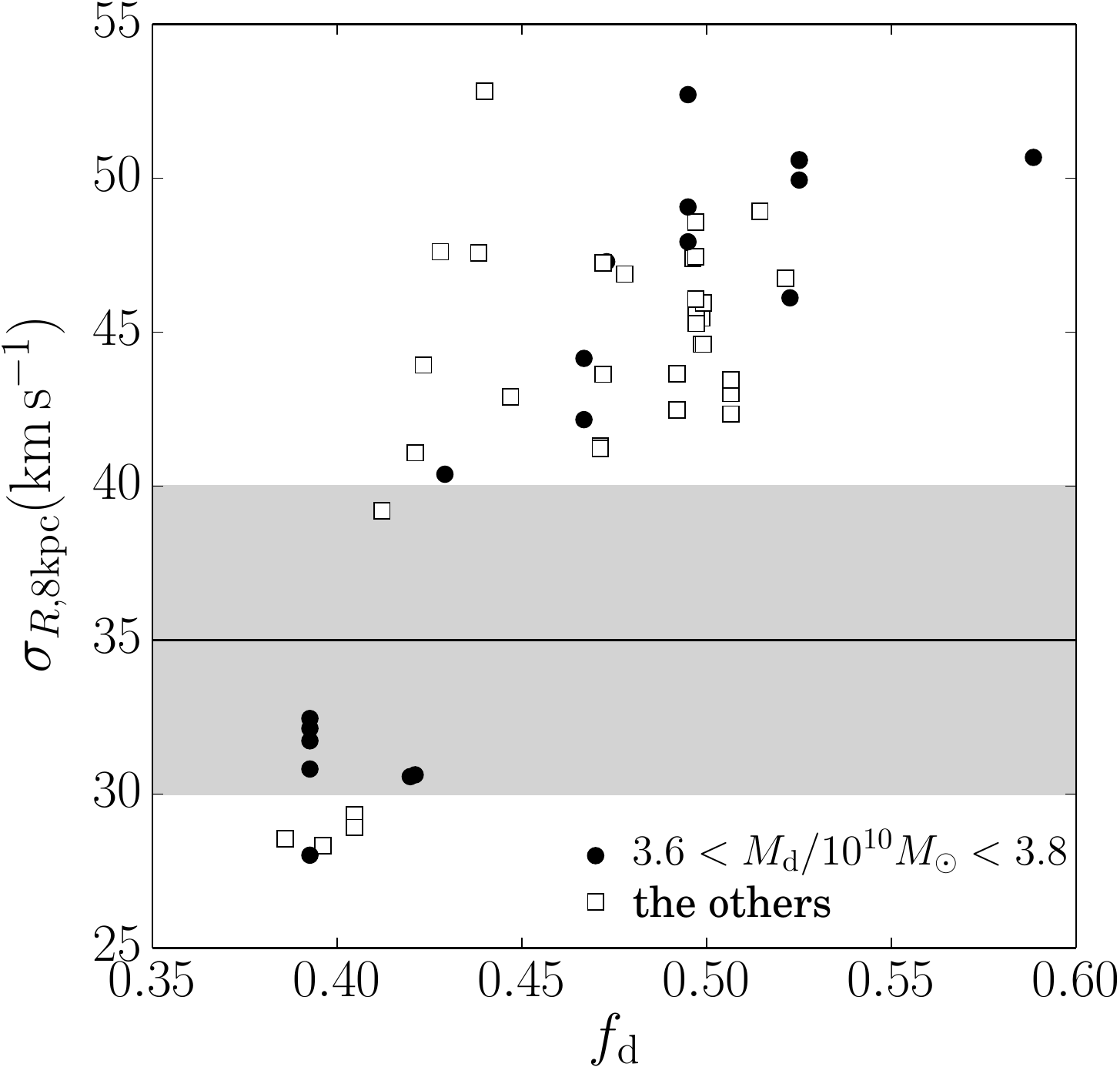}%
\caption{Relation between the disc mass fraction at $2.2R_{\rm d}$ ($f_{\rm d}$) and 
$\sigma_{R}$ at 8\,kpc. Black line and grey shaded region indicate the observed value; 
 $35\pm 5$ km\,s$^{-1}$\label{fig:sigmaR_fd}.}
\end{figure}

We find a correlation between $\sigma_R$ at 8\,kpc ($\sigma_{R, {\rm 8kpc}}$) and 
the disc mass fraction ($f_{\rm d}$) with a correlation coefficient of 0.85
for models with $3.6 \times10^{10}M_{\odot} <M_{\rm d}<3.8 \times 10^{10}M_{\odot}$ and 0.80 for all other models.
This correlation is presented in Fig.~\ref{fig:sigmaR_fd}.
A larger $f_{\rm d}$ results in earlier bar formation~\citep{2018MNRAS.477.1451F},
which causes the velocity dispersion to increase as $f_{\rm d}$ increases.
The results of Fig.~\ref{fig:sigmaR_fd} indicate that $0.4\lesssim f_{\rm d} \lesssim 0.45$ 
should be chosen to best agree with observations. 

We also find an anti-correlation between $K_z$ and $f_{\rm d}$. 
In panel (e) we plot $K_{\rm z}$ \citep{1991ApJ...367L...9K} at 
the end of the simulation ($t=10$\,Gyr, red line). 
The black circles, with 
error bars are observed values taken from~\citep{2013ApJ...779..115B}.
The vertical dotted line marks the location of the Sun at 8\,kpc from the 
Galactic centre. We take $K_{\rm z}/2\pi G=70\pm 5$\,($M_{\odot}\,{\rm pc}^{-2}$) at 8\,kpc
as the observed value \citep{2016ARA&A..54..529B}.

\begin{figure}
\includegraphics[width=0.9\columnwidth]{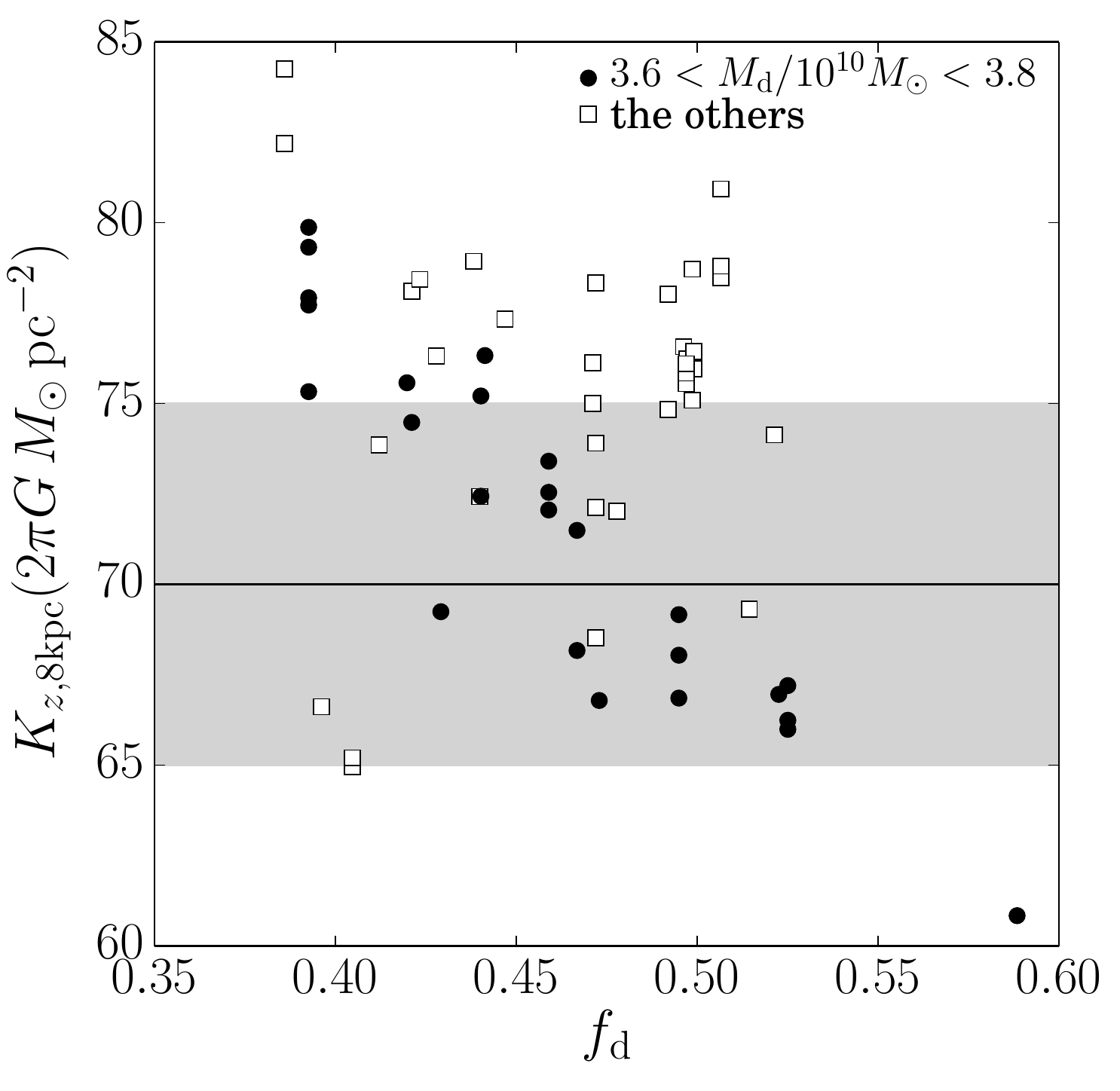}%
\caption{Relation between the disc mass fraction at $2.2R_{\rm d}$ ($f_{\rm d}$) and 
$K_z$ at 8\,kpc. Black line and grey shaded region indicate the observed value; 
$K_{\rm z}/2\pi G=70\pm 5$\,(km\,s$^{-1}$). \label{fig:Kz_fd}}
\end{figure}

The value of $K_Z$ should depend on both the disc and halo structure. 
We evaluated the correlation between $K_z$ and all initial parameters 
in Fig.~\ref{fig:Kz_fd}.
When we look at the models with similar disc mass 
($3.6 \times 10^{10}M_{\odot} <M_{\rm d}<3.8 \times 10^{10}M_{\odot}$)
we find an anti-correlation of -0.86 between $K_z$ and the disc mass fraction ($f_{\rm d}$).
This plot suggests that $0.40<f_{\rm d}<0.55$ results in a good fit to the observations.
Thus, both $\sigma_{R}$ and $K_{z}$ at 8\,kpc suggest $f_{\rm d}\sim 0.4$--0.45.

We measure the disc scale height by taking the root mean square of 
the disc particles $z$ coordinate. The results are presented 
in panel (f) for $t=0$ (black line) and $t=10$\,Gyr (red line).
The scale height of the MW disc is measured to be $0.30\pm0.05$\,kpc 
\citep{2016ARA&A..54..529B} and indicated by the grey shaded area.
Galactic discs thicken due to the heating induced by the bar and spiral arms and 
in the inner regions where the bar develops the scale height is significantly larger.
We find that if the initial scale height is set to 0.2--0.3\,kpc, the final disc scale 
height, for models with a halo spin, fall within the observed range at $\sim 8$\,kpc.
For the models without halo spin, the bar's strong dynamical heating causes the scale
height to become too high, even in the bar's outer region.
For the same reason as for the vertical velocity dispersion of the disc, we do not
consider matching the disc scale height with the observations when we tune our models.

Considering the above results, we conclude that 
$M_{\rm d}\sim 3.7\times 10^{10}M_{\odot}$,
$r_{\rm b}\sim0.7$--1.0\,kpc,
$\sigma_{b}\sim 300$\,km\,s$^{-1}$,
and $f_{\rm d}\sim 0.45$ 
are necessary for the MW model.
The values of $\Sigma_{\rm 8kpc}$, $\sigma_{R,{\rm 8kpc}}$, $\sigma_{z,{\rm 8kpc}}$,
$K_z$, and $V_{\rm c, 8kpc}$ are summarized in Tables \ref{tb:results} and 
\ref{tb:A3}.

\subsection{Comparison with bulge kinematics observations}

The kinematics of the bulge region is another data point that can be compared with 
available observational data.
We, therefore, compare our simulations with the bulge kinematics obtained from
the BRAVA 
observations~\citep{2012AJ....143...57K}. This gives 
the line-of-sight velocity and the velocity dispersion at the Galactic 
center for $b=-4^{\circ}$, $-6^{\circ}$, and $-8^{\circ}$ as a function of $l$.
We ``observe'' our simulated galaxies at $t=10$\,Gyr. To perform these
observations we set the observation angle with respect to the bar to $\phi_{\rm b}=25^{\circ}$.
In panels (g) and (h), we present the line-of-sight velocity and its 
dispersion as a function of $l$ for both the simulations and observations.
The BRAVA data is
represented by the yellow squares, orange circles, and cyan triangles. 
The simulation data is presented by the green squares, red circles and 
blue triangles. 
The observation angle to the bar ($\phi_{\rm b}=25^{\circ}$) is
motivated by observations that put it at $\sim 20^{\circ}$--$40^{\circ}$
\citep{2005A&A...439..107L,2013MNRAS.434..595C,2013MNRAS.435.1874W,2016ARA&A..54..529B}.
However, recent analysis suggests an angle closer to $40^{\circ}$~\citep{2017MNRAS.471.3988C}.
We, therefore, performed additional analysis where we used $30^{\circ}$ and $40^{\circ}$ 
angles, but the differences were minor and consistent with those of \citet{2010ApJ...720L..72S}.

In order to evaluate the match between observations and simulations,
we adopt the method used in \citet{2017MNRAS.470.1526A} and \citet{2014MNRAS.438.3275G},
in which the $N$-body simulations are compared with the BRAVA observations.
We measure $\chi^2$ for $v_{\rm los}$ and $\sigma_{los}$ for $b=-4^{\circ}$, $-6^{\circ}$, and $-8^{\circ}$, respectively. 
Here we calculate
\begin{eqnarray}
\chi^2_{a} = \sum^{N_{\rm data}}_{i} \left( \frac{a_{{\rm sim}, i}-a_{{\rm obs}, i}}{\sigma_{{\rm obs}, i}} \right)^2,
 \label{eq:chi_sq}
\end{eqnarray}
where $a_{\rm sim}$ and $a_{\rm obs}$ are data points obtained from simulations
and observations, and $\sigma_{\rm obs}$ is the observational error for
$a_{\rm obs}$. 
The $\chi^2$ values for $v_{\rm los}$ and $\sigma_{\rm los}$ are shown in
Fig.~\ref{fig:chi_sq_results}, where the value is normalized by the
number of the data points.

\begin{figure}
\includegraphics[width=0.9\columnwidth]{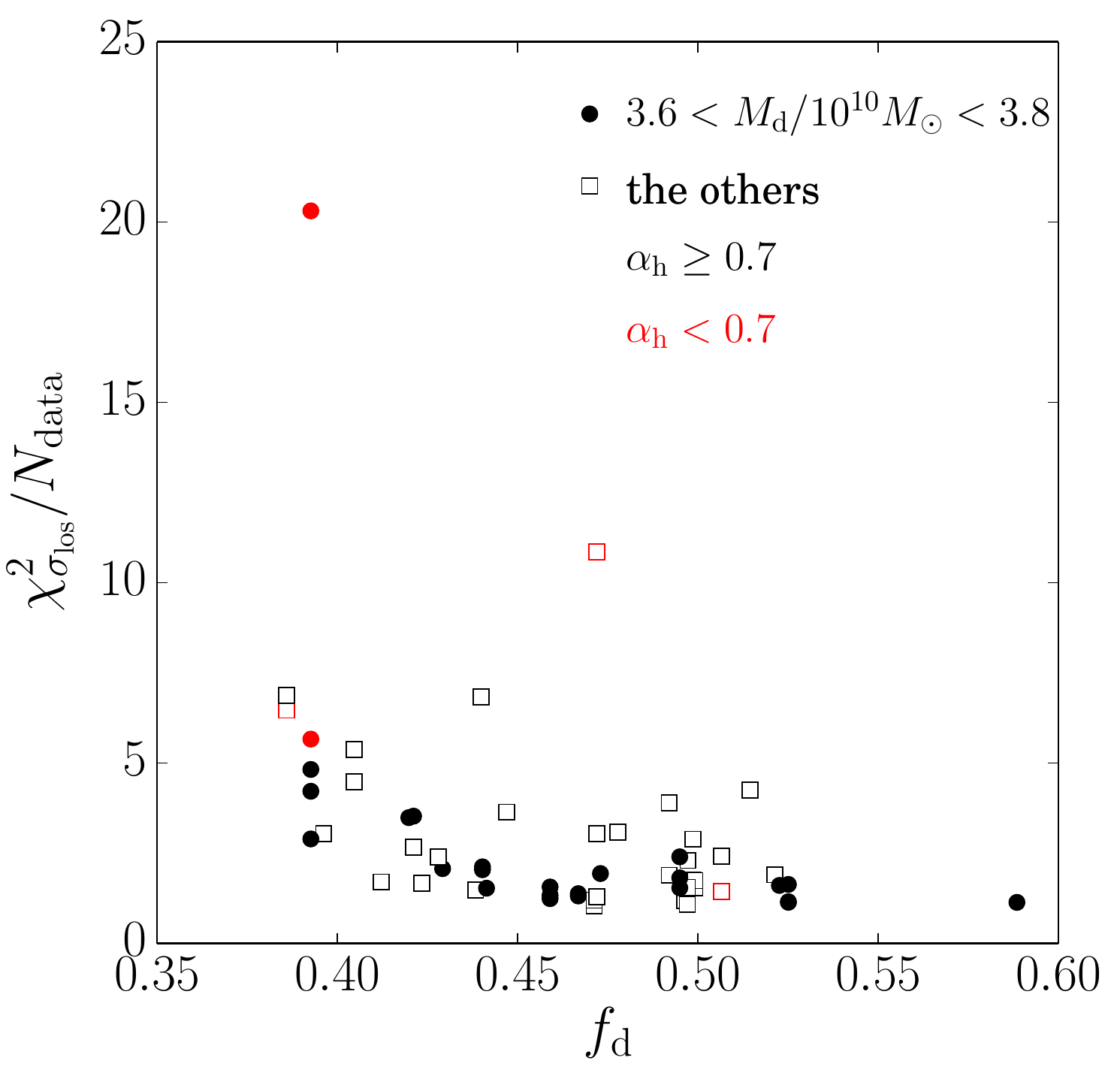}%
\caption{$\chi^2$ for $\sigma_{\rm los}$ as a function of $f_{\rm d}$. 
    The value of $\chi^2_{\sigma_{\rm los}}$ for model MWc0.5 is 91.4, and therefore 
    not shown in this panel.\label{fig:chi_sq_sigma_fd}}
\end{figure}

We did not find any strong correlations with the initial condition parameters, but 
we found that a large disc mass fraction fits better with the observation of $\sigma _{\rm los}$.
In Fig.~\ref{fig:chi_sq_sigma_fd} we present $\chi^2$ for $\sigma_{\rm los}$
(sum of $\chi^2$ for $b=-4^{\circ}$, $-6^{\circ}$, and $-8^{\circ}$ between $l=-20^{\circ}$ to $l=20^{\circ}$ normalized
by the number of the data points, $N_{\rm data}=63$) as a function of $f_{\rm d}$.
This figure suggests that $f_{\rm d}\gtrsim 0.4$. 
This is simply because a stronger bar with a smaller $f_{\rm d}$ heats the bulge as well as 
the disc. Bulge line-of-sight velocity, on the other hand, shows the rotation speed, 
which is barely affected by the heating induced by the bar.

\subsection{Bar length and pattern speed}

The bar's length and pattern speed are also important values to characterize 
the structure of a galactic disc. We use the same method 
as~\citet{2018MNRAS.477.1451F} to measure the length of the bar
\citep[see also][]{2015PASJ...67...63O}. 
We first perform a Fourier decomposition of the disc's surface density:
\begin{eqnarray}
\frac{\Sigma(R,\phi)}{\Sigma_0(R)}=\sum_{m=0}^{\infty}A_m(R){\rm e}^{{\rm i}m[\phi-\phi_m(R)]}
\label{eq:Fourier}
\end{eqnarray}
where $A_m(R)$ is the Fourier amplitude and $\phi_m(R)$ is the phase angle 
for the $m$-th mode. 
Using the $m=2$ component and $\Delta R=1$ kpc radial bins we compute the 
phase angle ($\phi_{2}$) and amplitude ($A_{2}$) of the bar for each radius up to 20 kpc. 
The radius with the highest amplitude value ($A_{\rm 2, max}$) 
is defined as the bar's phase angle ($\phi_{\rm 2, max}$). 
Then, starting at the radius for which we obtained $A_{\rm 2, max}$, we measure the 
phase angle to each outer radial bin. The bin for which the difference 
with $\phi_{\rm 2, max}$ is $\Delta \phi>0.05{\rm \pi}$ defines the bar 
length ($R_{\rm b}$). The result is shown using the red line in panel (i)
of Figs.~\ref{fig:a5B}--\ref{fig:c0.5}. 
In this panel, the grey shaded region indicates the co-rotation radius 
(4.5--7.0\,kpc) as suggested by observations \citep{2016ARA&A..54..529B}. 
Because a bar cannot be longer than the co-rotation 
radius~\citep{1980A&A....81..198C,1992MNRAS.259..328A}, it must be shorter 
than the grey shaded region.

As is suggested by previous simulations \citep{2014ApJ...783L..18L},
the halo spin parameter influences the bar length and the pattern 
speed the most. For the models with the strongest halo 
spin ($\alpha_{\rm h}=0.8$), the final bar length ($\sim 4$\,kpc) is 
slightly shorter than the co-rotation radius.
For the models with a weak halo spin ($\alpha_{\rm h}=0.65$) the final bar length 
is slightly longer and reaches the lower limit of the co-rotation radius ($4.5$\,kpc). 
In the models without halo spin, the bar continues to develop and 
the final length exceeds 6\,kpc. This is much longer than what observations suggest.
Estimates for the MW bar length, obtained from 
a model fit to near-IR star count, are $5.0\pm0.2$\,kpc~\citep{2015MNRAS.450.4050W,2016ARA&A..54..529B}
and $\sim 4$\,kpc \citep{2017MNRAS.471.3988C}.
Our measured bar length ($\sim 4$\,kpc) is consistent with these 
observations. We note that the used measurement method affect 
the resulting bar lengths. In a method that uses the $m=2$ mode, 
the obtained bar length tends to be shorter compared to other radial
profile methods~\citep{2015MNRAS.450.4050W}.

The bar's pattern speed is presented in panel (j) of
Figs.~\ref{fig:a5B}--\ref{fig:c0.5}. This is obtained 
by computing the phase of the bar ($\phi_{\rm 2, max}$), for each snapshot,
using the Fourier decomposition (Eq.~\ref{eq:Fourier}). The angular speed 
is then determined by computing the difference between the snapshots
and presented using the red curve.
In this panel, the grey shaded region marks the observed pattern 
speed of the Galactic bar, 
$\Omega_{\rm p}=43\pm9$km\,s$^{-1}$~\citep{2016ARA&A..54..529B},
although the pattern speed is still under debate 
\citep{2017MNRAS.466L.113M,2017ApJ...840L...2P,2018arXiv180401920H}.
For models with $\alpha_{\rm h}=0.8$, the final pattern speed stabilizes 
within the grey area, but for models with $\alpha_{\rm h}=0.65$ (moderate spin) 
and 
0.5 (no spin), the pattern speed drops below the observational constraints.
In Appendix C, we show more extreme cases (negative and maximum spins).
With a negative spin the final pattern speed is even slower than the case 
without spin. With the maximum spin, the pattern speed was comparable to 
the case with $\alpha_{\rm h}=0.8$ or even larger.
Thus, the pattern speed also indicates that $\alpha_{\rm h}\sim0.8$.
For our models in which the bulge kinematics fit the observations, the final
speed of the bar was 40--50 km\,s$^{-1}$. 
As far as we tested, no model has a pattern speed faster than $\sim$50 km\,s$^{-1}$.

We define the final pattern speed as the averaged pattern speed over the 
last 10--20 snapshots. The results are 45, 38, and 40\,km\,s$^{-1}$ for
models MWa5B, b6B, and c7B, respectively.
The time-scale for the 10--20 snapshots covers the oscillation period 
of the bar's pattern speed. 
The measured pattern speeds are summarized in Tables~\ref{tb:results} and \ref{tb:A3}.
We discuss the oscillation of the bar's pattern speed further in Sect.~\ref{Sect:discussion}.

When we compare the final pattern speed of the bar for models with a different 
number of particles ($N$) we see that it is not exactly the same (Table~\ref{tb:results}).
This could be caused by resolution differences or run-to-run variations~\citep{2009MNRAS.398.1279S}.
In order to determine if the difference is caused by $N$ or by run-to-run variations,
we perform four runs for the same initial parameters. The runs differ in $N$ and 
in the random seed used to generate the initial particle positions and velocities. 
The results are summarized in Appendix~\ref{Sect:Resolution}.
For our tests the bar's pattern speed varied by a few km\,s$^{-1}$ for the models 
with $N\sim$100M. For the $N\sim$1B models the run-to-run variation is smaller, $\sim$1\,km\,s$^{-1}$.

\begin{figure}
\includegraphics[width=0.9\columnwidth]{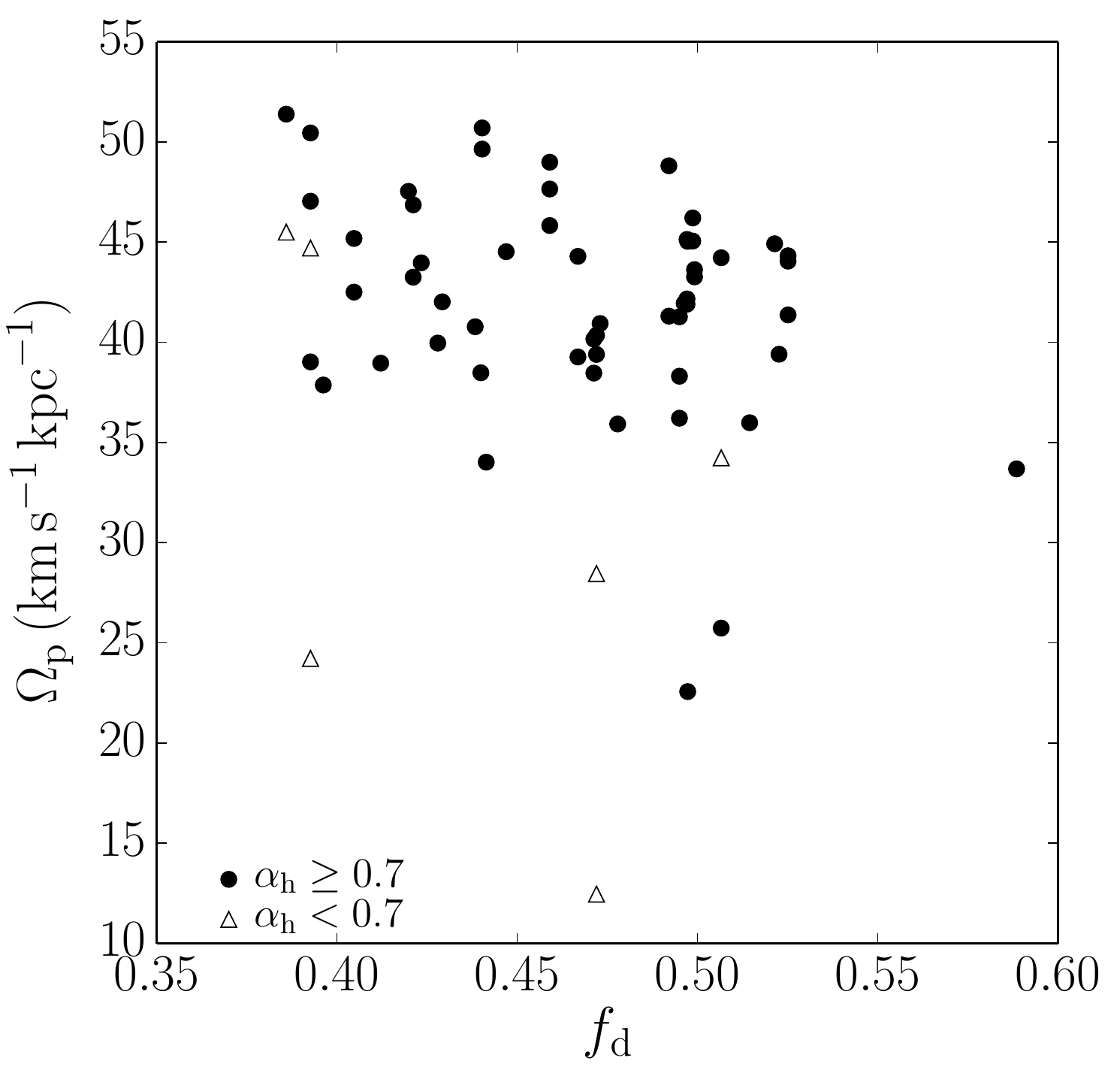}%
\caption{Relation between the initial disc mass fraction and the bar's pattern speed at $t=10$\,Gyr. \label{fig:fd_ps}}
\end{figure}

Except for the halo spin, it is unclear which parameter directly influences the 
pattern speed of the bar. In Fig.~\ref{fig:fd_ps}, we observe a weak dependence on the disc-mass to
total-mass fraction, $f_{\rm d}$. We find that the bar speed drops slightly when
$f_{\rm d}$ increases.
For $f_{\rm d}\sim 0.45$, the expected pattern speed is 40--50\,km\,s$^{-1}$.
In addition, we did not see any model with a bar pattern speed faster than
$\sim 50$\,km\,s$^{-1}$ in our simulations.

We further measure the co-rotation radius ($R_{\rm CR}$), and the inner ($R_{\rm ILR}$)
and outer ($R_{\rm OLR}$) Lindblad resonance radii. These values
are presented in columns 8, 9 and 10 of Tables~\ref{tb:results} and 
\ref{tb:A3}.
Models MWa and a5B have a relatively fast rotating bar, and therefore the $R_{\rm OLR}$ is $\sim 9$\,kpc, 
but for models MWb, b6B, c0.8, and c7B we find $R_{\rm OLR}\sim 10$\,kpc.
For both models, the location of the outer Lindblad resonances was further out than the 
Galactic radius of the Sun.

\subsection{Best-fitting models}

\begin{figure}
\includegraphics[width=0.9\columnwidth]{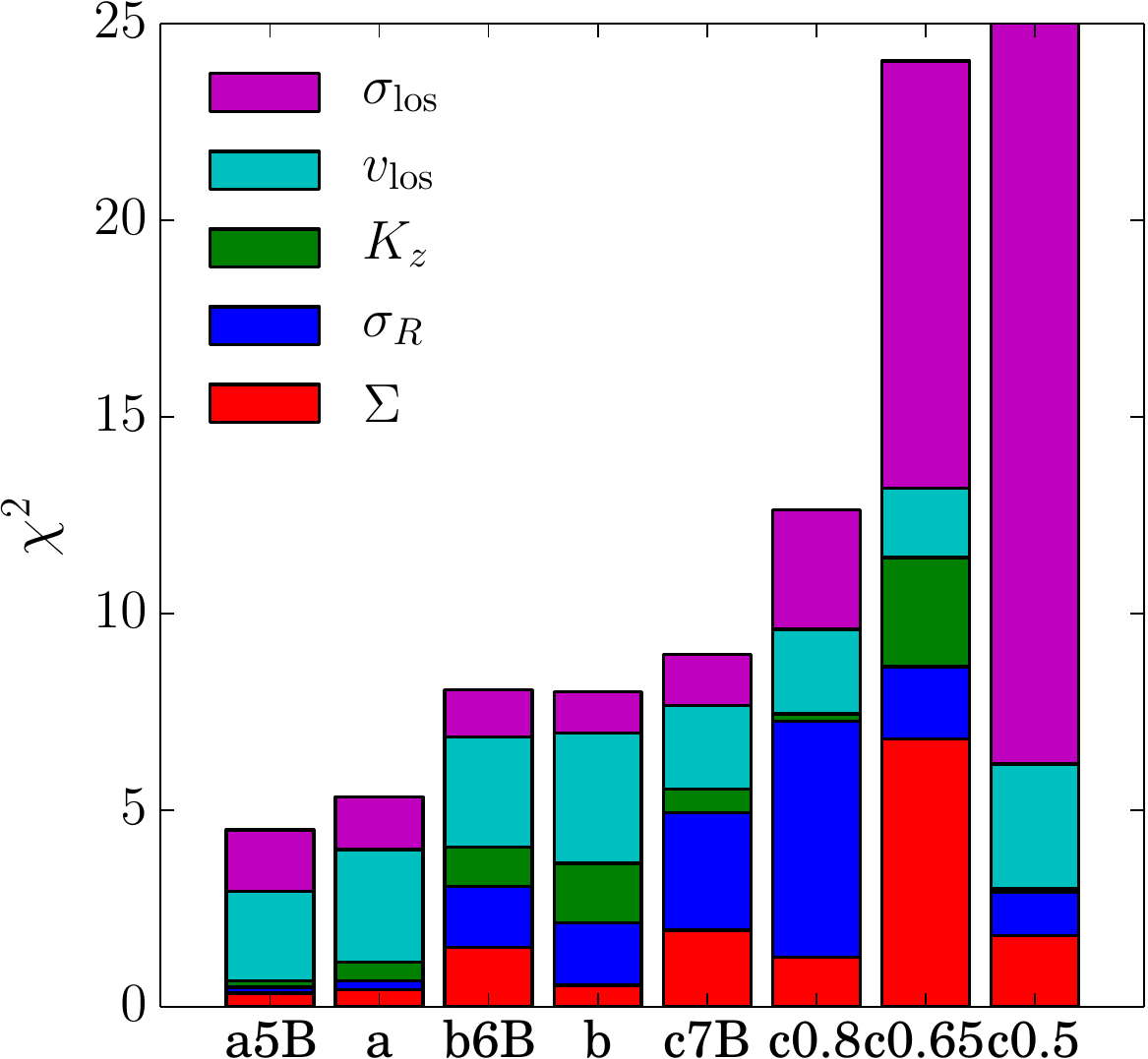}%
\caption{$\chi^2$ for all models listed in Table~\ref{tb:params}. The total $\chi^2$ value for model MWc0.5 is 97.6. \label{fig:chi_sq_results}.}
\end{figure}

In order to evaluate the comparison between models and observations,
we also calculate $\chi^2$ (see Eq.~\ref{eq:chi_sq}) 
for $\Sigma$, $\sigma_{R}$, $K_z$ at 8\,kpc, which are
$\chi_{\Sigma}^2$, $\chi_{\Sigma_{R}}^2$, and $\chi_{K_z}^2$.
The results for the best fitting models are summarized in Table~\ref{tb:chi_sq} 
and Fig.~\ref{fig:chi_sq_results}. The results for all models are presented in Table~\ref{tb:A4}.
Models MWa, b, and c0.8 are models for which the sum of all $\chi^2$ is relatively small,
where MWa/a5B have the smallest $\chi^2$ value. 
The total $\chi^2$ of model MWb/b6B is not the smallest, but the $\chi^2$ for $\sigma _{\rm los}$
is the smallest.
For model MWc0.8/MWc7B, the sum of $\chi_{\Sigma}^2$,
$\chi_{\Sigma_{R}}^2$, and $\chi_{K_z}^2$ is $\sim 5.5$, but
the sum of $\chi^2_{v_{\rm los}}$ and $\chi^2_{\sigma_{\rm los}}$ is the smallest.
Without halo spin, the $\chi^2$ values are very large (see model MWc0.5).
This is the reason we showed these models in detail in Figs.~\ref{fig:a5B}--\ref{fig:c7B}.
In the discussion section, we describe these models in more detail.

Summarizing the results of the previous sections we conclude that the following parameters for the
initial conditions lead to the best-fitting models. 
For the disc mass, the best fit value is $M_{\rm d}\sim 3.70 \times 10^{10}M_{\odot}$. 
With an initial $Q$ at $2.5R_{\rm d}$ of 1.2, this leads to a central velocity dispersion for the disc
of $\sigma_{R0}\sim 90$\,km\,s$^{-1}$.
Compared to the disc the parameters for the bulge are less clear. We find that 
$r_{\rm b}\sim0.7$--1.0\,kpc and $\sigma_{b}\sim 300$\,km\,s$^{-1}$ fit the observations 
and result in $M_{\rm b}=3$--$5 \times 10^{9}M_{\odot}$, assuming a Hernquist model. 
If we calculate the bulge-to-disc mass ratio, it is $\sim0.07$--0.15.  
In contrast to the result of \citet{2010ApJ...720L..72S}, we find that it is not necessary to have a bulge-to-disc 
mass ratio that is smaller than 0.08.

The halo spin is a very important parameter.
For the best-fitting models, the halo spin parameter was $\alpha_{\rm h}=0.8$,
which corresponds to $\lambda\sim0.06$. This is larger than the 
median value obtained from cosmological $N$-body 
simulations, $\lambda=0.03$--0.04 \citep{2007MNRAS.376..215B}. 
However, the spin parameters obtained from cosmological simulations have a significant
amount of dispersion. 
The found $\lambda\sim0.06$ is, therefore, not an extremely large value.
All models with a smaller or even no preferential spin give a bar pattern speed
that is too slow (at 10\,Gyr). This result suggests that the 
MW halo initially had a significant amount of rotation.
The final value of the spin parameter was $\sim 0.07$ for models MWa5B, 
MWb6B, and MWc7B. This value is slightly higher than the observationally suggested spin parameter for short barred galaxies ($\lambda = 0.061$), but still comparable \citep{2013ApJ...775...19C}. 
In Appendix C, we summarized the final spin parameters for the other models.

We further find that disc-to-total mass fraction ($f_{\rm d}$) is an important parameter. 
This parameter controls the bar formation epoch, which corresponds to the opposition against 
bar instability \citep{2018MNRAS.477.1451F}. If $f_{\rm d}$ is larger, the bar forms
earlier and grows stronger. 
Since $\sigma_{R,{\rm 8kpc}}$ and $K_{z, {\rm 8kpc}}$ correlate and anti-correlate with $f_{\rm d}$,
the preferable value of $f_{\rm d}$ is $\sim 0.45$. 
The line-of-sight velocity dispersion in the bulge region also indicates $\sim 0.45$ as
preferable value.

\begin{table*}
\scriptsize
\raggedright
\caption{Disc properties for the simulated galaxies at 10\,Gyr\label{tb:results}}
\begin{tabular}{lccccccccc}

\hline
Model   & $\Sigma_{\rm 8kpc}$ & $\sigma_{R,{\rm 8kpc}}$ & $\sigma_{z,{\rm 8kpc}}$ & $K_{z,{\rm 8kpc}}$ & $V_{\rm c,8kpc}$ & $\Omega _{\rm b}$ & $R_{\rm CR}$ & $R_{\rm OLR}$ & $R_{\rm ILR}$ \\
   & ($M_{\odot}\,{\rm kpc}^{-2}$) & (km\,s$^{-1}$) & (km\,s$^{-1}$) & ($2\pi G \, M_{\odot}\,{\rm kpc}^{-2}$) & (km\,s$^{-1}$) & (km\,s$^{-1}$\,kpc$^{-1}$) & (kpc) & (kpc) & (kpc) \\
\hline \hline
MWa5B & 45.1 & 36.9 & 14.6 & 72.1 & 241 & 46 & 4.7 & 9.1 & 1.7 \\ 
MWa & 44.8 & 37.4 & 14.9 & 73.4 & 241 & 48 & 4.4 & 8.5 & 1.6   \\ 
MWb6B & 51.3 & 41.2 & 15.8 & 75.0 & 230 & 38 & 5.6 & 10.1 & 1.9  \\ 
MWb & 49.6 & 41.3 & 15.9 & 76.1 & 230 & 40 & 5.4 & 9.8 & 1.8  \\ 
MWc7B & 51.9 & 43.6 & 16.3 & 73.9 & 228 & 40 & 5.2 & 9.8 & 1.9  \\ 
MWc & 50.9 & 47.3 & 17.2 & 72.1 & 226 & 39 & 5.3 & 10.1 & 2.0  \\ 
MWc0.65 & 56.0 & 41.8 & 17.2 & 78.3 & 226 & 28 & 7.9 & 15.7 & 2.7  \\ 
MWc0.5 & 51.7 & 40.2 & 18.8 & 68.5 & 214 & 12 & 18.6 & 30.0 & 5.5  \\ 
\hline
\end{tabular}
\newline
\medskip
{ \scriptsize The first column indicates the name of the model. Columns 2--6 give the surface
density, radial and vertical velocity dispersion, vertical force at $z=1.1$\,kpc, and circular velocity at $R=8$\,kpc. 
The seventh column gives the pattern speed of the bar. 
Column 8--10 show the corotation, outer and inner Lindblad resonance radii.
}

\end{table*}

\begin{table}
\scriptsize
\caption{$\chi^2$ for $\Sigma_{\rm 8kpc}$, $\sigma_{R,{\rm 8kpc}}$, $K_{z,{\rm 8kpc}}$, $v_{\rm los}$, and $\sigma_{\rm los}$  at 10\,Gyr\label{tb:chi_sq}}
\begin{tabular}{lcccccc}
\hline
Model   & ${\Sigma_{\rm 8kpc}}$ & $\sigma_{R,{\rm 8kpc}}$ & $K_{z,{\rm 8kpc}}$ & $v_{\rm los}$ & $\sigma_{\rm los}$ & total\\
\hline \hline
MWa5B & 0.4 & 0.1 & 0.2 & 2.3 & 1.6 & 4.5  \\ 
MWa & 0.4 & 0.2 & 0.5 & 2.9 & 1.3 & 5.3  \\ 
MWb6B & 1.5 & 1.5 & 1.0 & 2.8 & 1.2 & 8.1  \\ 
MWb & 0.6 & 1.6 & 1.5 & 3.3 & 1.0 & 8.0  \\ 
MWc7B & 2.0 & 3.0 & 0.6 & 2.1 & 1.3 & 9.0  \\ 
MWc0.8 & 1.3 & 6.0 & 0.2 & 2.2 & 3.0 & 12.6  \\ 
MWc0.65 & 6.8 & 1.8 & 2.8 & 1.8 & 10.9 & 24.1  \\ 
MWc0.5 & 1.8 & 1.1 & 0.1 & 3.2 & 91.4 & 97.6  \\ 
\hline
\end{tabular}
\end{table}

\section{Discussion}\label{Sect:discussion}

Here we ``observe'' our best-fitting models, and discuss
the origin of structures observed in the MW galaxy.

\subsection{Inner bar structure}
\begin{figure*}
\epsscale{.3}
\plotone{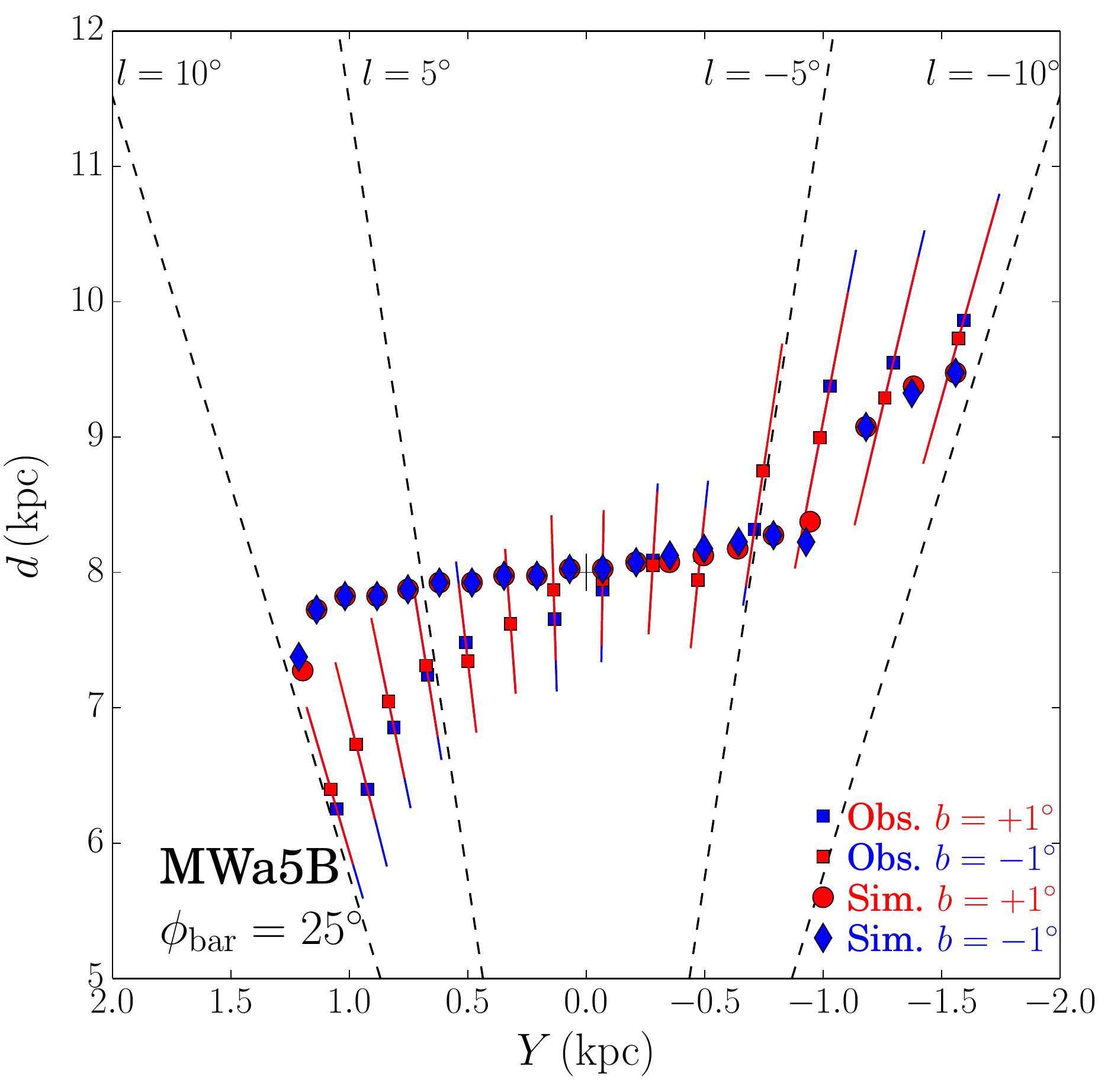}\plotone{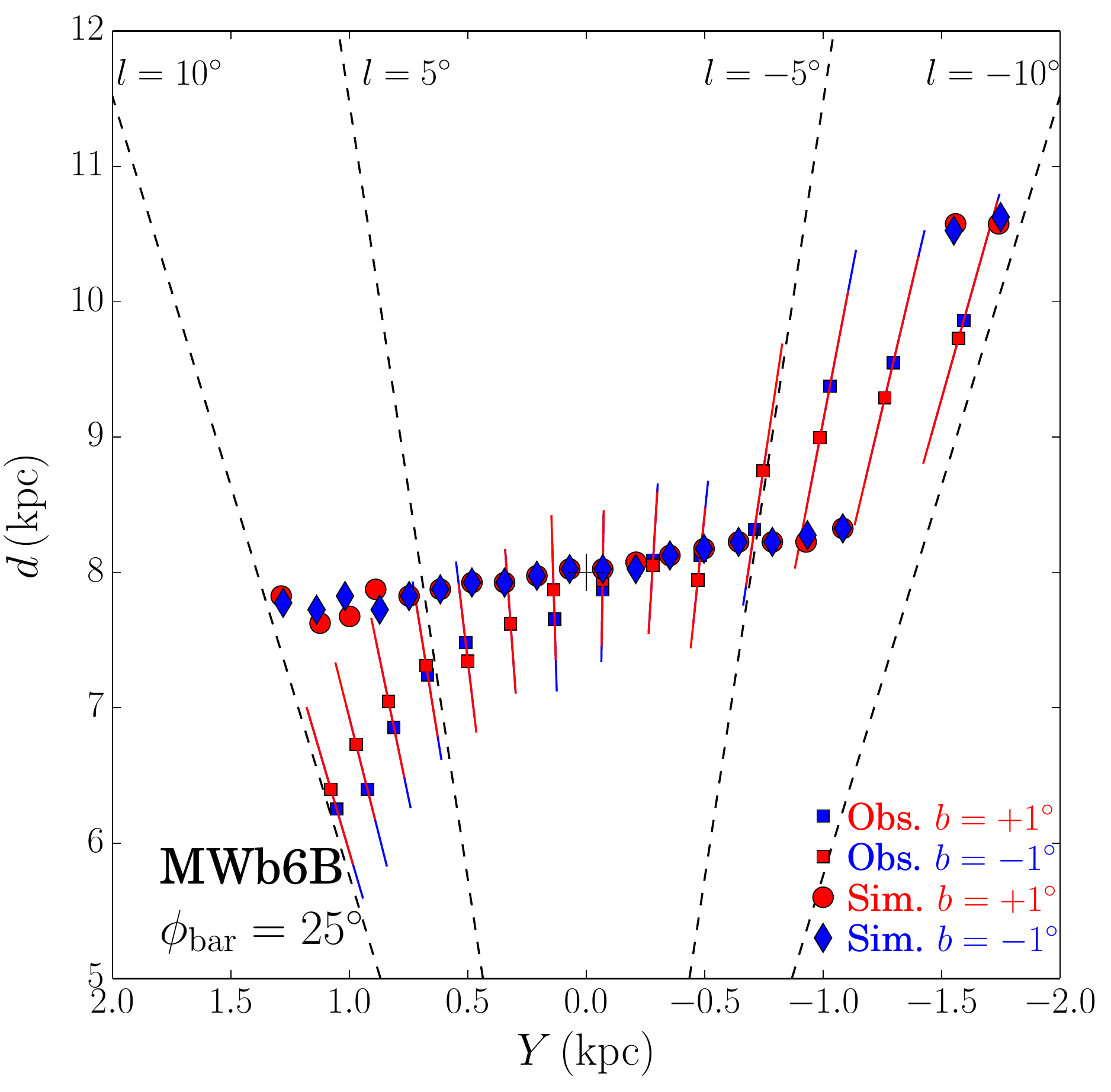}\plotone{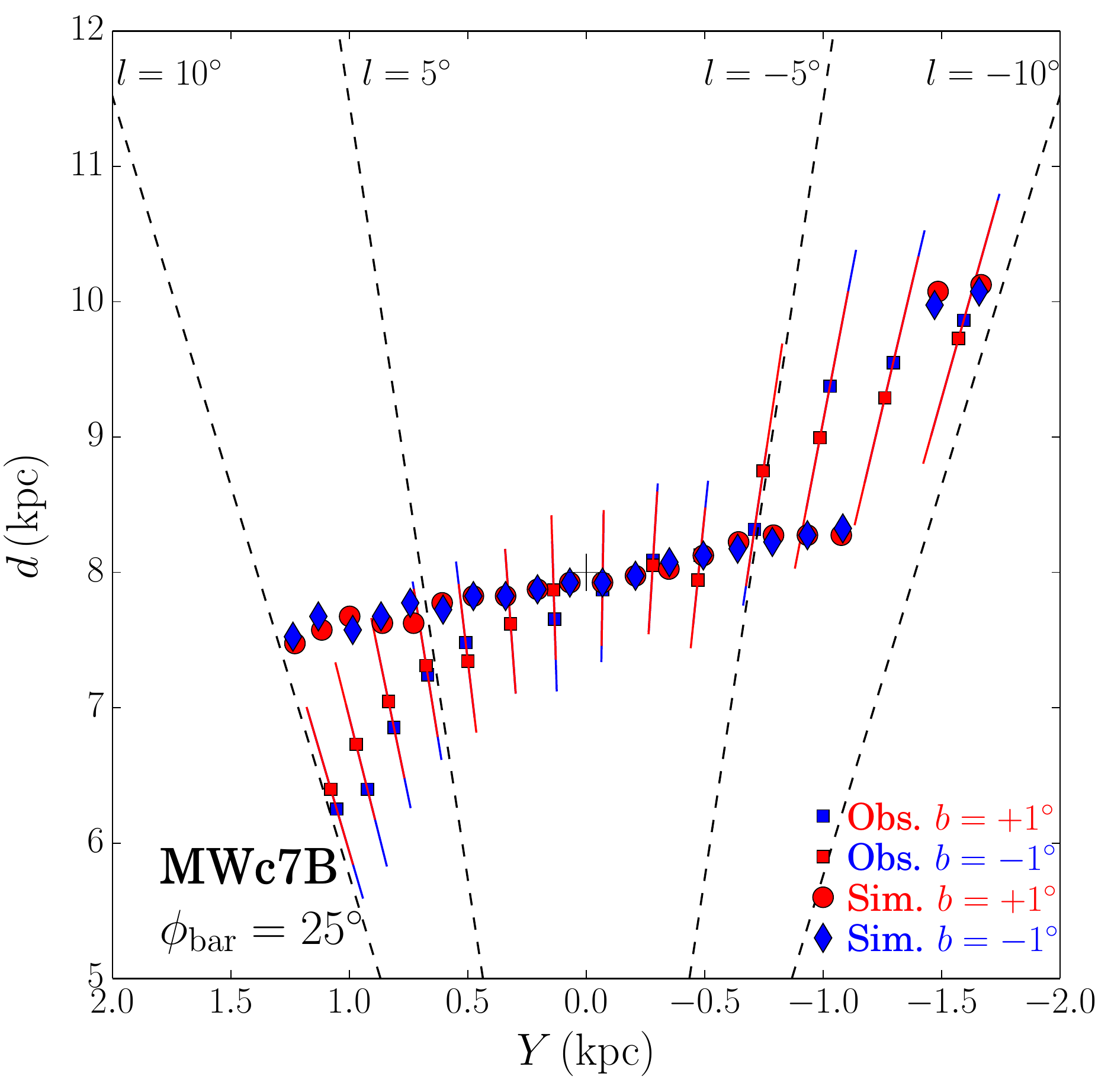}\\
\plotone{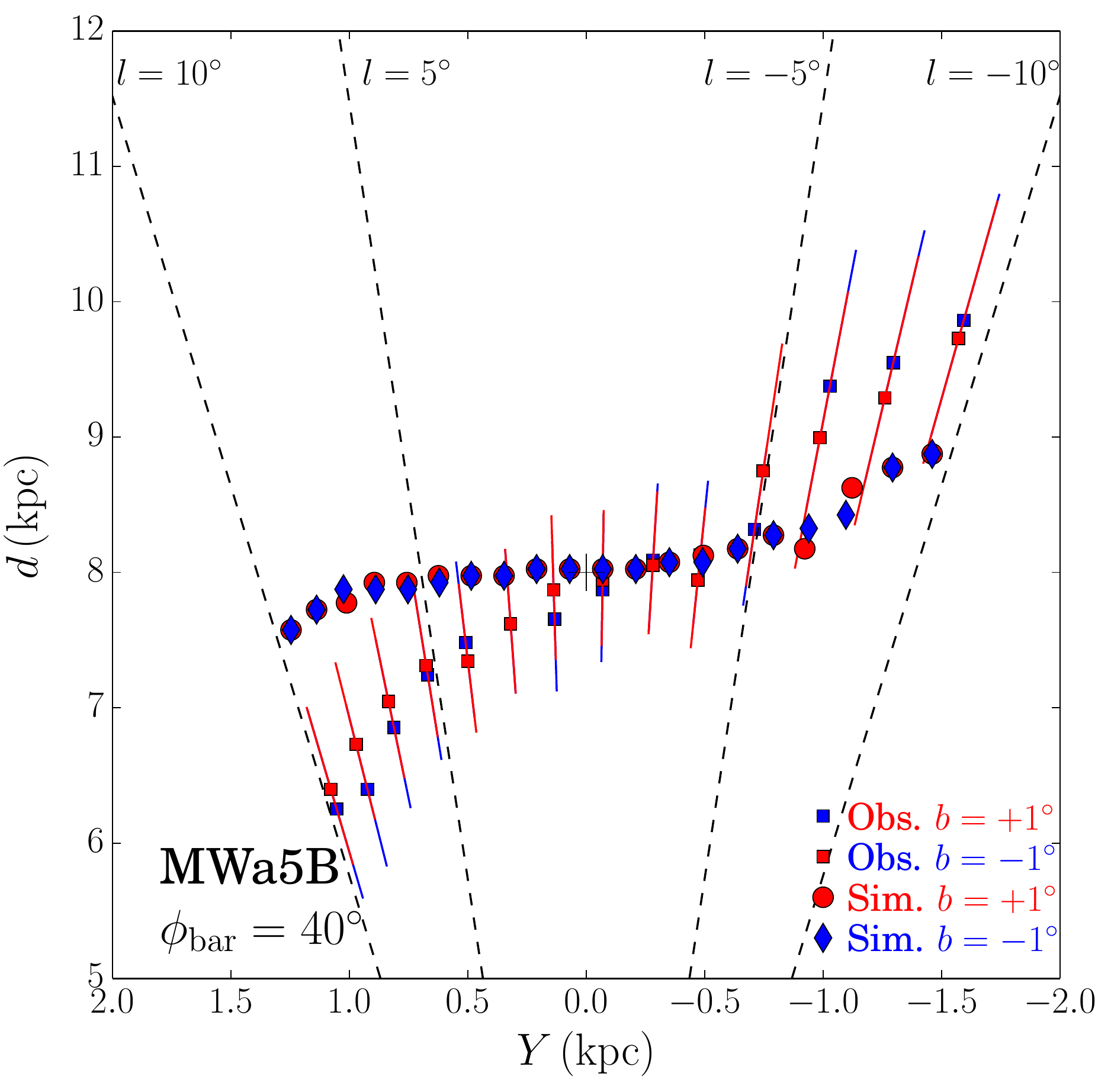}\plotone{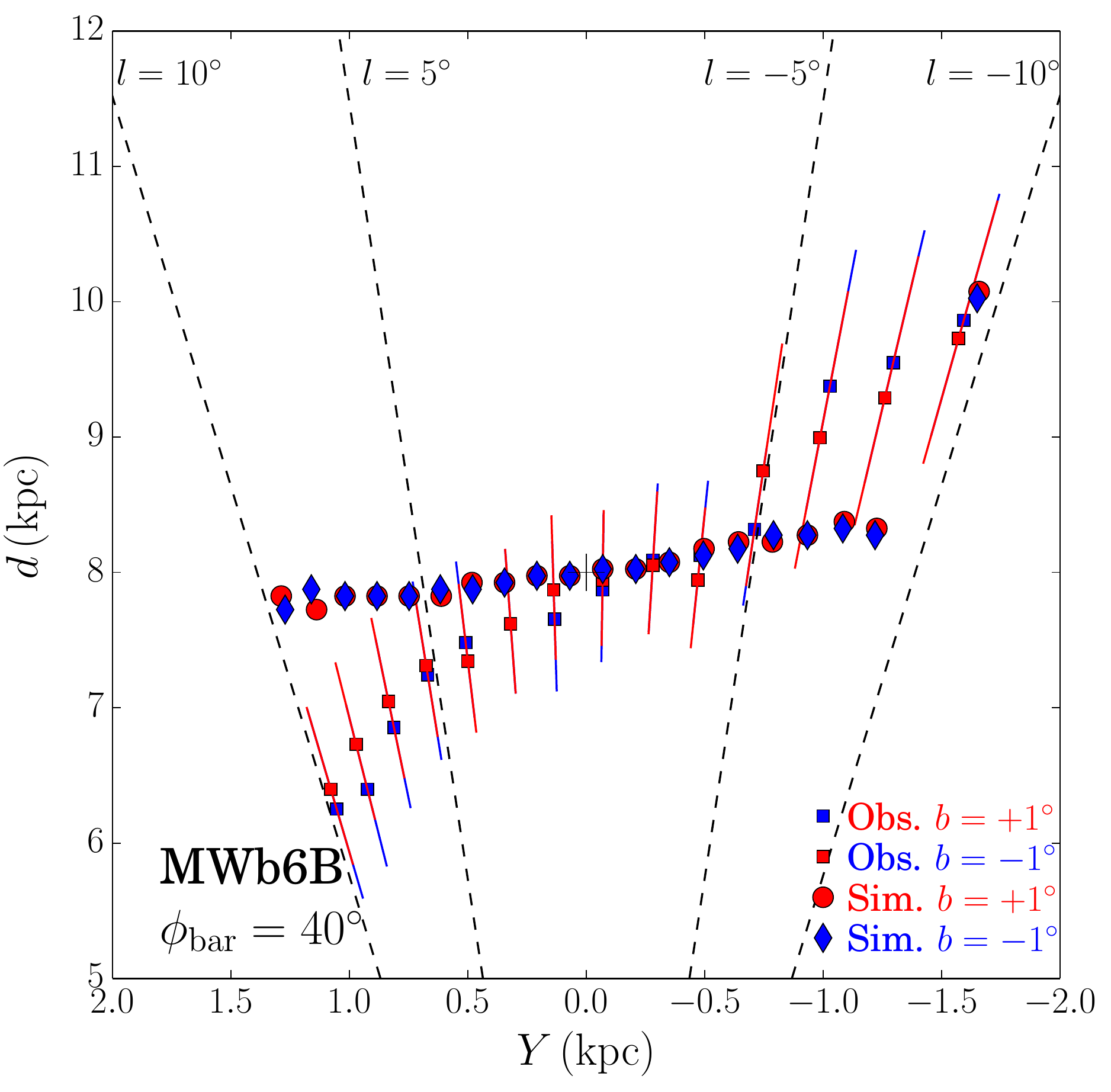}\plotone{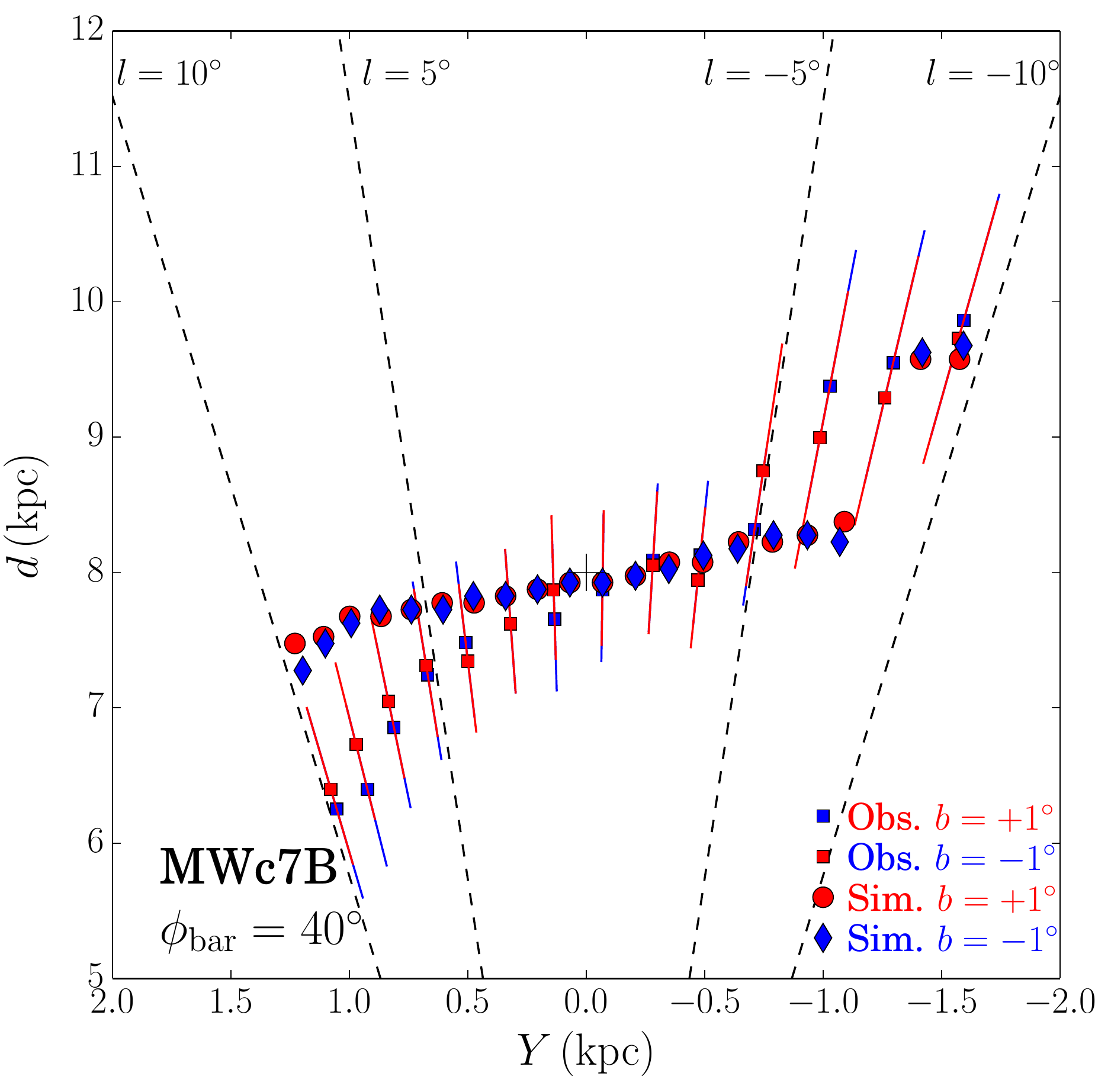}\\
\caption{Position of the Galactic bar with respect to the Sun, traced via the density 
  peaks \citep[similar to Fig.~3 of ][]{2011A&A...534L..14G,2016PASA...33...25Z}
  for models MWa5B (left), MWb6B (middle), and MWc7B (right).
  The bar angle is set to be $\phi_{\rm bar}=25^{\circ}$ (top) and $40^{\circ}$ (bottom).
  The Sun is located on the $y$-axis.\label{fig:bar_rc}}  
\end{figure*}

Using the ESO Vista Variables from the Via Lactea survey (VVV), \citet{2011A&A...534L..14G}
traced the bar structure of the Galactic centre.
They counted the number of clumps with red stars for 
each galactic longitude bin ($|l|<\pm10^{\circ}$) for $b=\pm 1^{\circ}$ as a function of the $K$ magnitude.
They converted the magnitude into the distance from the Sun and used this to measure the distance to 
the peak of the distribution of red clumps, and assumed that the peaks trace the Galactic bar. 
In order to reproduce this measurement, we use disc and bulge particles instead of red clumps. 
In each ($l$, $b$) bin, we count the number of stars as a function of the distance from the Sun 
($d$) and measure the peak distance. In Fig.~\ref{fig:bar_rc} we present the results for 
the $\phi_{\rm bar}=25^{\circ}$ and $40^{\circ}$ view angles. Similar to the results of 
\citet{2011A&A...534L..14G} and also \citet{2016PASA...33...25Z}, we ``observe'' bars 
much more inclined than the assumed bar angle and the bar angle changes between 
$|l|=$5--10$^{\circ}$. This is due to the projection effect and the existence of the bulge 
which was nicely shown, using $N$-body simulations, by \citet{2012ApJ...744L...8G}.
The points in which the bar breaks (i.e., the outer edge of the ``inner bar'') 
only barely depends on the bar's viewing angle ($\phi_{\rm bar}$).

\subsection{Hercules stream in simulated galaxies}

There are some known structures in the velocity space distribution 
of solar neighbourhood stars.
One of the most significant structures is the Hercules stream
\citep{1998AJ....115.2384D,2000AJ....119..800D}:
a co-moving group of stars with $30$--$50$\,km\,s$^{-1}$ slower than
the velocity of the local standard of rest ($V_{\rm LSR}$)
and with $U<0$\,km\,s$^{-1}$ \citep{2000AJ....119..800D,2018MNRAS.478..228Q}.
In a stellar number distribution, as a function of $V$,
the Hercules stream appears as a second peak at $-30$ to $-50$\,km\,s$^{-1}$
from the main peak \citep{2009NewA...14..615F,2017MNRAS.466L.113M}.

The origin of the Hercules stream is not yet completely clear. 
\citet{2000AJ....119..800D} suggested that it is caused by the
outer Lindblad resonance (OLR) 
\citep[see also][]{2014A&A...563A..60A,2017MNRAS.466L.113M,2018MNRAS.474...95H}.
For this model, a fast bar ($\gtrsim 50$\,km\,s$^{-1}$\,kpc$^{-1}$) is
required to bring the OLR radius near the Sun. 
\citet{2017ApJ...840L...2P}, on the other hand, suggested that the Hercules stream
is caused by resonant stars between the co-rotation and OLR radii. 
For this model a slower bar ($\sim 40$\,km\,s$^{-1}$\,kpc$^{-1}$)
is suggested.
In addition, \citet{2018arXiv180401920H} recently suggested a model in which a slow 
bar (36\,km\,s$^{-1}$) combined with spiral arms can 
reproduce the Hercules stream. However, they also found that just a fast bar, or 
both a fast bar and spirals can reproduce the Hercules stream.
\citet{2018MNRAS.477.3945H} explained the Hercules stream using a 4:1 
OLR of a slow bar. The above studies assume a fixed potential for the bar and spiral
arms. \citet{2018MNRAS.tmp.2421H}, on the other hand, showed that it can be 
reproduced by transient spirals only.
Using `live' $N$-body simulations, \citet{2011MNRAS.417..762Q}
studied the existence of the Hercules stream. They concluded that 
a Hercules-like stream appears outside of the OLR and with 
$\phi_{\rm b}=45^{\circ}$ when the spiral arm configuration is 
similar to the one observed. In this section we investigate
Hercules-like streams in our $N$-body models.

\begin{figure*}
\epsscale{1.0}
\plotone{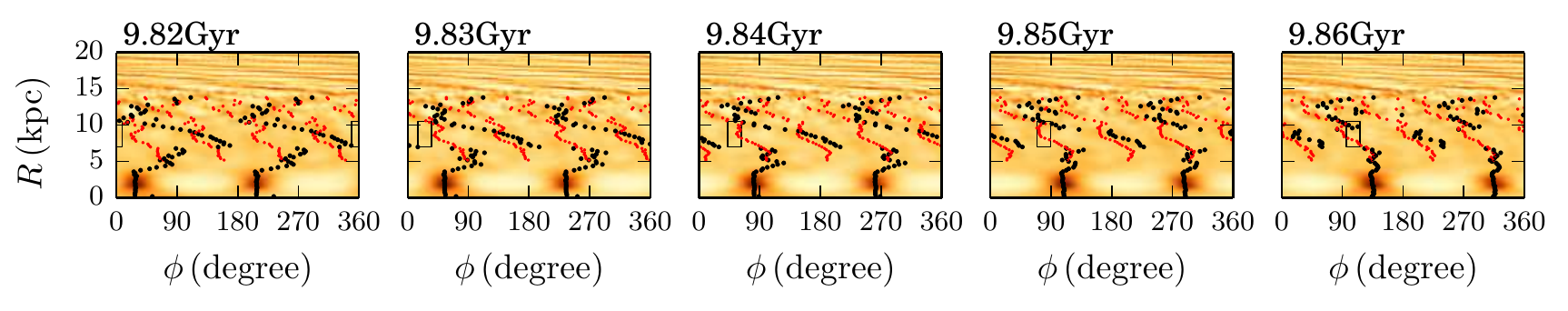}\\
\plotone{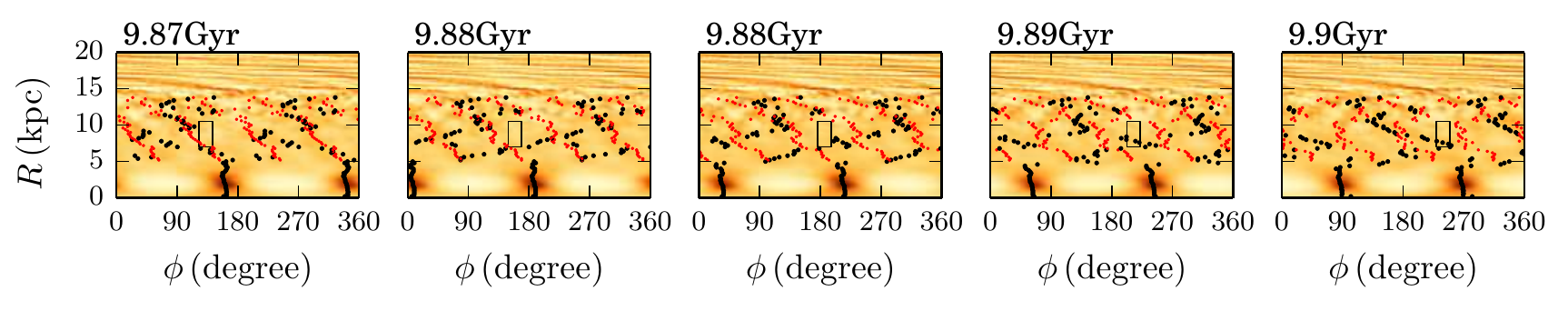}\\
\plotone{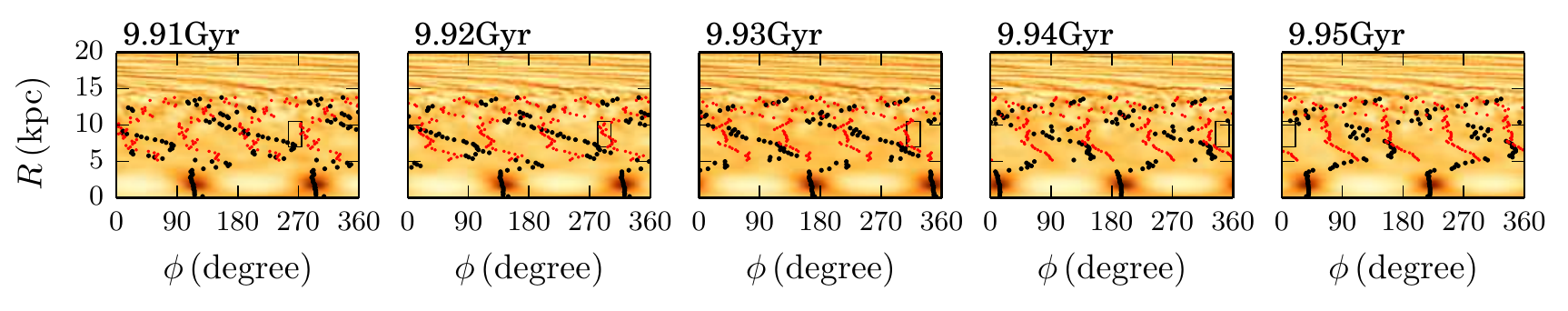}\\
\plotone{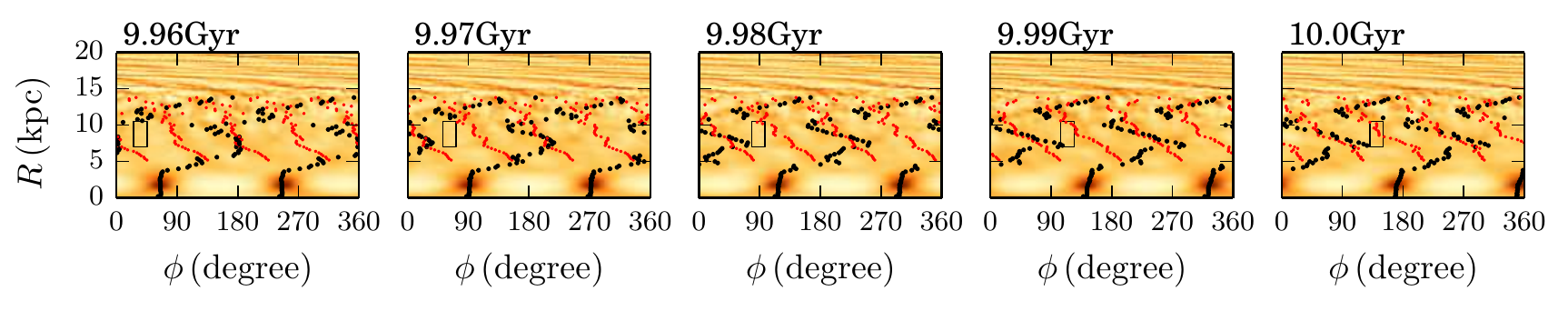}\\
\caption{Surface density of model MWa5B at $t=9.82$--10.00\,Gyr. 
    The surface density is normalized by the average at each radius. 
    The rectangle in each panel marks the region with $R=7$--10.5\,kpc 
    and $\phi_{\rm bar}=20$--40$^{\circ}$.
    Black and red dots indicate the phase, determined using Fourier
    decomposition (Eq.~\ref{eq:Fourier}), for $m=2$ and $m=4$, respectively.
    \label{fig:MWa5B_Rp_10Gyr}}
\end{figure*}

Our MWa5B model has the fastest bar pattern speed (45\,km\,s$^{-1}$\,kpc$^{-1}$)
and the resulting OLR radius is 9.1\,kpc.
For this model, we take the position of the Sun every 0.5\,kpc 
between 7.5--11.5\,kpc.
The bar's view angle is not fully determined from the observations,
we, therefore, take angles between 20--40$^{\circ}$ \citep{2016ARA&A..54..529B}.
In Fig.~\ref{fig:MWa5B_Rp_10Gyr} we present the surface densities
in the $R$-$\phi$ plane for $t=9.82$--10.00\,Gyr. In this figure, 
the surface density is normalized within each radial bin 
(i.e., the mean density at the radius is set to be 1). 
We also perform a Fourier decomposition (see equation~\ref{eq:Fourier}) at each radius
and include the $m=2$ and $m=4$ phases in the figure. 
These phases roughly trace the spiral arm positions.
As is seen in the figure, the spiral structure changes significantly over time. 
The position of the Sun, $R\sim 8$\,kpc and $\phi_{\rm b}\sim 20^{\circ}$--40$^{\circ}$,
falls mostly in the inter-arm regions. 
 
\begin{figure*}
\epsscale{1.0}
\plotone{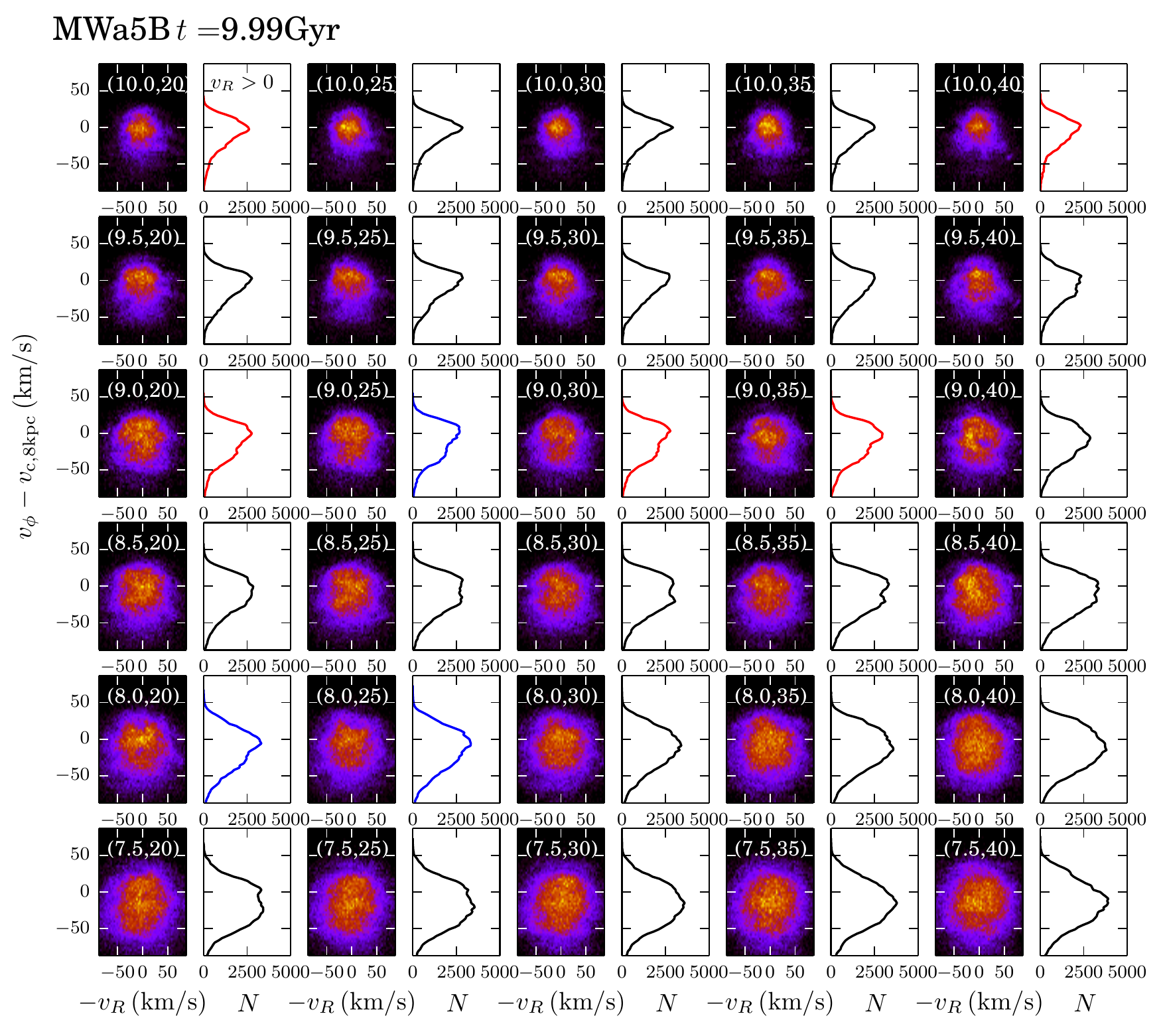}\\
\caption{Velocity distribution of model MWa5B at $t=9.99$\,Gyr. 
Hercules-like streams, detected using the least-mean-square fitting method, 
are marked by the red (in arm location i.e., higher surface density than the average at the radius) and blue lines (inter-arm i.e., lower surface density than the average), black 
lines indicate no detection. 
\label{fig:Rpn1023}}
\end{figure*}

In Fig.~\ref{fig:Rpn1023}, we present the velocity distribution of
stars that are within 0.5\,kpc from the assumed position of the Sun.
Here we take 
 $R=7.5$ to $R=10$\,kpc for every 0.5\,kpc and from $\phi_{\rm bar}=20^{\circ}$
to $\phi_{\rm bar}=40^{\circ}$ for every $5^{\circ}$ as the position of the Sun.
We show $t=9.99$\,Gyr, in which the configuration of the spiral arms is similar to those observed in 
the MW disc.
The Perseus arm at $\sim 10$\,kpc
and Scutum or Sagittarius arm at $\sim 6$\,kpc from the Galactic centre
\citep{2014ApJ...783..130R,2017NewAR..79...49V}. 
In simulations $-v_{R}$, where $v_{R}$ is the radial velocity,
is equivalent to $U$ in observations. 
For the tangential velocity ($v_{\phi}$), $v_{\phi}-V_{\rm LSR}$ is
equivalent with $V$.
We use the circular velocity at 8\,kpc ($v_{\rm circ, 8kpc}=241$\,km\,s$^{-1}$ for this model) 
as $V_{\rm LSR}$.
Hercules-like structures are unclear when only looking at the phase-space maps, and therefore 
we also present the number of
stars for $v_{\rm R}>0$ as a function of $v_{\phi}-v_{\rm c, 8kpc}$.

We detect the Hercules-like 
stream using a least-mean-square method. We fit the sum of two Gaussian 
functions to the distribution of stars with $v_{\rm R}>0$ using,
\begin{eqnarray}
f(v_{R}) = N_1\exp \left[-\frac{(v_R-v_1)^2}{2\sigma_1^2}\right] + N_2\exp \left[-\frac{(v_R-v_2)^2}{2\sigma_2^2}\right],
\end{eqnarray}
where $N_1$, $N_2$, $v_1$, $v_2$, $\sigma_1$, and $\sigma_2$
are fitting parameters, with $N_1>N_2$. 
We assume that a Hercules-like stream is detected when this function gives a better fit than a single Gaussian function 
and when the following conditions are satisfied:
\begin{enumerate}
\item $N_2>0.1N_1$ 
\item $-20<v_1<20$\,km\,s$^{-1}$
\item $-60<v_2<-20$\,km\,s$^{-1}$
\item $20<v_1-v_2<60$\,km\,s$^{-1}$.
\end{enumerate}
Using this function, we find two-peak structures when the Sun is located on the 
outer Lindblad resonance ($\sim 9$\,kpc).  
We perform this fitting procedure for 40 different Sun locations ($7.0\leq R\leq 10.5$\,kpc every 0.5\,kpc and 
$20^{\circ}\leq \phi_{\rm bar}\leq 40^{\circ}$ every $5^{\circ}$).
The locations in which a stream is found are indicated using the red and blue lines in Fig.~\ref{fig:Rpn1023}.

\begin{figure*}
\epsscale{1.0}
\plotone{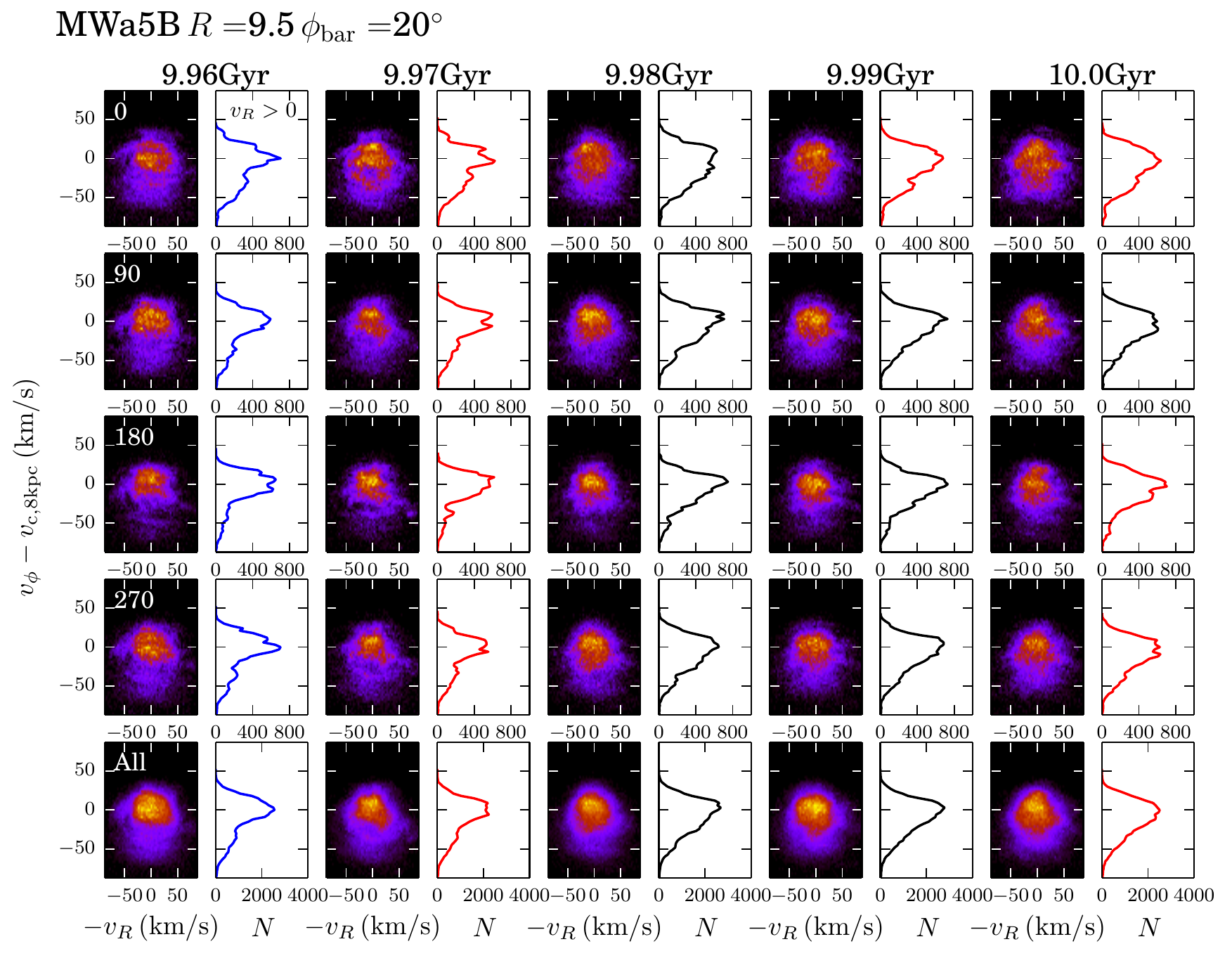}\\
\caption{Time evolution of the velocity distribution for model MWa5B at $R=9.5$\,kpc and $\phi_{\rm bar}=20^{\circ}$ for $t=9.96$--10.00\,Gyr. 
From top to bottom, velocity distribution for $l=0$, 90, 180, $270\pm45^{\circ}$, and all (total of the above). Red, blue and black lines are the same as those in Fig.~\ref{fig:Rpn1023}\label{fig:R9.5p20n1020}}
\end{figure*}

These structures, however, are not always seen.
In the top panels of Fig.~\ref{fig:R9.5p20n1020}
we present the velocity distribution of model MWa5B, assuming the position of the Sun is at 
$R=9.5$\,kpc and $\phi_{\rm b}=20^{\circ}$ for $t=9.96$--10.00\,Gyr. At this location
we often detected the stream structure.
Although Hercules-like structures are seen in $t=9.96$--9.98\,Gyr,
they are not seen at slightly different times such as $t=9.99$ and 10.00\,Gyr,
in which the location of the Sun is still between two major arms.

\begin{figure*}
\epsscale{.5}
\plotone{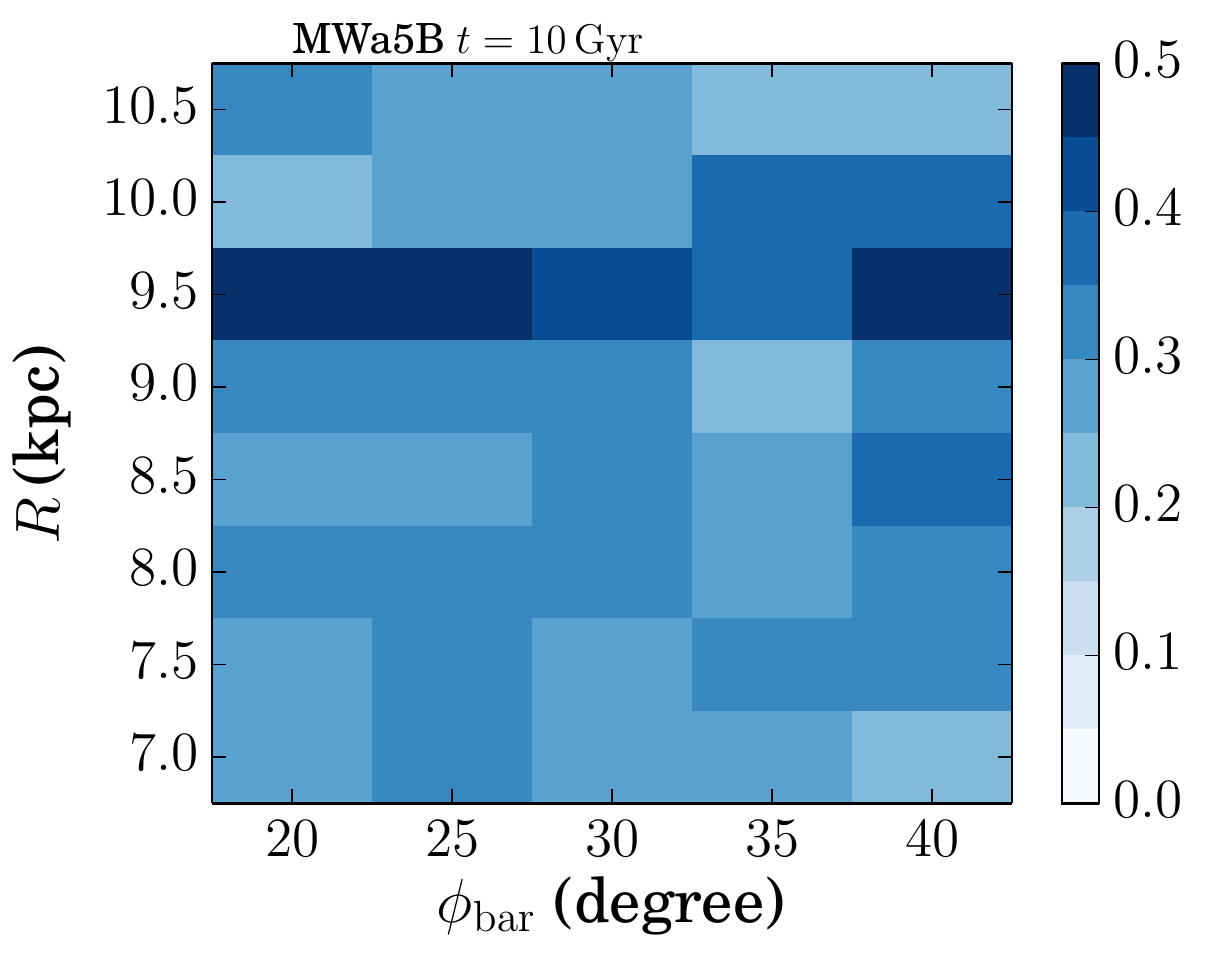}
\plotone{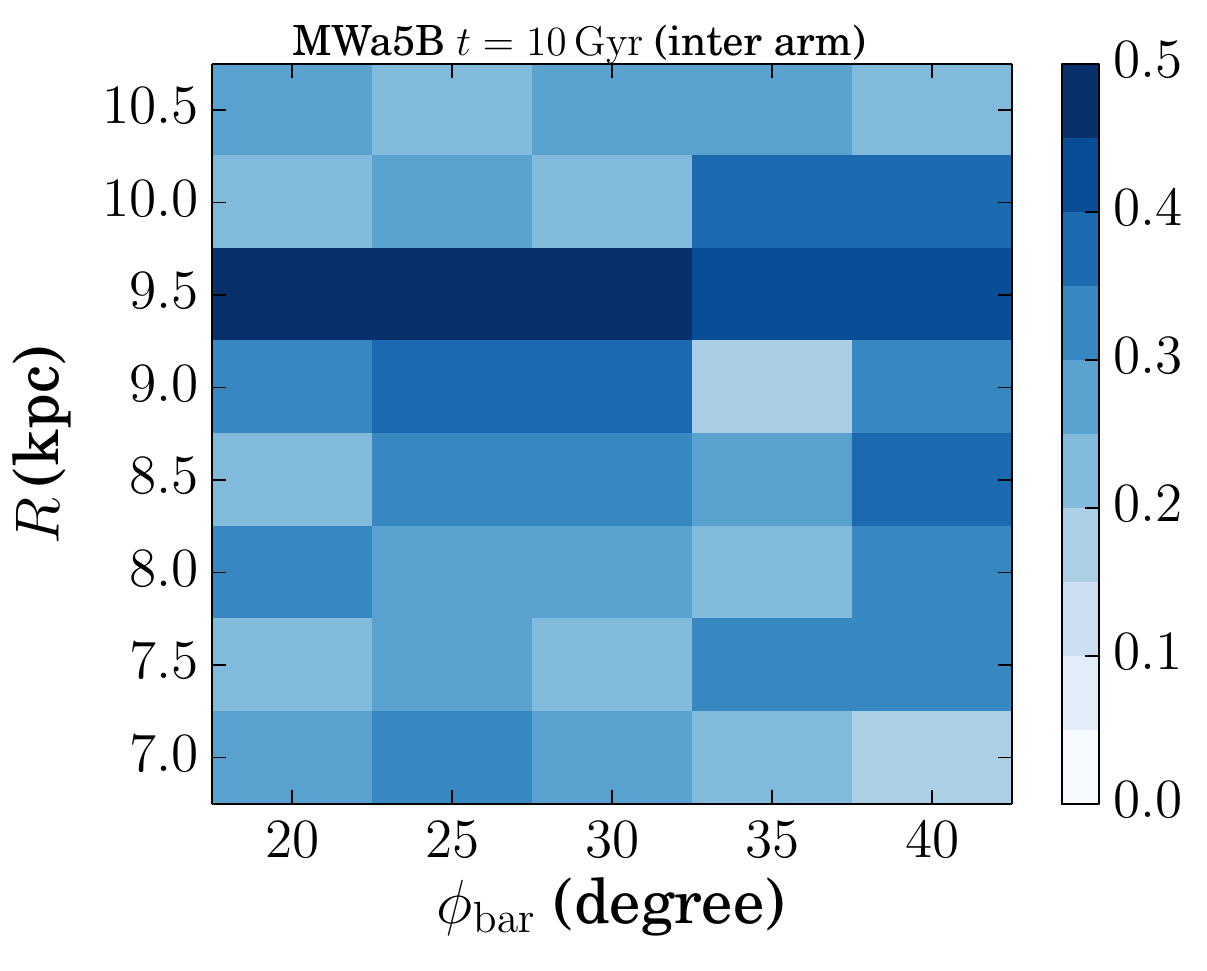}\\
\caption{Stream frequency for the last 100 snapshots ($t=9.0$--$10.0$\,Gyr) for model MWa5B. Left: irrespective of the position of spiral arms. Right: only for inter arm regions. \label{fig:stream_freq_MWa_10Gyr}}
\end{figure*}

\begin{figure*}
\epsscale{.5}
\plotone{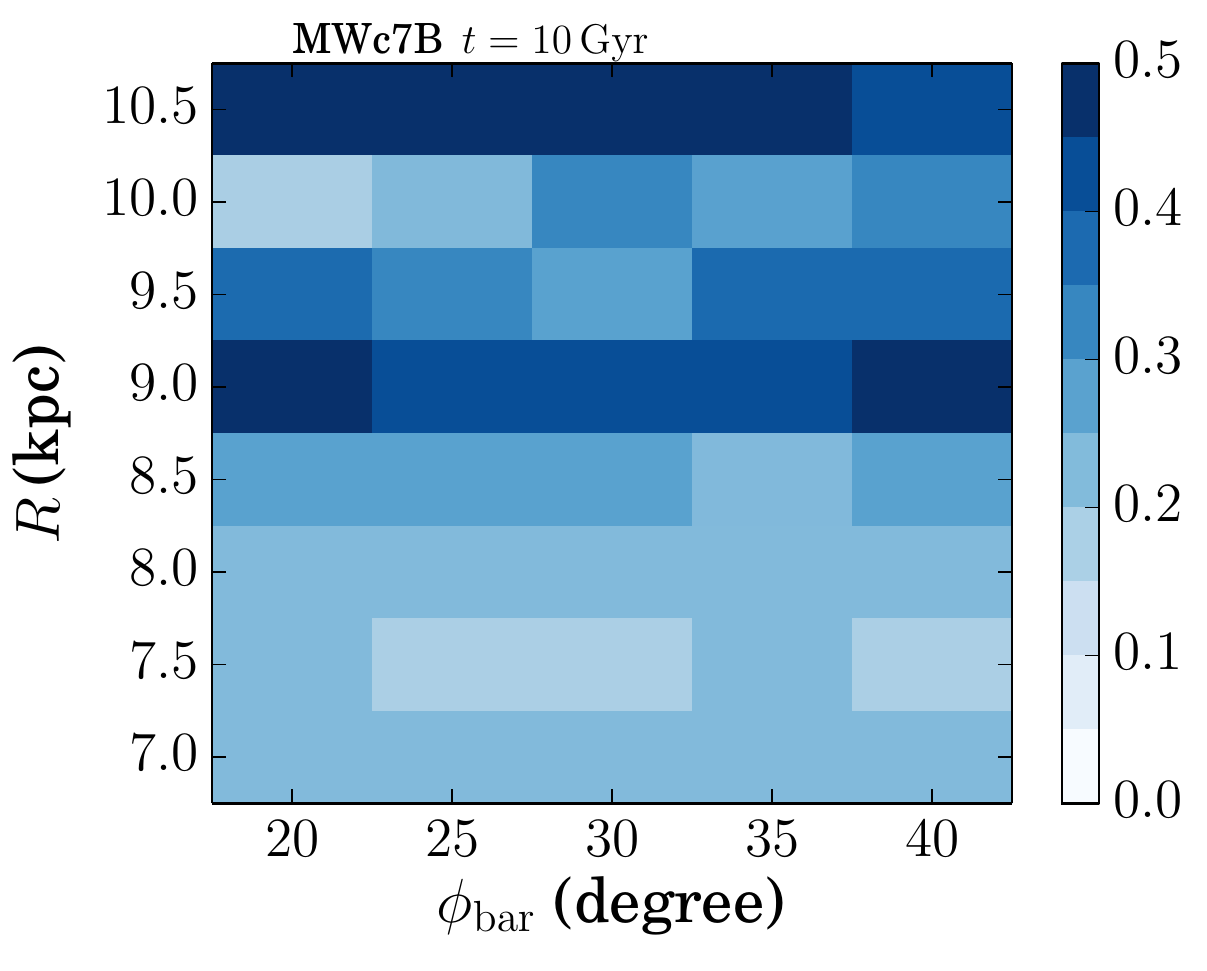}
\plotone{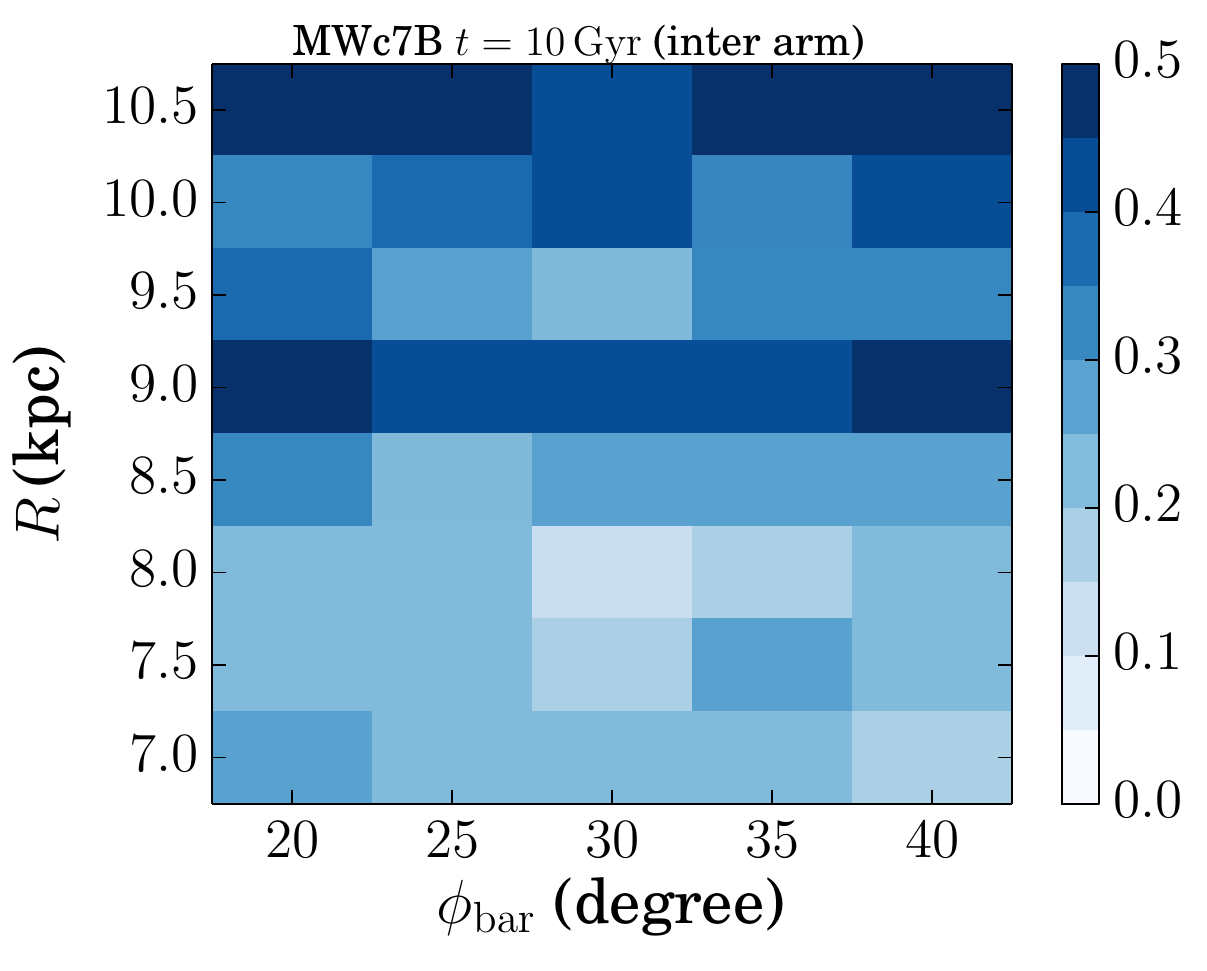}\\
\caption{Same as Fig.~\ref{fig:stream_freq_MWa_10Gyr}, but for model MWc7B. \label{fig:stream_freq_MWc_10Gyr}}
\end{figure*}

In order to quantify the frequency with which we ``observe'' the
Hercules-like stream, we perform this fitting for the last 100 snapshots and 
present the frequency of the Hercules-like stream for models MWa5B and MWc7B in 
Figs.~\ref{fig:stream_freq_MWa_10Gyr} and \ref{fig:stream_freq_MWc_10Gyr}, respectively. 
We did not perform this analysis for model MWb6B because the OLR radius is very similar to that of 
model MWc7B. 
We find an enhancement of the streams frequency at a radius slightly further out than the OLR radius,
namely at $R\sim9$ and 10\,kpc for models MWa5B and MWc7B, respectively.
At this radius, we detect a Hercules-like stream in at most $\sim 50$\,\% of the snapshots. 
For model MWc7B, we further find a high stream frequency at $R=9$\,kpc, which is 
between the co-rotation and OLR radius.

We further tested a situation where the Sun is located in an inter-arm region
\citep{2016ARA&A..54..529B}, i.e.,
the local surface density is smaller than the mean density at the Galactocentric distance ($R$). 
The fraction of the inter-arm regions among the distributions in which a Hercules-like stream is 
detected is $\sim 60$\,\% and $\sim 70$\,\% for models MWa5B and MWc7B, respectively. 
Hercules-like streams, again, frequently appear at the OLR radii, and the maximum frequency
is $\sim 50$\,\% among inter-arm regions for all the simulation times that we analyzed.

We also test the recent observation that finds the Hercules stream at 
the Galactic longitude of $l=270^{\circ}$~\citep{2018MNRAS.478..228Q}. 
Fig.~\ref{fig:R9.5p20n1020} therefore also shows the velocity distribution 
for model MWa5B at $R=9.5$\,kpc where $l=0$, 90, 180, and $270\pm 45^{\circ}$ and 
$\phi_{\rm bar}=20^{\circ}$.
These are the locations in which the Hercules-like streams most often appear 
(see Fig.~\ref{fig:stream_freq_MWa_10Gyr}). 
Indeed, the strength of the Hercules-like structure is different for each $l$, but
there is no sign that a Hercules-like feature is seen more often for $l=270\pm 45^{\circ}$.

\begin{figure*}
\epsscale{.5}
\plotone{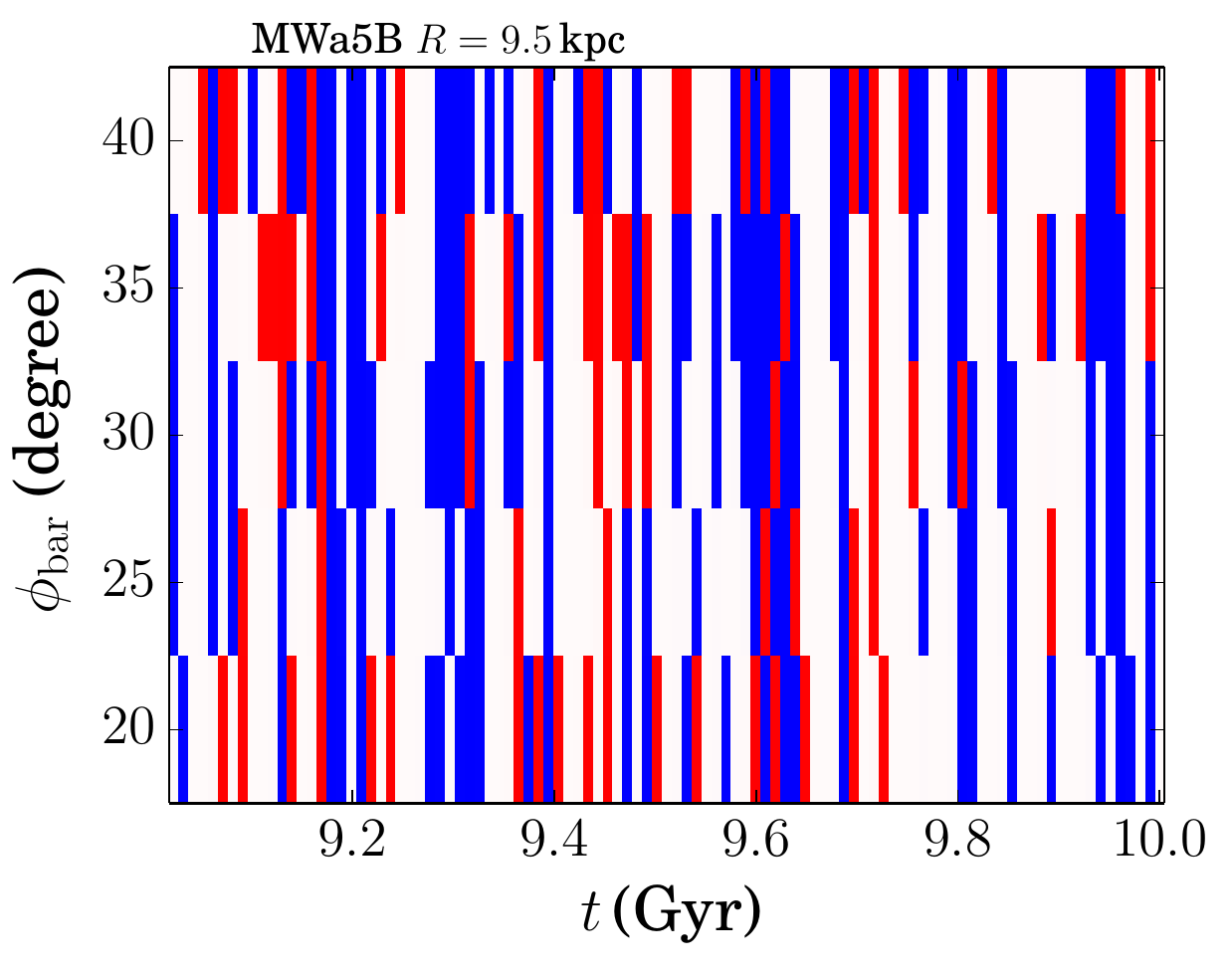}
\plotone{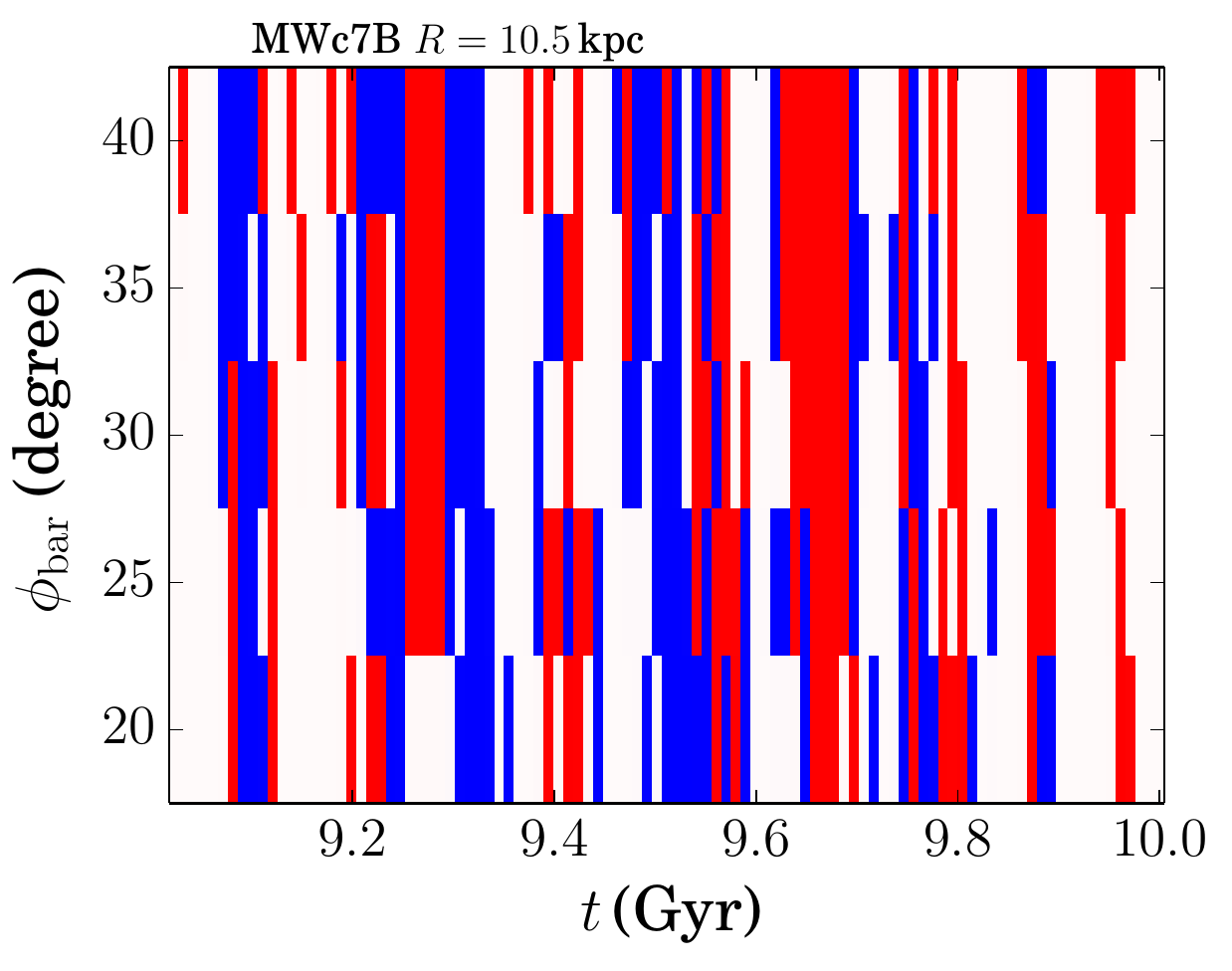}\\
\caption{Stream appearances as a function of time for Models MWa5B $R=9.5$\,kpc (left) and MWc7B $R=10.5$\,kpc (right). The time interval of each snapshot is $\sim 10$\,Myr.
    Red and blue indicate timings in which we detected a stream in the arm and inter-arm regions, respectively. \label{fig:stream_time}}
\end{figure*}

For the radii in which the stream frequency is high, we investigated the time dependence. 
In Fig. ~\ref{fig:stream_time}, we present the time at which we detected a
Hercules-like stream for the last 100 snapshot ($\sim 1$\,Gyr).
The timing for which we ``observe''
Hercules-like streams is not periodically, but once they appear, they tend to continue
for 20--30\,Myr and sometimes even for $\sim100$\,Myr.

We repeat the above analysis for $t\sim5$\,Gyr, when the bar rotates faster 
than at $t=10$\,Gyr and the pattern speed oscillates between 48--56\,km\,s$^{-1}$. 
The corresponding OLR occurs at $R\sim7.5$--9\,kpc.

\begin{figure*}
\epsscale{1.0}
\plotone{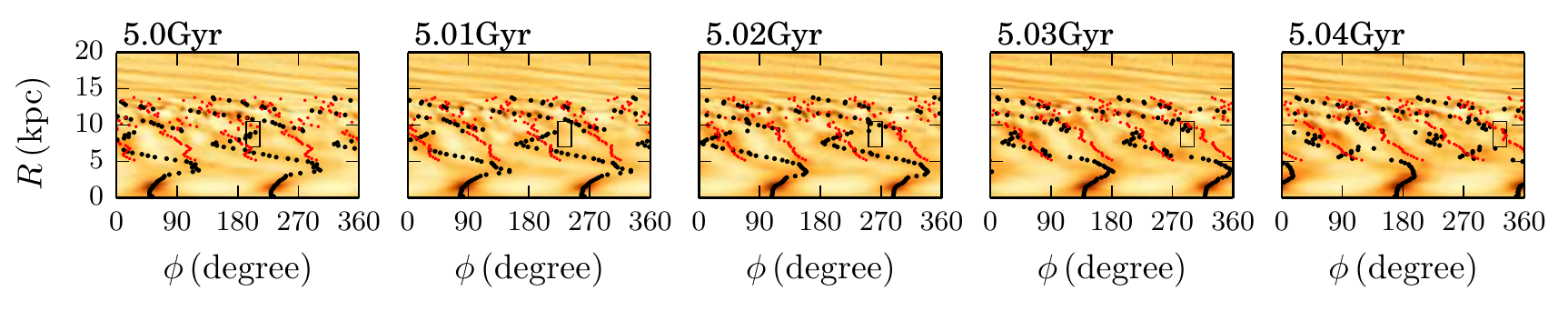}\\
\plotone{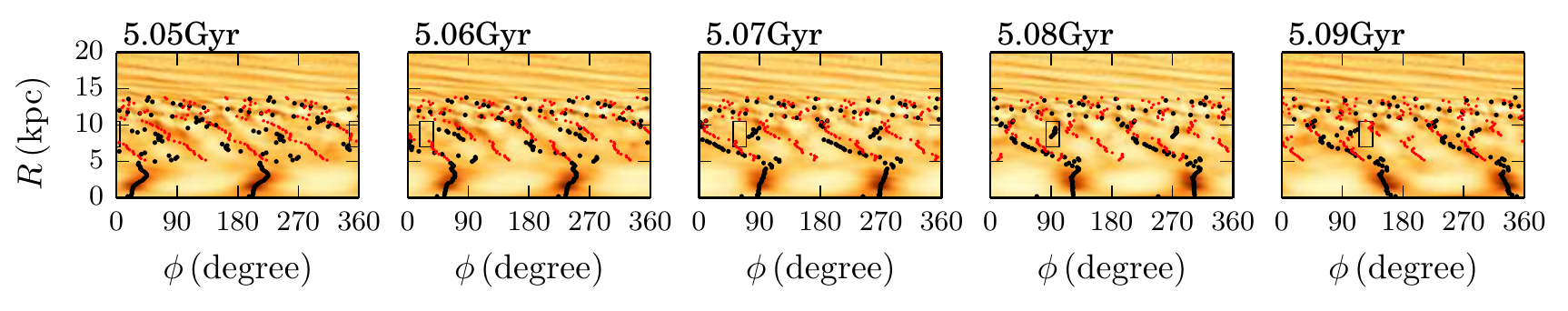}\\
\plotone{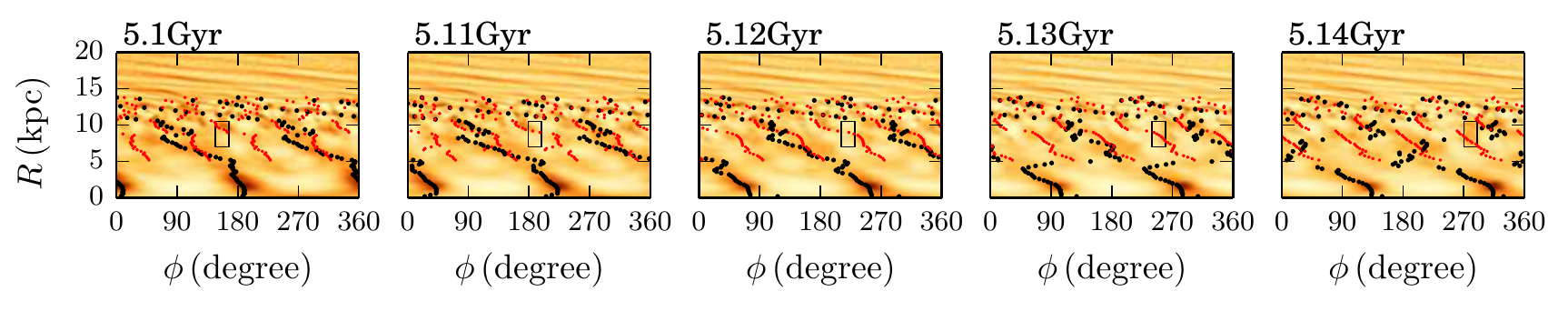}\\
\plotone{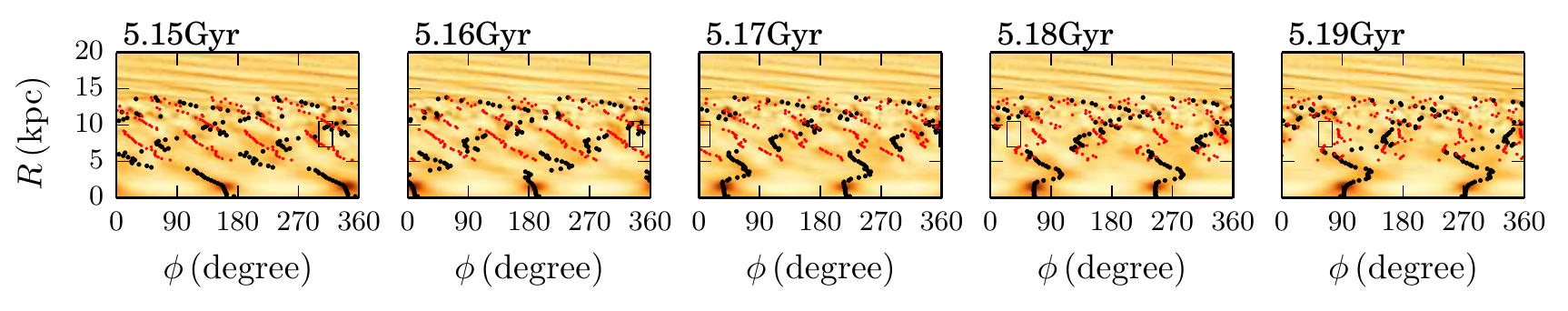}\\
\caption{Same as Fig.~\ref{fig:MWa5B_Rp_10Gyr}, but for $t=5.00$--$5.19$\,Gyr.\label{fig:MWa5B_Rp_5Gyr}}
\end{figure*}

In Fig.~\ref{fig:MWa5B_Rp_5Gyr} we present the surface density plots in the 
$R$-$\phi$ plane of model MWa5B for $t=5.00$--5.19\,Gyr. 
In this figure, we see that the bar oscillates back and forth over a period 
of $\sim 200$\,Myr.
This oscillation corresponds to the oscillation in the pattern speed of the bar.
The pattern speed of the bar reaches a minimum at $t=5$\,Gyr.
Then it gradually
speeds up and reaches the local fastest velocity (56\,km\,s$^{-1}$) 
at $t=5.08$\,Gyr before it slows down again.
At $t=5.16$\,Gyr, the pattern speed again reaches the slowest phase (48\,km\,s$^{-1}$). 
This oscillation seems to be due to the interaction with spiral arms
as suggested by \citet{1988MNRAS.231P..25S} and also 
\citet{2015MNRAS.454.2954B}. 
Since the spiral arms outside the bar have a pattern speed slower than that 
of the bar, 
they repeatedly connect and disconnect. 
When the bar catches up with 
the outer spiral arms, the bar's pattern speed is accelerated 
($t=5.00$--5.08\,Gyr) and reaches its maximum pattern speed at $t=5.08$\,Gyr.
Since the spiral arms move slower than the bar, 
the spirals start to get behind ($t=5.09$--5.15\,Gyr) and 
detach from the bar at $t=5.16$\,Gyr. During this process,
the bar slows down and reaches the slowest pattern speed at 
$t=5.16$\,Gyr. After that the bar connects to the spiral arms which are 
ahead of the bar ($t=5.17$--5.19\,Gyr).
This oscillation becomes weaker with time, but continues until the end of the 
simulation. We therefore also see 
this oscillation around 10\,Gyr (see Fig.~\ref{fig:MWa5B_Rp_10Gyr}),
although it is much weaker than at 5\,Gyr.

\begin{figure*}
\epsscale{1.0}
\plotone{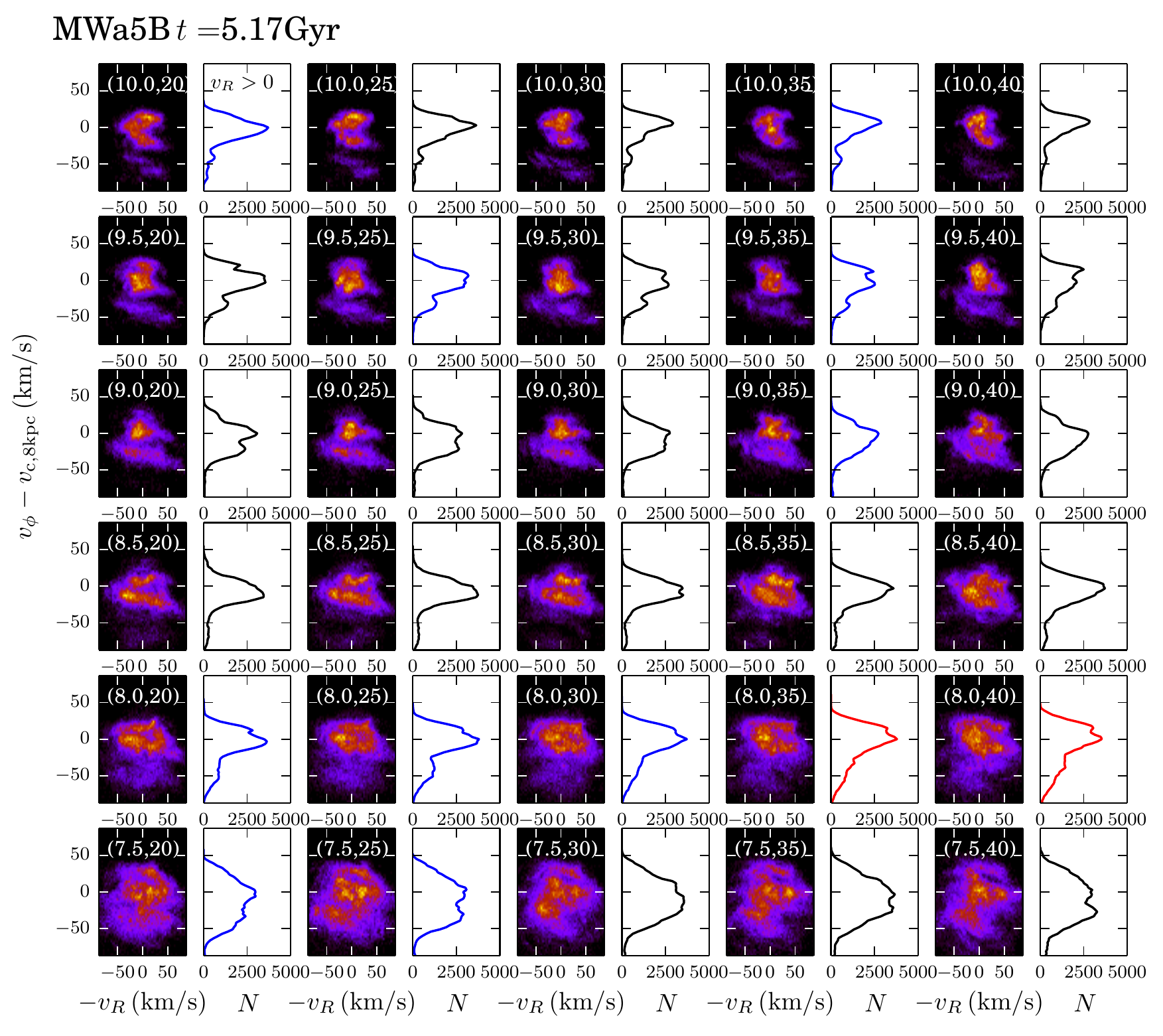}\\
\caption{Same as Fig.~\ref{fig:Rpn1023}, but for $t=5.17$\,Gyr. \label{fig:Rpn529}}
\end{figure*}

In Fig.~\ref{fig:Rpn529},
we present the velocity distribution at $t=5.17$\,Gyr for 
$R=7.5$--10\,kpc and $\phi_{\rm bar}=20$--40$^{\circ}$.
At these times, the position of the Sun is between two major spiral arms.
The figures contain more structures in the velocity distribution
than those seen at $\sim 10$\,Gyr.
The difference appears to be caused by the existence of stronger spiral structures at
$\sim 5$\,Gyr. \citet{2011MNRAS.417..762Q} showed that the velocity distribution
observed in their $N$-body simulation changes on a short time-scale and 
suggested that this change is caused by the spiral arms.

\begin{figure*}
\epsscale{.5}
\plotone{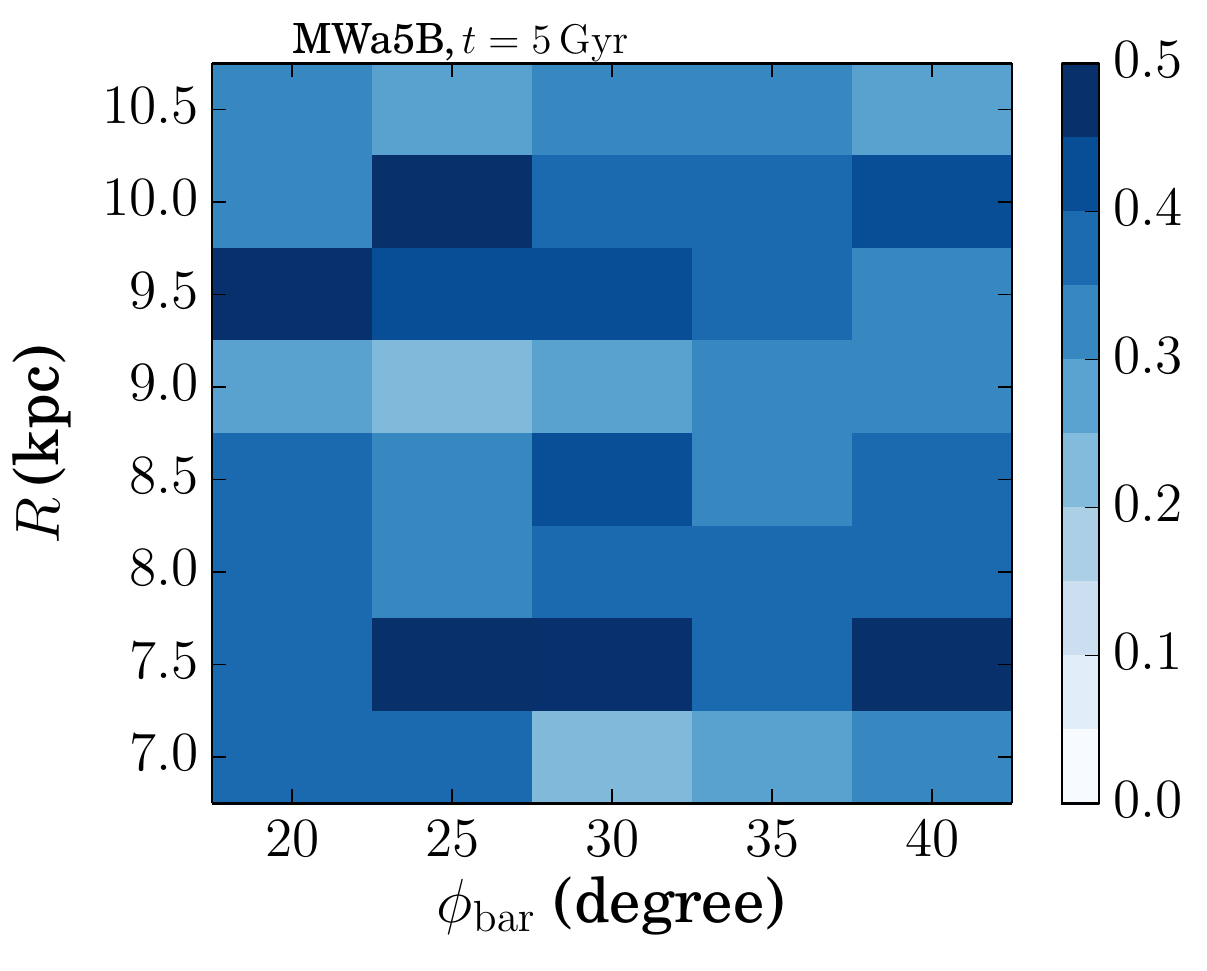}
\plotone{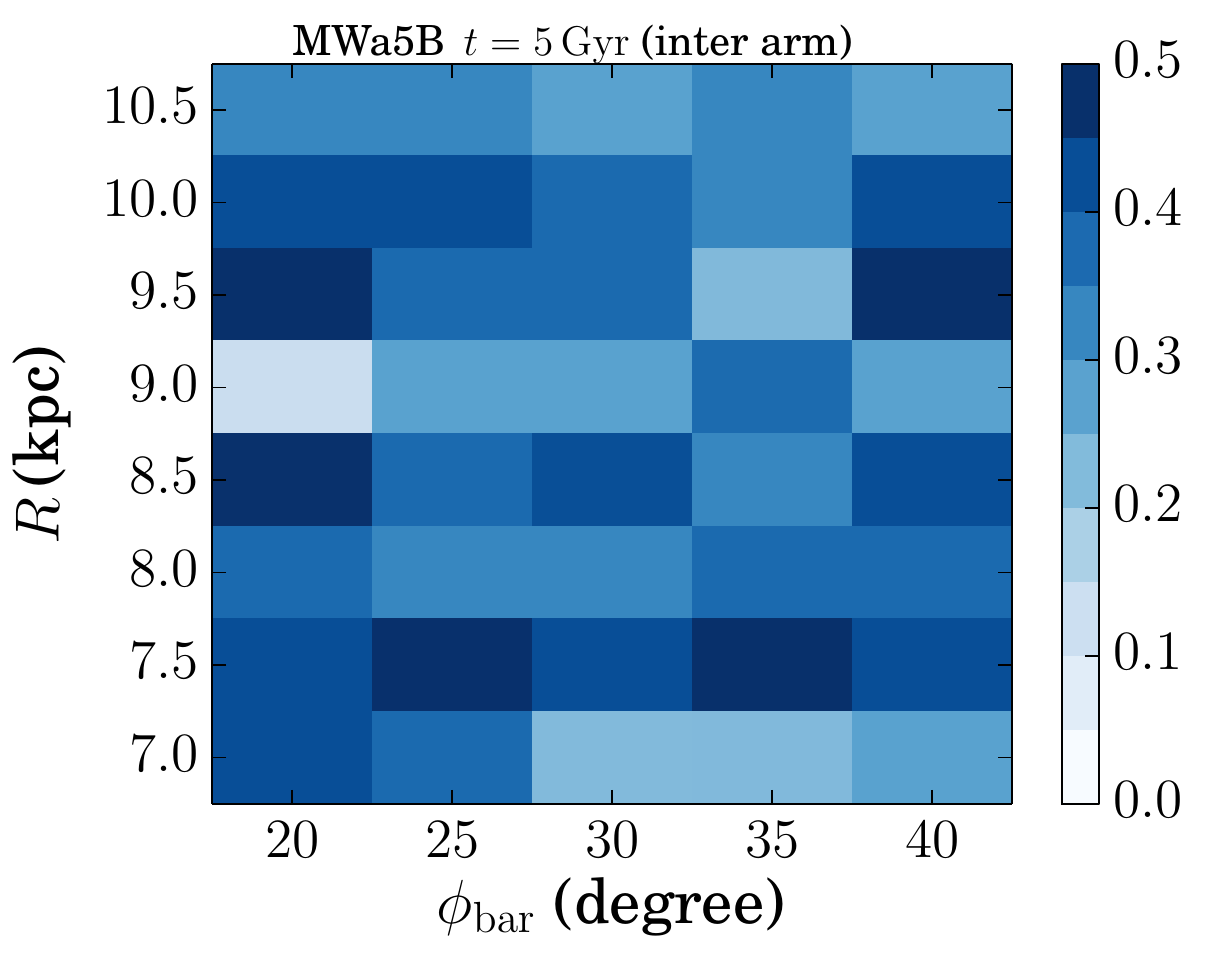}\\
\caption{Same as Fig.~\ref{fig:stream_freq_MWa_10Gyr}, but for $t=5$\,Gyr. \label{fig:stream_freq_MWa_5Gyr}}
\end{figure*}

We also present the frequency of the Hercules-like stream for $t=5.00$--5.05\,Gyr in 
Fig.~\ref{fig:stream_freq_MWa_5Gyr}. Among the detected streams, $\sim 60$\,\% are in an inter-arm region. 
Hercules-like streams most frequently appear at $R=7.5$\,kpc, while the OLR radii oscillate
between  
$R\sim7.5$ and 9\,kpc because of the oscillation of the pattern speed of the bar.

In $N$-body simulations the velocity distribution changes significantly on a 20--30\,Myr time-scale. 
Thus it is difficult to connect the pattern speed of the bar with the local velocity distributions 
and to determine the location of the Sun with respect to the resonance and the angle to the bar.
One reason is that in $N$-body simulations both the bar and spiral arms 
change their shape within a few tens of Myr. 
In contrast to studies in which the bar and spirals are modeled by analytic potentials
\citep{2017ApJ...840L...2P,2018arXiv180401920H},
the orbits of stars moving in the potential are chaotic, and therefore
it may be difficult to obtain a time independent result.

However, we find that if the Sun is located close to the OLR radii and 
in an inter-arm region, we were able to observe a Hercules-like stream in $\sim50$\,\% of the cases.
With more observational data obtained by Gaia
\citep{2016A&A...595A...1G,2018arXiv180409365G} and future simulations with an even higher resolution, 
we would be able to compare the simulations with observations 
of a larger region of the Galactic disc than just the solar neighbourhood.
Then we might be able to understand the dynamical origin of the velocity distribution structure.

\section{Summary}

We performed a series of $N$-body simulations of Milky Way models
and compared the final products at $t=10$\,Gyr with Milky Way Galaxy 
observations.
We found the set of parameters for the initial conditions which 
are important to reproduce the observed Milky Way galaxy. 
The stellar disc in our best-fitting model has a mass of
 $M_{\rm d}\sim 3.7 \times 10^{10}M_{\odot}$ and a central velocity dispersion 
of $\sigma_{R0}\sim 90$\,km\,s$^{-1}$. We configured the initial $Q$ value 
to be 1.2
and the initial disc scale height was set to 0.2\,kpc.
For the bulge, the initial bulge characteristic velocity of 
$\sigma_{\rm b}\sim 300$\,km\,s$^{-1}$ 
and bulge scale length of 1\,kpc.
The bulge mass is not a critical parameter, but if we assume a 
Hernquist model with the above scale length and velocity we get 
$M_{\rm b}=3$--$5 \times 10^{9}M_{\odot}$.
The mass ratio of the bulge-to-disc for the best fitting model is $\sim0.06$--0.15, 
which is somewhat higher than previous results, which found that the bulge mass 
should be less than 8\,\% of the disc mass 
in order to reproduce the observed bulge kinematics \citep{2010ApJ...720L..72S}.

In our simulations, the angular momentum of the halo is one of the most crucial parameters for 
reproducing the bar of the Milky Way. 
The best match with the Milky Way is obtained for a spin parameter of $\lambda\sim0.06$, which is 
larger than the mean spin-parameters measured for Milky-Way size dark matter
halos of cosmological $N$-body simulations, $\lambda=0.03$--0.04 
\citep{2007MNRAS.376..215B}, but consistent with the estimated value
for short-barred galaxies in SDSS DR7 \citep{2013ApJ...775...19C}.
Without an initial halo spin, we find that the bar that develops is too long and it 
rotates too slow compared to the bar in the Milky Way.

We also find that the final state of the galactic disc is sensitive to the fraction  of disc mass 
to the total mass of the galaxy ($f_{\rm d}$) in the initial conditions.
This parameter controls the bar instability in the sense that a 
larger value of $f_{\rm d}$ causes the bar to form earlier \citep{2018MNRAS.477.1451F}.
This parameter correlates with the radial velocity
dispersion of the disc at $R=8$\,kpc but anti-correlates with the total surface
density ($K_{z}/s\pi G$) at  $R=8$\,kpc. Comparing the bulge kinematics with observations
would suggest a larger value of $f_{\rm d}$ when we fix the disc mass and halo spin parameters.
Based on these results our models suggest that $f_{\rm d}\sim 0.45$ is preferable
for modelling the Milky Way galaxy.

The pattern speed of the bar for our best matching models at an age of $t=10$\,Gyr
is 40--45\,km\,s$^{-1}$\,kpc$^{-1}$.
This is similar to the velocity of a `slow' bar-model
in which the Sun is located between the co-rotation and outer Lindblad 
resonance radii \citep{2017ApJ...840L...2P}.
Except for the halo spin, the pattern speed of the bar also depends on $f_{\rm d}$. 
As $f_{\rm d}$ becomes smaller, the pattern speed at $t=10$\,Gyr becomes faster. 
For $f_{\rm d}\sim 0.45$, the expected pattern speed is 40--50\,km\,s$^{-1}$.
In addition, we did not see any model in which the bar's pattern speed 
was faster than $52$\,km\,s$^{-1}$ at $t=10$\,Gyr.

Based on the ``observations'' of our best-fitting models, we confirm
that we observe the inner-bar structure due to the projection effect
and the existence of the bulge \citep{2012ApJ...744L...8G}.

The velocity distribution in the solar neighbourhood experiences large
changes on a time scale of 20--30\,Myr. In at most 50\,\% of the snapshots, we
observed a velocity structure similar to the Hercules stream.  
The radius at which we most frequently see the Hercules-like stream 
is slightly further out than the outer Lindblad resonance.
If we assume that the Sun is located in an inter-arm region, 
then the frequency is at most $\sim 50$\,\%. 
These variations over time are the result of the complicated interactions between the spiral 
and bar structure as seen in $N$-body simulations of galactic discs 
\citep[e.g.,][]{2011MNRAS.417..762Q,2015MNRAS.454.2954B}.
It is thus difficult to give an estimate of the entire galactic disc structure 
based only on the phase-space distribution of solar neighbourhood stars.
Data covering a larger region would be required to make a more precise 
estimate of the Milky-Way Galaxy structure.

\section*{Acknowledgments}

We thank the referee for their useful comments.
This work was supported by JSPS KAKENHI Grant Number 26800108, 
17F17764, 18K03711, 
HPCI Strategic Program Field 5 'The origin of matter and the universe,' 
Initiative on Promotion of Supercomputing for Young or Women Researchers
Supercomputing Division Information Technology Center The University of Tokyo,
and the Netherlands Research School for Astronomy (NOVA).
Simulations are performed using GPU clusters, HA-PACS at the
University of Tsukuba, Piz Daint at CSCS and Little Green Machine II
(621.016.701).
This work was use of the {\tt AMUSE} framework. Initial development has been done using the Titan
computer Oak Ridge National Laboratory.  This work was supported by a
grant from the Swiss National Supercomputing Centre (CSCS) under
project ID s548 and s716.  This research used resources of the Oak Ridge
Leadership Computing Facility at the Oak Ridge National Laboratory,
which is supported by the Office of Science of the U.S. Department of
Energy under Contract No. DE-AC05-00OR22725 and by the European
Union's Horizon 2020 research and innovation programme under grant
agreement No 671564 (COMPAT project).

\appendix

\appendix

\section{Model parameters for all models}\label{Sect:AllModel}

We summarize the parameters for the initial conditions and final 
outcome for all models except for the models shown in the main text. 
We summarize the parameter sets for {\tt GalacICS} in Table~\ref{tb:A1} and 
the resulting mass and radius of the initial conditions and the number of 
particles in Table~\ref{tb:A2}.
Table \ref{tb:A3} contains the disc properties such as velocity 
dispersion, surface density, the pattern speed of the bar, and 
the resulting resonance radii at 10\,Gyr. 
In Table~\ref{tb:A4} we summarize the $\chi^2$ value for all models.

\begin{table*}
\scriptsize
\caption{Model parameters\label{tb:A1}}
\begin{tabular}{lccccccccccc}
\hline
           &  \multicolumn{4}{l}{Halo} &  \multicolumn{4}{l}{Disc} &  \multicolumn{3}{l}{Bulge} \\
Model &  $a_{\rm h}$ & $\sigma_{\rm h}$ & $\epsilon_{\rm h}$ & $\alpha_{\rm h}$& $M_{\rm d}$ & $R_{\rm d}$ & $z_{\rm d}$ & $\sigma_{\rm R0}$  & $a_{\rm b}$ & $\sigma_{\rm b}$ & $\epsilon_{\rm b}$ \\ 
   &  (kpc) & ($\kms$) &  &  & $(10^{10}M_{\odot})$ & (kpc) & (kpc) & ($\kms$)  & (kpc) & $(\kms)$\\
\hline \hline

0 & 12.5 & 400 & 0.8 & 0.8 & 4.1 & 2.6 & 0.3 & 94 & 0.73 & 333 & 0.95  \\ 
1 & 10.0 & 400 & 0.85 & 0.7 & 3.1 & 2.6 & 0.3 & 70 & 0.75 & 300 & 0.99  \\ 
2 & 12. & 409 & 0.8 & 0.8 & 4.1 & 2.6 & 0.2 & 91 & 0.75 & 300 & 0.95  \\ 
3 & 10.0 & 400 & 0.85 & 0.8 & 3.1 & 2.6 & 0.3 & 70 & 0.75 & 300 & 0.99  \\ 
4 & 10.0 & 400 & 0.85 & 0.8 & 3.1 & 2.6 & 0.3 & 70 & 0.75 & 300 & 0.99  \\ 
5 & 12. & 409 & 0.8 & 0.8 & 3.61 & 2.6 & 0.2 & 83 & 0.75 & 300 & 0.95  \\ 
6 & 12. & 390 & 0.8 & 0.8 & 4.1 & 2.6 & 0.2 & 90 & 1.2 & 280 & 0.95  \\ 
7 & 12. & 430 & 0.8 & 0.8 & 4.1 & 2.6 & 0.2 & 91 & 0.4 & 300 & 0.95  \\ 
8 & 14. & 420 & 0.82 & 0.8 & 3.61 & 2.6 & 0.2 & 85 & 0.78 & 280 & 0.95  \\ 
9 & 22. & 500 & 0.7 & 0.8 & 3.61 & 2.6 & 0.2 & 90 & 0.78 & 280 & 0.95  \\ 
10 & 12. & 450 & 0.8 & 0.8 & 4.1 & 2.6 & 0.2 & 91 & 0.4 & 330 & 0.8  \\ 
11 & 12. & 450 & 0.8 & 0.7 & 4.1 & 2.6 & 0.2 & 91 & 0.4 & 330 & 0.8  \\ 
12 & 10. & 440 & 0.85 & 0.6 & 4.1 & 2.6 & 0.2 & 87 & 0.2 & 400 & 0.8  \\ 
13 & 10. & 440 & 0.85 & 0.8 & 4.1 & 2.6 & 0.2 & 87 & 0.2 & 400 & 0.8  \\ 
14 & 10. & 440 & 0.85 & 0.8 & 4.1 & 2.6 & 0.2 & 87 & 0.2 & 400 & 0.8  \\ 
15 & 10.0 & 440 & 0.85 & 0.6 & 3.61 & 2.6 & 0.2 & 74 & 0.75 & 300 & 0.99  \\ 
16 & 10.0 & 440 & 0.85 & 0.7 & 3.61 & 2.6 & 0.2 & 74 & 0.75 & 300 & 0.99  \\ 
17 & 10.0 & 440 & 0.85 & 0.7 & 3.61 & 2.6 & 0.2 & 80 & 0.75 & 300 & 0.99  \\ 
18 & 10.0 & 440 & 0.85 & 0.8 & 3.61 & 2.6 & 0.2 & 74 & 0.75 & 300 & 0.99  \\ 
19 & 10.0 & 440 & 0.85 & 0.65 & 3.61 & 2.6 & 0.3 & 74 & 0.75 & 300 & 0.99  \\ 
20 & 12.0 & 500 & 0.8 & 0.65 & 3.84 & 2.6 & 0.3 & 75 & 0.75 & 300 & 0.95  \\ 
21 & 12.0 & 500 & 0.8 & 0.8 & 3.84 & 2.6 & 0.3 & 75 & 0.75 & 300 & 0.95  \\ 
22 & 10.0 & 440 & 0.85 & 0.8 & 3.61 & 2.4 & 0.3 & 80 & 0.75 & 300 & 0.99  \\ 
23 & 10.0 & 400 & 0.85 & 0.8 & 3.61 & 2.3 & 0.3 & 85 & 0.75 & 330 & 0.99  \\ 
24 & 10.0 & 440 & 0.85 & 0.8 & 3.61 & 2.3 & 0.3 & 85 & 0.75 & 270 & 0.99  \\ 
25 & 12.0 & 430 & 0.85 & 0.8 & 4.1 & 3.0 & 0.2 & 76 & 0.75 & 300 & 0.99  \\ 
26 & 18.0 & 530 & 0.75 & 0.8 & 4.1 & 3.0 & 0.2 & 75 & 0.75 & 300 & 0.99  \\ 
27 & 12.0 & 440 & 0.82 & 0.8 & 4.05 & 3.0 & 0.2 & 76 & 0.7 & 350 & 0.99  \\ 
28 & 18.0 & 550 & 0.75 & 0.8 & 4.1 & 3.0 & 0.2 & 75 & 0.75 & 360 & 0.99  \\ 
29 & 12.0 & 480 & 0.82 & 0.8 & 3.91 & 2.8 & 0.2 & 76 & 0.6 & 365 & 0.99  \\ 
30 & 12. & 450 & 0.8 & 0.8 & 4.08 & 2.6 & 0.2 & 88 & 0.4 & 350 & 0.8  \\ 
31 & 12. & 450 & 0.8 & 0.8 & 4.08 & 2.6 & 0.2 & 91 & 0.4 & 350 & 0.8  \\ 
32 & 12. & 450 & 0.8 & 0.8 & 3.89 & 2.6 & 0.2 & 85 & 0.4 & 370 & 0.8  \\ 
33 & 12. & 450 & 0.8 & 0.8 & 3.89 & 2.6 & 0.2 & 92 & 0.4 & 370 & 0.8  \\ 
34 & 12. & 450 & 0.8 & 0.8 & 4.01 & 2.6 & 0.2 & 87 & 0.4 & 360 & 0.8  \\ 
35 & 12. & 450 & 0.8 & 0.8 & 4.01 & 2.6 & 0.2 & 94 & 0.4 & 360 & 0.8  \\ 
36 & 12. & 459 & 0.8 & 0.8 & 3.91 & 2.6 & 0.2 & 86 & 0.4 & 400 & 0.8  \\ 
37 & 12. & 459 & 0.8 & 0.8 & 3.91 & 2.6 & 0.2 & 92 & 0.4 & 400 & 0.8  \\ 
38 & 12. & 459 & 0.8 & 0.8 & 3.91 & 2.6 & 0.2 & 92 & 0.4 & 400 & 0.8  \\ 
39 & 12. & 400 & 0.8 & 0.8 & 3.96 & 3.2 & 0.2 & 72 & 0.9 & 300 & 0.97  \\ 
40 & 12. & 400 & 0.8 & 0.8 & 3.96 & 2.9 & 0.2 & 80 & 0.9 & 300 & 0.8  \\ 
41 & 12. & 400 & 0.8 & 0.8 & 3.59 & 2.2 & 0.2 & 96 & 0.9 & 300 & 0.97  \\ 
42 & 12. & 400 & 0.8 & 0.8 & 3.59 & 2.6 & 0.2 & 82 & 0.9 & 300 & 0.97  \\ 
43 & 12. & 400 & 0.8 & 0.9 & 3.59 & 2.2 & 0.2 & 96 & 0.9 & 300 & 0.97  \\ 
44 & 12. & 400 & 0.8 & 0.9 & 3.59 & 2.2 & 0.2 & 105 & 0.9 & 300 & 0.97  \\ 
45 & 12. & 400 & 0.8 & 0.9 & 3.59 & 2.6 & 0.2 & 89 & 0.9 & 300 & 0.97  \\ 
46 & 12. & 400 & 0.8 & 0.8 & 3.59 & 2.4 & 0.2 & 82 & 0.9 & 300 & 0.97  \\ 
47 & 12. & 400 & 0.8 & 0.9 & 3.59 & 2.4 & 0.2 & 82 & 0.9 & 300 & 0.97  \\ 
48 & 12. & 400 & 0.8 & 0.8 & 3.59 & 2.4 & 0.2 & 82 & 0.9 & 300 & 0.97  \\ 
49 & 10. & 440 & 0.85 & 0.8 & 3.61 & 2.3 & 0.2 & 85 & 0.75 & 330 & 0.99  \\ 
50 & 10. & 420 & 0.85 & 0.8 & 3.61 & 2.3 & 0.2 & 88 & 0.75 & 330 & 0.99  \\ 
51 & 10. & 440 & 0.85 & 0.8 & 3.61 & 2.3 & 0.2 & 92 & 0.75 & 340 & 0.99  \\ 
\hline
\end{tabular}
\end{table*}

\begin{table*}
\scriptsize
\caption{Mass, radius, and the number of particles\label{tb:A2}}
\begin{tabular}{lcccccccccccc}
\hline
Model    & $M_{\rm d}$ & $M_{\rm b}$ & $M_{\rm h}$ & $M_{\rm b}/M_{\rm d}$ & $R_{\rm d, t}$ & $r_{\rm b, t}$ & $r_{\rm h, t}$ & $Q_0$  & $N_{\rm d}$ & $N_{\rm b}$ & $N_{\rm h}$ & $f_{\rm d}$\\ 
   & ($10^{10}M_{\odot}$) & ($10^{10}M_{\odot}$) & ($10^{10}M_{\odot}$) & &  (kpc) & (kpc) & (kpc) &  &   &  & & \\ 
\hline \hline
0 & 4.19 & 0.646 & 90.7 & 0.15 & 31.6 & 3.3 & 223.0 & 1.2  & 8.3M  & 1.3M  & 179M  & 0.514  \\ 
1 & 3.22 & 0.491 & 74.6 & 0.15 & 31.6 & 3.29 & 246.0 & 1.2  & 8.3M  & 1.3M  & 192M  & 0.396  \\ 
2 & 4.19 & 0.444 & 83.6 & 0.11 & 31.6 & 2.89 & 223.0 & 1.2  & 8.3M  & 0.9M  & 164M  & 0.496  \\ 
3 & 3.2 & 0.472 & 74.8 & 0.15 & 31.6 & 3.28 & 245.0 & 1.2  & 8.3M  & 1.3M  & 192M  & 0.405  \\ 
4 & 3.2 & 0.472 & 74.8 & 0.15 & 31.6 & 3.28 & 245.0 & 1.2  & 8.3M  & 1.3M  & 192M  & 0.405  \\ 
5 & 3.7 & 0.469 & 83.2 & 0.13 & 31.6 & 3.03 & 219.0 & 1.2  & 8.3M  & 0.7M  & 159M  & 0.473  \\ 
6 & 4.2 & 0.366 & 73.1 & 0.09 & 31.6 & 3.13 & 234.0 & 1.2  & 8.3M  & 0.8M  & 162M  & 0.478  \\ 
7 & 4.17 & 0.348 & 92.0 & 0.08 & 31.6 & 2.25 & 219.0 & 1.2  & 8.3M  & 0.7M  & 151M  & 0.521  \\ 
8 & 3.67 & 0.401 & 105.8 & 0.11 & 31.4 & 2.99 & 280.0 & 1.2  & 8.3M  & 0.6M  & 169M  & 0.523  \\ 
9 & 3.66 & 0.436 & 154.4 & 0.12 & 31.4 & 3.25 & 244.0 & 1.2  & 8.3M  & 1.0M  & 349M  & 0.588  \\ 
10 & 4.18 & 0.35 & 100.7 & 0.08 & 31.6 & 1.63 & 209.0 & 1.2  & 8.3M  & 0.7M  & 151M  & 0.492  \\ 
11 & 4.18 & 0.35 & 100.7 & 0.08 & 31.6 & 1.63 & 209.0 & 1.2  & 8.3M  & 0.7M  & 151M  & 0.492  \\ 
12 & 4.18 & 0.325 & 105.0 & 0.08 & 31.6 & 1.13 & 209.0 & 1.2  & 8.3M  & 0.7M  & 199M  & 0.507  \\ 
13 & 4.18 & 0.325 & 105.0 & 0.08 & 31.6 & 1.13 & 209.0 & 1.2  & 8.3M  & 0.7M  & 199M  & 0.507  \\ 
14 & 4.18 & 0.325 & 105.0 & 0.08 & 31.6 & 1.13 & 209.0 & 1.2  & 8.3M  & 0.7M  & 199M  & 0.507  \\ 
15 & 3.72 & 0.427 & 86.4 & 0.11 & 31.6 & 2.95 & 257.0 & 1.2  & 8.3M  & 1.0M  & 192M  & 0.393  \\ 
16 & 3.72 & 0.427 & 86.4 & 0.11 & 31.6 & 2.95 & 257.0 & 1.2  & 8.3M  & 1.0M  & 192M  & 0.393  \\ 
17 & 3.72 & 0.427 & 86.4 & 0.11 & 31.6 & 2.95 & 257.0 & 1.2  & 8.3M  & 1.0M  & 192M  & 0.393  \\ 
18 & 3.72 & 0.427 & 86.4 & 0.11 & 31.6 & 2.95 & 257.0 & 1.2  & 8.3M  & 1.0M  & 192M  & 0.393  \\ 
19 & 3.75 & 0.446 & 86.1 & 0.12 & 31.6 & 2.96 & 259.0 & 1.2  & 8.3M  & 1.0M  & 192M  & 0.393  \\ 
20 & 3.98 & 0.447 & 109.9 & 0.11 & 31.6 & 2.72 & 227.0 & 1.2  & 8.3M  & 1.0M  & 192M  & 0.386  \\ 
21 & 3.98 & 0.447 & 109.9 & 0.11 & 31.6 & 2.72 & 227.0 & 1.2  & 8.3M  & 1.0M  & 192M  & 0.386  \\ 
22 & 3.74 & 0.412 & 88.3 & 0.11 & 31.6 & 2.75 & 256.0 & 1.2  & 8.3M  & 0.9M  & 197M  & 0.42  \\ 
23 & 3.74 & 0.56 & 92.2 & 0.15 & 31.6 & 3.11 & 242.0 & 1.2  & 8.3M  & 0.9M  & 197M  & 0.429  \\ 
24 & 3.71 & 0.289 & 85.4 & 0.08 & 31.6 & 2.41 & 272.0 & 1.2  & 8.3M  & 0.6M  & 190M  & 0.421  \\ 
25 & 4.2 & 0.488 & 101.7 & 0.12 & 31.6 & 3.39 & 299.0 & 1.2  & 8.3M  & 0.9M  & 197M  & 0.421  \\ 
26 & 4.19 & 0.526 & 159.4 & 0.13 & 31.6 & 3.64 & 252.0 & 1.2  & 8.3M  & 0.9M  & 197M  & 0.438  \\ 
27 & 4.17 & 0.756 & 105.2 & 0.18 & 31.6 & 3.98 & 234.0 & 1.2  & 8.3M  & 1.5M  & 209M  & 0.428  \\ 
28 & 4.2 & 0.889 & 185.2 & 0.21 & 31.6 & 4.44 & 242.0 & 1.2  & 8.3M  & 1.0M  & 314M  & 0.44  \\ 
29 & 4.03 & 0.762 & 123.2 & 0.19 & 31.6 & 3.7 & 229.0 & 1.2  & 8.3M  & 1.5M  & 209M  & 0.423  \\ 
30 & 4.16 & 0.406 & 104.4 & 0.1 & 31.6 & 1.69 & 204.0 & 1.2  & 8.3M  & 0.8M  & 208M  & 0.499  \\ 
31 & 4.16 & 0.406 & 104.4 & 0.1 & 31.6 & 1.69 & 204.0 & 1.2  & 8.3M  & 0.8M  & 208M  & 0.499  \\ 
32 & 3.97 & 0.469 & 108.3 & 0.12 & 31.6 & 1.76 & 199.0 & 1.2  & 8.3M  & 1.0M  & 226M  & 0.497  \\ 
33 & 3.97 & 0.469 & 108.3 & 0.12 & 31.6 & 1.76 & 199.0 & 1.2  & 8.3M  & 1.0M  & 226M  & 0.497  \\ 
34 & 4.09 & 0.436 & 106.4 & 0.11 & 31.6 & 1.73 & 202.0 & 1.2  & 8.3M  & 0.9M  & 216M  & 0.499  \\ 
35 & 4.09 & 0.436 & 106.4 & 0.11 & 31.6 & 1.73 & 202.0 & 1.2  & 8.3M  & 0.9M  & 216M  & 0.499  \\ 
36 & 4.0 & 0.564 & 118.0 & 0.14 & 31.6 & 1.83 & 195.0 & 1.2  & 8.3M  & 1.2M  & 245M  & 0.497  \\ 
37 & 4.0 & 0.564 & 118.0 & 0.14 & 31.6 & 1.83 & 195.0 & 1.2  & 8.3M  & 1.2M  & 245M  & 0.497  \\ 
38 & 4.0 & 0.564 & 118.0 & 0.14 & 31.6 & 1.83 & 195.0 & 1.2  & 8.3M  & 1.2M  & 245M  & 0.497  \\ 
39 & 4.22 & 0.55 & 76.1 & 0.13 & 31.6 & 3.73 & 221.0 & 1.2  & 8.3M  & 1.1M  & 150M  & 0.412  \\ 
40 & 4.27 & 0.514 & 77.7 & 0.12 & 31.6 & 3.52 & 220.0 & 1.2  & 8.3M  & 0.8M  & 100M  & 0.447  \\ 
41 & 3.66 & 0.434 & 82.5 & 0.12 & 31.6 & 3.05 & 217.0 & 1.2  & 8.3M  & 1.0M  & 187M  & 0.525  \\ 
42 & 3.68 & 0.497 & 79.5 & 0.13 & 31.6 & 3.41 & 220.0 & 1.2  & 8.3M  & 1.1M  & 179M  & 0.467  \\ 
43 & 3.66 & 0.434 & 82.5 & 0.12 & 31.6 & 3.05 & 217.0 & 1.2  & 8.3M  & 1.0M  & 187M  & 0.525  \\ 
44 & 3.66 & 0.434 & 82.5 & 0.12 & 31.6 & 3.05 & 217.0 & 1.2  & 8.3M  & 1.0M  & 187M  & 0.525  \\ 
45 & 3.68 & 0.497 & 79.5 & 0.13 & 31.6 & 3.41 & 220.0 & 1.2  & 8.3M  & 1.1M  & 179M  & 0.467  \\ 
46 & 3.67 & 0.467 & 80.9 & 0.13 & 31.6 & 3.24 & 219.0 & 1.2  & 8.3M  & 1.1M  & 182M  & 0.495  \\ 
47 & 3.67 & 0.467 & 80.9 & 0.13 & 31.6 & 3.24 & 219.0 & 1.2  & 8.3M  & 1.1M  & 182M  & 0.495  \\ 
48 & 3.67 & 0.467 & 80.9 & 0.13 & 31.6 & 3.24 & 219.0 & 1.2  & 8.3M  & 1.1M  & 182M  & 0.495  \\ 
49 & 3.71 & 0.535 & 92.5 & 0.14 & 31.6 & 3.09 & 242.0 & 1.2  & 8.3M  & 1.2M  & 206M  & 0.44  \\ 
50 & 3.71 & 0.535 & 92.5 & 0.14 & 31.6 & 3.09 & 242.0 & 1.2  & 8.3M  & 1.2M  & 206M  & 0.44  \\ 
51 & 3.71 & 0.588 & 94.0 & 0.16 & 31.6 & 3.21 & 238.0 & 1.2  & 8.3M  & 1.3M  & 209M  & 0.441  \\ 
\hline
\end{tabular}
\end{table*}

\begin{table*}
\scriptsize
\caption{Disc properties for the simulated galaxies at 10\,Gyr\label{tb:A3}}
\begin{tabular}{lccccccccc}
\hline

Model   & $\Sigma_{\rm 8kpc}$ & $\sigma_{R,{\rm 8kpc}}$ & $\sigma_{z,{\rm 8kpc}}$ & $K_{z,{\rm 8kpc}}$ & $V_{\rm c,8kpc}$ & $\Omega _{\rm b}$ & $R_{\rm CR}$ & $R_{\rm OLR}$ & $R_{\rm ILR}$\\
   & ($M_{\odot}\,{\rm kpc}^{-2}$) & (km\,s$^{-1}$) & (km\,s$^{-1}$) & ($2\pi G \, M_{\odot}\,{\rm kpc}^{-2}$) & (km\,s$^{-1}$) & (km\,s$^{-1}$\,kpc$^{-1}$) & (kpc) & (kpc) & (kpc) \\
\hline \hline
0 & 53.6 & 48.9 & 20.9 & 69.3 & 212.9 & 36.0 & 5.6 & 11.2 & 2.2  \\ 
1 & 40.3 & 28.3 & 16.1 & 66.6 & 226.7 & 37.9 & 5.5 & 10.7 & 1.7  \\ 
2 & 56.3 & 47.4 & 17.4 & 76.6 & 223.5 & 41.9 & 4.8 & 9.6 & 1.9  \\ 
3 & 41.2 & 29.3 & 16.2 & 65.0 & 226.6 & 45.2 & 4.4 & 8.8 & 1.4  \\ 
4 & 40.2 & 28.9 & 16.0 & 65.2 & 226.8 & 42.5 & 4.7 & 9.4 & 1.5  \\ 
5 & 50.6 & 47.3 & 16.5 & 66.8 & 216.5 & 40.9 & 4.7 & 9.6 & 1.9  \\ 
6 & 48.8 & 46.9 & 16.9 & 72.0 & 224.8 & 35.9 & 6.0 & 11.0 & 2.2  \\ 
7 & 55.6 & 46.7 & 16.7 & 74.1 & 220.2 & 44.9 & 4.3 & 8.6 & 1.7  \\ 
8 & 48.8 & 46.1 & 16.6 & 67.0 & 211.2 & 39.4 & 4.8 & 10.1 & 2.0  \\ 
9 & 48.8 & 50.7 & 17.7 & 60.8 & 193.0 & 33.7 & 5.3 & 12.0 & 2.3  \\ 
10 & 54.9 & 43.6 & 16.6 & 74.8 & 233.1 & 48.8 & 4.0 & 8.4 & 1.6  \\ 
11 & 56.2 & 42.5 & 16.4 & 78.0 & 234.2 & 41.3 & 4.8 & 10.7 & 1.8  \\ 
12 & 58.2 & 42.3 & 16.9 & 78.5 & 231.4 & 34.2 & 6.1 & 14.0 & 2.1  \\ 
13 & 60.1 & 43.0 & 17.2 & 80.9 & 231.8 & 25.7 & 9.5 & 19.4 & 2.3  \\ 
14 & 57.8 & 43.5 & 17.2 & 78.8 & 231.0 & 44.2 & 4.4 & 9.9 & 1.7  \\ 
15 & 46.5 & 32.1 & 14.9 & 77.7 & 243.8 & 24.2 & 10.4 & 18.5 & 3.0  \\ 
16 & 47.1 & 30.8 & 14.4 & 79.9 & 248.2 & 47.0 & 4.6 & 9.4 & 1.4  \\ 
17 & 45.9 & 31.7 & 14.2 & 75.3 & 248.2 & 39.0 & 5.9 & 11.3 & 1.8  \\ 
18 & 46.0 & 32.5 & 14.6 & 79.3 & 247.1 & 50.5 & 4.3 & 8.7 & 1.4  \\ 
19 & 45.8 & 28.0 & 17.0 & 77.9 & 249.4 & 44.7 & 5.0 & 9.8 & 1.4  \\ 
20 & 50.6 & 28.5 & 17.5 & 84.2 & 259.9 & 45.5 & 5.0 & 10.6 & 1.5  \\ 
21 & 50.7 & 28.6 & 17.8 & 82.2 & 260.0 & 51.4 & 4.3 & 9.1 & 1.3  \\ 
22 & 44.5 & 30.6 & 17.1 & 75.6 & 250.3 & 47.5 & 4.7 & 9.3 & 1.4  \\ 
23 & 44.9 & 40.4 & 18.9 & 69.2 & 232.1 & 42.0 & 5.1 & 10.0 & 1.9  \\ 
24 & 42.1 & 30.6 & 17.1 & 74.5 & 251.1 & 46.9 & 4.9 & 9.2 & 1.3  \\ 
25 & 55.9 & 41.1 & 16.5 & 78.1 & 229.8 & 43.3 & 4.7 & 9.6 & 1.7  \\ 
26 & 61.0 & 47.6 & 18.2 & 78.9 & 223.1 & 40.8 & 4.8 & 10.8 & 1.9  \\ 
27 & 56.5 & 47.6 & 17.5 & 76.3 & 223.4 & 40.0 & 5.0 & 10.3 & 2.0  \\ 
28 & 54.2 & 52.8 & 18.9 & 72.4 & 217.5 & 38.5 & 5.2 & 12.2 & 2.2  \\ 
29 & 56.6 & 43.9 & 17.2 & 78.4 & 235.0 & 44.0 & 4.7 & 10.0 & 1.8  \\ 
30 & 58.0 & 44.6 & 17.1 & 75.1 & 230.6 & 45.1 & 4.3 & 9.4 & 1.7  \\ 
31 & 57.1 & 45.5 & 16.9 & 78.7 & 231.0 & 46.2 & 4.1 & 9.3 & 1.6  \\ 
32 & 57.9 & 45.6 & 17.6 & 76.2 & 224.4 & 22.6 & 10.7 & 21.4 & 2.8  \\ 
33 & 55.0 & 45.3 & 16.7 & 76.1 & 226.5 & 45.0 & 4.2 & 9.1 & 1.7  \\ 
34 & 58.4 & 45.9 & 17.5 & 76.4 & 228.4 & 43.6 & 4.4 & 9.6 & 1.8  \\ 
35 & 58.4 & 44.6 & 17.2 & 76.0 & 228.4 & 43.3 & 4.5 & 9.9 & 1.7  \\ 
36 & 58.7 & 48.6 & 17.5 & 75.6 & 225.5 & 42.2 & 4.5 & 10.0 & 1.9  \\ 
37 & 56.3 & 47.4 & 17.0 & 75.8 & 226.6 & 45.1 & 4.2 & 8.8 & 1.7  \\ 
38 & 58.3 & 46.1 & 17.0 & 76.1 & 226.9 & 41.9 & 4.6 & 10.1 & 1.9  \\ 
39 & 53.1 & 39.2 & 15.9 & 73.9 & 218.2 & 39.0 & 5.1 & 10.6 & 1.8  \\ 
40 & 53.6 & 42.9 & 16.2 & 77.3 & 224.3 & 44.5 & 4.4 & 9.2 & 1.7  \\ 
41 & 46.8 & 49.9 & 16.8 & 66.0 & 220.1 & 41.4 & 4.8 & 29.9 & 2.0  \\ 
42 & 48.6 & 44.1 & 16.5 & 68.2 & 217.4 & 39.3 & 5.0 & 9.9 & 1.9  \\ 
43 & 46.9 & 50.6 & 17.5 & 67.2 & 220.1 & 44.1 & 4.5 & 8.6 & 1.9  \\ 
44 & 45.7 & 50.6 & 16.9 & 66.2 & 220.3 & 44.3 & 4.6 & 29.6 & 1.9  \\ 
45 & 49.0 & 42.2 & 15.8 & 71.5 & 219.0 & 44.3 & 4.4 & 8.5 & 1.7  \\ 
46 & 47.6 & 49.1 & 16.8 & 68.0 & 217.6 & 36.2 & 5.5 & 10.7 & 2.2  \\ 
47 & 46.5 & 52.7 & 17.5 & 66.9 & 215.6 & 41.3 & 4.8 & 9.1 & 2.2  \\ 
48 & 48.5 & 47.9 & 17.0 & 69.2 & 218.1 & 38.3 & 5.2 & 29.8 & 2.1  \\ 
49 & 45.5 & 36.4 & 14.8 & 72.4 & 247.8 & 49.6 & 4.4 & 8.9 & 1.6  \\ 
50 & 46.3 & 41.9 & 15.7 & 72.5 & 238.9 & 47.7 & 4.5 & 9.1 & 1.7  \\ 
51 & 45.4 & 36.4 & 14.9 & 76.3 & 247.5 & 34.0 & 7.1 & 13.4 & 1.9  \\ 
\hline
\end{tabular}
\end{table*}

\begin{table*}
\scriptsize
\caption{$\chi^2$ for $\Sigma_{\rm 8kpc}$, $\sigma_{R,{\rm 8kpc}}$, $K_{z,{\rm 8kpc}}$, $v_{\rm los}$, and $\sigma_{\rm los}$ \label{tb:A4}}
\begin{tabular}{lcccccc}
\hline
Model   & ${\Sigma_{\rm 8kpc}}$ & $\sigma_{R,{\rm 8kpc}}$ & $K_{z,{\rm 8kpc}}$ & $v_{\rm los}$ & $\sigma_{\rm los}$ & total\\
\hline \hline
0 & 3.6 & 7.8 & 0.02 & 2.0 & 4.3 & 17.6  \\ 
1 & 4.0 & 1.8 & 0.5 & 1.9 & 3.0 & 11.3  \\ 
2 & 7.4 & 6.2 & 1.7 & 2.3 & 1.2 & 18.7  \\ 
3 & 3.0 & 1.3 & 1.0 & 3.1 & 5.4 & 13.8  \\ 
4 & 4.2 & 1.5 & 0.9 & 2.6 & 4.5 & 13.7  \\ 
5 & 1.0 & 6.0 & 0.4 & 1.9 & 1.9 & 11.4  \\ 
6 & 0.3 & 5.7 & 0.2 & 2.3 & 3.1 & 11.5  \\ 
7 & 6.2 & 5.5 & 0.7 & 2.2 & 1.9 & 16.5  \\ 
8 & 0.3 & 4.9 & 0.4 & 2.2 & 1.6 & 9.3  \\ 
9 & 0.2 & 9.8 & 3.4 & 2.0 & 1.1 & 16.6  \\ 
10 & 5.3 & 3.0 & 0.9 & 2.4 & 3.9 & 15.5  \\ 
11 & 7.2 & 2.2 & 2.6 & 1.9 & 1.9 & 15.7  \\ 
12 & 10.6 & 2.2 & 2.9 & 2.0 & 1.4 & 19.2  \\ 
13 & 14.6 & 2.6 & 4.8 & 2.0 & 2.4 & 26.4  \\ 
14 & 9.8 & 2.9 & 3.1 & 2.0 & 2.4 & 20.2  \\ 
15 & 0.03 & 0.3 & 2.4 & 2.2 & 20.3 & 25.3  \\ 
16 & 0.0 & 0.7 & 3.9 & 3.4 & 4.2 & 12.2  \\ 
17 & 0.1 & 0.4 & 1.1 & 1.8 & 2.9 & 6.4  \\ 
18 & 0.1 & 0.3 & 3.5 & 3.1 & 4.8 & 11.7  \\ 
19 & 0.1 & 2.0 & 2.5 & 2.6 & 5.7 & 12.9  \\ 
20 & 1.1 & 1.7 & 8.1 & 2.4 & 6.5 & 19.7  \\ 
21 & 1.1 & 1.7 & 5.9 & 3.3 & 6.9 & 18.9  \\ 
22 & 0.6 & 0.8 & 1.2 & 5.8 & 3.5 & 11.9  \\ 
23 & 0.4 & 1.2 & 0.02 & 3.0 & 2.1 & 6.7  \\ 
24 & 2.2 & 0.8 & 0.8 & 6.0 & 3.5 & 13.3  \\ 
25 & 6.7 & 1.5 & 2.6 & 2.1 & 2.7 & 15.6  \\ 
26 & 16.6 & 6.3 & 3.2 & 2.1 & 1.5 & 29.7  \\ 
27 & 7.6 & 6.4 & 1.6 & 1.9 & 2.4 & 19.8  \\ 
28 & 4.3 & 12.7 & 0.2 & 1.9 & 6.8 & 26.0  \\ 
29 & 7.9 & 3.2 & 2.8 & 2.0 & 1.7 & 17.6  \\ 
30 & 10.3 & 3.7 & 1.0 & 2.1 & 1.8 & 18.9  \\ 
31 & 8.7 & 4.4 & 3.0 & 2.1 & 2.9 & 21.1  \\ 
32 & 10.1 & 4.5 & 1.6 & 2.1 & 2.3 & 20.6  \\ 
33 & 5.4 & 4.2 & 1.5 & 2.1 & 2.3 & 15.6  \\ 
34 & 11.1 & 4.8 & 1.7 & 2.1 & 1.5 & 21.1  \\ 
35 & 11.0 & 3.7 & 1.4 & 2.1 & 1.7 & 20.0  \\ 
36 & 11.7 & 7.4 & 1.2 & 2.1 & 1.1 & 23.5  \\ 
37 & 7.3 & 6.2 & 1.4 & 2.0 & 1.6 & 18.5  \\ 
38 & 10.8 & 4.9 & 1.5 & 2.4 & 1.1 & 20.7  \\ 
39 & 3.1 & 0.7 & 0.6 & 1.9 & 1.7 & 8.0  \\ 
40 & 3.7 & 2.5 & 2.2 & 3.3 & 3.6 & 15.2  \\ 
41 & 0.01 & 8.9 & 0.6 & 2.1 & 1.6 & 13.4  \\ 
42 & 0.2 & 3.3 & 0.1 & 2.0 & 1.4 & 7.1  \\ 
43 & 0.0 & 9.7 & 0.3 & 2.5 & 1.1 & 13.6  \\ 
44 & 0.2 & 9.7 & 0.6 & 2.7 & 1.2 & 14.3  \\ 
45 & 0.3 & 2.1 & 0.1 & 2.2 & 1.3 & 6.0  \\ 
46 & 0.02 & 7.9 & 0.2 & 2.1 & 2.4 & 12.6  \\ 
47 & 0.03 & 12.6 & 0.4 & 2.1 & 1.5 & 16.6  \\ 
48 & 0.2 & 6.7 & 0.03 & 1.9 & 1.8 & 10.7  \\ 
49 & 0.2 & 0.1 & 0.2 & 2.4 & 2.0 & 5.0  \\ 
50 & 0.1 & 1.9 & 0.3 & 3.5 & 1.2 & 7.0  \\ 
51 & 0.3 & 0.1 & 1.6 & 2.3 & 1.5 & 5.8  \\ 
\hline
\end{tabular}
\end{table*}


\section{The effect of the particle resolution}\label{Sect:Resolution}

It is known that the number of particles ($N$) affects the results. 
\citet{2009ApJ...697..293D} showed that the bar formation epoch is delayed if 
the number of particles increases, because a bar develops from particle noise 
which in turn depends on $N$. In simulations of spiral galaxies, the maximum 
and final spiral amplitude depends on $N$ because of the same reason~\citep{2011ApJ...730..109F}.

Even if the number of particles increases, the dynamical evolution of galactic 
discs may change when we use a different random seed to generate the initial 
conditions~\citep{2009MNRAS.398.1279S}. We, therefore, test the convergence of the results 
by performing multiple simulations for the same Galaxy model. For this model 
we vary the number of particles and the random seed. 
The model parameters for this test are summarized in Tables~\ref{tb:B1} and \ref{tb:B2}. 
We use $N\sim 0.8$M, 8M, and 80M particles for the disc. The total number of particles 
for each model then becomes $\sim 12$M, 120M, and 1.2B, respectively. For each $N$ we
use four different random seeds to generate four different realizations.

\begin{figure*}
\epsscale{.4}
\plotone{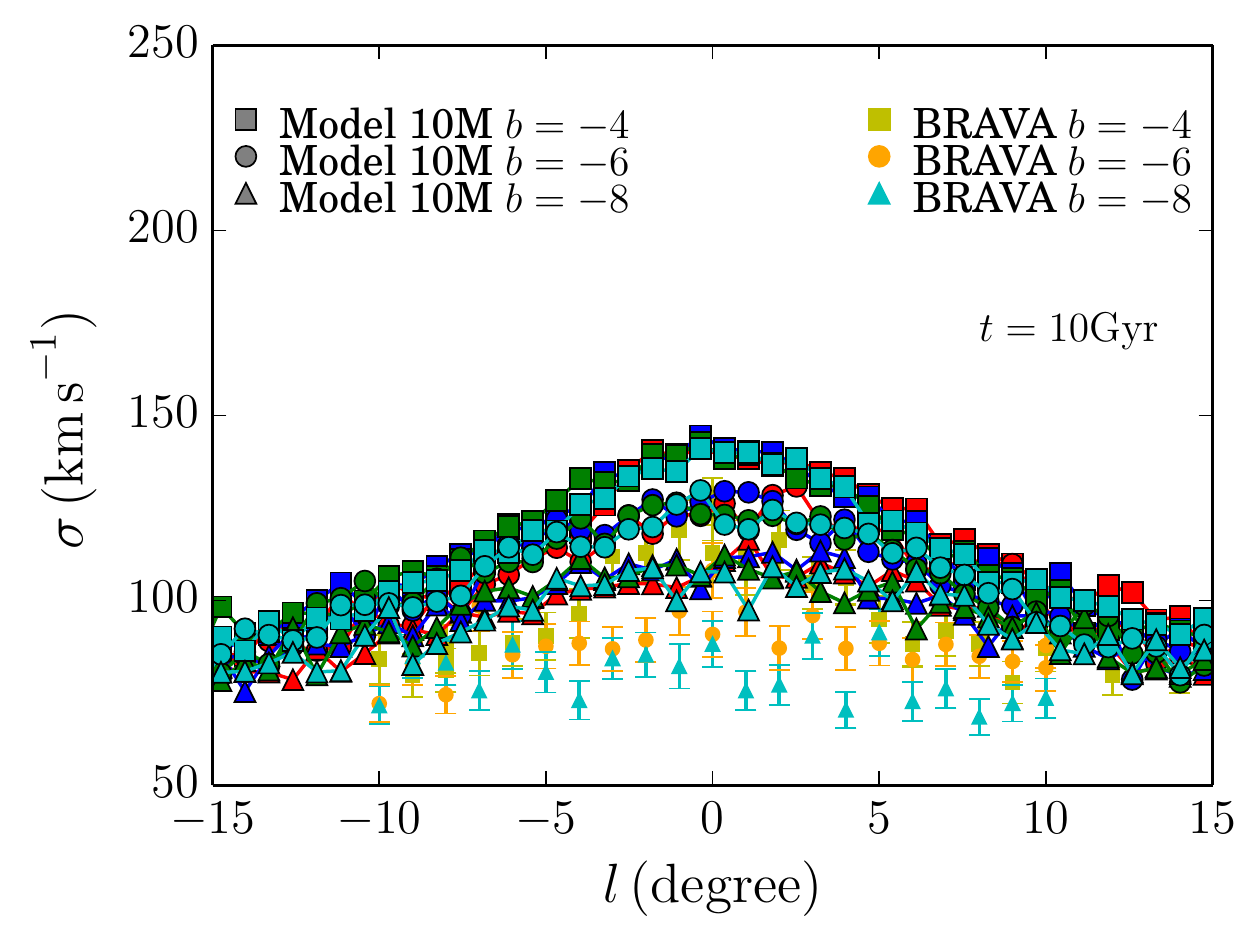}\plotone{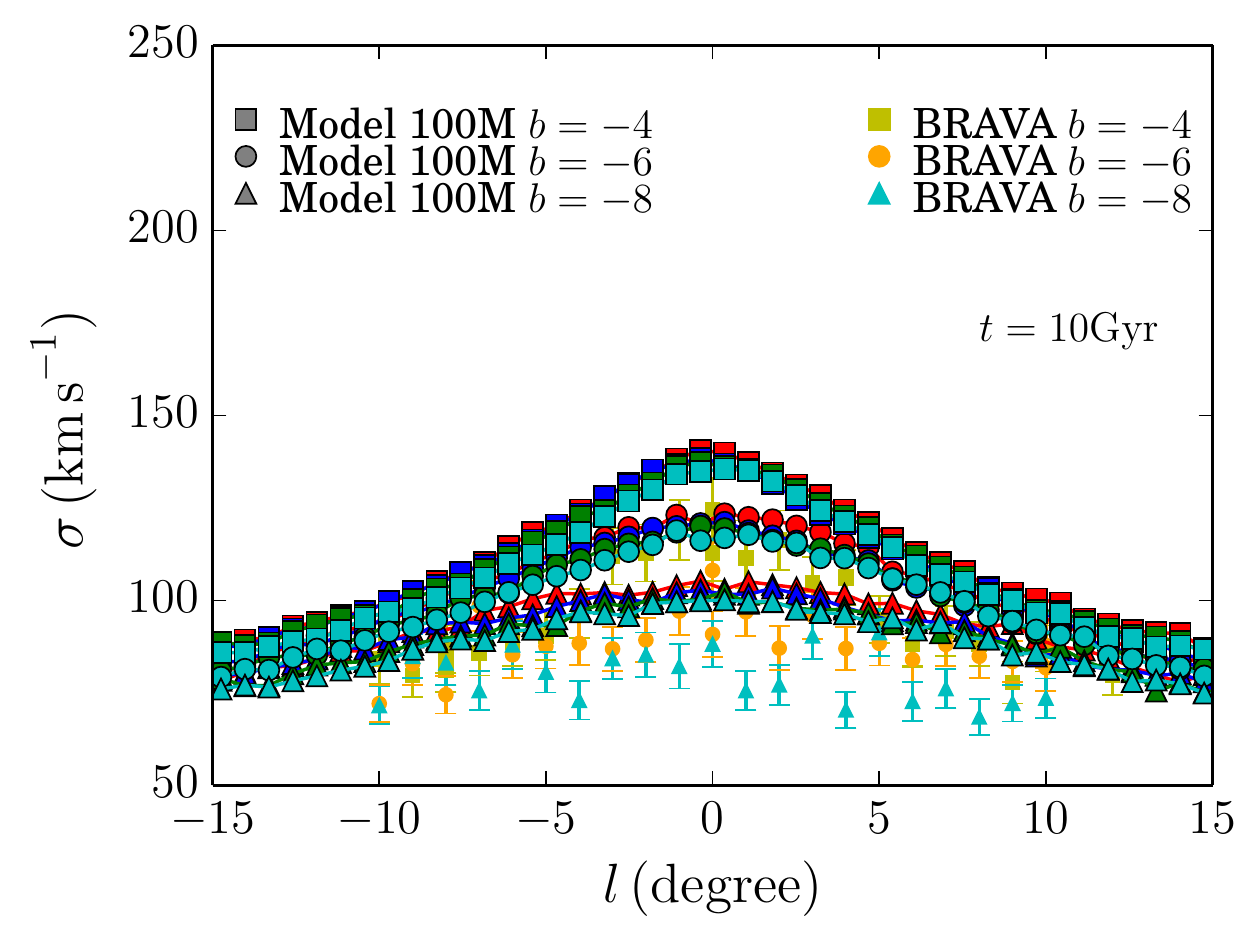}\plotone{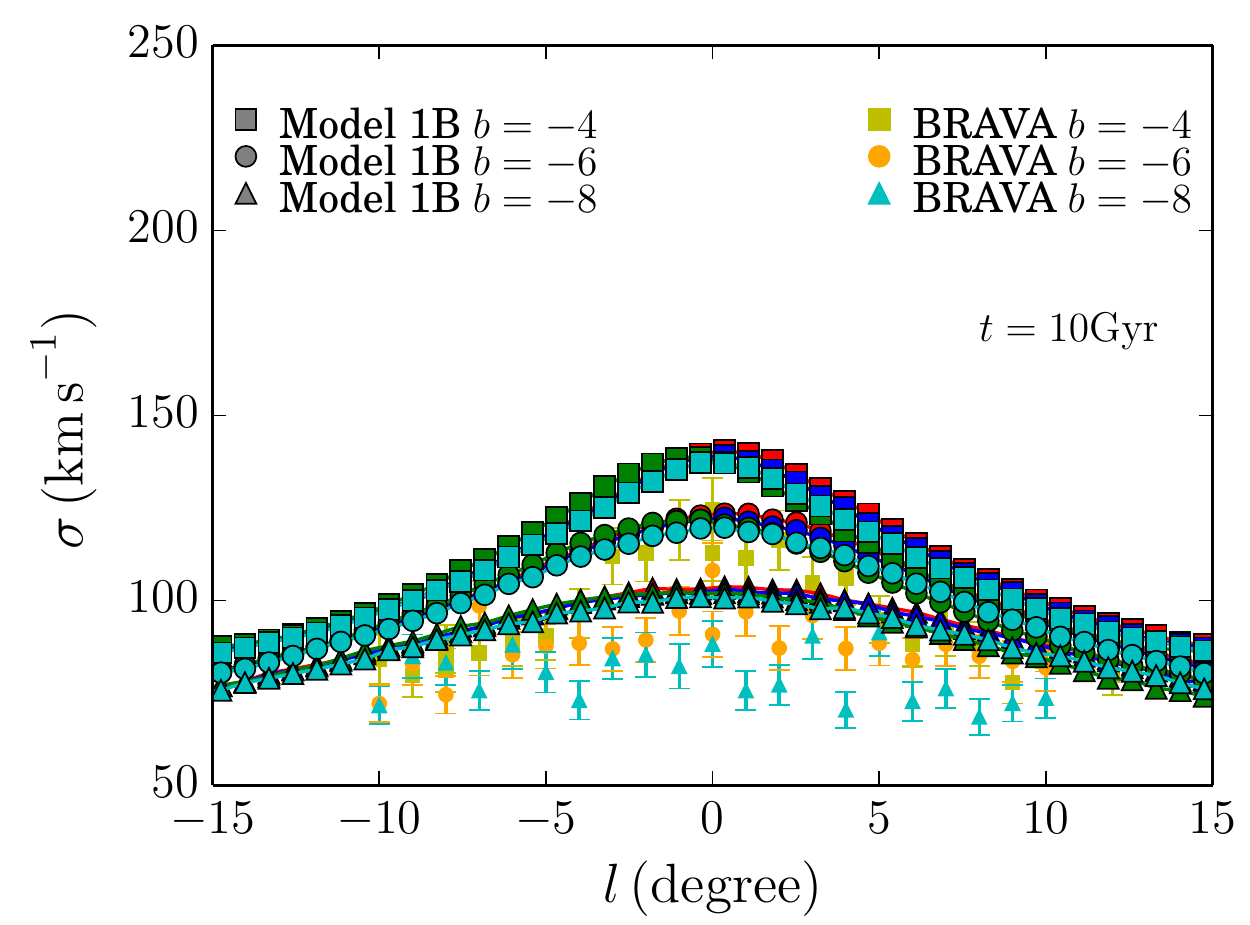}\\
\plotone{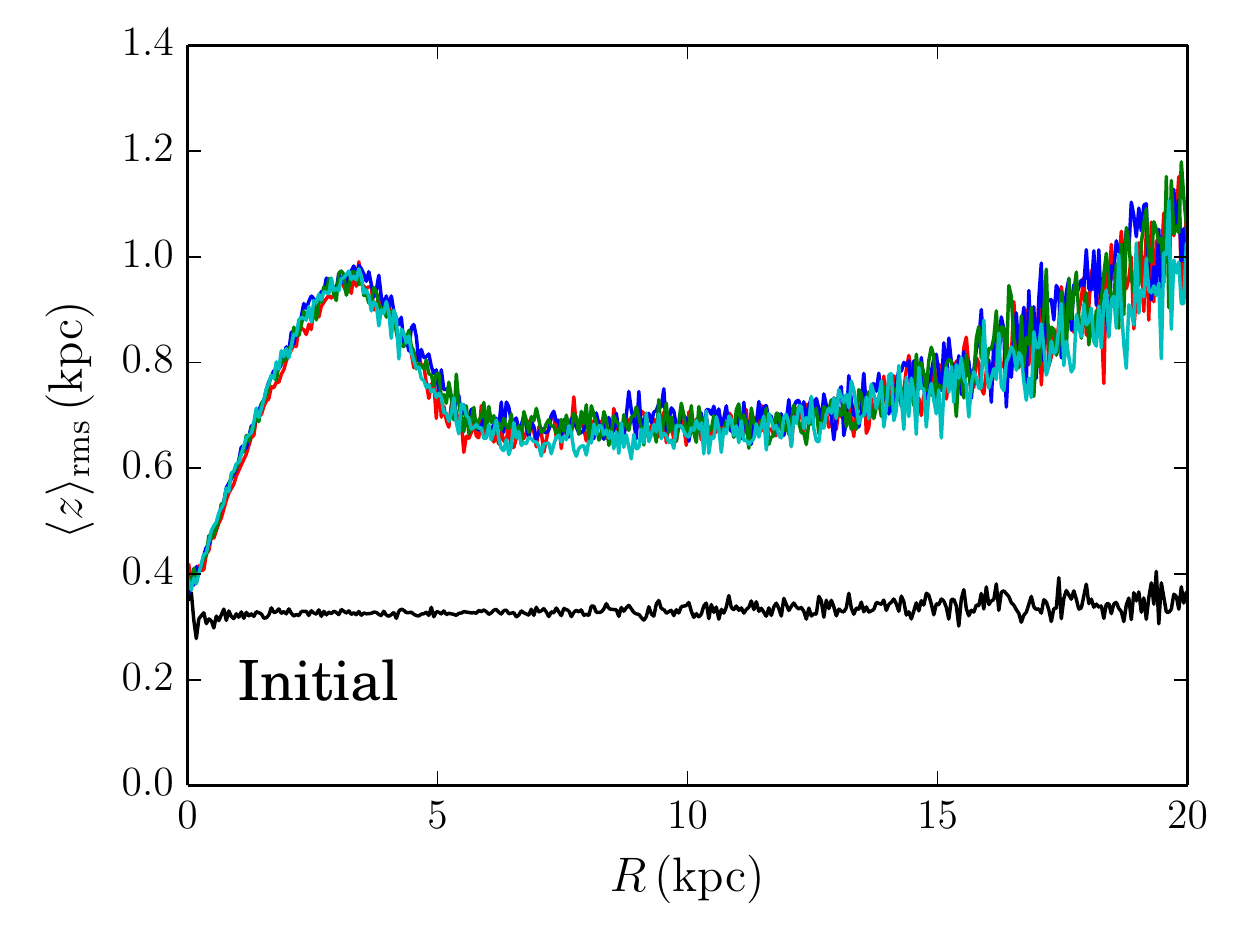}\plotone{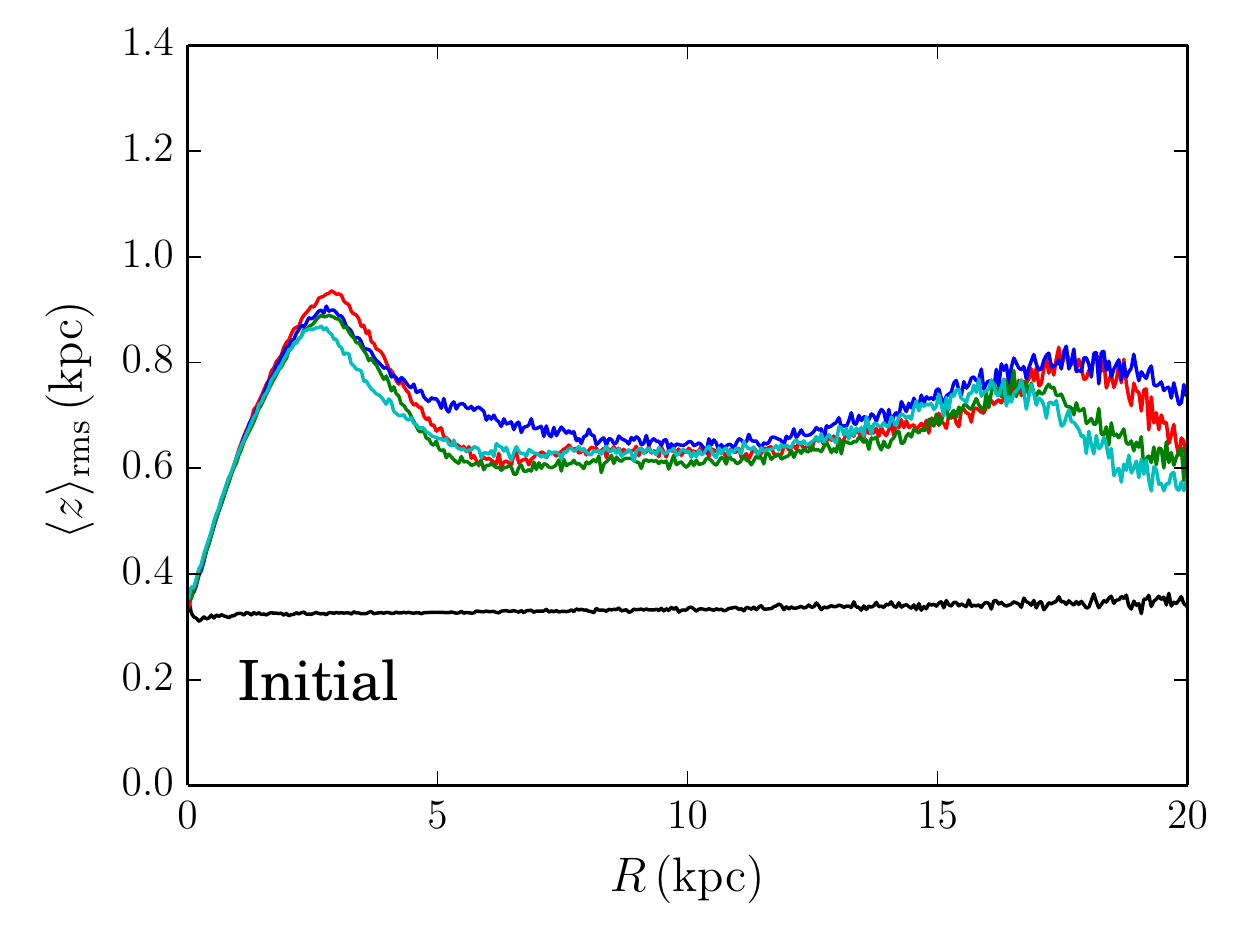}\plotone{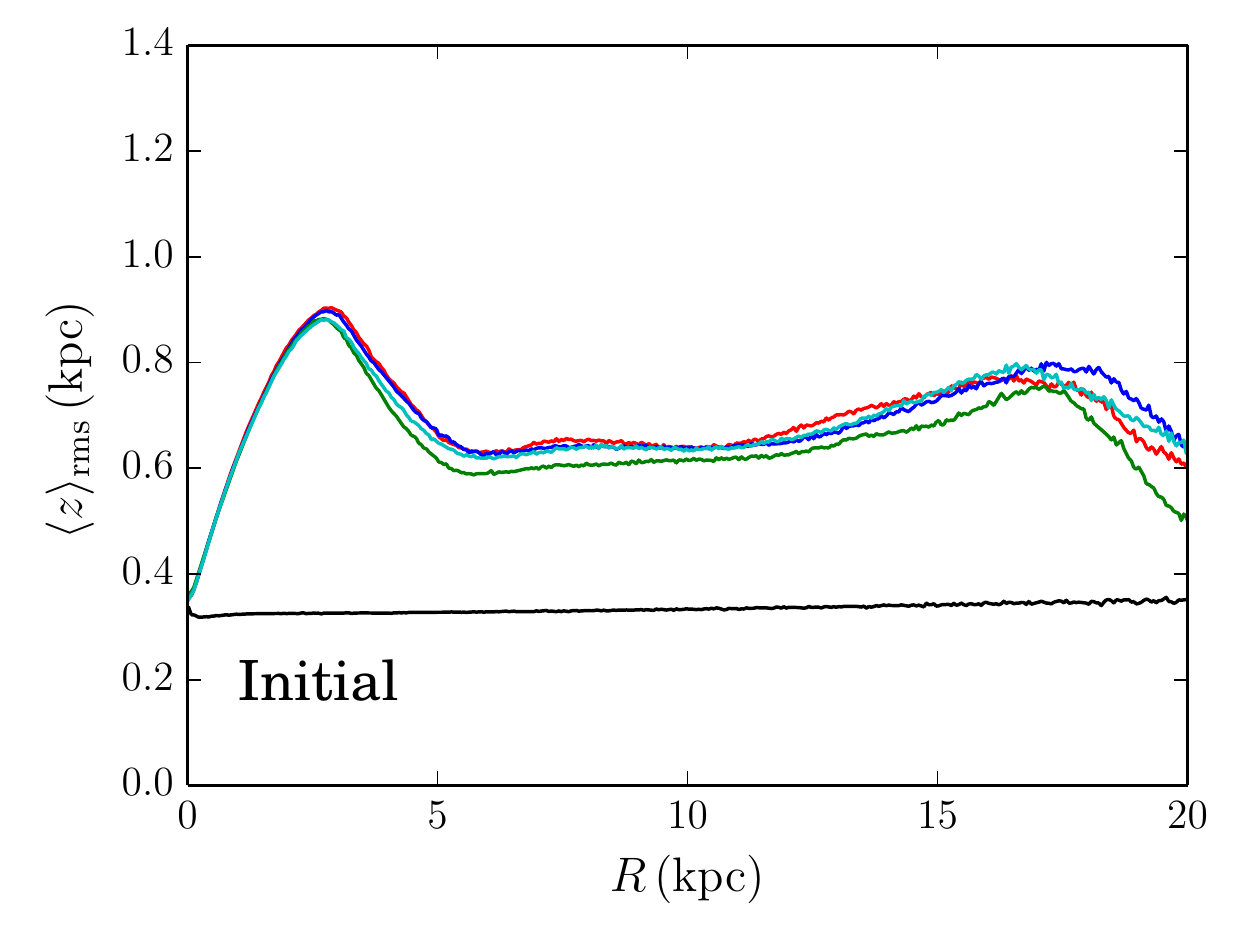}\\
\plotone{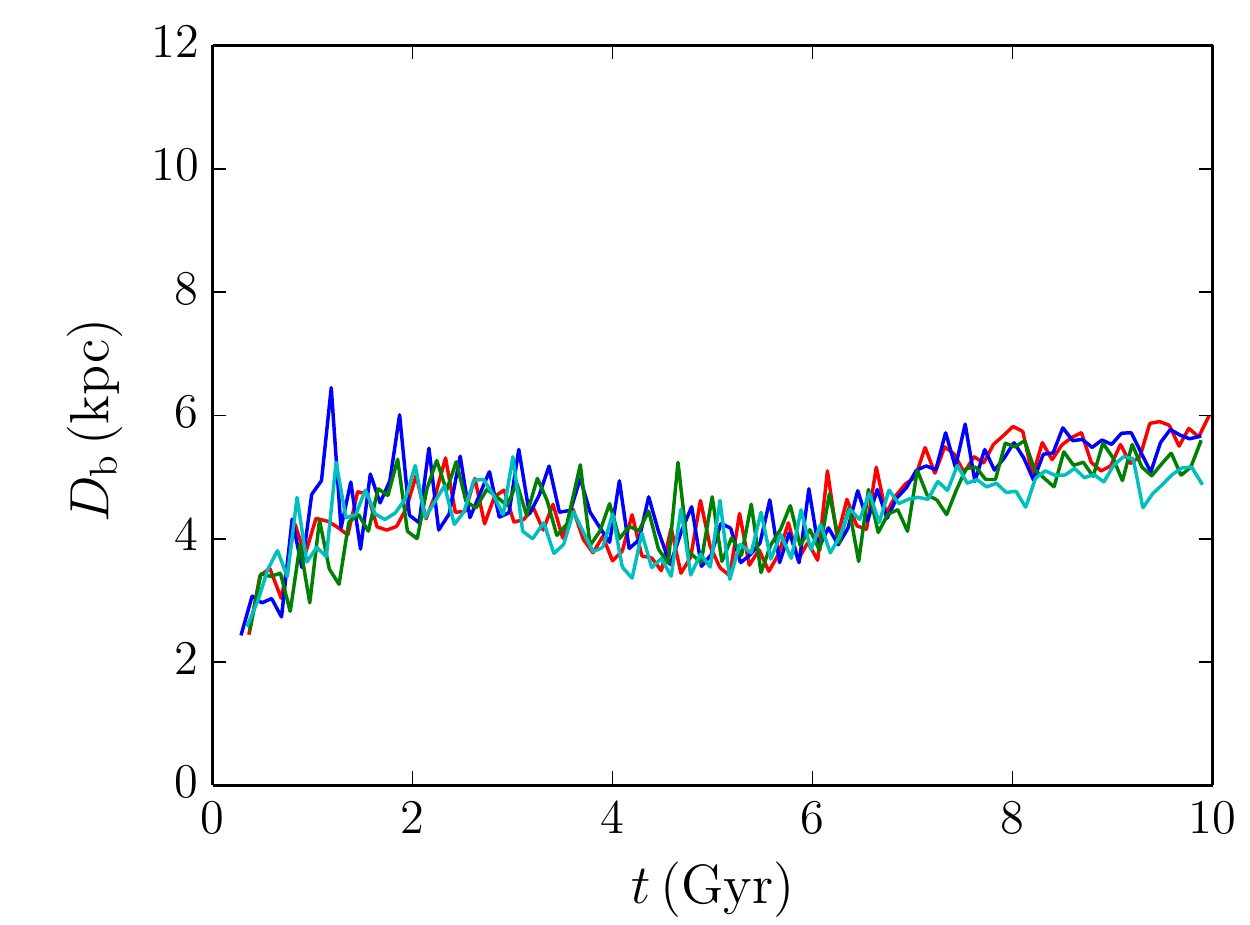}\plotone{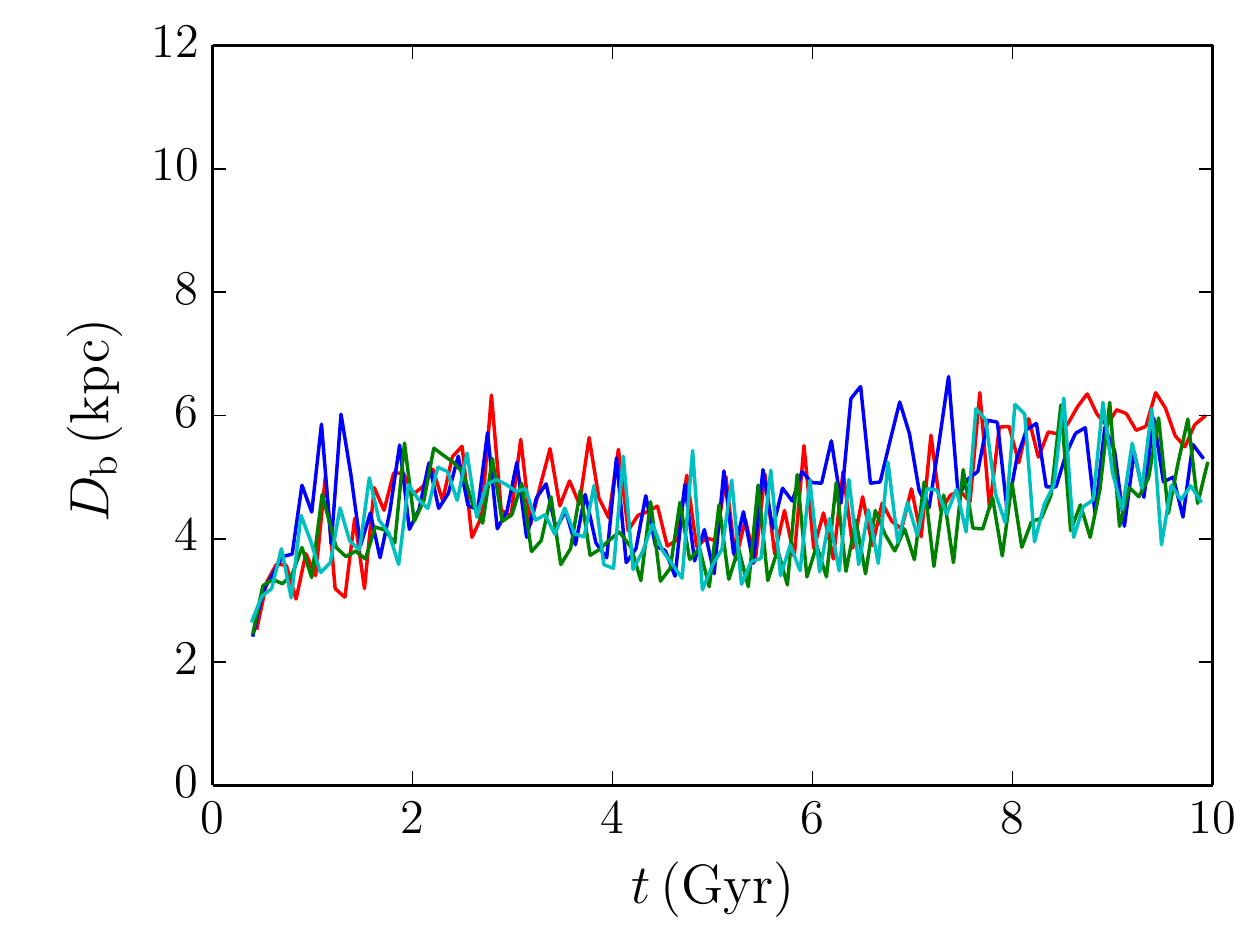}\plotone{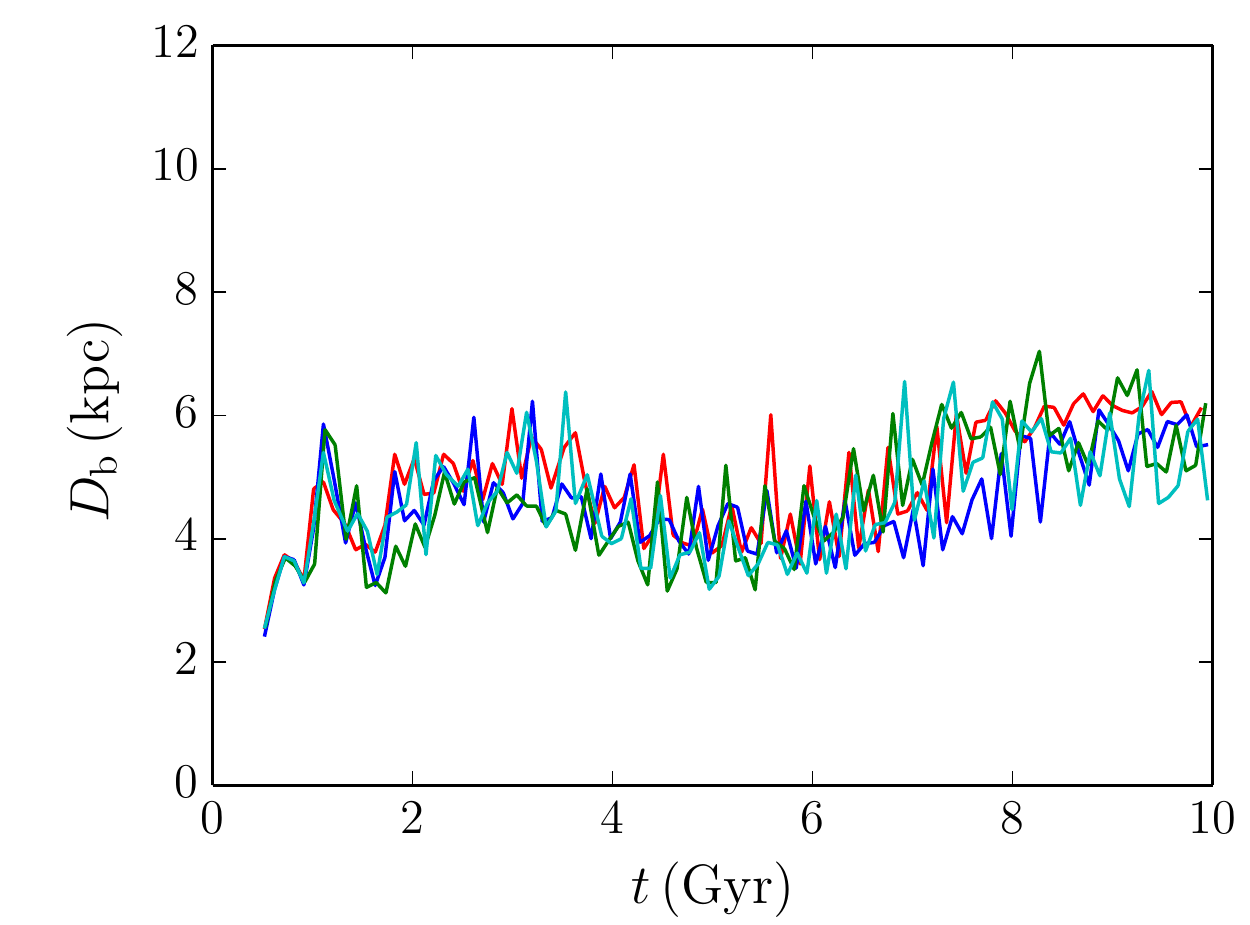}\\
\plotone{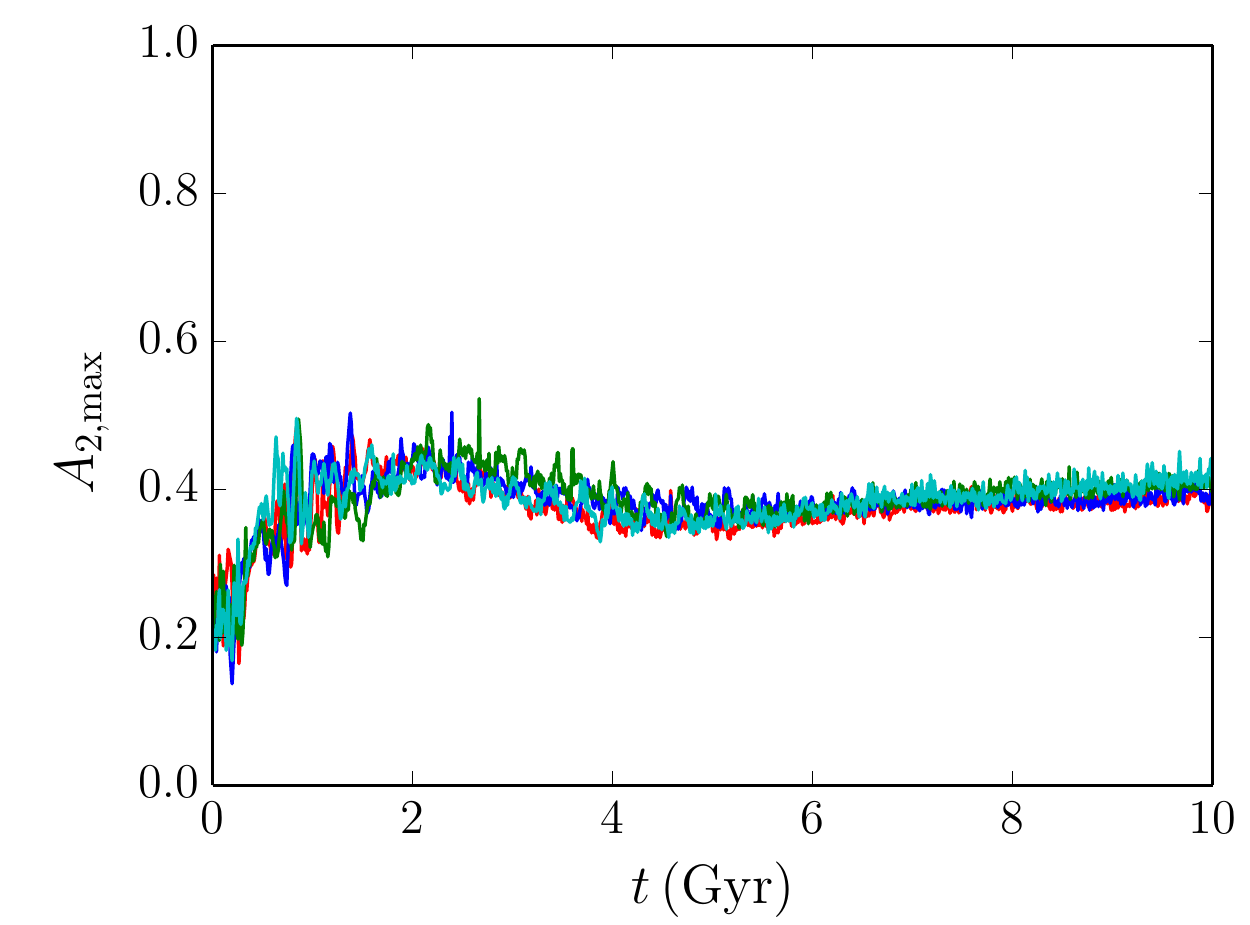}\plotone{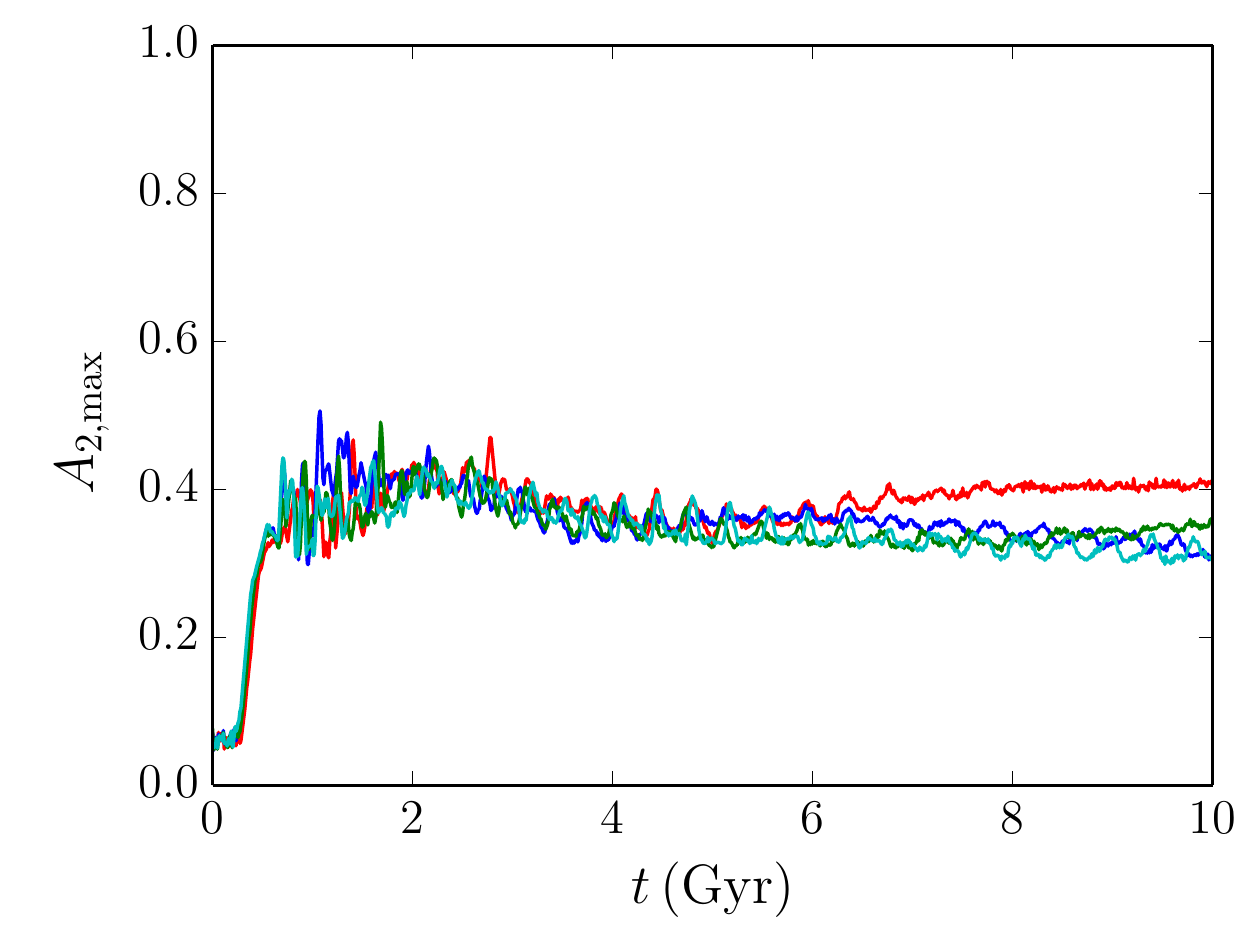}\plotone{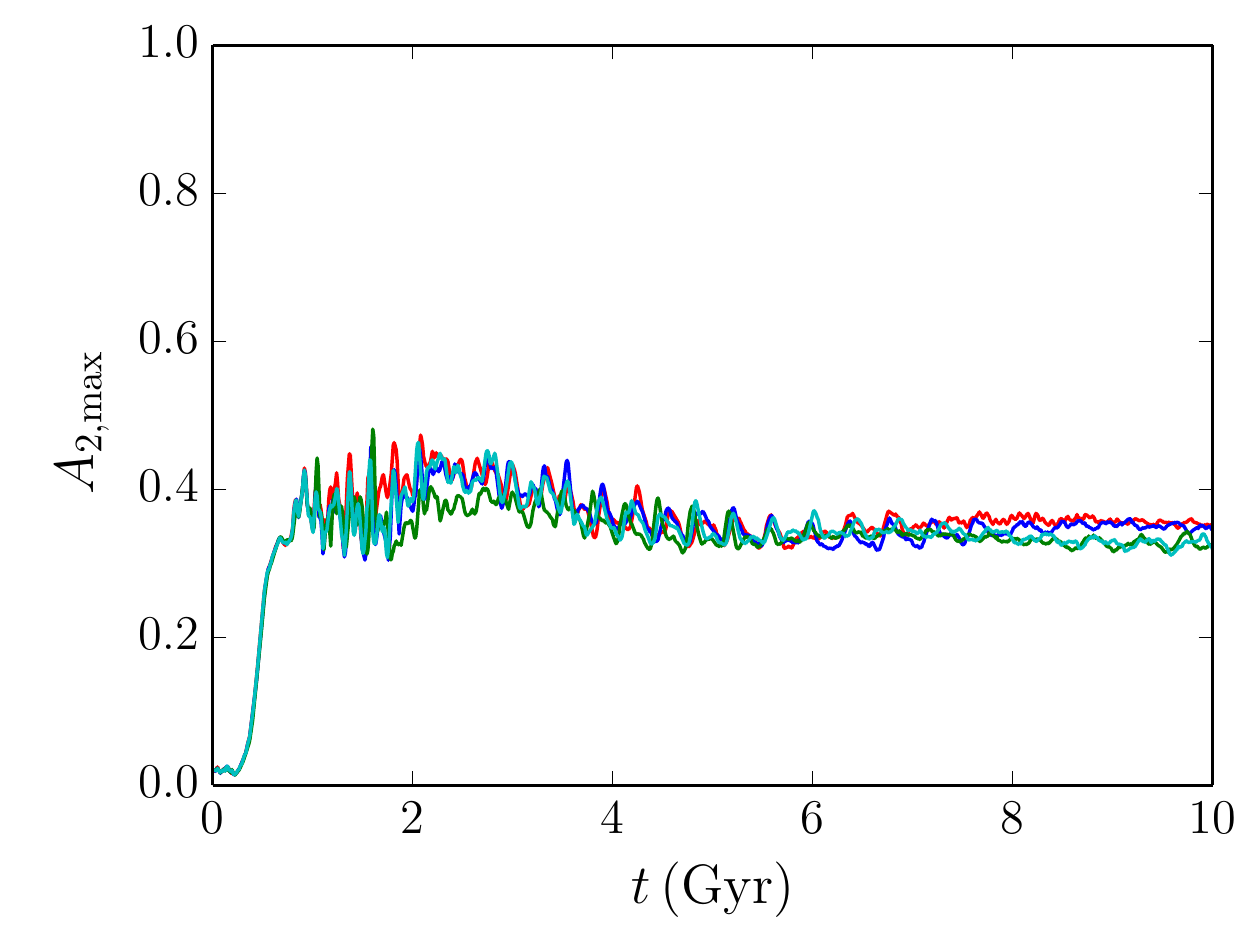}\\
\plotone{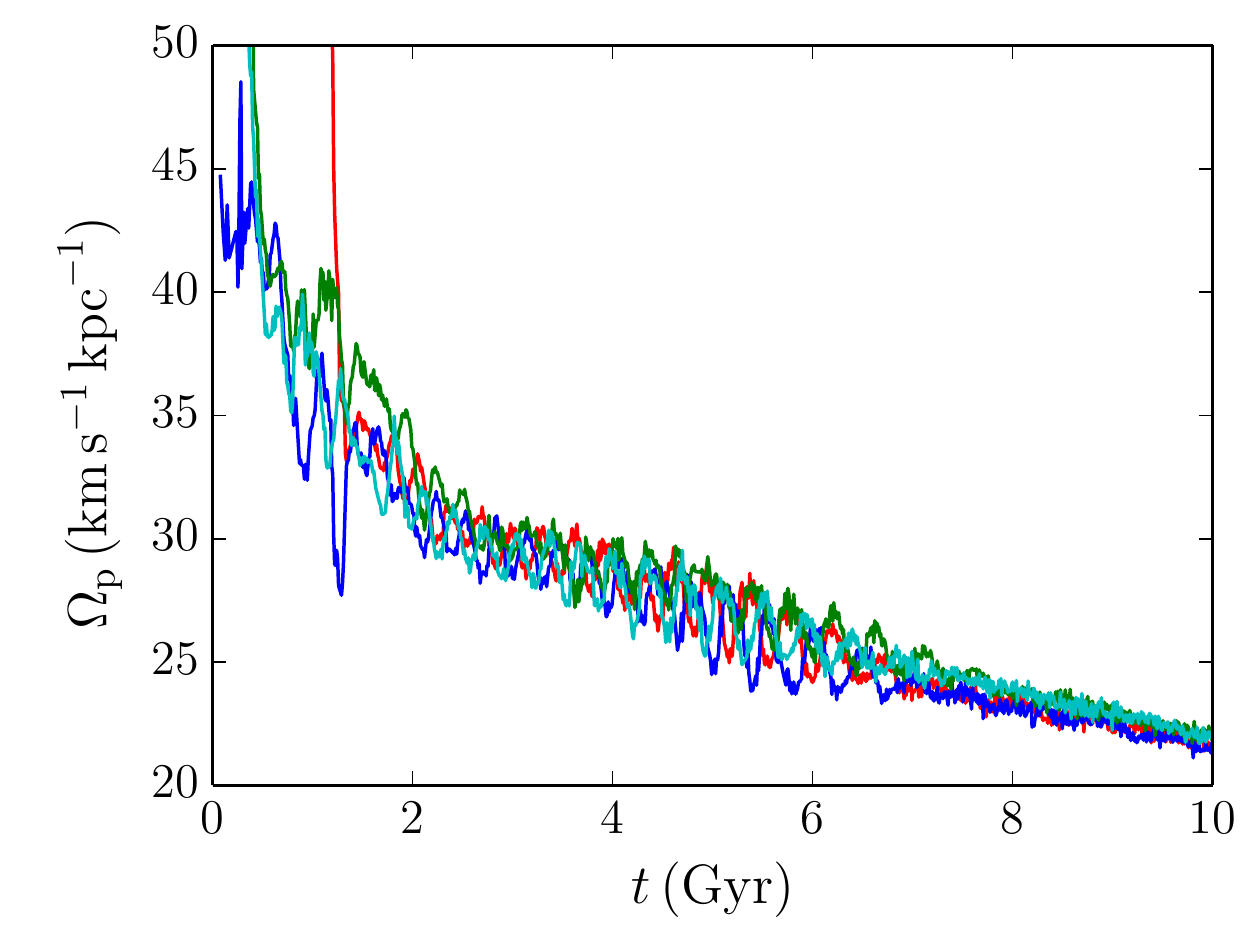}\plotone{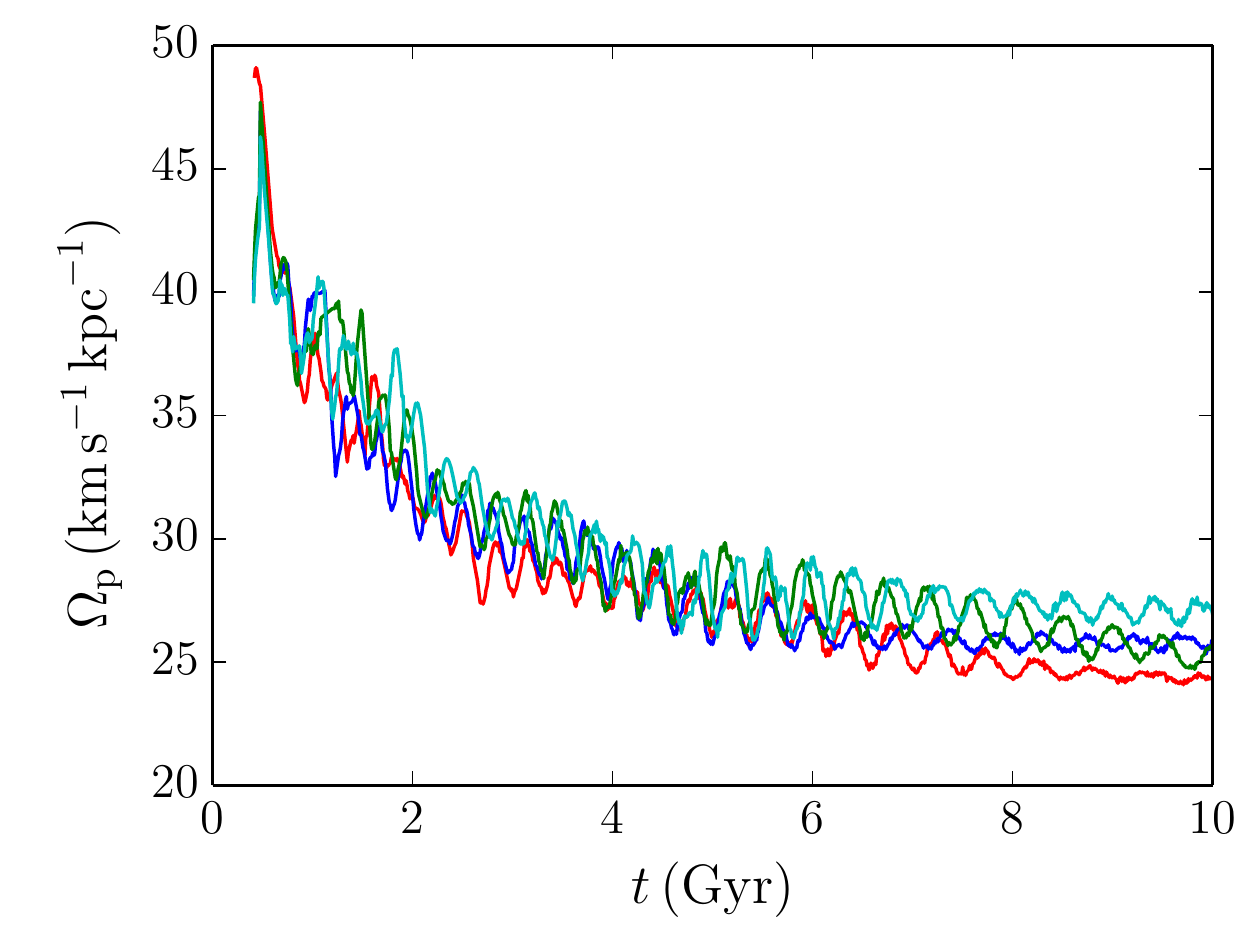}\plotone{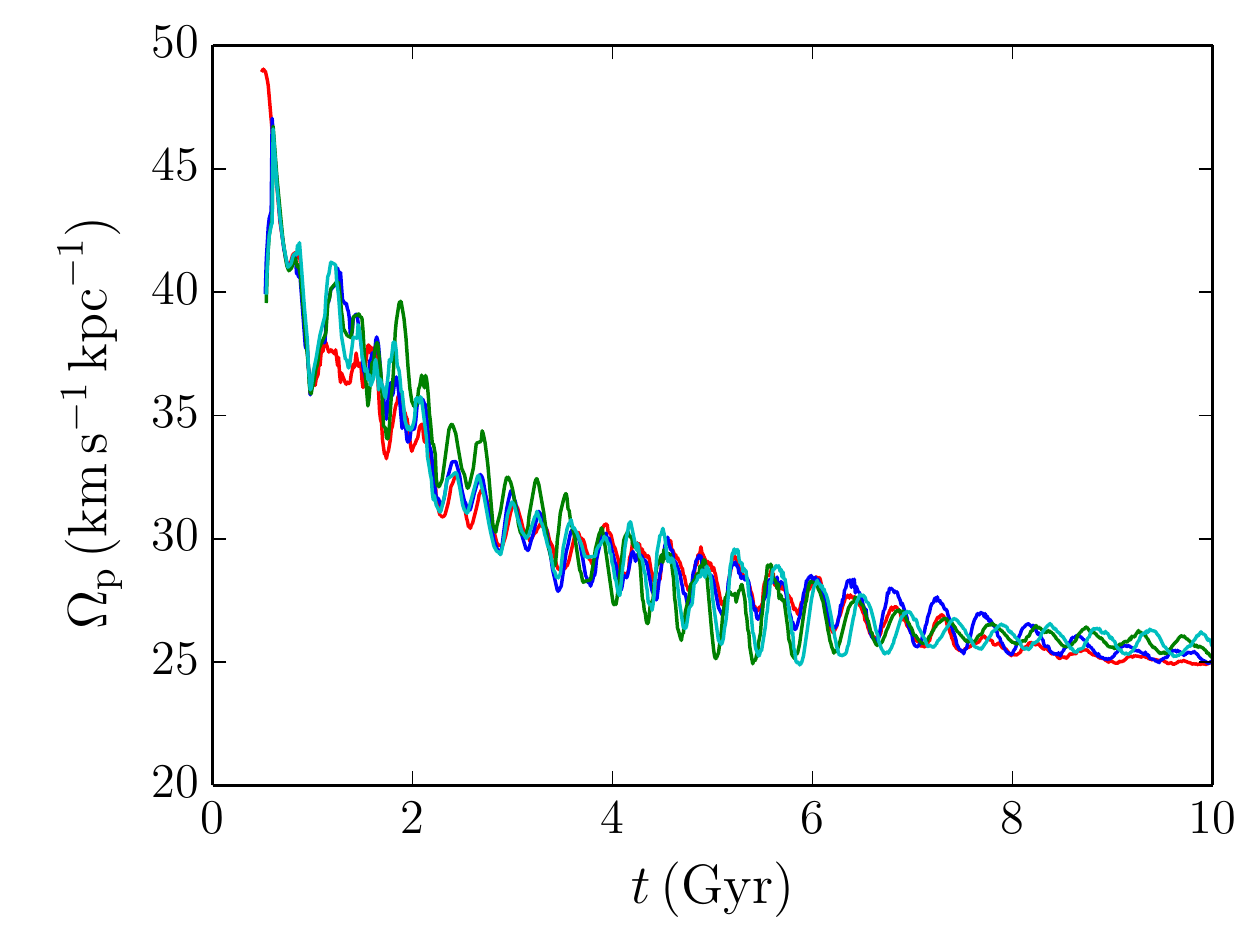}\\
\caption{Results for models with different $N$ (12M, 120M, and 1.2B from left to right). From top to bottom the panels show,  bulge line-of-sight velocity, 
bulge velocity dispersion, disc scale height, the evolution of the bar length and the bar amplitude. Colors indicate different random seeds. \label{fig:N}}
\end{figure*}

In Fig.~\ref{fig:N}, we present the following results (from top to bottom): 
bulge radial velocity, velocity dispersion, disc scale height, bar length, 
bar amplitude, and bar pattern speed. 
From left to right we show the results for $N\sim 12$M, 120M, and 1.2B.
In each panel, the results for the four different runs are over-plotted using 
different colours. For the top two panels, we also show the 
BRAVA observations~\citep{2012AJ....143...57K}.

The convergence of the bulge velocity distribution and disc scale height improves 
as $N$ increases.
The length of the bar changes with time and the fluctuation of the bar length,
between the different random seeds, becomes larger as $N$ increases. 
This may be because finer structures, which affect the bar length,  appear in the higher
resolution simulations. Indeed, the bar repeatedly connects and disconnects
with spiral arms (see Figure \ref{fig:MWa5B_Rp_5Gyr}). 
Evolution over a longer time scale is, however, similar for all $N$ and 
the different random seeds only have minor influence. 

In the fourth row, we can see that the bar formation epoch is delayed,
but the final bar amplitude converges as $N$ increases.
The initial peak of the bar amplitude fluctuates for different random seeds, 
but the final amplitude settles down to similar values. 
In the bottom row, we present the evolution of the pattern speed of the bar. 
With the lowest resolution, the pattern speed continuously drops, in contrast
for the higher resolutions, it settled down at $t\sim 6$\,Gyr.
In the highest resolution (1.2B), the results look converged, even for 
the pattern speed oscillation at $t\sim 5$\,Gyr.

Based on the results of these tests, we decided to use models with 
a total number of particle of $\sim 100$M 
($\sim 8$M for discs) to find parameters fitting to the Milky Way observations. 
We performed additional simulations for some of the best-fitting models, with up to 8
billion particles in total.

\begin{table*}
\scriptsize
\raggedright 
\caption{Model parameters\label{tb:B1}}
\begin{tabular}{lccccccccccc}
\hline
           & \multicolumn{4}{l}{Halo} &  \multicolumn{4}{l}{Disc} &  \multicolumn{3}{l}{Bulge} \\
Parameters &  $a_{\rm h}$ & $\sigma_{\rm h}$ & $\epsilon_{\rm h}$ & $\alpha_{\rm h}$& $M_{\rm d}$ & $R_{\rm d}$ & $z_{\rm d}$ & $\sigma_{\rm R0}$  & $a_{\rm b}$ & $\sigma_{\rm b}$ & $\epsilon_{\rm b}$ \\ 
&  (kpc) & ($\kms$) &  &  & $(10^{10}M_{\odot})$ & (kpc) & (kpc) & ($\kms$)  & (kpc) & $(\kms)$\\
\hline \hline
MWtest  & 15.5 &  320 & 0.76 & 0.8 & $4.66$ & 2.6 & 0.36 & 117  & 0.78 & 255 & 0.99 \\
\hline
\end{tabular}
\end{table*}

\begin{table*}
\scriptsize
\raggedright 
\caption{Mass, radius, and the number of particles\label{tb:B2}}
\begin{tabular}{lccccccccccc}
\hline
Model    & $M_{\rm d}$ & $M_{\rm b}$ & $M_{\rm h}$ & $M_{\rm b}/M_{\rm d}$ & $R_{\rm d, t}$ & $r_{\rm b, t}$ & $r_{\rm h, t}$ & $Q_0$  & $N_{\rm d}$ & $N_{\rm b}$ & $N_{\rm h}$\\ 
   & ($10^{10}M_{\odot}$) & ($10^{10}M_{\odot}$) & ($10^{10}M_{\odot}$) & &  (kpc) & (kpc) & (kpc) &  &   &  & \\ 
\hline \hline
MWtest120M  & 4.68 &  0.329 & 66.6 & 0.070 & 31.6 & 2.95 & 254 & 1.2 & 8.3M & 0.57M & 120M\\
\hline
\end{tabular}
\newline
{ \scriptsize
The number of particles for models MWtest12M and MWtest1.2B are 0.1 and 10 times that of model MWtest120M, respectively.
}

\end{table*}

\section{The Effect of Halo Spin}\label{Sect:HaloSpin}

We investigate the effect of halo spin on the evolution of the disk, 
and especially the bar structure.
In \citet{2014ApJ...783L..18L}, they perform a series of $N$-body simulations
changing the initial spin parameter of the halo ($\lambda$).
They tested five parameters from 0 to 0.09 and found a sequential change
in bar amplitude and pattern speed after 10\,Gyr. They also found that the time 
at which the bar amplitude reaches its maximum is delayed when the halo spin decreases.

\citet{2013MNRAS.434.1287S} also performed similar simulations, but for a timescale 
shorter than the simulations of \citet{2014ApJ...783L..18L}. They, however, also 
tested a case with a negative initial halo spin and also found a sequential change
when negative spin parameters were included.

Here, we perform a series of $N$-body simulations for model MWc where we change the initial
halo spin parameter. In addition to models MWc0.5, MWc0.65, and MWc0.8, we performed 
simulations for the same models but with $\alpha_{\rm h}=0.3$ (model MWc0.3) and 1.0 
(model MWc1.0). The initial spin parameters are $\lambda=-0.053$ and $0.14$ for 
models MWc0.3 and MWc1.0, respectively. Using {\tt GalactICS}, $\alpha=1.0$ is 
the maximum spin we can set. 

\begin{figure*}
\epsscale{.45}
\plotone{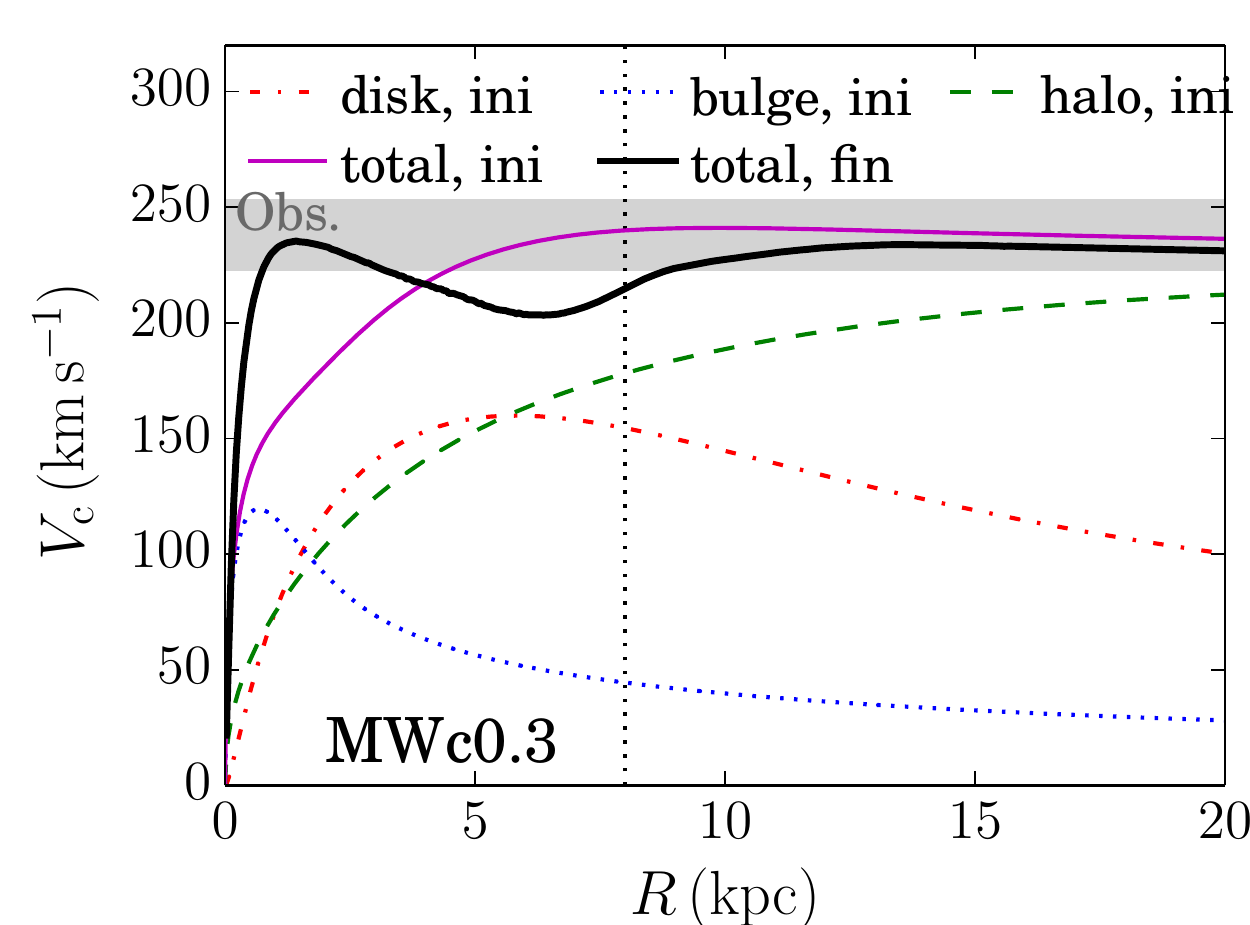}\plotone{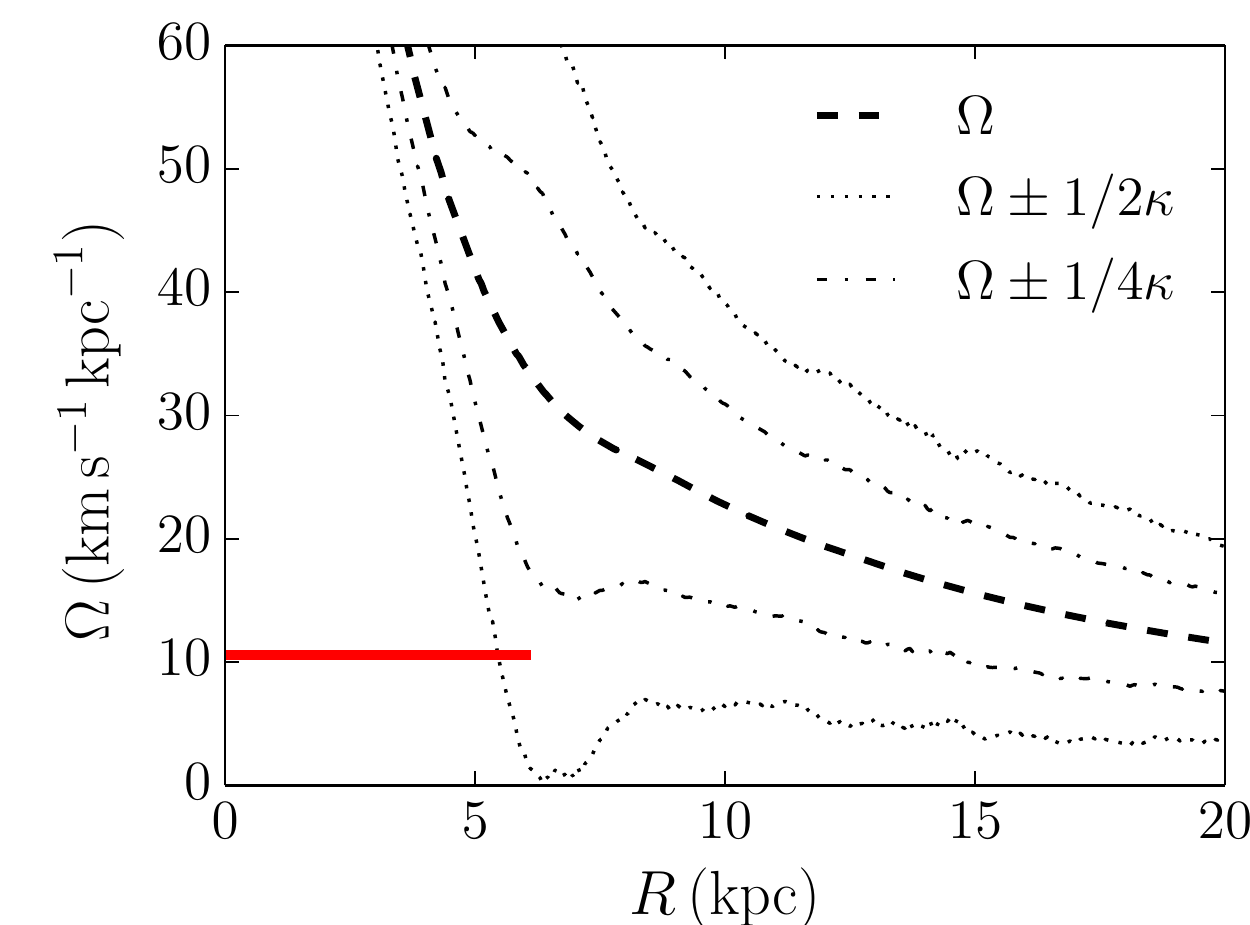}\\
\plotone{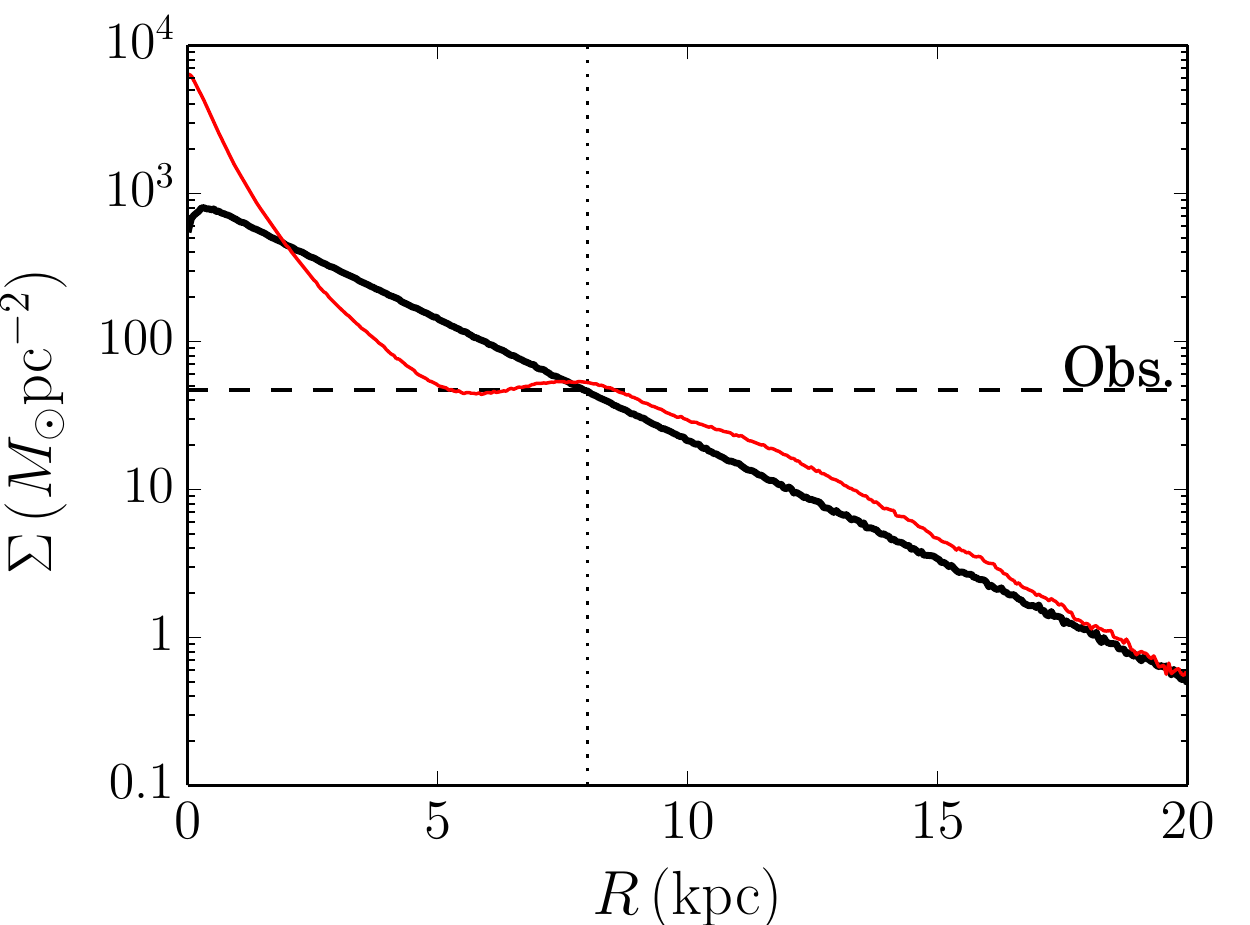}\plotone{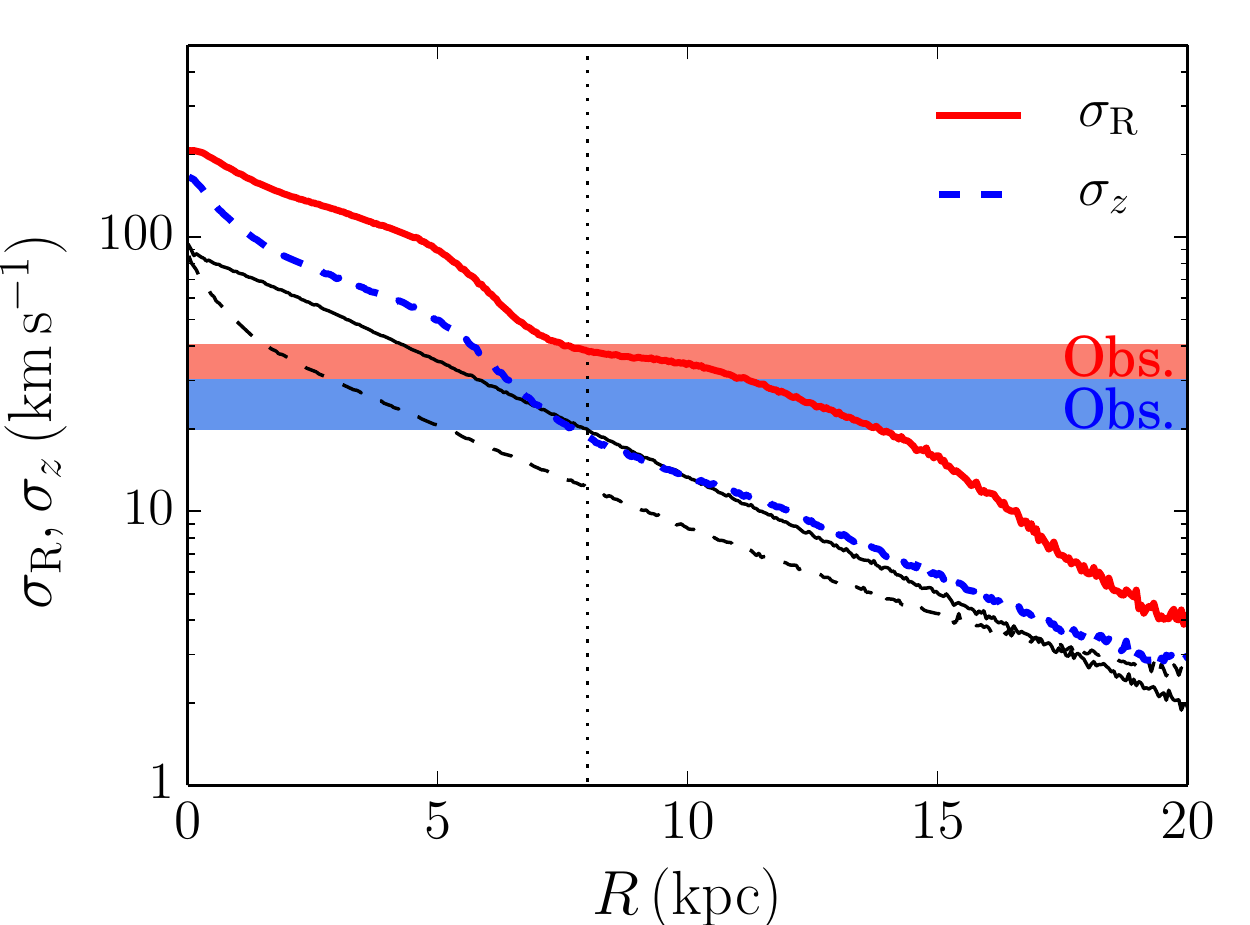}\\
\plotone{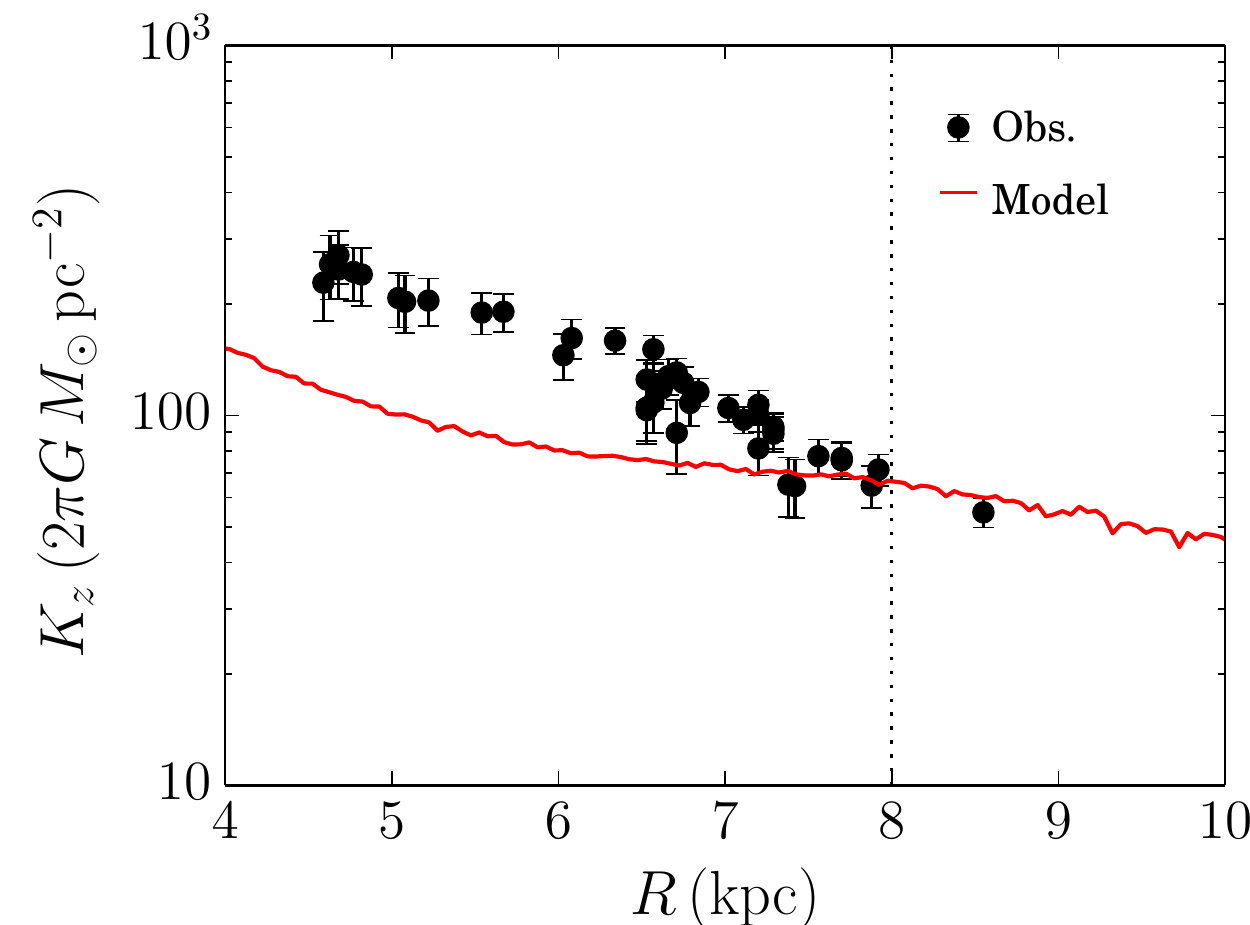}\plotone{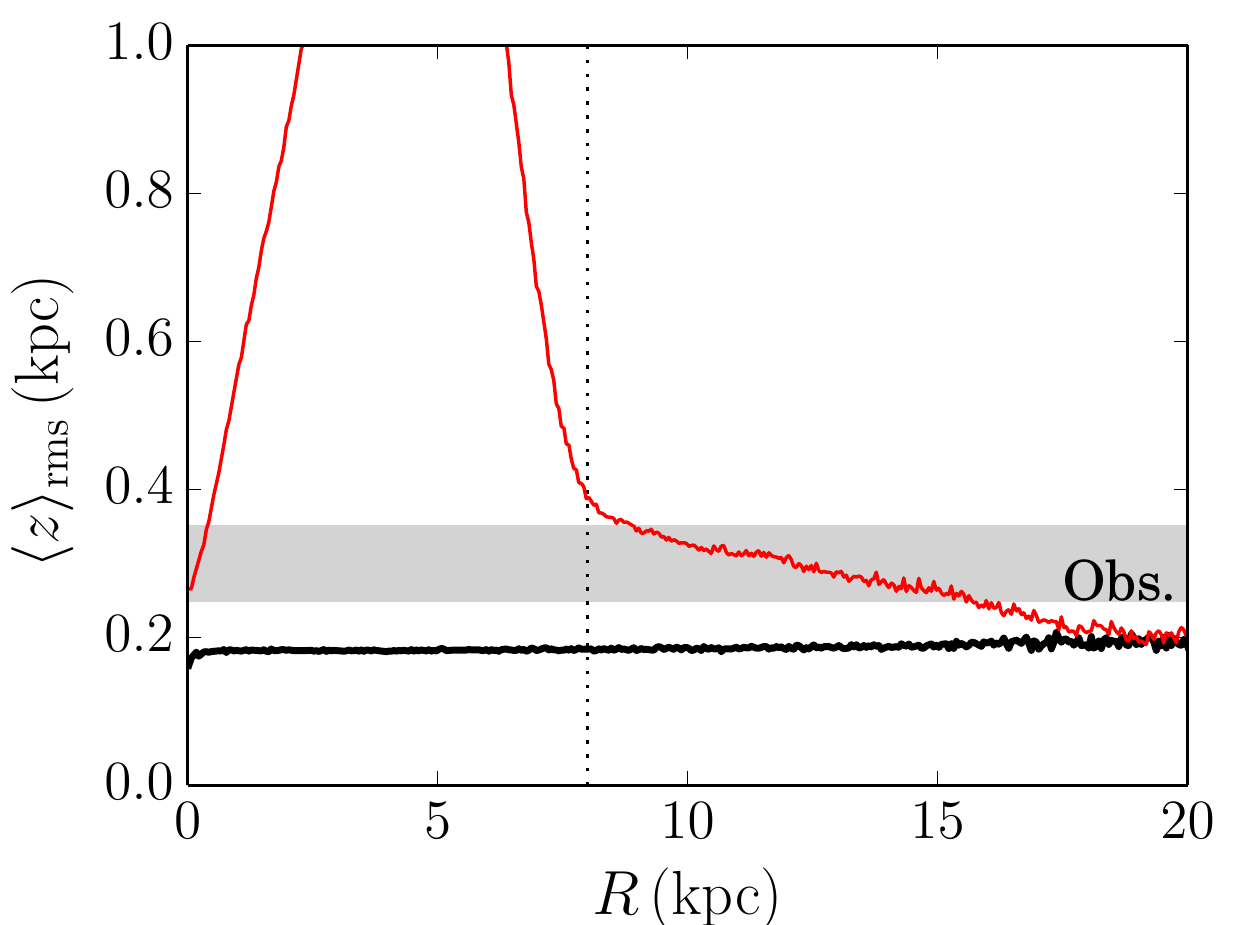}\\
\plotone{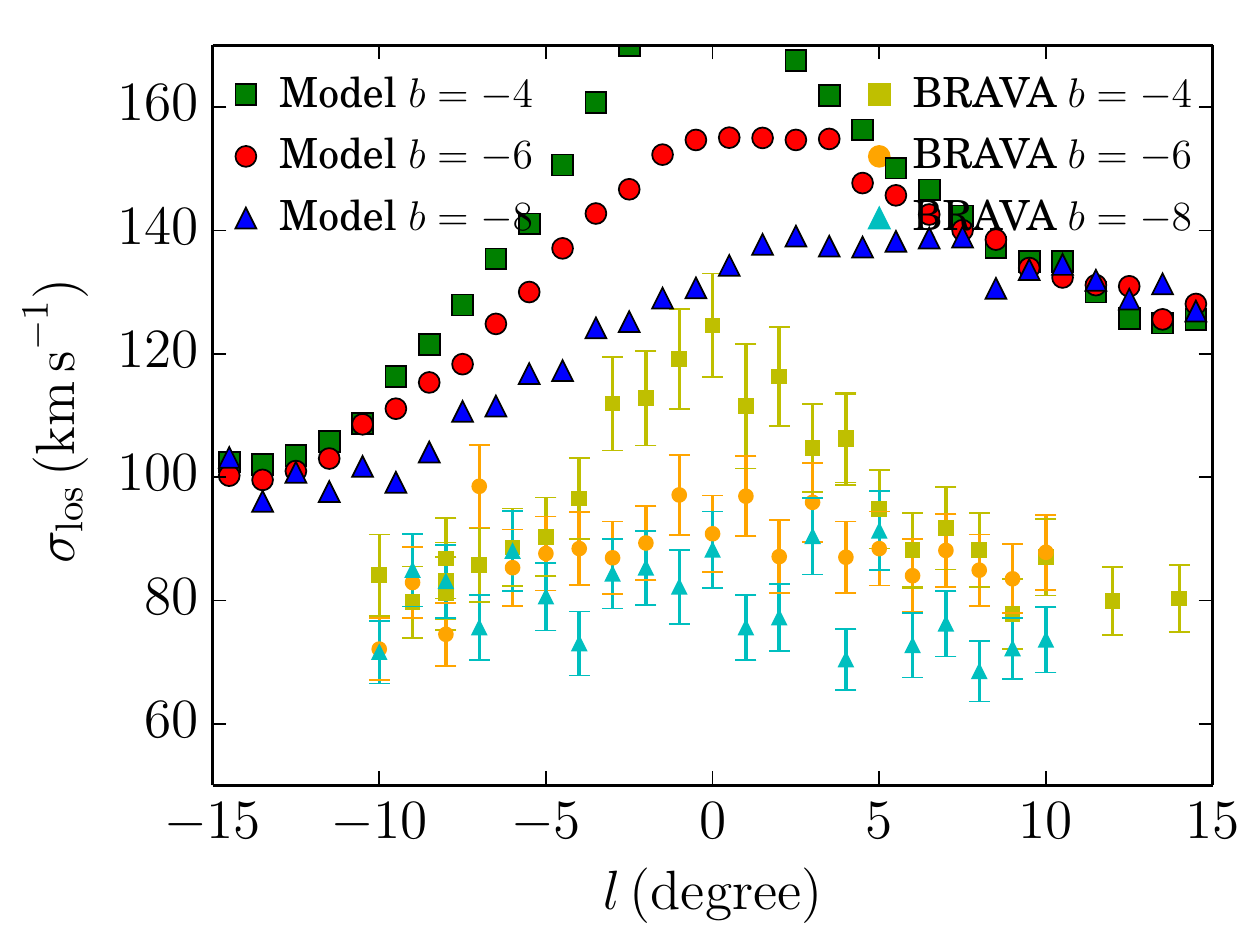}\plotone{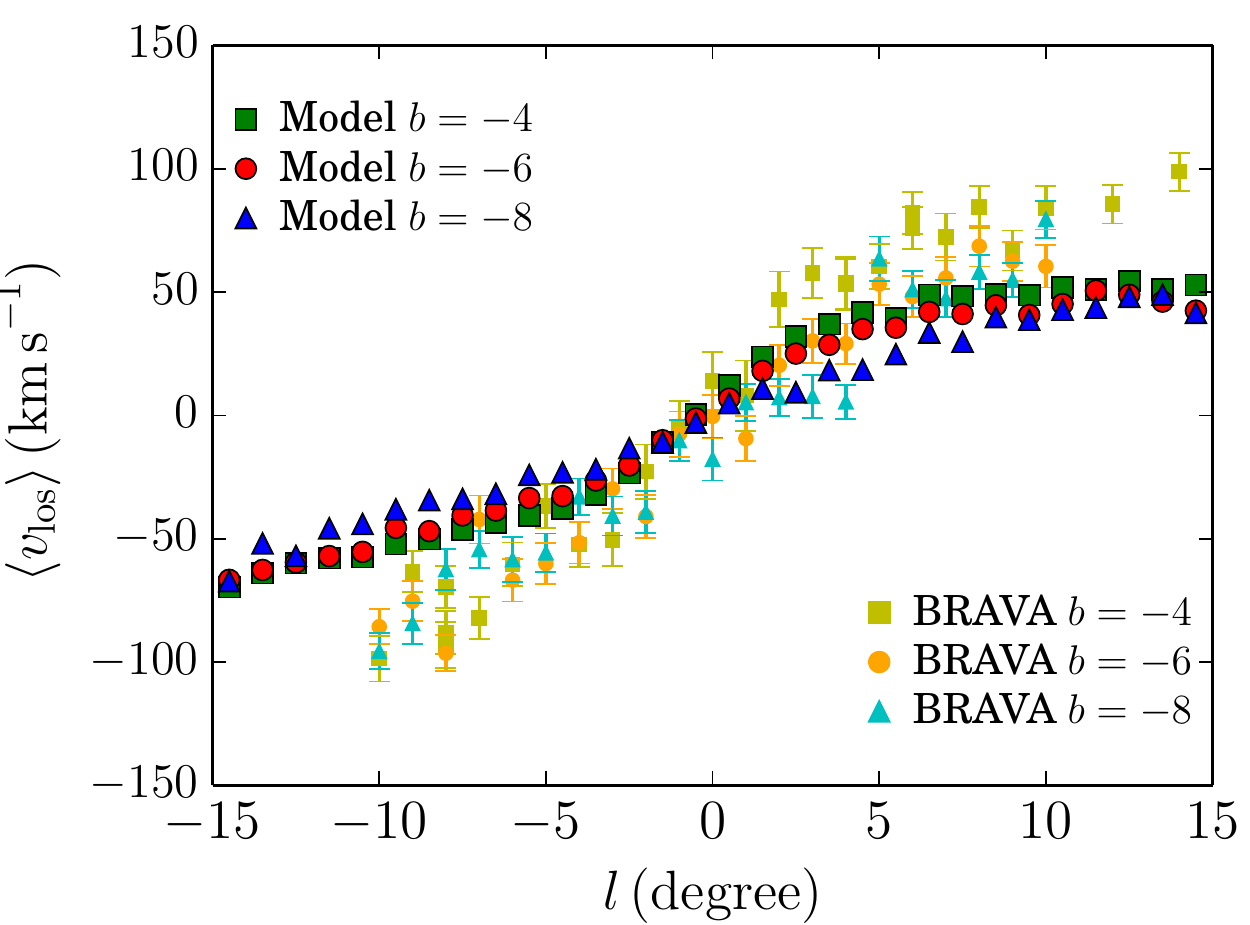}\\
\plotone{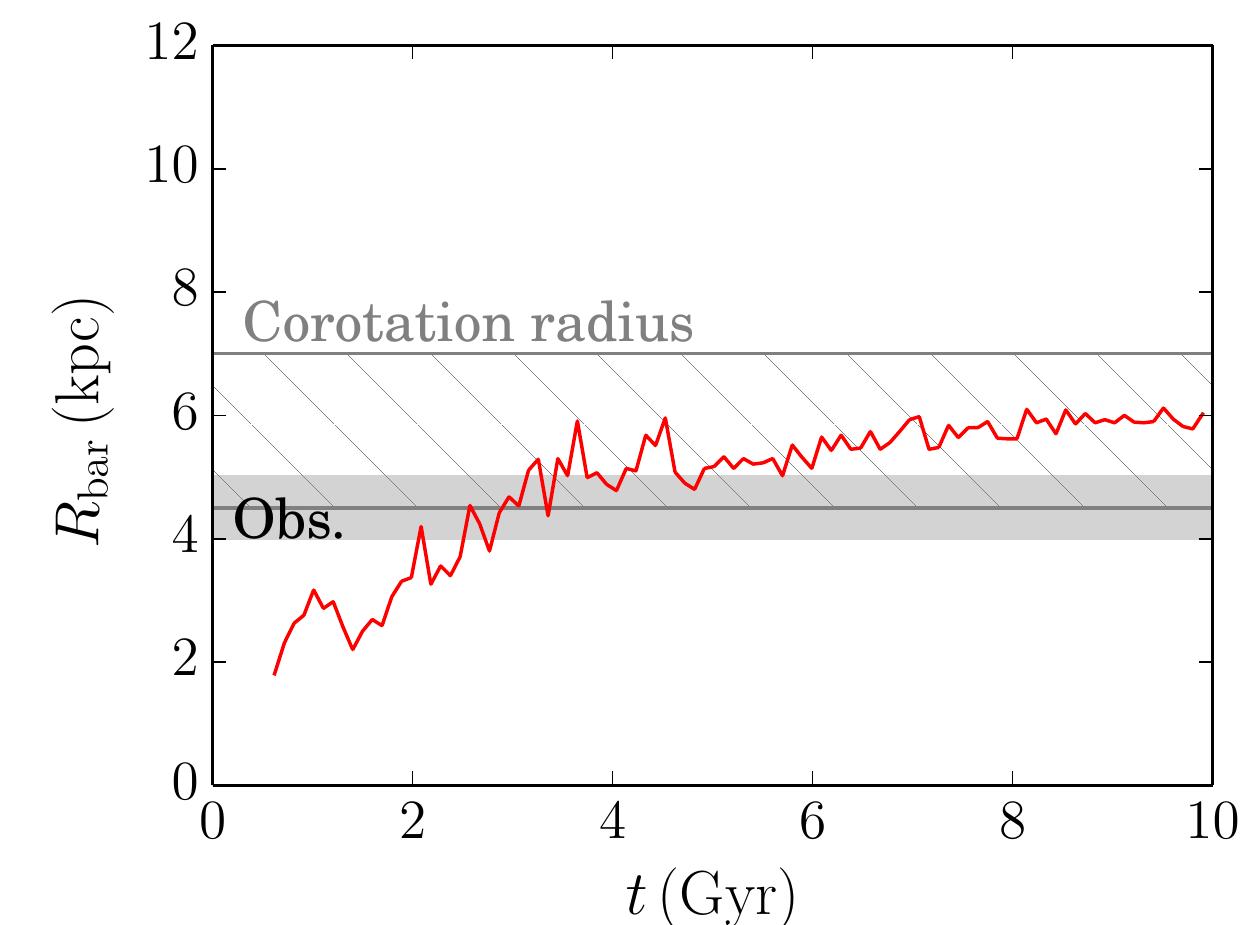}\plotone{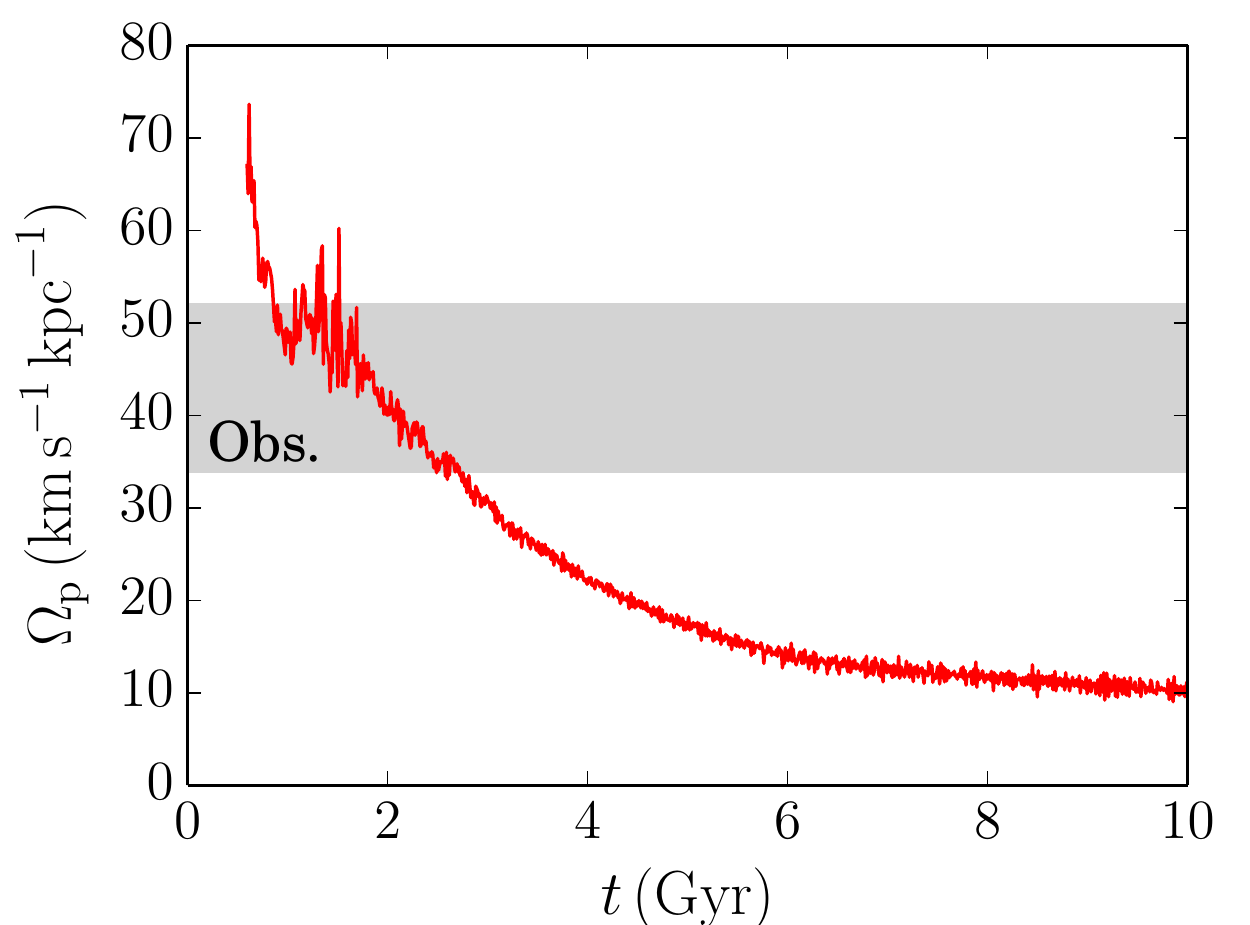}\\
\caption{Same as Fig.\ref{fig:a5B}, but for model MWc0.3. \label{fig:c0.3}}
\end{figure*}

\begin{figure*}
\epsscale{.45}
\plotone{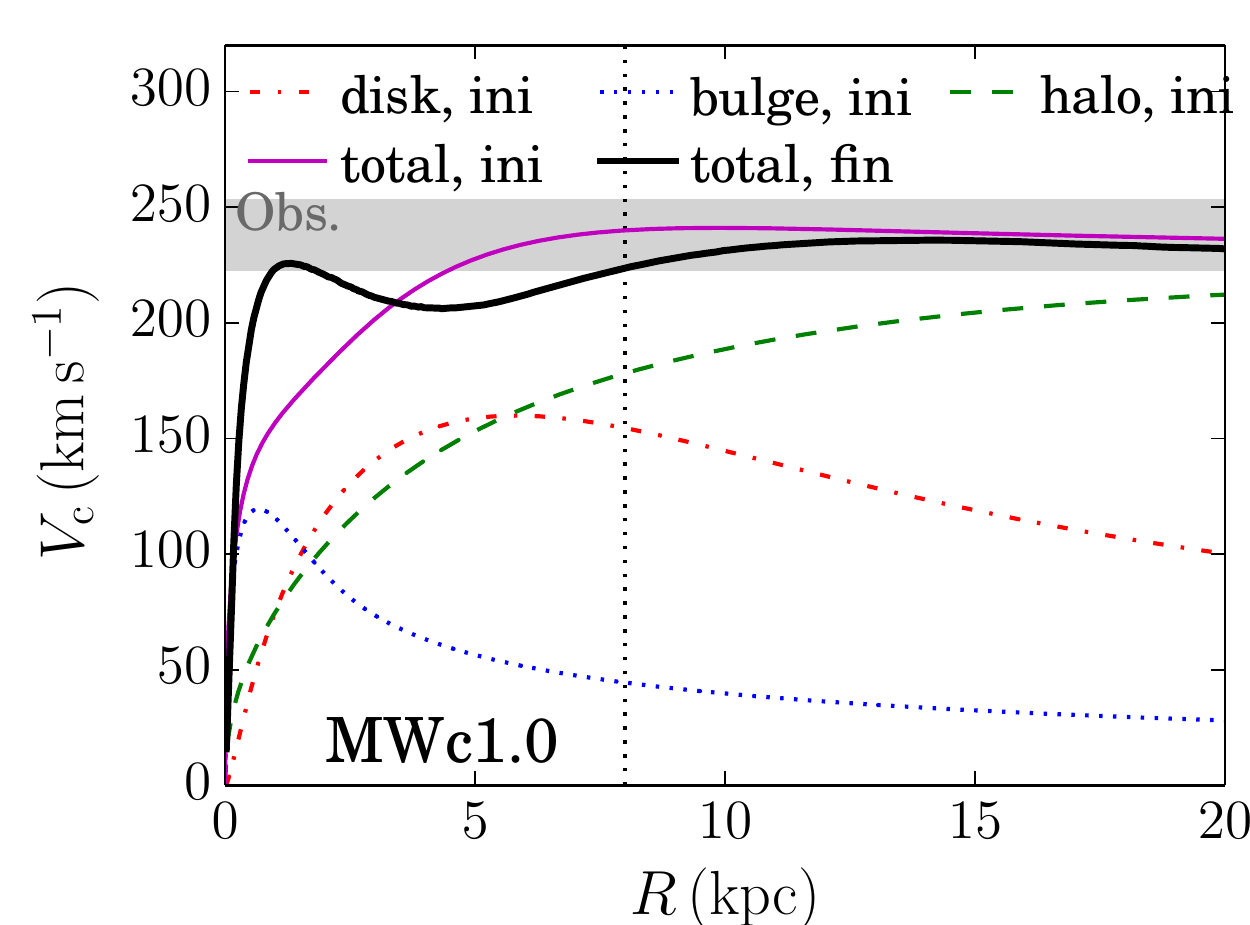}\plotone{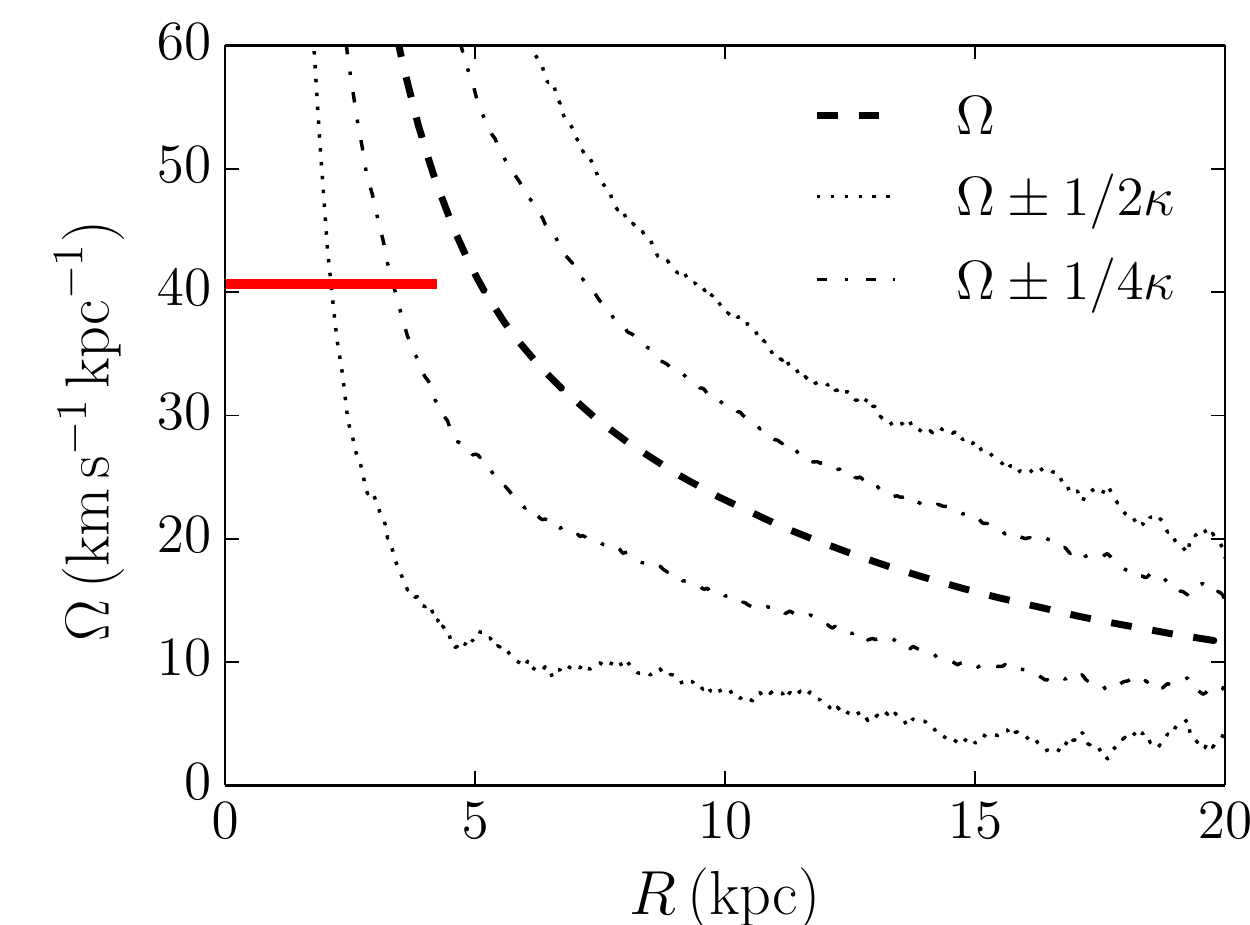}\\
\plotone{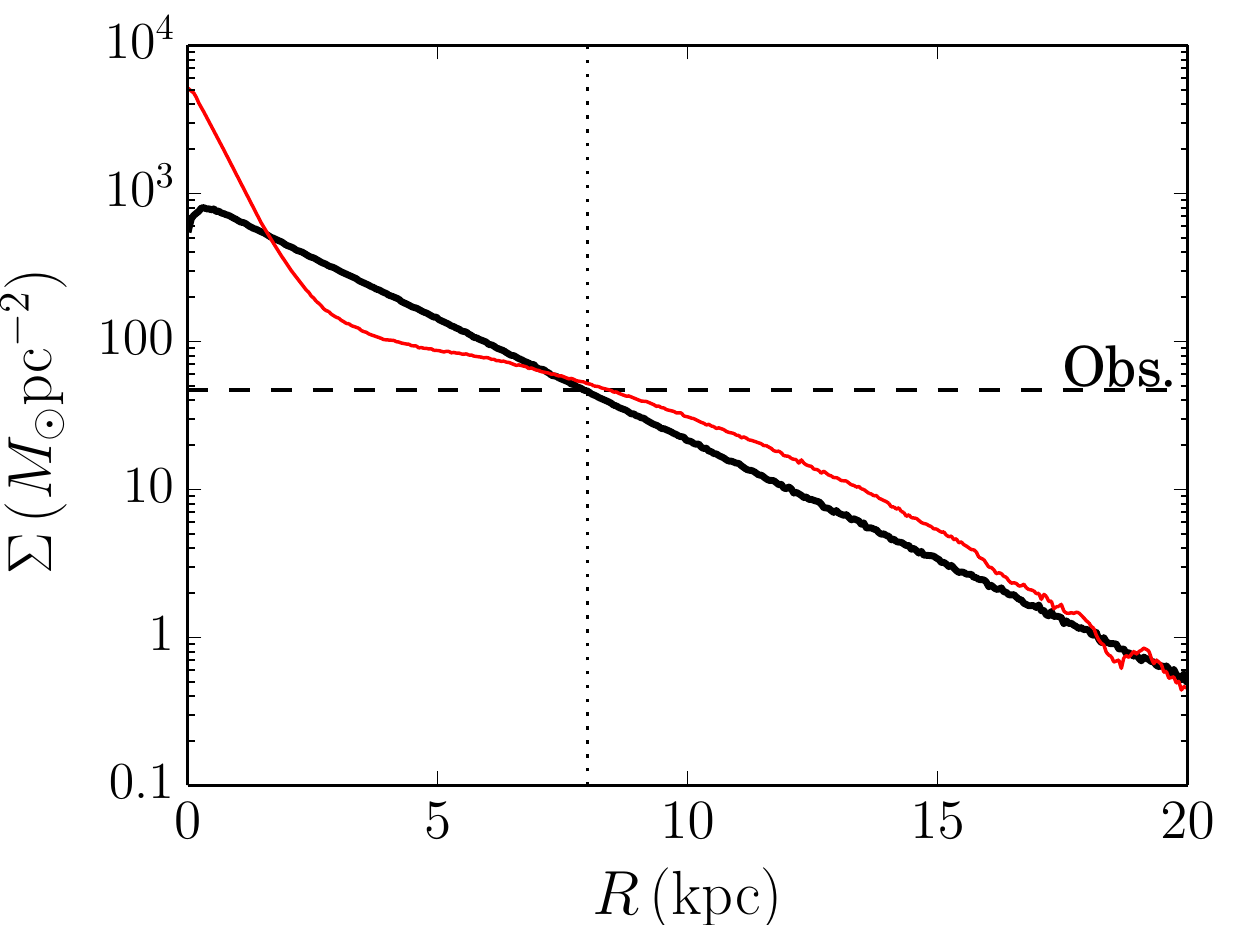}\plotone{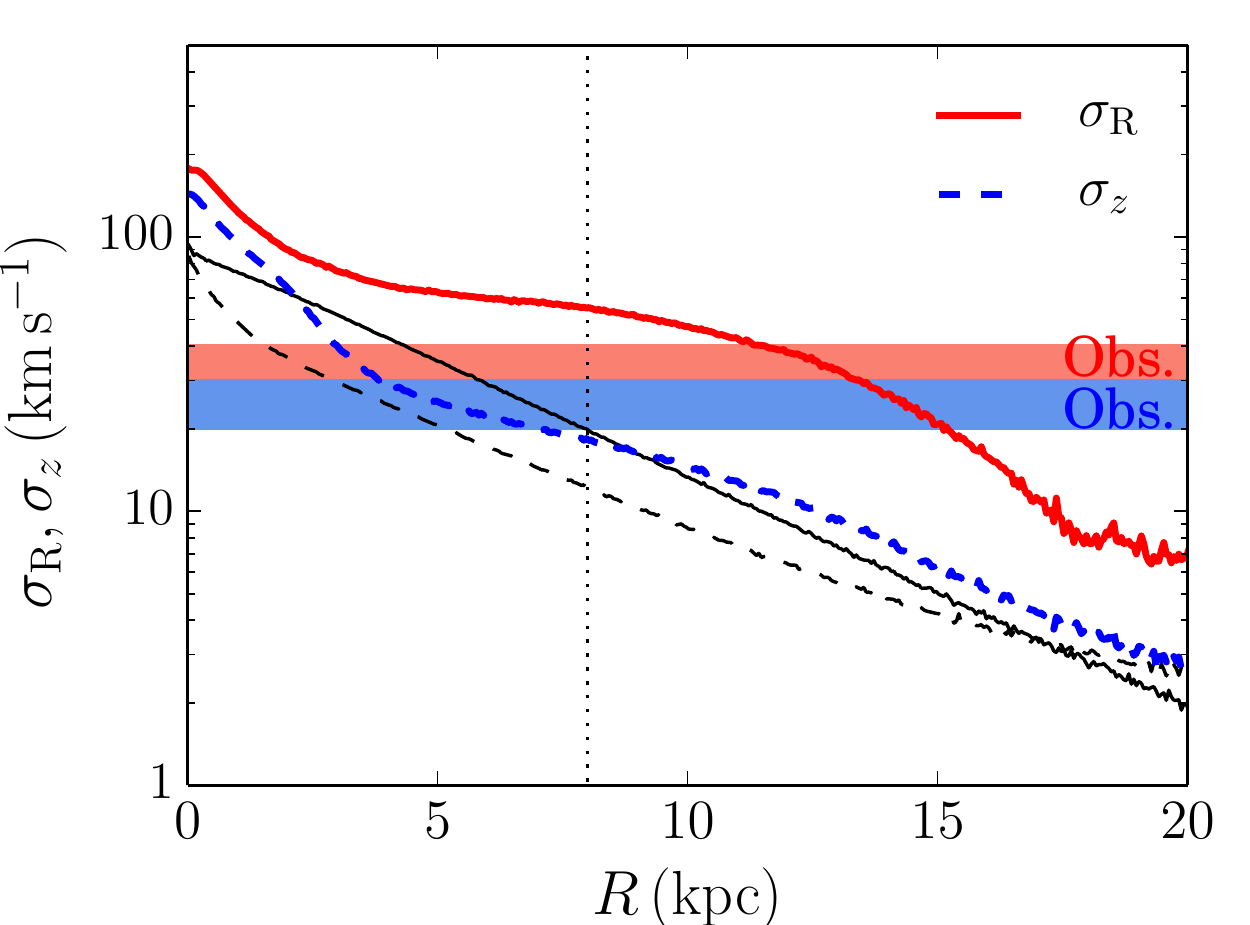}\\
\plotone{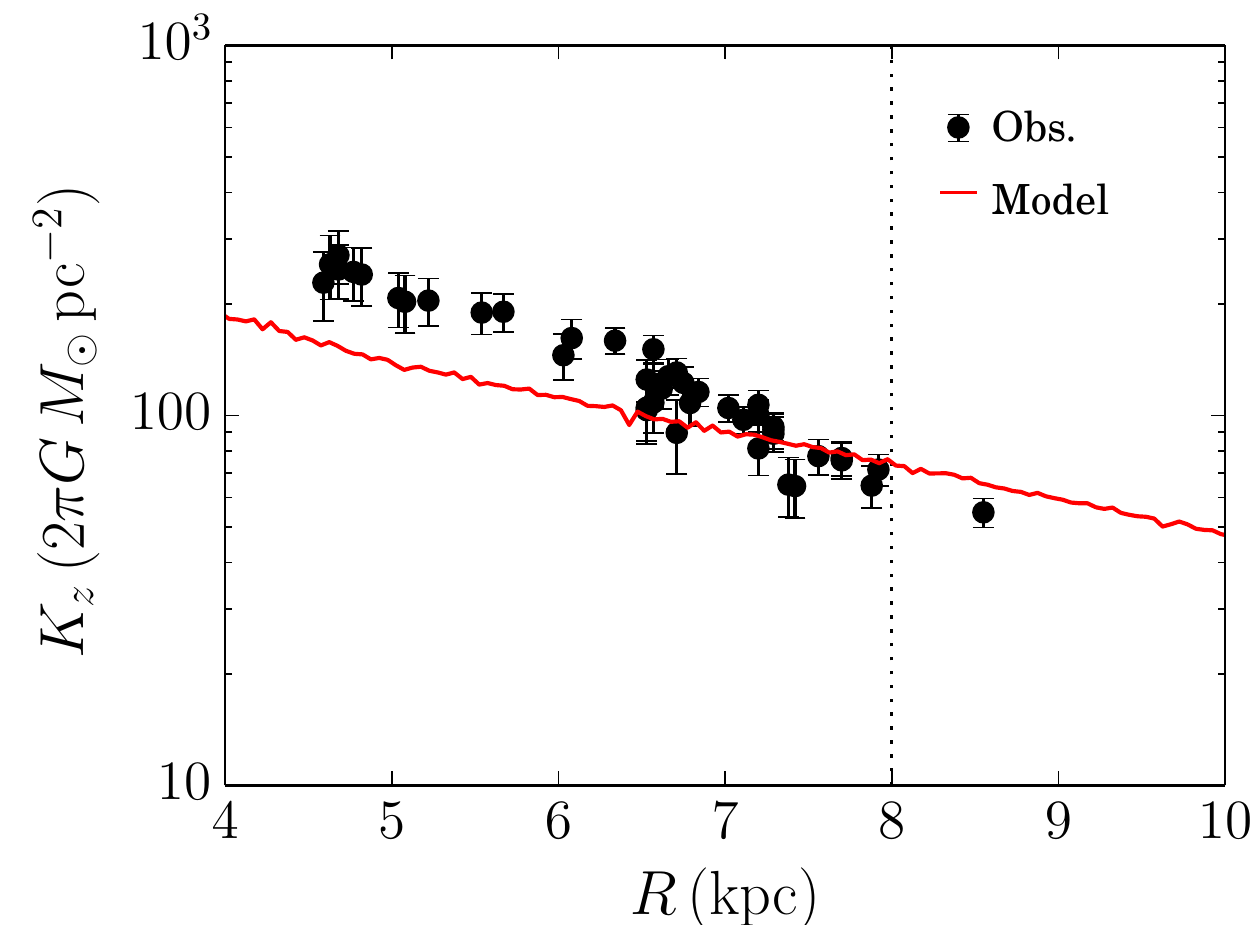}\plotone{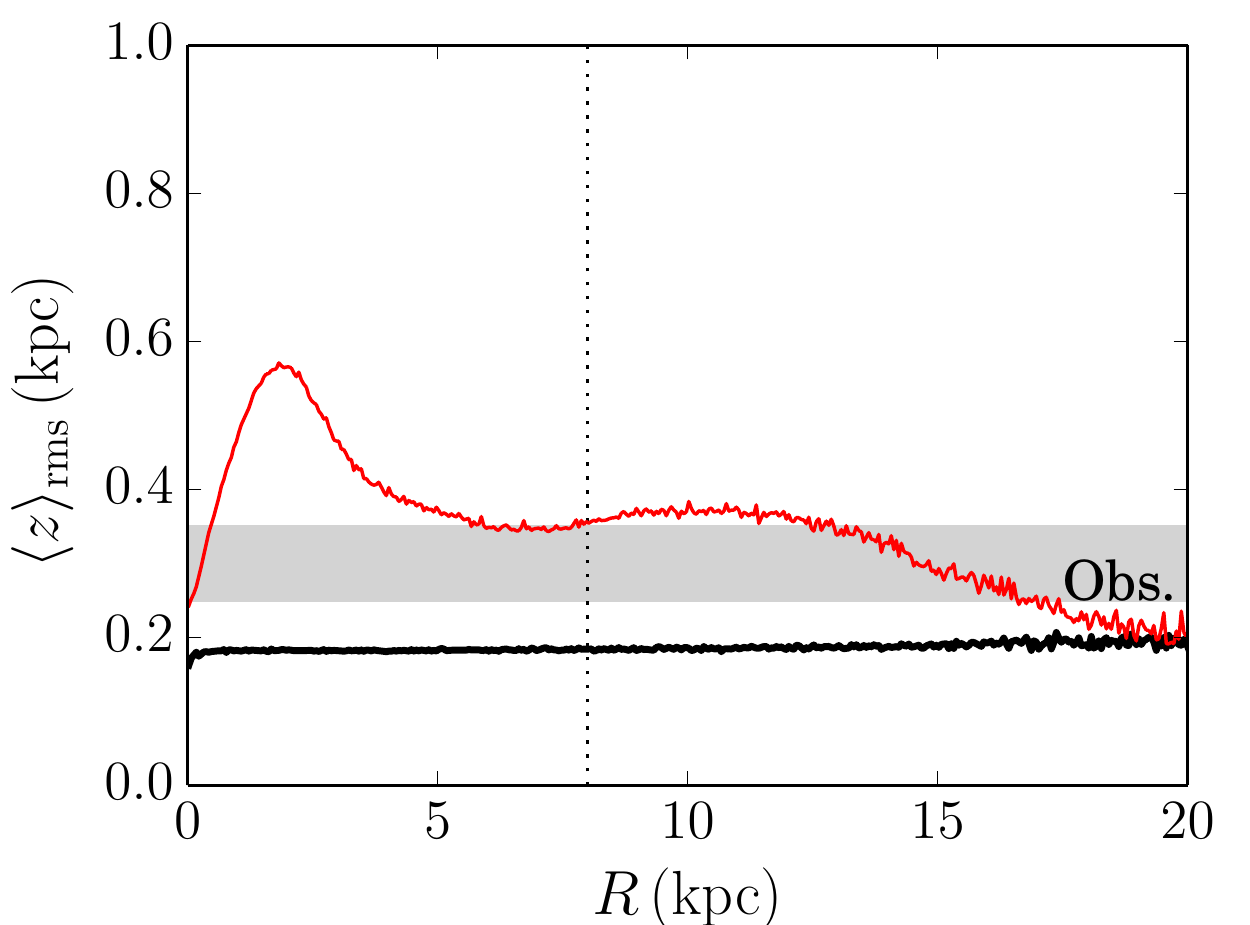}\\
\plotone{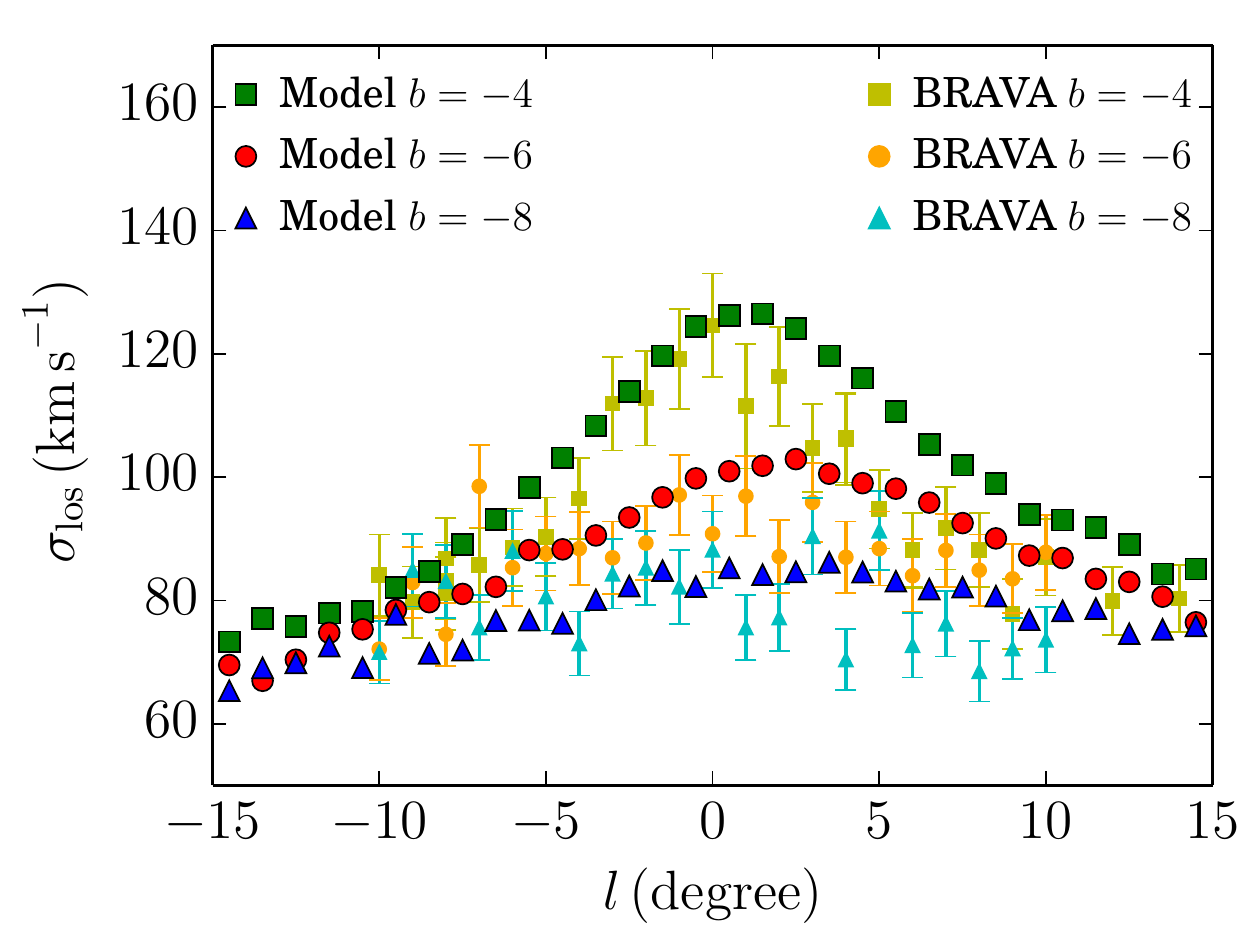}\plotone{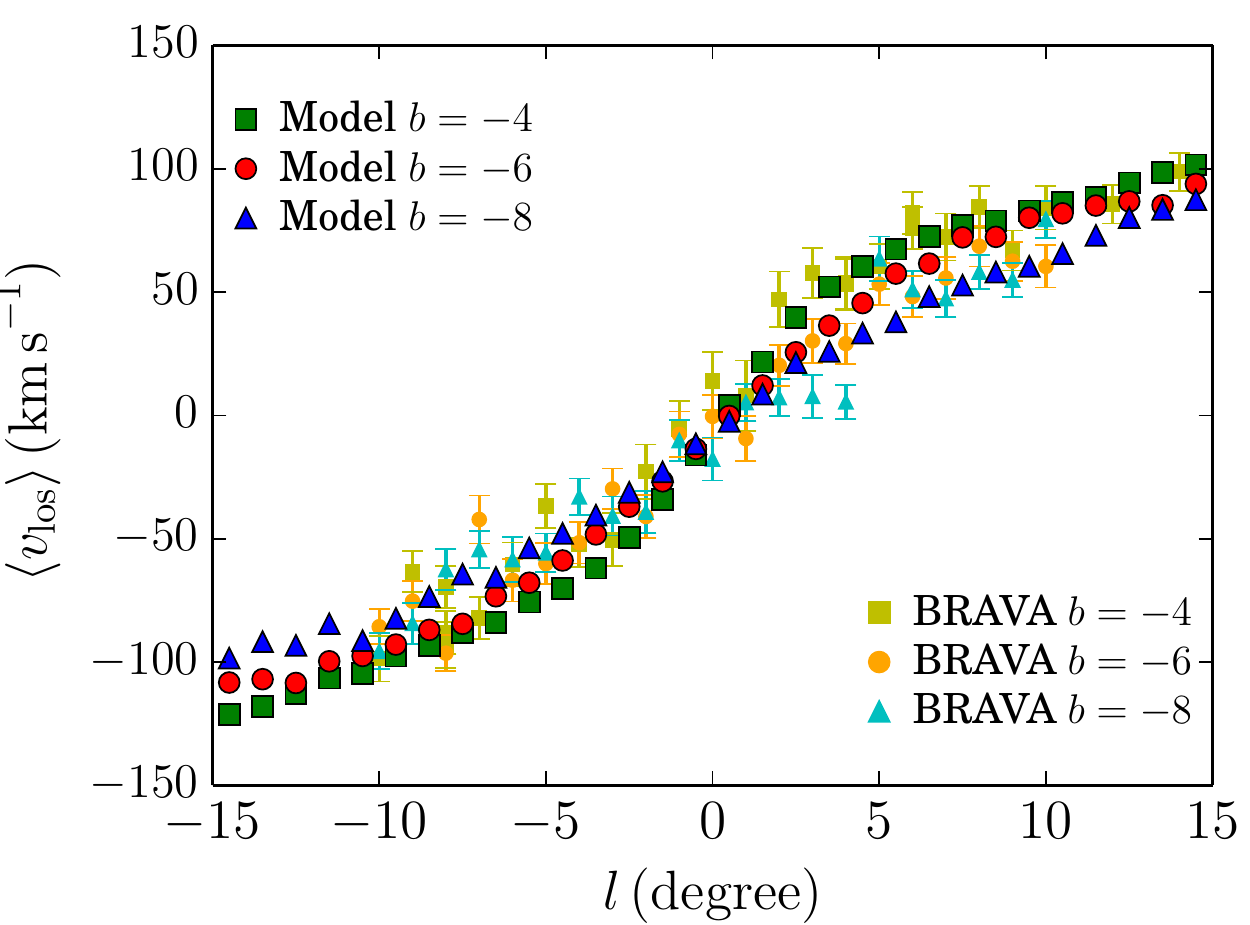}\\
\plotone{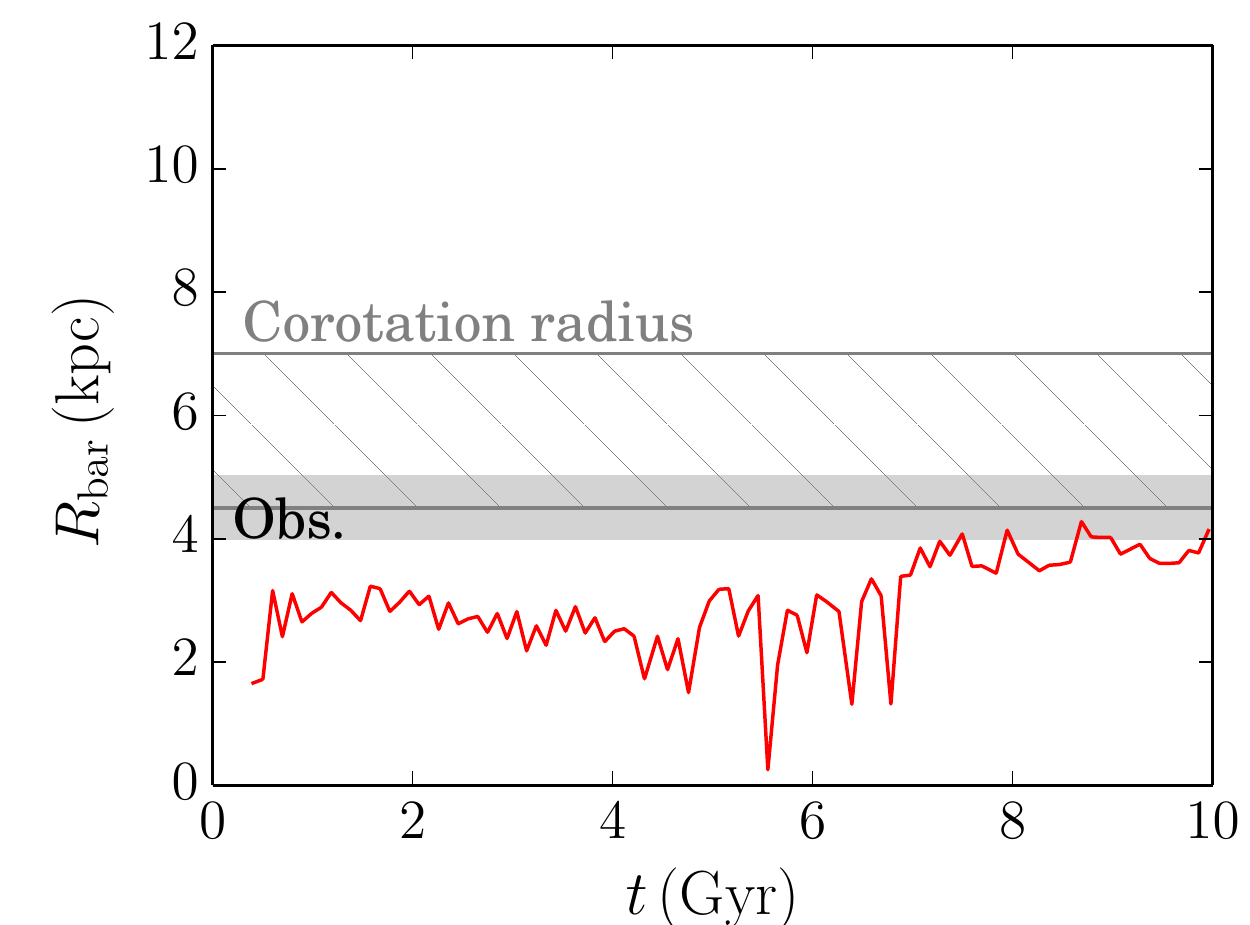}\plotone{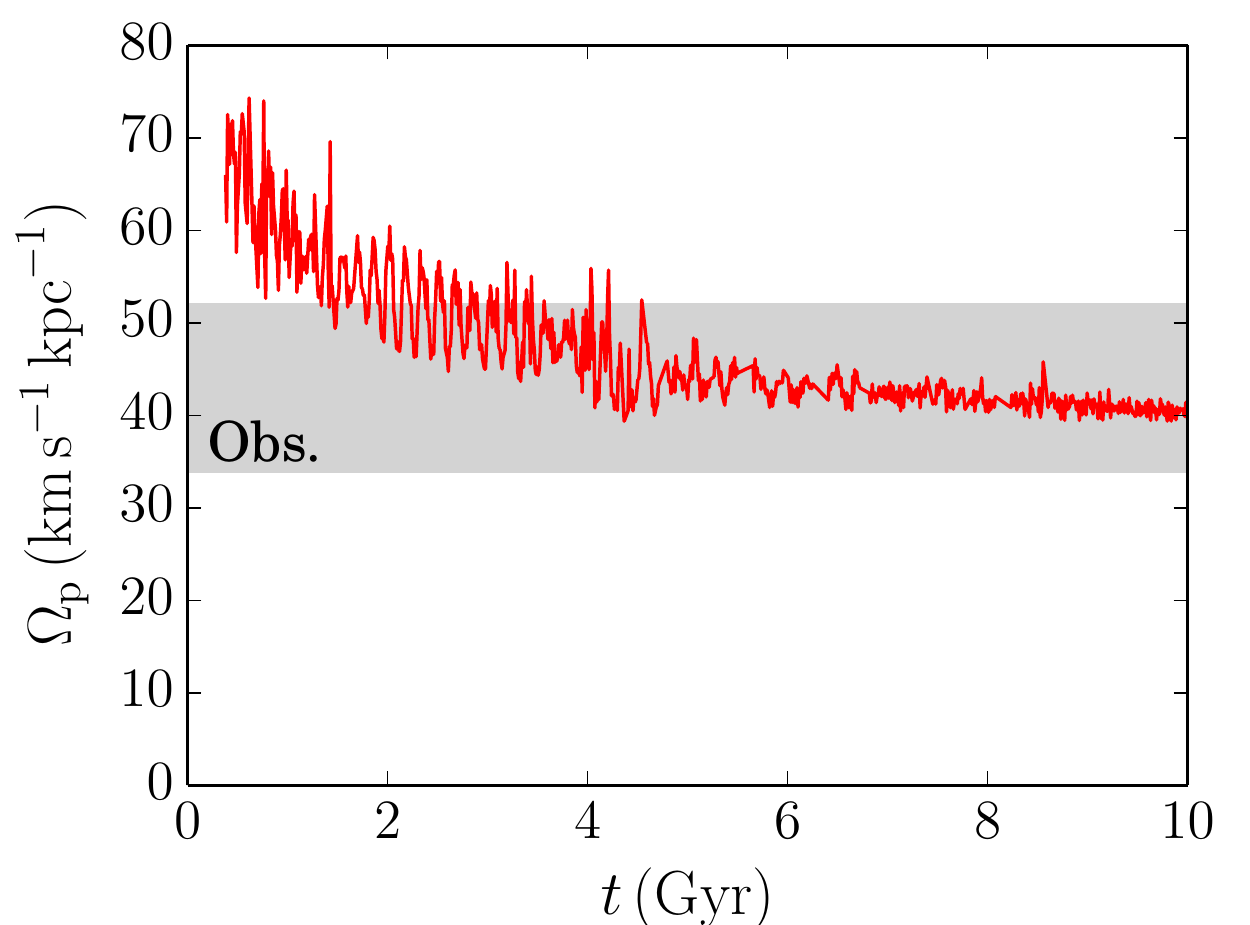}\\
\caption{Same as Fig.\ref{fig:a5B}, but for model MWc1.0. \label{fig:c1.0}}
\end{figure*}

\begin{figure*}
\raggedright
\epsscale{.35}
\plotone{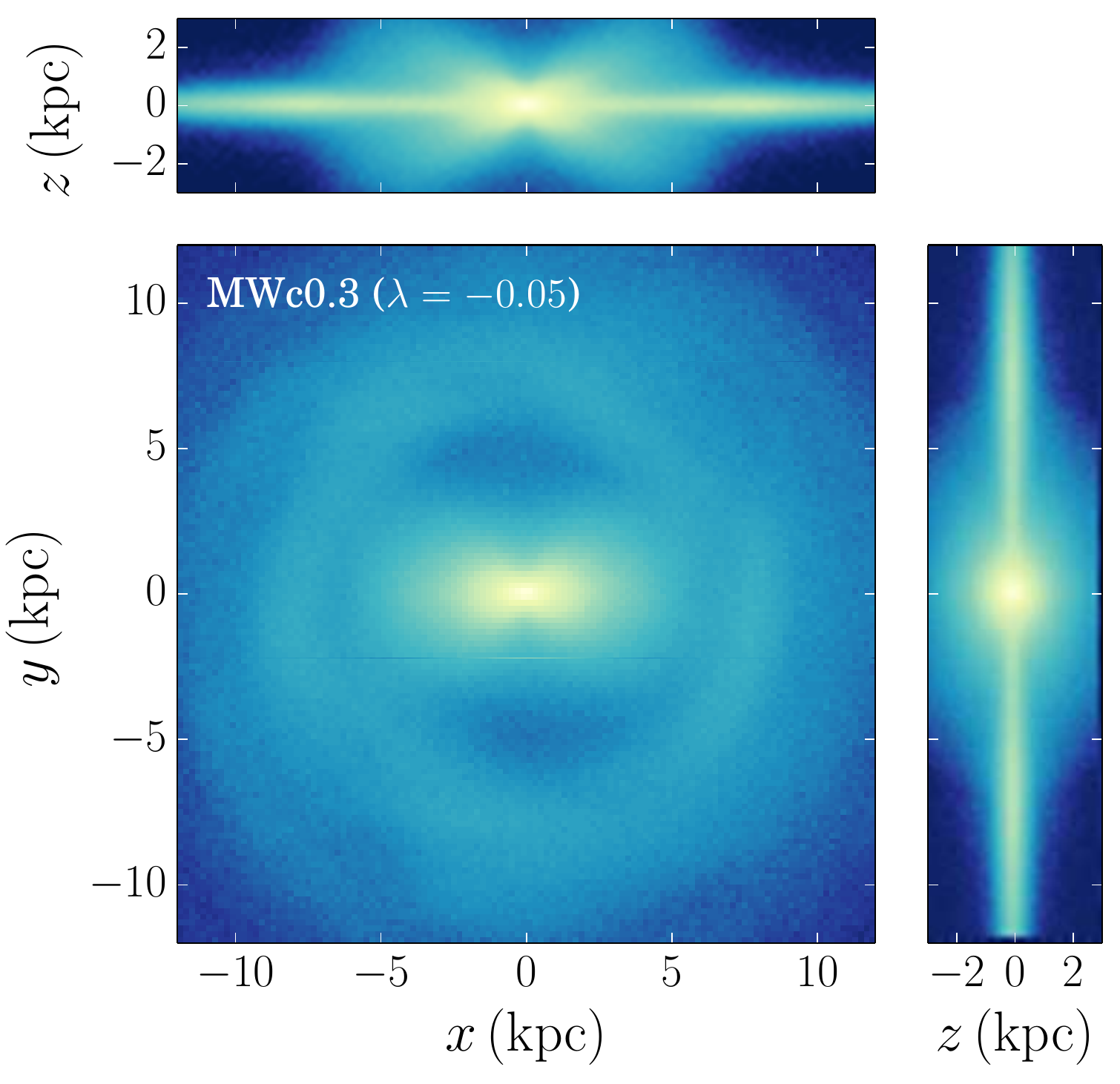}
\plotone{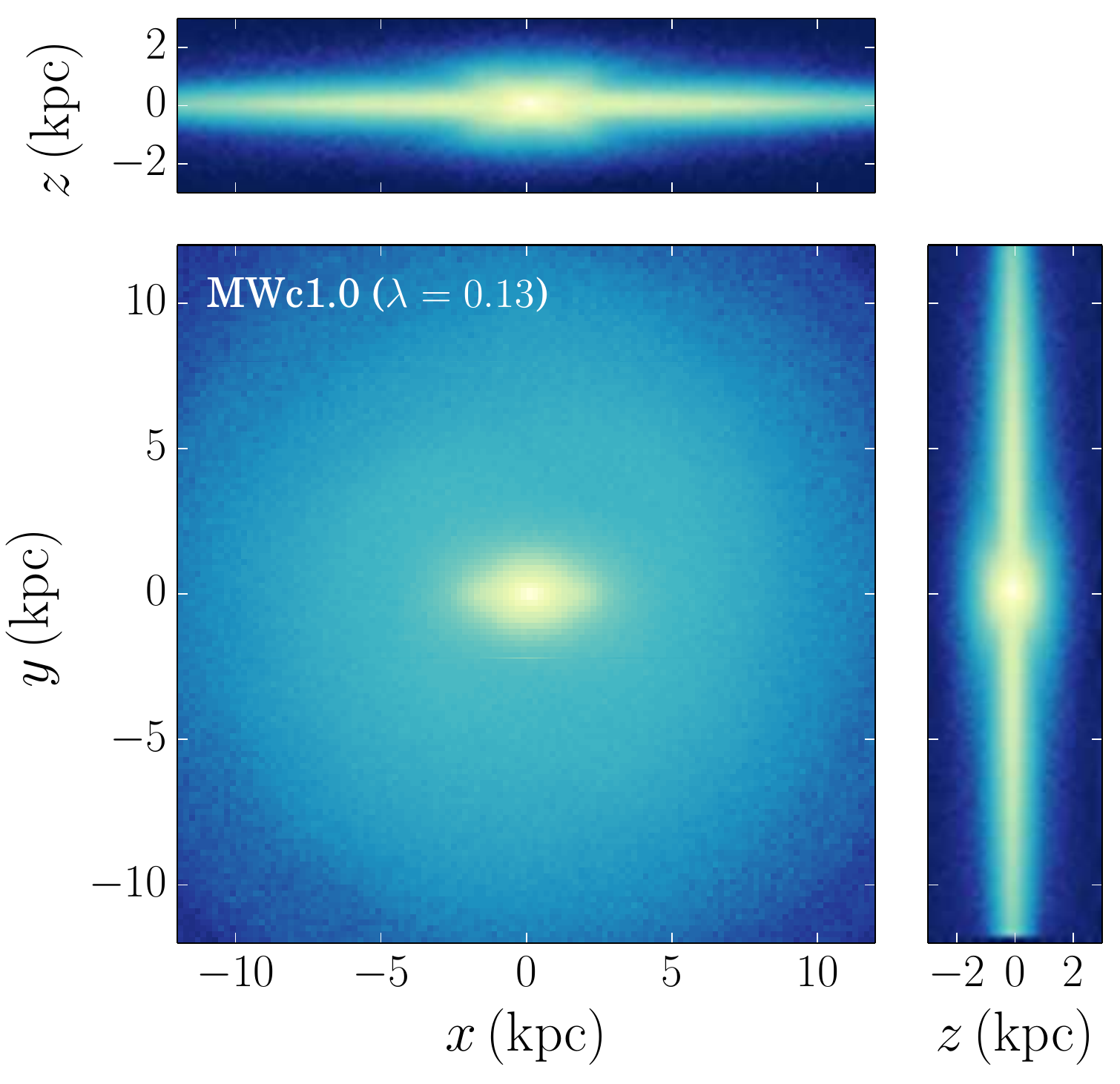}\\
\caption{Surface density maps at $t=10$\,Gyr for models MWc0.3 and MWc01.0. \label{fig:snap_c_spin}}
\end{figure*}

\begin{figure}
\raggedright
\epsscale{.5}
\plotone{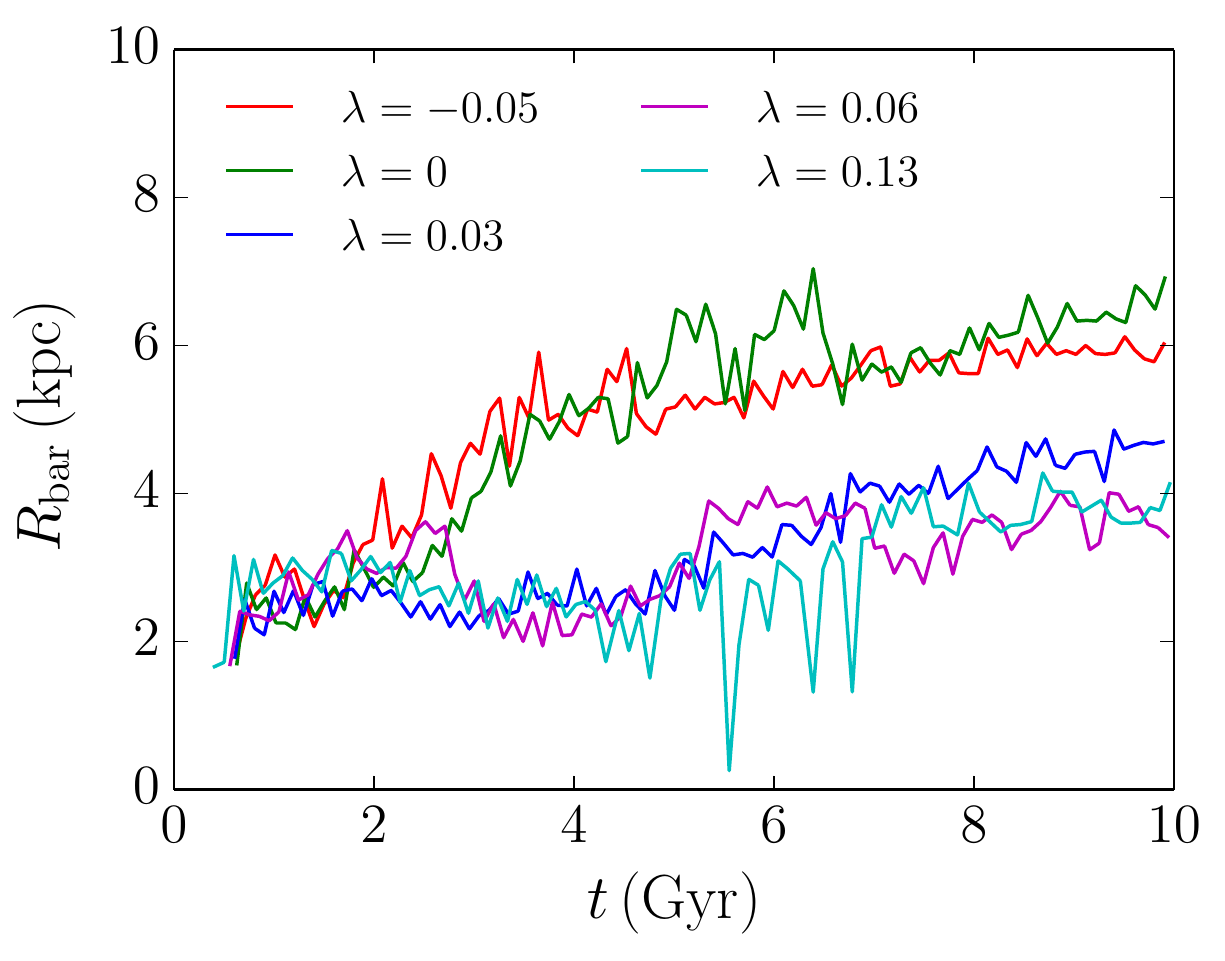}\\
\plotone{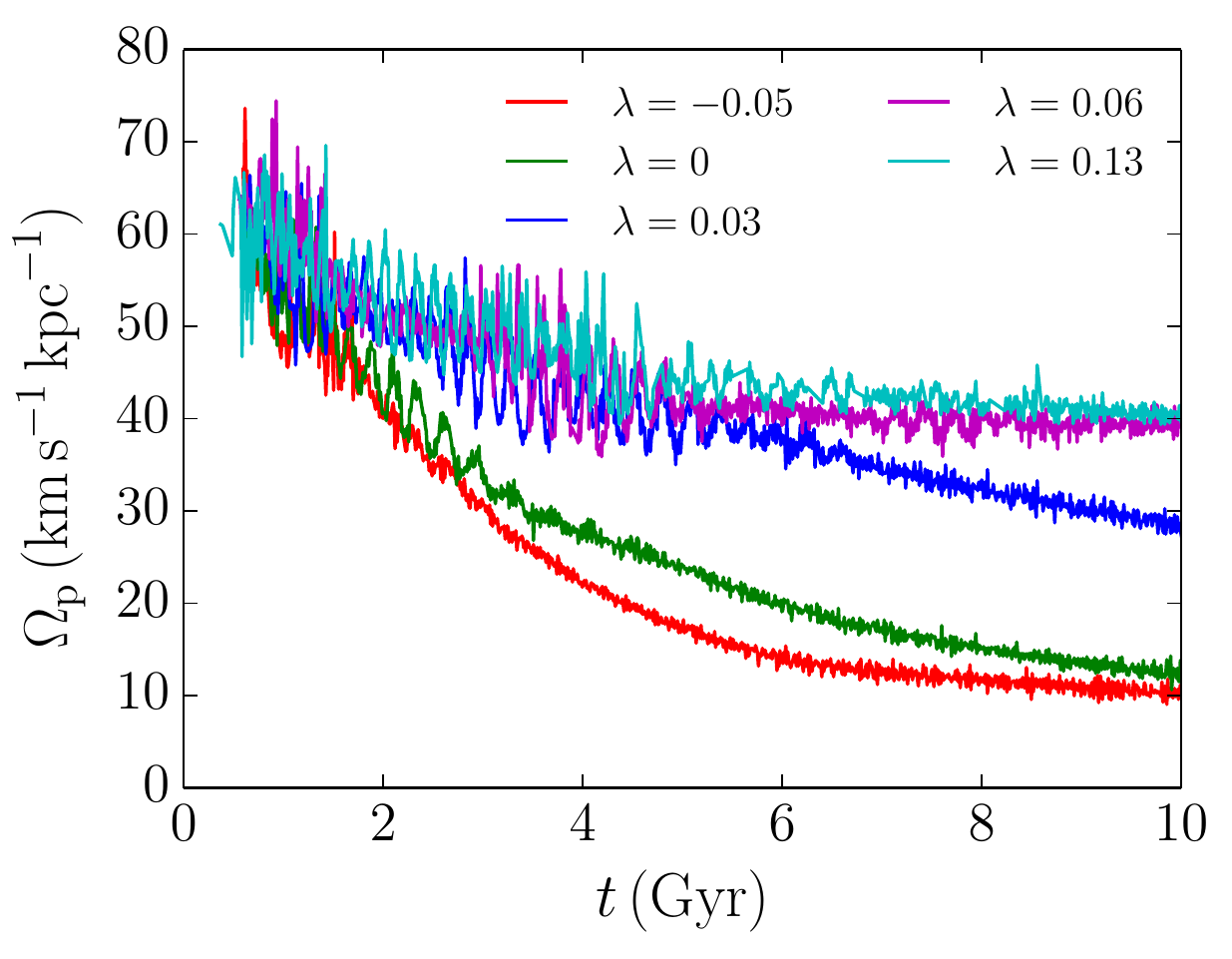}\\
\caption{Bar length (top) and pattern speed (bottom) as a function of time for models MWc0.3 ($\lambda=-0.05$), MWc0.5 ($\lambda=0.0$),  MWc0.65 ($\lambda=0.03$),  MWc0.8 ($\lambda=0.06$), and MWc1.0 ($\lambda=0.13$).  \label{fig:bar_spin}}
\end{figure}

The evolution and the final structures of these models are summarized in Figs.
\ref{fig:c0.3} and \ref{fig:c1.0}. 
In Fig. \ref{fig:snap_c_spin}, we present the final snapshots of models MWc0.3 and MWc1.0.
Model MWc0.3 (negative spin) forms a bar stronger than model MWc0.5 (see Fig. \ref{fig:snap_c}). 
The bar of model MWc1.0 is similar to that of model MWc7B (MWc0.8).
In the top panel of Fig. \ref{fig:bar_spin}, we summarized the time evolution of the 
bar length for all model MWc variations. The measured bar length is comparable for 
models MWc0.3 and MWc0.5 and models MWc0.8 and MWc1.0. The evolution of the bar length
for model MWc0.65 is similar to that of the high spin models, but the bar becomes slightly longer.
We also present the time evolution of the bar pattern speed in the bottom 
panel of Fig. \ref{fig:bar_spin}.
Although model MWc1.0 has a spin parameter that is larger than used in  model MWc0.8, the time
evolution and the final pattern speed of the bar is very similar to that of 
model MWc0.8. For model MWc0.3, which initially has a negative spin parameter, 
the pattern speed drops most rapidly, and the final pattern speed of the bar is 
slower than that of model MWc0.5. 

\begin{figure}
\raggedright
\epsscale{.5}
\plotone{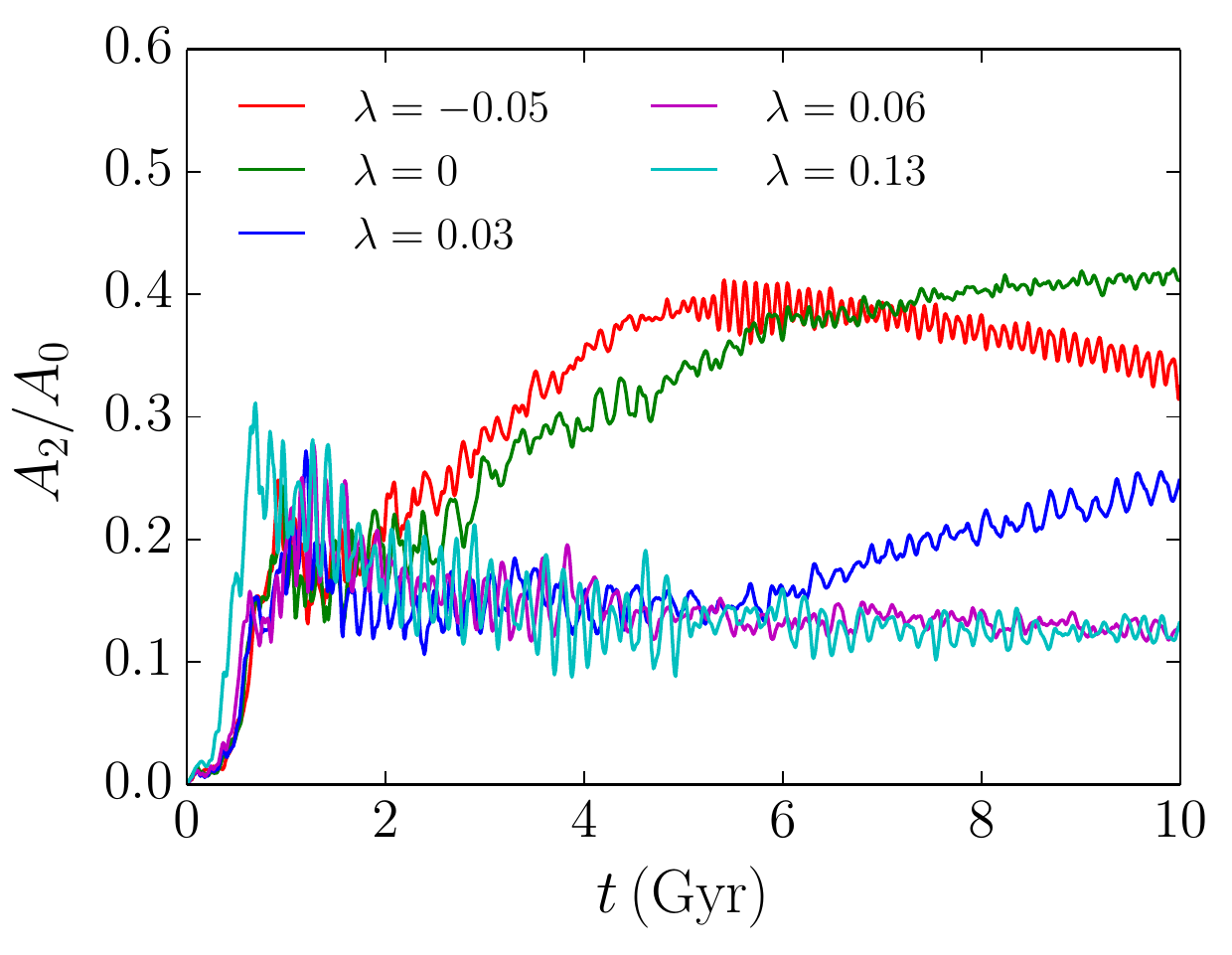}\\
\plotone{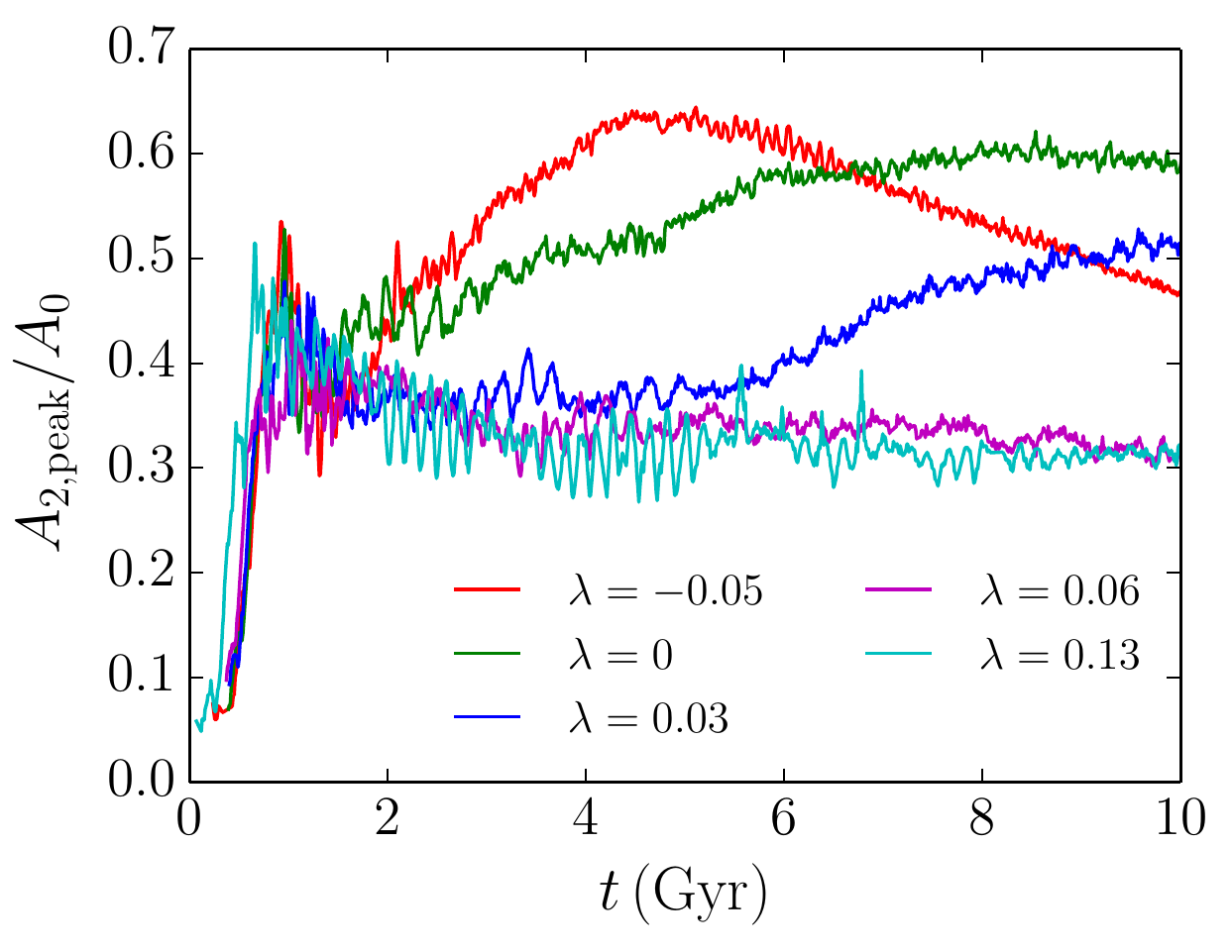}\\
\caption{Bar amplitude as a function of time for models MWc0.3 ($\lambda=-0.05$), MWc0.5 ($\lambda=0.0$),  MWc0.65 ($\lambda=0.03$),  MWc0.8 ($\lambda=0.06$), and MWc1.0 ($\lambda=0.13$). Top panel shows the mean amplitude for $R<0.65$\,kpc. Bottom panel shows the peak amplitude measured at each galactic radius.\label{fig:amp_spin}}
\end{figure}

We also investigate the evolution of the bar amplitude. We adopt two different definitions
for the amplitude. We first measure the average amplitude of the $m=2$ Fourier amplitude 
(eq. \ref{eq:Fourier}) for $R<2.5R_{\rm d}$ ($=6.5$\,kpc). This is the same as that used 
in \citet{2014ApJ...783L..18L} and `Integrated $A_2$' in \citet{2013MNRAS.434.1287S}.
The result is presented in the top panel of Fig. \ref{fig:amp_spin}.
We see features which are also seen in previous studies; for a larger spin model 
(1) the initial peak amplitude is larger and (2) the peak time is earlier. 
In \citet{2014ApJ...783L..18L}, the amplitude decreases initially, but increases more
than the initial peak for their model without initial halo spin. We see the 
same evolution behaviour in model MWc0.5. With a negative spin (model MWc0.3), the amplitude 
decreased and increased again similar as model MWc0.5, but then decreased again after 
$\sim 5$\,Gyr. 
For positive spin models, the amplitude increased again if the model has a moderate
spin (model MWc0.65), but not for the stronger spin models (MWc0.8 and MWc1.0).
These results are consistent with previous studies \citep{2014ApJ...783L..18L}.
We also measure the maximum amplitude for amplitudes measured at each galactic radius.
This is equivalent with 'Peak of $A_2$' in \citet{2013MNRAS.434.1287S}.
The results are presented in the bottom panel of Fig. \ref{fig:amp_spin}. The evolution of 
the peak amplitude is similar to that of the integrated amplitude.

We also measure the final halo spin ($\lambda_{\rm fin}$) and the spin change 
($\Delta \lambda = \lambda_{\rm fin}-\lambda_{\rm ini}$). The results are summarized in 
Table \ref{tb:spin}. For models with a negative to moderate spin the final spin parameter
increased. For the highest spin model (MWc1.0) the final
spin parameter was smaller than the initial one.

Thus, the initial halo spin sequencially changes the length and pattern
speed of the bar. As the halo spin increases, the bar becomes shorter
and the pattern speed increases. This is due to the angular momentum
transfer from the bar to the halo 
\citep{2002ApJ...569L..83A,2009ApJ...697..293D}.
This angular momentum transfer seems to saturate if we increase the initial
halo spin (maximum spin). This result implies that models with a fixed halo
potential would be similar to the maximum spin cases. 

\begin{table}
\scriptsize
\caption{Spin parameters.\label{tb:spin}}
\begin{tabular}{lcccc}
\hline
Model   & $\alpha_{\rm h}$ & $\lambda_{\rm ini}$ & $\lambda_{\rm fin}$ & $\Delta \lambda$\\
\hline \hline
MWc0.3 & 0.3 & -0.053 & -0.047 & 0.006   \\ 
MWc0.5 & 0.5 & 0.0 & 0.00050 & 0.0005   \\ 
MWc0.65 & 0.65 & 0.032 & 0.036 & 0.004   \\ 
MWc0.8 & 0.8 & 0.061 & 0.075 & 0.014   \\ 
MWc7B & 0.8 & 0.061 & 0.072 & 0.011   \\ 
MWc1.0 & 1.0 & 0.14 & 0.12 &  -0.02  \\ 
\hline
\end{tabular}
\newline
{ \scriptsize
 $\lambda_{\rm ini}$ and $\lambda_{\rm fin}$ are the initial and final halo spin parameters, and $\Delta \lambda=\lambda_{\rm fin}-\lambda_{\rm ini}$.
}
\end{table}

\bibliographystyle{mnras}
\bibliography{reference}

\begin{thebibliography}{}
\makeatletter
\relax
\def\mn@urlcharsother{\let\do\@makeother \do\$\do\&\do\#\do\^\do\_\do\%\do\~}
\def\mn@doi{\begingroup\mn@urlcharsother \@ifnextchar [ {\mn@doi@}
  {\mn@doi@[]}}
\def\mn@doi@[#1]#2{\def\@tempa{#1}\ifx\@tempa\@empty \href
  {http://dx.doi.org/#2} {doi:#2}\else \href {http://dx.doi.org/#2} {#1}\fi
  \endgroup}
\def\mn@eprint#1#2{\mn@eprint@#1:#2::\@nil}
\def\mn@eprint@arXiv#1{\href {http://arxiv.org/abs/#1} {{\tt arXiv:#1}}}
\def\mn@eprint@dblp#1{\href {http://dblp.uni-trier.de/rec/bibtex/#1.xml}
  {dblp:#1}}
\def\mn@eprint@#1:#2:#3:#4\@nil{\def\@tempa {#1}\def\@tempb {#2}\def\@tempc
  {#3}\ifx \@tempc \@empty \let \@tempc \@tempb \let \@tempb \@tempa \fi \ifx
  \@tempb \@empty \def\@tempb {arXiv}\fi \@ifundefined
  {mn@eprint@\@tempb}{\@tempb:\@tempc}{\expandafter \expandafter \csname
  mn@eprint@\@tempb\endcsname \expandafter{\@tempc}}}

\bibitem[\protect\citeauthoryear{{Abbott}, {Valluri}, {Shen}  \&
  {Debattista}}{{Abbott} et~al.}{2017}]{2017MNRAS.470.1526A}
{Abbott} C.~G.,  {Valluri} M.,  {Shen} J.,   {Debattista} V.~P.,  2017, \mn@doi
  [\mnras] {10.1093/mnras/stx1262}, \href
  {http://adsabs.harvard.edu/abs/2017MNRAS.470.1526A} {470, 1526}

\bibitem[\protect\citeauthoryear{{Antoja} et~al.,}{{Antoja}
  et~al.}{2014}]{2014A&A...563A..60A}
{Antoja} T.,  et~al., 2014, \mn@doi [\aap] {10.1051/0004-6361/201322623}, \href
  {http://adsabs.harvard.edu/abs/2014A%26A...563A..60A} {563, A60}

\bibitem[\protect\citeauthoryear{{Athanassoula}}{{Athanassoula}}{1992}]{1992MNRAS.259..328A}
{Athanassoula} E.,  1992, \mn@doi [\mnras] {10.1093/mnras/259.2.328}, \href
  {http://adsabs.harvard.edu/abs/1992MNRAS.259..328A} {259, 328}

\bibitem[\protect\citeauthoryear{{Athanassoula}}{{Athanassoula}}{2002}]{2002ApJ...569L..83A}
{Athanassoula} E.,  2002, \mn@doi [\apjl] {10.1086/340784}, \href
  {http://adsabs.harvard.edu/abs/2002ApJ...569L..83A} {569, L83}

\bibitem[\protect\citeauthoryear{{Baba}}{{Baba}}{2015}]{2015MNRAS.454.2954B}
{Baba} J.,  2015, \mn@doi [\mnras] {10.1093/mnras/stv2220}, \href
  {http://adsabs.harvard.edu/abs/2015MNRAS.454.2954B} {454, 2954}

\bibitem[\protect\citeauthoryear{{Baba}, {Saitoh}  \& {Wada}}{{Baba}
  et~al.}{2010}]{2010PASJ...62.1413B}
{Baba} J.,  {Saitoh} T.~R.,   {Wada} K.,  2010, \mn@doi [\pasj]
  {10.1093/pasj/62.6.1413}, \href
  {http://adsabs.harvard.edu/abs/2010PASJ...62.1413B} {62, 1413}

\bibitem[\protect\citeauthoryear{{B{\'e}dorf}, {Gaburov}  \& {Portegies
  Zwart}}{{B{\'e}dorf} et~al.}{2012}]{2012JCoPh.231.2825B}
{B{\'e}dorf} J.,  {Gaburov} E.,   {Portegies Zwart} S.,  2012, \mn@doi [Journal
  of Computational Physics] {10.1016/j.jcp.2011.12.024}, \href
  {http://adsabs.harvard.edu/abs/2012JCoPh.231.2825B} {231, 2825}

\bibitem[\protect\citeauthoryear{{B{\'e}dorf}, {Gaburov}, {Fujii}, {Nitadori},
  {Ishiyama}  \& {Portegies Zwart}}{{B{\'e}dorf}
  et~al.}{2014}]{2014hpcn.conf...54B}
{B{\'e}dorf} J.,  {Gaburov} E.,  {Fujii} M.~S.,  {Nitadori} K.,  {Ishiyama} T.,
    {Portegies Zwart} S.,  2014, in Proceedings of the International Conference
  for High Performance Computing, Networking, Storage and Analysis, p. 54-65.
  pp 54--65 (\mn@eprint {arXiv} {1412.0659}), \mn@doi{10.1109/SC.2014.10OTHER:
  http://dl.acm.org/citation.cfm?id=2683600}

\bibitem[\protect\citeauthoryear{{Bett}, {Eke}, {Frenk}, {Jenkins}, {Helly}  \&
  {Navarro}}{{Bett} et~al.}{2007}]{2007MNRAS.376..215B}
{Bett} P.,  {Eke} V.,  {Frenk} C.~S.,  {Jenkins} A.,  {Helly} J.,   {Navarro}
  J.,  2007, \mn@doi [\mnras] {10.1111/j.1365-2966.2007.11432.x}, \href
  {http://adsabs.harvard.edu/abs/2007MNRAS.376..215B} {376, 215}

\bibitem[\protect\citeauthoryear{{Binney} \& {Tremaine}}{{Binney} \&
  {Tremaine}}{2008}]{2008gady.book.....B}
{Binney} J.,  {Tremaine} S.,  2008, {Galactic Dynamics: Second Edition}.
Princeton University Press

\bibitem[\protect\citeauthoryear{{Bland-Hawthorn} \&
  {Gerhard}}{{Bland-Hawthorn} \& {Gerhard}}{2016}]{2016ARA&A..54..529B}
{Bland-Hawthorn} J.,  {Gerhard} O.,  2016, \mn@doi [\araa]
  {10.1146/annurev-astro-081915-023441}, \href
  {http://adsabs.harvard.edu/abs/2016ARA%26A..54..529B} {54, 529}

\bibitem[\protect\citeauthoryear{{Boekholt} \& {Portegies Zwart}}{{Boekholt} \&
  {Portegies Zwart}}{2015}]{2015ComAC...2....2B}
{Boekholt} T.,  {Portegies Zwart} S.,  2015, \mn@doi [Computational
  Astrophysics and Cosmology] {10.1186/s40668-014-0005-3}, \href
  {http://adsabs.harvard.edu/abs/2015ComAC...2....2B} {2, 2}

\bibitem[\protect\citeauthoryear{{Bovy} \& {Rix}}{{Bovy} \&
  {Rix}}{2013}]{2013ApJ...779..115B}
{Bovy} J.,  {Rix} H.-W.,  2013, \mn@doi [\apj] {10.1088/0004-637X/779/2/115},
  \href {http://adsabs.harvard.edu/abs/2013ApJ...779..115B} {779, 115}

\bibitem[\protect\citeauthoryear{{Cao}, {Mao}, {Nataf}, {Rattenbury}  \&
  {Gould}}{{Cao} et~al.}{2013}]{2013MNRAS.434..595C}
{Cao} L.,  {Mao} S.,  {Nataf} D.,  {Rattenbury} N.~J.,   {Gould} A.,  2013,
  \mn@doi [\mnras] {10.1093/mnras/stt1045}, \href
  {http://adsabs.harvard.edu/abs/2013MNRAS.434..595C} {434, 595}

\bibitem[\protect\citeauthoryear{{Carlberg}}{{Carlberg}}{1987}]{1987ApJ...322...59C}
{Carlberg} R.~G.,  1987, \mn@doi [\apj] {10.1086/165702}, \href
  {http://adsabs.harvard.edu/abs/1987ApJ...322...59C} {322, 59}

\bibitem[\protect\citeauthoryear{{Cervantes-Sodi}, {Li}, {Park}  \&
  {Wang}}{{Cervantes-Sodi} et~al.}{2013}]{2013ApJ...775...19C}
{Cervantes-Sodi} B.,  {Li} C.,  {Park} C.,   {Wang} L.,  2013, \mn@doi [\apj]
  {10.1088/0004-637X/775/1/19}, \href
  {http://adsabs.harvard.edu/abs/2013ApJ...775...19C} {775, 19}

\bibitem[\protect\citeauthoryear{{Ciambur}, {Graham}  \&
  {Bland-Hawthorn}}{{Ciambur} et~al.}{2017}]{2017MNRAS.471.3988C}
{Ciambur} B.~C.,  {Graham} A.~W.,   {Bland-Hawthorn} J.,  2017, \mn@doi
  [\mnras] {10.1093/mnras/stx1823}, \href
  {http://adsabs.harvard.edu/abs/2017MNRAS.471.3988C} {471, 3988}

\bibitem[\protect\citeauthoryear{{Contopoulos}}{{Contopoulos}}{1980}]{1980A&A....81..198C}
{Contopoulos} G.,  1980, \aap, \href
  {http://adsabs.harvard.edu/abs/1980A%26A....81..198C} {81, 198}

\bibitem[\protect\citeauthoryear{{Correa}, {Wyithe}, {Schaye}  \&
  {Duffy}}{{Correa} et~al.}{2015}]{2015MNRAS.452.1217C}
{Correa} C.~A.,  {Wyithe} J.~S.~B.,  {Schaye} J.,   {Duffy} A.~R.,  2015,
  \mn@doi [\mnras] {10.1093/mnras/stv1363}, \href
  {http://adsabs.harvard.edu/abs/2015MNRAS.452.1217C} {452, 1217}

\bibitem[\protect\citeauthoryear{{Dehnen}}{{Dehnen}}{1998}]{1998AJ....115.2384D}
{Dehnen} W.,  1998, \mn@doi [\aj] {10.1086/300364}, \href
  {http://adsabs.harvard.edu/abs/1998AJ....115.2384D} {115, 2384}

\bibitem[\protect\citeauthoryear{{Dehnen}}{{Dehnen}}{2000}]{2000AJ....119..800D}
{Dehnen} W.,  2000, \mn@doi [\aj] {10.1086/301226}, \href
  {http://adsabs.harvard.edu/abs/2000AJ....119..800D} {119, 800}

\bibitem[\protect\citeauthoryear{{Dubinski}, {Berentzen}  \&
  {Shlosman}}{{Dubinski} et~al.}{2009}]{2009ApJ...697..293D}
{Dubinski} J.,  {Berentzen} I.,   {Shlosman} I.,  2009, \mn@doi [\apj]
  {10.1088/0004-637X/697/1/293}, \href
  {http://adsabs.harvard.edu/abs/2009ApJ...697..293D} {697, 293}

\bibitem[\protect\citeauthoryear{{Francis} \& {Anderson}}{{Francis} \&
  {Anderson}}{2009}]{2009NewA...14..615F}
{Francis} C.,  {Anderson} E.,  2009, \mn@doi [\na]
  {10.1016/j.newast.2009.03.004}, \href
  {http://adsabs.harvard.edu/abs/2009NewA...14..615F} {14, 615}

\bibitem[\protect\citeauthoryear{{Fujii}, {Baba}, {Saitoh}, {Makino}, {Kokubo}
  \& {Wada}}{{Fujii} et~al.}{2011}]{2011ApJ...730..109F}
{Fujii} M.~S.,  {Baba} J.,  {Saitoh} T.~R.,  {Makino} J.,  {Kokubo} E.,
  {Wada} K.,  2011, \mn@doi [\apj] {10.1088/0004-637X/730/2/109}, \href
  {http://adsabs.harvard.edu/abs/2011ApJ...730..109F} {730, 109}

\bibitem[\protect\citeauthoryear{{Fujii}, {B{\'e}dorf}, {Baba}  \& {Portegies
  Zwart}}{{Fujii} et~al.}{2018}]{2018MNRAS.477.1451F}
{Fujii} M.~S.,  {B{\'e}dorf} J.,  {Baba} J.,   {Portegies Zwart} S.,  2018,
  \mn@doi [\mnras] {10.1093/mnras/sty711}, \href
  {http://adsabs.harvard.edu/abs/2018MNRAS.477.1451F} {477, 1451}

\bibitem[\protect\citeauthoryear{{Fux}}{{Fux}}{1997}]{1997A&A...327..983F}
{Fux} R.,  1997, \aap, \href {http://ads.nao.ac.jp/abs/1997A%26A...327..983F}
  {327, 983}

\bibitem[\protect\citeauthoryear{{Gaia Collaboration} et~al.,}{{Gaia
  Collaboration} et~al.}{2016}]{2016A&A...595A...1G}
{Gaia Collaboration} et~al., 2016, \mn@doi [\aap]
  {10.1051/0004-6361/201629272}, \href
  {http://adsabs.harvard.edu/abs/2016A%26A...595A...1G} {595, A1}

\bibitem[\protect\citeauthoryear{{Gaia Collaboration}, {Brown}, {Vallenari},
  {Prusti}, {de Bruijne}, {Babusiaux}  \& {Bailer-Jones}}{{Gaia Collaboration}
  et~al.}{2018}]{2018arXiv180409365G}
{Gaia Collaboration} {Brown} A.~G.~A.,  {Vallenari} A.,  {Prusti} T.,  {de
  Bruijne} J.~H.~J.,  {Babusiaux} C.,   {Bailer-Jones} C.~A.~L.,  2018,
  preprint, \href {http://adsabs.harvard.edu/abs/2018arXiv180409365G} {}
  (\mn@eprint {arXiv} {1804.09365})

\bibitem[\protect\citeauthoryear{{Gardner}, {Debattista}, {Robin},
  {V{\'a}squez}  \& {Zoccali}}{{Gardner} et~al.}{2014}]{2014MNRAS.438.3275G}
{Gardner} E.,  {Debattista} V.~P.,  {Robin} A.~C.,  {V{\'a}squez} S.,
  {Zoccali} M.,  2014, \mn@doi [\mnras] {10.1093/mnras/stt2430}, \href
  {http://adsabs.harvard.edu/abs/2014MNRAS.438.3275G} {438, 3275}

\bibitem[\protect\citeauthoryear{{Gerhard} \& {Martinez-Valpuesta}}{{Gerhard}
  \& {Martinez-Valpuesta}}{2012}]{2012ApJ...744L...8G}
{Gerhard} O.,  {Martinez-Valpuesta} I.,  2012, \mn@doi [\apjl]
  {10.1088/2041-8205/744/1/L8}, \href
  {http://adsabs.harvard.edu/abs/2012ApJ...744L...8G} {744, L8}

\bibitem[\protect\citeauthoryear{{Gonzalez}, {Rejkuba}, {Minniti}, {Zoccali},
  {Valenti}  \& {Saito}}{{Gonzalez} et~al.}{2011}]{2011A&A...534L..14G}
{Gonzalez} O.~A.,  {Rejkuba} M.,  {Minniti} D.,  {Zoccali} M.,  {Valenti} E.,
  {Saito} R.~K.,  2011, \mn@doi [\aap] {10.1051/0004-6361/201117959}, \href
  {http://adsabs.harvard.edu/abs/2011A%26A...534L..14G} {534, L14}

\bibitem[\protect\citeauthoryear{{Hattori}, {Gouda}, {Yano}, {Sakai}, {Tagawa},
  {Baba}  \& {Kumamoto}}{{Hattori} et~al.}{2018}]{2018arXiv180401920H}
{Hattori} K.,  {Gouda} N.,  {Yano} T.,  {Sakai} N.,  {Tagawa} H.,  {Baba} J.,
  {Kumamoto} J.,  2018, preprint, \href
  {http://adsabs.harvard.edu/abs/2018arXiv180401920H} {} (\mn@eprint {arXiv}
  {1804.01920})

\bibitem[\protect\citeauthoryear{{Hernquist}}{{Hernquist}}{1990}]{1990ApJ...356..359H}
{Hernquist} L.,  1990, \mn@doi [\apj] {10.1086/168845}, \href
  {http://adsabs.harvard.edu/abs/1990ApJ...356..359H} {356, 359}

\bibitem[\protect\citeauthoryear{{Howard}, {Rich}, {Reitzel}, {Koch}, {De
  Propris}  \& {Zhao}}{{Howard} et~al.}{2008}]{2008ApJ...688.1060H}
{Howard} C.~D.,  {Rich} R.~M.,  {Reitzel} D.~B.,  {Koch} A.,  {De Propris} R.,
   {Zhao} H.,  2008, \mn@doi [\apj] {10.1086/592106}, \href
  {http://adsabs.harvard.edu/abs/2008ApJ...688.1060H} {688, 1060}

\bibitem[\protect\citeauthoryear{{Huang} et~al.,}{{Huang}
  et~al.}{2016}]{2016MNRAS.463.2623H}
{Huang} Y.,  et~al., 2016, \mn@doi [\mnras] {10.1093/mnras/stw2096}, \href
  {http://adsabs.harvard.edu/abs/2016MNRAS.463.2623H} {463, 2623}

\bibitem[\protect\citeauthoryear{{Hunt} \& {Bovy}}{{Hunt} \&
  {Bovy}}{2018}]{2018MNRAS.477.3945H}
{Hunt} J.~A.~S.,  {Bovy} J.,  2018, \mn@doi [\mnras] {10.1093/mnras/sty921},
  \href {http://adsabs.harvard.edu/abs/2018MNRAS.477.3945H} {477, 3945}

\bibitem[\protect\citeauthoryear{{Hunt}, {Hong}, {Bovy}, {Kawata}  \&
  {Grand}}{{Hunt} et~al.}{2018a}]{2018MNRAS.tmp.2421H}
{Hunt} J.~A.~S.,  {Hong} J.,  {Bovy} J.,  {Kawata} D.,   {Grand} R.~J.~J.,
  2018a, \mn@doi [\mnras] {10.1093/mnras/sty2532}, \href
  {http://adsabs.harvard.edu/abs/2018MNRAS.tmp.2421H} {}

\bibitem[\protect\citeauthoryear{{Hunt} et~al.,}{{Hunt}
  et~al.}{2018b}]{2018MNRAS.474...95H}
{Hunt} J.~A.~S.,  et~al., 2018b, \mn@doi [\mnras] {10.1093/mnras/stx2777},
  \href {http://adsabs.harvard.edu/abs/2018MNRAS.474...95H} {474, 95}

\bibitem[\protect\citeauthoryear{{Jenkins} \& {Binney}}{{Jenkins} \&
  {Binney}}{1990}]{1990MNRAS.245..305J}
{Jenkins} A.,  {Binney} J.,  1990, \mnras, \href
  {http://adsabs.harvard.edu/abs/1990MNRAS.245..305J} {245, 305}

\bibitem[\protect\citeauthoryear{{Klypin}, {Zhao}  \& {Somerville}}{{Klypin}
  et~al.}{2002}]{2002ApJ...573..597K}
{Klypin} A.,  {Zhao} H.,   {Somerville} R.~S.,  2002, \mn@doi [\apj]
  {10.1086/340656}, \href {http://adsabs.harvard.edu/abs/2002ApJ...573..597K}
  {573, 597}

\bibitem[\protect\citeauthoryear{{Kuijken} \& {Dubinski}}{{Kuijken} \&
  {Dubinski}}{1995}]{1995MNRAS.277.1341K}
{Kuijken} K.,  {Dubinski} J.,  1995, \mn@doi [\mnras]
  {10.1093/mnras/277.4.1341}, \href
  {http://adsabs.harvard.edu/abs/1995MNRAS.277.1341K} {277, 1341}

\bibitem[\protect\citeauthoryear{{Kuijken} \& {Gilmore}}{{Kuijken} \&
  {Gilmore}}{1991}]{1991ApJ...367L...9K}
{Kuijken} K.,  {Gilmore} G.,  1991, \mn@doi [\apjl] {10.1086/185920}, \href
  {http://adsabs.harvard.edu/abs/1991ApJ...367L...9K} {367, L9}

\bibitem[\protect\citeauthoryear{{Kunder} et~al.,}{{Kunder}
  et~al.}{2012}]{2012AJ....143...57K}
{Kunder} A.,  et~al., 2012, \mn@doi [\aj] {10.1088/0004-6256/143/3/57}, \href
  {http://adsabs.harvard.edu/abs/2012AJ....143...57K} {143, 57}

\bibitem[\protect\citeauthoryear{{Long}, {Shlosman}  \& {Heller}}{{Long}
  et~al.}{2014}]{2014ApJ...783L..18L}
{Long} S.,  {Shlosman} I.,   {Heller} C.,  2014, \mn@doi [\apjl]
  {10.1088/2041-8205/783/1/L18}, \href
  {http://adsabs.harvard.edu/abs/2014ApJ...783L..18L} {783, L18}

\bibitem[\protect\citeauthoryear{{L{\'o}pez-Corredoira}, {Cabrera-Lavers}  \&
  {Gerhard}}{{L{\'o}pez-Corredoira} et~al.}{2005}]{2005A&A...439..107L}
{L{\'o}pez-Corredoira} M.,  {Cabrera-Lavers} A.,   {Gerhard} O.~E.,  2005,
  \mn@doi [\aap] {10.1051/0004-6361:20053075}, \href
  {http://adsabs.harvard.edu/abs/2005A%26A...439..107L} {439, 107}

\bibitem[\protect\citeauthoryear{{McKee}, {Parravano}  \& {Hollenbach}}{{McKee}
  et~al.}{2015}]{2015ApJ...814...13M}
{McKee} C.~F.,  {Parravano} A.,   {Hollenbach} D.~J.,  2015, \mn@doi [\apj]
  {10.1088/0004-637X/814/1/13}, \href
  {http://adsabs.harvard.edu/abs/2015ApJ...814...13M} {814, 13}

\bibitem[\protect\citeauthoryear{{McMillan}}{{McMillan}}{2017}]{2017MNRAS.465...76M}
{McMillan} P.~J.,  2017, \mn@doi [\mnras] {10.1093/mnras/stw2759}, \href
  {http://adsabs.harvard.edu/abs/2017MNRAS.465...76M} {465, 76}

\bibitem[\protect\citeauthoryear{{Monari}, {Kawata}, {Hunt}  \&
  {Famaey}}{{Monari} et~al.}{2017}]{2017MNRAS.466L.113M}
{Monari} G.,  {Kawata} D.,  {Hunt} J.~A.~S.,   {Famaey} B.,  2017, \mn@doi
  [\mnras] {10.1093/mnrasl/slw238}, \href
  {http://adsabs.harvard.edu/abs/2017MNRAS.466L.113M} {466, L113}

\bibitem[\protect\citeauthoryear{{Navarro}, {Frenk}  \& {White}}{{Navarro}
  et~al.}{1997}]{1997ApJ...490..493N}
{Navarro} J.~F.,  {Frenk} C.~S.,   {White} S.~D.~M.,  1997, \apj, \href
  {http://adsabs.harvard.edu/abs/1997ApJ...490..493N} {490, 493}

\bibitem[\protect\citeauthoryear{{Okamoto}, {Isoe}  \& {Habe}}{{Okamoto}
  et~al.}{2015}]{2015PASJ...67...63O}
{Okamoto} T.,  {Isoe} M.,   {Habe} A.,  2015, \mn@doi [\pasj]
  {10.1093/pasj/psv037}, \href
  {http://adsabs.harvard.edu/abs/2015PASJ...67...63O} {67, 63}

\bibitem[\protect\citeauthoryear{{Peebles}}{{Peebles}}{1969}]{1969ApJ...155..393P}
{Peebles} P.~J.~E.,  1969, \mn@doi [\apj] {10.1086/149876}, \href
  {http://adsabs.harvard.edu/abs/1969ApJ...155..393P} {155, 393}

\bibitem[\protect\citeauthoryear{{Peebles}}{{Peebles}}{1971}]{1971A&A....11..377P}
{Peebles} P.~J.~E.,  1971, \aap, \href
  {http://adsabs.harvard.edu/abs/1971A%26A....11..377P} {11, 377}

\bibitem[\protect\citeauthoryear{{Pelupessy}, {van Elteren}, {de Vries},
  {McMillan}, {Drost}  \& {Portegies Zwart}}{{Pelupessy}
  et~al.}{2013}]{2013A&A...557A..84P}
{Pelupessy} F.~I.,  {van Elteren} A.,  {de Vries} N.,  {McMillan} S.~L.~W.,
  {Drost} N.,   {Portegies Zwart} S.~F.,  2013, \mn@doi [\aap]
  {10.1051/0004-6361/201321252}, \href
  {http://adsabs.harvard.edu/abs/2013A%26A...557A..84P} {557, A84}

\bibitem[\protect\citeauthoryear{{P{\'e}rez-Villegas}, {Portail}, {Wegg}  \&
  {Gerhard}}{{P{\'e}rez-Villegas} et~al.}{2017}]{2017ApJ...840L...2P}
{P{\'e}rez-Villegas} A.,  {Portail} M.,  {Wegg} C.,   {Gerhard} O.,  2017,
  \mn@doi [\apjl] {10.3847/2041-8213/aa6c26}, \href
  {http://adsabs.harvard.edu/abs/2017ApJ...840L...2P} {840, L2}

\bibitem[\protect\citeauthoryear{{Portegies Zwart} \& Boekholt}{{Portegies
  Zwart} \& Boekholt}{2018}]{PORTEGIESZWART2018160}
{Portegies Zwart} S.~F.,  Boekholt T.~C.,  2018, \mn@doi [Communications in
  Nonlinear Science and Numerical Simulation]
  {https://doi.org/10.1016/j.cnsns.2018.02.002}, 61, 160

\bibitem[\protect\citeauthoryear{{Portegies Zwart} \& {McMillan}}{{Portegies
  Zwart} \& {McMillan}}{2018}]{AMUSE}
{Portegies Zwart} S.,  {McMillan} S.,  2018, {Astrophysical Recipes: the Art of
  AMUSE}.
AAS IOP Astronomy

\bibitem[\protect\citeauthoryear{{Portegies Zwart}, {McMillan}, {van Elteren},
  {Pelupessy}  \& {de Vries}}{{Portegies Zwart}
  et~al.}{2013}]{2013CoPhC.183..456P}
{Portegies Zwart} S.,  {McMillan} S.~L.~W.,  {van Elteren} E.,  {Pelupessy} I.,
    {de Vries} N.,  2013, \mn@doi [Computer Physics Communications]
  {10.1016/j.cpc.2012.09.024}, \href
  {http://adsabs.harvard.edu/abs/2013CoPhC.183..456P} {183, 456}

\bibitem[\protect\citeauthoryear{{Quillen}, {Dougherty}, {Bagley}, {Minchev}
  \& {Comparetta}}{{Quillen} et~al.}{2011}]{2011MNRAS.417..762Q}
{Quillen} A.~C.,  {Dougherty} J.,  {Bagley} M.~B.,  {Minchev} I.,
  {Comparetta} J.,  2011, \mn@doi [\mnras] {10.1111/j.1365-2966.2011.19349.x},
  \href {http://adsabs.harvard.edu/abs/2011MNRAS.417..762Q} {417, 762}

\bibitem[\protect\citeauthoryear{{Quillen} et~al.,}{{Quillen}
  et~al.}{2018}]{2018MNRAS.478..228Q}
{Quillen} A.~C.,  et~al., 2018, \mn@doi [\mnras] {10.1093/mnras/sty865}, \href
  {http://adsabs.harvard.edu/abs/2018MNRAS.478..228Q} {478, 228}

\bibitem[\protect\citeauthoryear{{Reid} et~al.,}{{Reid}
  et~al.}{2014}]{2014ApJ...783..130R}
{Reid} M.~J.,  et~al., 2014, \mn@doi [\apj] {10.1088/0004-637X/783/2/130},
  \href {http://adsabs.harvard.edu/abs/2014ApJ...783..130R} {783, 130}

\bibitem[\protect\citeauthoryear{{Saha} \& {Naab}}{{Saha} \&
  {Naab}}{2013}]{2013MNRAS.434.1287S}
{Saha} K.,  {Naab} T.,  2013, \mn@doi [\mnras] {10.1093/mnras/stt1088}, \href
  {http://adsabs.harvard.edu/abs/2013MNRAS.434.1287S} {434, 1287}

\bibitem[\protect\citeauthoryear{{Sellwood} \& {Debattista}}{{Sellwood} \&
  {Debattista}}{2009}]{2009MNRAS.398.1279S}
{Sellwood} J.~A.,  {Debattista} V.~P.,  2009, \mn@doi [\mnras]
  {10.1111/j.1365-2966.2009.15219.x}, \href
  {http://adsabs.harvard.edu/abs/2009MNRAS.398.1279S} {398, 1279}

\bibitem[\protect\citeauthoryear{{Sellwood} \& {Sparke}}{{Sellwood} \&
  {Sparke}}{1988}]{1988MNRAS.231P..25S}
{Sellwood} J.~A.,  {Sparke} L.~S.,  1988, \mn@doi [\mnras]
  {10.1093/mnras/231.1.25P}, \href
  {http://adsabs.harvard.edu/abs/1988MNRAS.231P..25S} {231, 25P}

\bibitem[\protect\citeauthoryear{{Shen}, {Rich}, {Kormendy}, {Howard}, {De
  Propris}  \& {Kunder}}{{Shen} et~al.}{2010}]{2010ApJ...720L..72S}
{Shen} J.,  {Rich} R.~M.,  {Kormendy} J.,  {Howard} C.~D.,  {De Propris} R.,
  {Kunder} A.,  2010, \mn@doi [\apjl] {10.1088/2041-8205/720/1/L72}, \href
  {http://adsabs.harvard.edu/abs/2010ApJ...720L..72S} {720, L72}

\bibitem[\protect\citeauthoryear{{Sofue}}{{Sofue}}{2012}]{2012PASJ...64...75S}
{Sofue} Y.,  2012, \mn@doi [\pasj] {10.1093/pasj/64.4.75}, \href
  {http://adsabs.harvard.edu/abs/2012PASJ...64...75S} {64, 75}

\bibitem[\protect\citeauthoryear{{Toomre}}{{Toomre}}{1964}]{1964ApJ...139.1217T}
{Toomre} A.,  1964, \mn@doi [\apj] {10.1086/147861}, \href
  {http://adsabs.harvard.edu/abs/1964ApJ...139.1217T} {139, 1217}

\bibitem[\protect\citeauthoryear{{Tremaine} et~al.,}{{Tremaine}
  et~al.}{2002}]{2002ApJ...574..740T}
{Tremaine} S.,  et~al., 2002, \mn@doi [\apj] {10.1086/341002}, \href
  {http://adsabs.harvard.edu/abs/2002ApJ...574..740T} {574, 740}

\bibitem[\protect\citeauthoryear{{Vall{\'e}e}}{{Vall{\'e}e}}{2017}]{2017NewAR..79...49V}
{Vall{\'e}e} J.~P.,  2017, \mn@doi [\nar] {10.1016/j.newar.2017.09.001}, \href
  {http://adsabs.harvard.edu/abs/2017NewAR..79...49V} {79, 49}

\bibitem[\protect\citeauthoryear{{Wegg} \& {Gerhard}}{{Wegg} \&
  {Gerhard}}{2013}]{2013MNRAS.435.1874W}
{Wegg} C.,  {Gerhard} O.,  2013, \mn@doi [\mnras] {10.1093/mnras/stt1376},
  \href {http://adsabs.harvard.edu/abs/2013MNRAS.435.1874W} {435, 1874}

\bibitem[\protect\citeauthoryear{{Wegg}, {Gerhard}  \& {Portail}}{{Wegg}
  et~al.}{2015}]{2015MNRAS.450.4050W}
{Wegg} C.,  {Gerhard} O.,   {Portail} M.,  2015, \mn@doi [\mnras]
  {10.1093/mnras/stv745}, \href
  {http://adsabs.harvard.edu/abs/2015MNRAS.450.4050W} {450, 4050}

\bibitem[\protect\citeauthoryear{{Widrow} \& {Dubinski}}{{Widrow} \&
  {Dubinski}}{2005}]{2005ApJ...631..838W}
{Widrow} L.~M.,  {Dubinski} J.,  2005, \mn@doi [\apj] {10.1086/432710}, \href
  {http://adsabs.harvard.edu/abs/2005ApJ...631..838W} {631, 838}

\bibitem[\protect\citeauthoryear{{Widrow}, {Pym}  \& {Dubinski}}{{Widrow}
  et~al.}{2008}]{2008ApJ...679.1239W}
{Widrow} L.~M.,  {Pym} B.,   {Dubinski} J.,  2008, \mn@doi [\apj]
  {10.1086/587636}, \href {http://adsabs.harvard.edu/abs/2008ApJ...679.1239W}
  {679, 1239}

\bibitem[\protect\citeauthoryear{{Zoccali} \& {Valenti}}{{Zoccali} \&
  {Valenti}}{2016}]{2016PASA...33...25Z}
{Zoccali} M.,  {Valenti} E.,  2016, \mn@doi [\pasa] {10.1017/pasa.2015.56},
  \href {http://adsabs.harvard.edu/abs/2016PASA...33...25Z} {33, e025}

\makeatother
\end{thebibliography}


\label{lastpage}
\end{document}